\documentclass[acmsmall,screen]{acmart}
\setcopyright{none}
\copyrightyear{2027}
\acmYear{2027}
\acmDOI{XXXXXXX.XXXXXXX}
\acmNumber{POPL}
\acmVolume{10}
\acmArticle{216}
\acmMonth{11}
\acmJournal{PACMPL}

\usepackage{tikz}
\usepackage{tabularx}
\usepackage{wrapfig2}

\usepackage{amsthm,proof,amssymb}
\usepackage{enumitem}
\usepackage{oktheorem,okgeneralmath,okcategory}
\usepackage{bbold}
\usepackage{epiolmec}
\usepackage{oksemantics}
\usepackage{mathpartir}
\usepackage{todonotes}
\usepackage{natbib}
\usepackage{fp}
\usepackage{xparse}
\usepackage{xspace}

\usepackage{contour}

\newboolean{WithSupplementary}
\setboolean{WithSupplementary}{true}
\newcommand\seclabel[1]{\label{sec:#1}}
\newcommand\secref[1]{\S\ref{sec:#1}}
\newcommand\applabel[1]{\label{app:#1}}
\newcommand\appref[2][]{Appendix~%
  \ifthenelse{\boolean{WithSupplementary}}%
             {\ref{app:#2}}%
             {#1}%
             }
\newcommand\applemmaref[2][]{%
  \ifthenelse{\boolean{WithSupplementary}}%
             {\lemmaref{#2}}%
             {Lemma~#1}%
}
\newcommand\apptheoremref[2][]{%
  \ifthenelse{\boolean{WithSupplementary}}%
             {\theoremref{#2}}%
             {Theorem~#1}%
}
\WithSuffix\newcommand\appref*[1]{\ref{app:#1}}
\newcommand\subseclabel[1]{\label{subsec:#1}}
\newcommand\subsecref[1]{\S\ref{subsec:#1}}

\newcommand\remarklabel[1]{\label{remark:#1}}
\newcommand\remarkref[1]{Remark~\ref{remark:#1}}

\newcommand\figlabel[1]{\label{fig:#1}}
\newcommand\figref[1]{Fig.~\ref{fig:#1}}

{
  \theoremstyle{definition}
  \newtheorem{remark}[definition]{Remark}
}

\newcommand{\citepos}[1]{\citeauthor{#1}'s~\cite{#1}}
\WithSuffix\newcommand\citepos*[2]{\citeauthor{#1}'s #2~\cite{#1}}
\WithSuffix\newcommand\citepos-[2]{\citeauthor{#1}#2~\cite{#1}}
\makeatletter

\newcommand\ibid[2][]{%
  \citetext{%
    \hyper@natlinkstart{#2}%
    \textit{ibid.}%
    \hyper@natlinkend%
    \ifx\ibid@empty#1\ibid@empty%
    \else%
    , #1%
    \fi%
  }%
}

\newcommand\citloc[1]{%
  \citetext{%
    \hyper@natlinkstart{#1}%
    \textit{cit.~loc.}%
    \hyper@natlinkend%
  }%
}
\newcommand\citlocpos[1]{%
  \citeauthor{#1}'s~\citloc{#1}%
}
\WithSuffix\newcommand\citlocpos*[2]{%
  \citeauthor{#1}'s #2~\citloc{#1}%
}
\makeatother

\newcommand\diagram[1]{\begin{center}
    \diagram*{#1}
\end{center}}
\WithSuffix\newcommand\diagram*[2][c]{\begin{tabular}[#1]{@{}c@{}}\includegraphics{generated-diagrams/diagram-#2.mps}\end{tabular}}

\usepackage{tcolorbox}
\newtcbox{\myfigbox}{%
  on line, arc=3pt, outer arc=3pt,
  colback=black!8, colframe=black!8,
  boxsep=0pt,
  left=2pt,right=2pt,top=1pt,bottom=1pt,%
  boxrule=0pt}
\makeatletter
\NewDocumentCommand\NewFontImport{O{\hbox}O{\tf@size}mmmmmm}{
\newcommand#3{\begingroup\mathchoice%
  {#1{\fontsize{#2}{#2}\usefont{#4}{#5}{#6}{#7}\char#8}}%
  {\hbox{\fontsize{\tf@size}{\tf@size}\usefont{#4}{#5}{#6}{#7}\char#8}}%
  {\hbox{\fontsize{\sf@size}{\sf@size}\usefont{#4}{#5}{#6}{#7}\char#8}}%
  {\hbox{\fontsize{\ssf@size}{\ssf@size}\usefont{#4}{#5}{#6}{#7}\char#8}}%
  \endgroup%
}}
\makeatother

\makeatletter
\newcommand\newmvar[3][]{%
  \newcommand#2[1][1]{#1{\ifcase ##1\or #3\or\else@ctrerr\fi}}%
}

\makeatother

\newcommand\newop[2]{\newcommand{#1}{\mathop{{\mathrm{#2}}}\nolimits}}

\newop{\return}{return}
\newop{\strengthName}{\mathbf{st}}
\newcommand{\strength}[1][]{\supimplicit[{#1}]{\strengthName}}

\newop{\shadow}{shadow}

\makeatletter
\let\llangle\@undefined
\let\rrangle\@undefined
\makeatother

\usepackage{mathtools}
\usepackage{pst-venn}
\usepackage{tikz-cd}
\usetikzlibrary{decorations.pathmorphing}

\usepackage[dvipsnames,svgnames,table]{xcolor}
\usepackage{listings}
\colorlet{code}{DarkBlue}
\colorlet{types}{Magenta}
\colorlet{implicit}{Gray}

\usepackage{scalerel,stackengine}
\stackMath
\newcommand\reallywidehat[1]{%
\savestack{\tmpbox}{\stretchto{%
  \scaleto{%
    \scalerel*[\widthof{\ensuremath{#1}}]{\kern-.6pt\bigwedge\kern-.6pt}%
    {\rule[-\textheight/2]{1ex}{\textheight}}
  }{\textheight}%
}{0.5ex}}%
\stackon[1pt]{#1}{\tmpbox}%
}

\newcommand\sem[1]{\scottbrackets{#1}}

\NewDocumentCommand\eval{oO{}}{\mathop{\mathrm{eval}}\nolimits_{\IfNoValueF{#1}{#1;}#2}}
\NewDocumentCommand\curry{oO{}}{\mathop{\mathrm{curry}}\nolimits_{\IfNoValueF{#1}{#1;}#2}}
\NewDocumentCommand\uncurry{oO{}}{\mathop{\mathrm{uncurry}}\nolimits_{\IfNoValueF{#1}{#1;}#2}}
\RenewDocumentCommand\Dom{oO{}}{\mathop{\mathrm{Dom}}\nolimits_{\IfNoValueF{#1}{#1;}#2}}
\newcommand\swap{\mathtt{F}}
\newcommand\stet{\mathtt{I}}
\newcommand\pto\rightharpoonup
\newcommand\xpto\xrightharpoonup

\DeclareFontFamily{U}{bskarr}{}
\DeclareFontShape{U}{bskarr}{m}{n}{<->bskarr10}{ }
\NewFontImport\rightarrowTriangle{U}{bskarr}{m}{n}{"FB}
\NewFontImport\rightarrowtriangle{U}{bskarr}{m}{n}{"FE}
\NewDocumentCommand\kto{O{}}{\mathrel{\rightarrowtriangle}_{%
  \ifthenelse{\equal{#1}{sig}}%
          {\sigma}{%
  \ifthenelse{\equal{#1}{prb}}{1}{%
  \ifthenelse{\equal{#1}{fin}}{\textrm{fin}}{%
  \ifthenelse{\equal{#1}{sub}}{\leq1}
             {#1}}}}%
}}

\DeclareFontEncoding{LS1}{}{}
\DeclareFontFamily{LS1}{stixarrows1}{ }
\DeclareFontSubstitution{LS1}{stix}{m}{n}
\DeclareFontShape{LS1}{stixarrows1}{m}{n} {<-> stix-mathsf}{}

\DeclareFontFamily{LS1}{stixintegrals}{ }
\DeclareFontShape{LS1}{stixintegrals}{m}{n} {<-> stix-mathcal}{}
\NewFontImport\intcupupop{LS1}{stixintegrals}{m}{n}{"C7}

\DeclareFontFamily{LS1}{stixletters}{ }
\DeclareFontShape{LS1}{stixletters}{m}{it}{<-> stix-mathit}{}
\NewFontImport{\scurel}{LS1}{stixletters}{m}{it}{"FE}
\DeclareFontFamily{LS1}{stixfrak}{ }
\DeclareFontShape{LS1}{stixfrak}{m}{n}{<-> stix-mathfrak}{}
\NewFontImport{\boxastSymb}{LS1}{stixfrak}{m}{n}{"B5}
\newcommand\mapmul{\mathbin{\boxastSymb}}

\newcommand\cotuple\coseq

\newcommand\vto[3]{%
  \begin{smallmatrix}%
        \hphantom{#2}#1     \\%
        #2\mathord\downarrow\\%
        \hphantom{#2}#3       %
  \end{smallmatrix}%
}
\WithSuffix\newcommand\vto+[3]{%
  \begin{smallmatrix}%
        #1     \\%
        \mathllap{#2}\mathord\downarrow\\%
        #3       %
  \end{smallmatrix}%
}
\WithSuffix\newcommand\vto*[3]{%
  \begin{smallmatrix}%
        #1\hphantom{#2}     \\%
        \mathord\downarrow#2\\%
        #3\hphantom{#2}       %
  \end{smallmatrix}%
}
\WithSuffix\newcommand\vto-[3]{%
  \begin{smallmatrix}%
        #1     \\%
        \mathord\downarrow\mathrlap{#2}\\%
        #3       %
  \end{smallmatrix}%
}
\newcommand\vtotwo[5]{%
  \begin{smallmatrix}%
        \hphantom{#2}#1     \\%
        #2\mathord\downarrow\\%
        \hphantom{#2}#3     \\%
        #4\mathord\downarrow\\%
        \hphantom{#4}#5       %
  \end{smallmatrix}%
}
\WithSuffix\newcommand\vtotwo+[5]{%
  \begin{smallmatrix}%
        #1                             \\%
        \mathllap{#2}\mathord\downarrow\\%
        #3                             \\%
        \mathllap{#4}\mathord\downarrow\\%
        #5
  \end{smallmatrix}%
}
\WithSuffix\newcommand\vtotwo*[5]{%
  \begin{smallmatrix}%
        #1\hphantom{#2}     \\%
        \mathord\downarrow#2\\%
        #3\hphantom{#2}     \\%
        \mathord\downarrow#4\\%
        #5\hphantom{#4}       %
  \end{smallmatrix}%
}

\newcommand\carrier[1]{\underline{#1}}

\newcommand\fun[1]{#1 \mapsto}

   {\begin{array}[t]{@{}l@{}}}
   {\end{array}}

\DeclareFontFamily{U}{bbold}{}
\DeclareFontShape{U}{bbold}{m}{n}{<->bbold10}{}
\DeclareFontSubstitution{U}{bbold}{m}{n}
\NewFontImport\SinSym{U}{bbold}{m}{n}{'017}

\newmvar\A{A \or B \or C \or D}
\newcommand\B{\A[2]}
\let\C\undefined
\newcommand\C{\A[3]}

\newmvar\X{X \or Y \or Z}
\newcommand\Y{\X[2]}

\newmvar\mX{X \or Y \or Z}
\newcommand\mY{\mX[2]}

\newmvar\mx{x \or y \or z}

\newmvar\E{E}
\newmvar\G{G \or H \or K}
\newmvar\M{M \or K \or L}
\let\H\undefined
\newcommand\H{\G[2]}

\let\P\undefined
\newmvar\P{P \or Q}

\newmvar\rep{\rho\or\sigma\or\tau}

\newcommand\implicit[2][]{\overset{{\color{implicit}#1}}{#2}}
\newcommand\inplicit[1]{{\color{implicit}#1}}
\newcommand\subimplicit[2][]{#2_{{\color{implicit}#1}}}
\newcommand\supimplicit[2][]{#2^{{\color{implicit}#1}}}

\newcommand\unit[1][]{\implicit[{#1}]{e}}
\newcommand\mul[1][]{\mathbin{\implicit[#1]{\ast}}}
\newcommand\ginv[2][]{\subimplicit[#1]{\inv{#2}}}
\WithSuffix\newcommand\ginv*[2][]{\subimplicit[#1]{\inv*{#2}}}

\newcommand\Action[1]{#1\text{-}\Category{Act}}
\WithSuffix\newcommand\Action*{\Category{Act}}

\newcommand\act[1][]{\mathbin{{\circledast}^{\inplicit{#1}}}}
\WithSuffix\newcommand\act'[1][]{\mathop{\implicit[#1]{\odot}}\nolimits}

\newcommand\iact[2][]{\mathbin{{\circledast}^{\inplicit{#1}}_{#2}}}
\WithSuffix\newcommand\iact*[2][]{\mathbin{{\circledast}^{\inplicit{#1}_{\inplicit{#2}}}}}
\WithSuffix\newcommand\iact+[2][]{\act[#1]}

\newcommand\Triv[1][]{\mathsf{Triv}_{#1}}

\newcommand\wlg{\mathop{\mathsf{wlog}}\nolimits}
\newcommand\equito{\mathrel{\raisebox{.25ex}{$\scriptscriptstyle\circledast$}}\joinrel\to}

\newcommand\types\vdash

\newcommand\Canon[2][]{\mathop{\mathsf{Can}}\nolimits_{#1}{#2}}
\newcommand\canoniser{c}

\newcommand\groth\oint

\newcommand\stabilise[2][\G]{#1_{#2}}
\newcommand\stabeq[1]{\overset{#1}\sim}
\newcommand\Fin{\mathop{\mathsf{\TConst{Fin}}}\nolimits}

\newcommand\Sym[1]{\mathbf{S}_{#1}}
\newcommand\Cyc[1]{\mathbf{C}_{#1}}
\newcommand\Dih[1]{\mathbf{D}_{#1}}

\newmvar\perm{\pi\or\sigma\or\tau}

\newcommand\sort{\mathop{\mathsf{sort}}\nolimits}

\newcommand\Prop{\mathbf{\TConst{Prop}}}
\newmvar\prop{\phi\or\psi\or\xi}

\newcommand\length\cardinality

\WithSuffix\newcommand\++{\mathbin{+\mspace{-8mu}{+}}}
\WithSuffix\newcommand\h+{\mathbin{{\color{red}+}\mspace{-8mu}{\color{blue}+}}}
\newcommand\preleq{\sqsubseteq}

\makeatletter
\long\def\ok@tree[#1][#2]{\mathop{\mathsf{T}}_{#1}^{#2}}
\newcommand\SearchTree{\optparams{\ok@tree}{[{}][{}]}}
\makeatother
\WithSuffix\newcommand\SearchTree*[1]{#1\text{-}\SearchTree}

\newmvar\btree{t\or s\or r}
\newcommand\mtree[1][1]{\BoundV{\btree[#1]}}

\WithSuffix\newcommand\leq?[1][\preleq]{\overset{?}{#1}}
\WithSuffix\newcommand\in?{\mathrel{\overset?\in}}
\WithSuffix\newcommand\=?{\mathrel{\overset?=}}
\WithSuffix\newcommand\in!{\mathrel{\overset!\in}}
\WithSuffix\newcommand\notin!{\mathrel{\overset!{\not\in}}}

\newmvar\Ctx{\Gamma\or\Delta\or\Xi\or\Psi\or\Phi}
\newcommand\Dtx{\Ctx[2]}
\newmvar\ctx{\gamma\or\delta\or\xi\or\psi\or\varphi}
\newcommand\dtx{\ctx[2]}
\newcommand\xtx{\ctx[3]}

\newcommand\preals{\mathbb{R}_+}
\newcommand\pnaturals{\mathbb{N}_+}

\newcommand\rulename\textsc

\newcommand\unitval\star
\newcommand\Board[1]{\TConst{Board}_{#1}}

\newcommand\fmap[1][]{\mathop{\subimplicit[{#1}]{F}}}

\newcommand\subst[2]{#1\bracks{#2}}
\newcommand\For[2]{#1{\mathrel{\color{DarkRed}\mapsto}}#2}

\newmvar\Grph{G\or H\or K}
\newcommand\walkto\leadsto

\newcommand\Row[1]{\TConst{Row}_{#1}}
\newcommand\Vct[2]{\TConst{Vect}_{#1}#2}
\newcommand\Mat[3]{\TConst{Mat}_{#1\times#2}#3}
\DeclareSymbolFont{okMnSymbolA}{U}{MnSymbolA}{m}{n}
\SetSymbolFont{okMnSymbolA}{bold}{U}{MnSymbolA}{b}{n}
\DeclareFontShape{U}{MnSymbolA}{b}{n}{
    <-6>  MnSymbolA-Bold5
   <6-7>  MnSymbolA-Bold6
   <7-8>  MnSymbolA-Bold7
   <8-9>  MnSymbolA-Bold8
   <9-10> MnSymbolA-Bold9
  <10-12> MnSymbolA-Bold10
  <12->   MnSymbolA-Bold12}{}
\DeclareMathSymbol{\rcurvearrownw}{\mathrel}{okMnSymbolA}{197}
\DeclareMathSymbol{\rcurvearrowleft}{\mathrel}{okMnSymbolA}{194}
\DeclareMathSymbol{\rcurvearrowright}{\mathrel}{okMnSymbolA}{192}
\DeclareMathSymbol{\lcurvearrowsw}{\mathrel}{okMnSymbolA}{190}
\DeclareMathSymbol{\lcurvearrowse}{\mathrel}{okMnSymbolA}{191}
\DeclareMathSymbol{\rcirclearrowright}{\mathrel}{okMnSymbolA}{248}
\DeclareMathSymbol{\downrsquigarrow}{\mathrel}{okMnSymbolA}{171}
\newcommand\Rotation{\mathord{\DConst{\boldsymbol\rcurvearrownw}}}
\newcommand\twiceRotation{\mathord{\DConst{\boldsymbol\rcurvearrowright}}}
\newcommand\unRotation{\mathord{\DConst{\boldsymbol\lcurvearrowse}}}
\newcommand\nonRotation{\mathord{\DConst{\boldsymbol\rcirclearrowright}}}
\newcommand\vplusstyle[3]{%
\mathbin{
  \begin{array}[c]{@{}>{#3}c@{}}
    \color{red}+\\[#1]\color{blue}+\\[#2]
  \end{array}
}}
\WithSuffix\newcommand\v+{%
\mathchoice%
        {\vplusstyle{-2.25ex}{.02ex}{\displaystyle}}%
        {\vplusstyle{-2.25ex}{.02ex}{\textstyle}}%
        {\vplusstyle{-2.25ex}{.2ex}{\scriptstyle}}%
        {\vplusstyle{-2.5ex}{.5ex}{\scriptscriptstyle}}%
}
\newcommand\Tile{\TConst{Tile}}

\newcommand\Direction{\TConst{Dir}}
\newcommand\Rt{\DConst{\mathord\Rightarrow}}
\newcommand\Lt{\DConst{\mathord\Leftarrow}}
\newcommand\Up{\DConst{\mathord\Uparrow}}
\newcommand\Dn{\DConst{\mathord\Downarrow}}
\newcommand\mkLt[1][\,]{\mathsf{mkLft}#1}

\newcommand\To\Rightarrow
\newcommand\mergeLeft[1][]{\merge_{\Lt}^{#1}}

\newcommand\merge{\mathsf{merge}}
\newcommand\minimal{\mathsf{indexMin}}
\newcommand\struct[1][]{\subimplicit[{#1}]{F}}
\usepackage[svgnames]{xcolor}
\usepackage{listings}
\colorlet{types}{DarkCyan}
\colorlet{code}{DarkRed}
\colorlet{bound}{MediumPurple}
\newcommand\TConst[1]{\mathsf{\textcolor{types}{#1}}}
\newcommand\DConst[1]{\mathsf{\textcolor{code}{#1}}}
\newcommand\BoundV[1]{#1}
\newcommand\Nil{\DConst{Nil}}
\newcommand\Leaf{\DConst{Leaf}}
\WithSuffix\newcommand\::{\mathbin{\DConst{::}}}
\newcommand\Node{\mathop{\DConst{Node}}\nolimits}

\newcommand\Maybe{\mathop{\TConst{Maybe}}}
\newcommand\Nothing{{\DConst{Nothing}}}
\newcommand\Just{\mathop{\DConst{Just}}}
\newcommand\Tree{\mathop{\TConst{Tree}}}

\newcommand\Or{\mathrel{\boldsymbol\vert}}
\newcommand\roll[1][]{\subimplicit[{#1}]{\mathop{\mathbf{roll}}\nolimits}}
\newcommand\unroll[1][]{\subimplicit[{#1}]{\mathop{\mathbf{unroll}}\nolimits}}
\newcommand\vecmap[1][]{\subimplicit[{#1}]{\mathop{\mathbf{map}}\nolimits}}
\newcommand\fold[1][]{\subimplicit[{#1}]{\mathop{\mathbf{fold}}\nolimits}}
\newcommand\pfold[1][]{\subimplicit[{#1}]{\mathop{\mathbf{pfold}}\nolimits}}

\newcommand\fix{\boldsymbol\mu}
\newcommand\TreeLabel{{\mathbf{Step}}}
\newcommand\TreePath{\TConst{Path}}
\newcommand\Lft{\DConst{L}}
\newcommand\Rgt{\DConst{R}}
\newenvironment{tree diagram}%
  {\scriptsize\begin{tikzcd}[ampersand replacement=\&,column sep=tiny,row sep=tiny, scale=0.3]}%
  {\end{tikzcd}}
\newcommand\treeLink[4]{\arrow["", from=#1-#2, to=#3-#4, no head, no tail]}
\newcommand\NodeV[3]{
 \begin{tree diagram}
      \& #2 \& \\
    \text{$#1$}\&    \& \text{$#3$}
   \treeLink1221
   \treeLink1223
   \end{tree diagram}
 }

\NewDocumentCommand\gstrength{O{}o}{
  \IfNoValueTF{#2}
    {\supimplicit[{#1}]{\strengthName}}
    {\supimplicit[{#1};{#2}]{\strengthName}}%
 }

\newcommand\WLOG{w.l.o.g.\xspace}
\newcommand\wLOG{W.l.o.g.\xspace}
\newcommand\lookup{\mathsf{lookup}}
\newcommand\lookupElim{\mathsf{lookup}^{\mathrm{Elim}}}
\newcommand\lookupCore{\mathsf{lookup}^{\mathrm{Core}}}
\newcommand\lookLeft{\mathsf{lookLeft}}

\newcommand\NodeInName{\mathsf{\TConst{InPath}}}
\newcommand\NodeIn{\mathop{\NodeInName}}
\newcommand\Type[1]{\mathop{\mathsf{\TConst{#1}}}}
\newmvar\msubst{\theta\or \xi\or\zeta}
\newcommand\OrbitOf[2][\G]{#1\act\bracks{#2}}

\newmvar\x{\BoundV x\or\BoundV y \or\BoundV z}
\newcommand\y{\x[2]}
\newcommand\z{\x[3]}
\newcommand\DPair[2]{\parent{#1 : #2}\mathbin{\times}}
\newcommand\dpair[1]{#1\mathbin{\DConst,}}
\newcommand\DFun[2]{\parent{#1 : #2}\mathbin{\to}}
\newcommand\EFun[2]{\parent{#1 : #2}\mathbin{\equito}}
\newcommand\Fun[1]{#1\mathbin{\to}}
\newcommand\Tru{\DConst{True}}
\newcommand\Fls{\DConst{False}}
\newmvar\mn{\BoundV m \or \BoundV n\or \BoundV k}
\newmvar\mm{\mn[2]}
\newmvar\mk{\mn[3]}
\newmvar\mt{\BoundV t}
\newmvar\mell{\BoundV \ell}
\newmvar\mr{\BoundV r}
\newmvar\ma{a\or b\or c}
\newcommand\mb{\ma[2]}
\newcommand\mc{\ma[3]}
\newcommand\mctx[1][1]{\BoundV{\ctx[#1]}}
\newmvar\g{g\or h\or k}
\newcommand\mg[1][1]{\BoundV{\g[#1]}}
\newcommand\Schur{\mathsf{Schur}}

\renewcommand\amalg{+}

\newcommand\keys{\mathbb K}

\newcommand\gast[1][\G]{#1\mathbin{\TConst\boxastSymb}}
\newcommand\lollybox[1][\G]{#1\mathrel{\TConst{\relbar\joinrel\mathrel{\boxastSymb}}}}
\newcommand\alg{\mathbf{alg}}
\newcommand\abs[1]{\left\lvert#1\right\rvert}
\newcommand\muast{\boxast}

\newcommand\Fam{\Category{Fam}}
\newop{\costrength}{costrength}
\newcommand\RED{\DConst{r}}
\newcommand\BLU{\DConst{b}}
\newcommand\lift{\mathop{\mathsf{\DConst{lift}}}\nolimits}
\newcommand\coerce{\mathop{\mathsf{coerce}}\nolimits}
\newcommand\goLeft{\mathsf{goLeft}}
\newcommand\flg[1][]{\subimplicit[{#1}]{\mathop{\mathbf{flog}}\nolimits}}
\newcommand\subtree{\mathsf{subtree}}
\newcommand\subtreeCore{\mathsf{subtree}^{\mathrm{Core}}}
\newcommand\subtreeAux{\mathsf{subtree}^{\mathrm{aux}}}
\newcommand\subtreeAuxCore{\mathsf{subtree}^{\mathrm{auxCore}}}
\newcommand\canonise[1][]{\mathop{\mathsf{canonise}}\nolimits_{#1}}

\setboolean{WithSupplementary}{true}

\begin{document}

\title[Mathematical foundations for symmetric programming and reasoning]{Mathematical foundations for symmetric programming and reasoning:
  eliminating symmetric cases with group actions}

\author{Ohad Kammar}
\email{ohad.kammar@ed.ac.uk}
\orcid{0000-0002-2071-0929}
\affiliation{%
  \institution{University of Edinburgh}
  \department{Laboratory for the Foundations of Computer Science, School of Informatics}
  \city{Edinburgh}
  \state{Scotland}
  \country{United Kingdom}
}
\author{Matija Pretnar}
\email{matija.pretnar@fmf.uni-lj.si}
\orcid{0000-0001-7755-2303}
\affiliation{%
  \institution{University of Ljubljana \& Institute of Mathematics, Physics and Mechanics}
  \city{Ljubljana}
  \country{Slovenia}
}

\renewcommand{\shortauthors}{O.~Kammar and M.~Pretnar}

\begin{abstract}
  We develop a mathematical abstraction for uniformly exploiting
symmetries in programming and reasoning based on the mathematical
proof principle `without-loss-of-generality' (w.l.o.g.). Given: group
actions that describe symmetries of the input and the output; and a
canoniser, which captures how arbitrary inputs get sent to canonical
ones, our construction extends any function from the canonical inputs
to the whole domain.

We characterise the conditions on the canonisers which guarantee that
the construction gives the unique equivariant (i.e.
symmetry-preserving) extension. Furthermore, we adapt the construction
to a dependently-typed setting, where it captures traditional
symmetric w.l.o.g. arguments via the Curry-Howard correspondence.
Finally, we extend the construction to algebraic data types, where the
symmetries of the constructors both uniquely determine the symmetries
of the data type and pave the way for efficient execution.

\end{abstract}




\maketitle


\section{Introduction}

When faced with repetition, programmers naturally turn to boilerplate or
parametrisation. In this paper, we focus on repetition that
arises from \emph{symmetry}. We develop
mathematical foundations that explicitly and uniformly capture
existing implicit patterns in programming and reasoning, and
define a construction that removes redundant symmetric cases from
functions and proofs.

Consider Gabriele Cirulli's famous game of 2048\footnote{Available to
play: \url{https://play2048.co}}, in which the player slides tiles
labelled by powers of $2$, merging tiles with identical labels. The
player can slide tiles in all four directions, and the game
source-code accommodates that ability through a general merging code
that takes the direction as a parameter, builds a sequence of matrix
indices to consider, and traverses over them\footnote{See
\texttt{GameManager.prototype.move}:
\url{https://github.com/gabrielecirulli/2048/blob/478b6ec/js/game_manager.js\#L129}.}.

Our anecdotal experience with half a dozen student projects
suggests this parametrisation can be difficult to implement
correctly. Here is an approach that is simpler to understand and to implement.
Define the merging logic only for a single direction, say left,
and exploit the inherent symmetry of the game for the other three
directions: rotate the board so that the task is to slide left,
perform the merge, and rotate back, restoring the original
orientation. Here is an example downwards merge:
\usetikzlibrary{arrows,arrows.meta}
\[
  \newcommand{\twofortyeight}[2]{\vcenter{\hbox{
        \begin{tikzpicture}[scale=0.6, tile/.style={rounded corners=0.05cm, minimum size=0.55cm, draw=white, thick, font=\sffamily}
          ]
          \foreach \x in {0,1,2,3} {
            \foreach \y in {0,1,2,3} {
              \node[tile, fill=gray!10] (\x\y) at (\x, \y) {};
            }
          }
          \foreach \row/\col/\num in {#1}
          \node[tile, fill=gray!10!orange!\num0] at (\col, 3 - \row) {\num};
          #2
  \end{tikzpicture}}}}
  \twofortyeight{0/1/2, 0/3/2, 1/3/2, 2/1/2, 2/3/4, 3/1/4, 3/3/8}
  {\draw[ double distance=1.5pt
      , arrows = {-Latex[]}
    ]  (02.center) node[above=-2pt]
    {
      \begin{tabular}{@{}>{\scriptsize\centering}p{.6cm}@{}}
        merge \\
        down
      \end{tabular}
    }
    -- (01.center)
    ;}
  \quad
  \explain\unRotation{rotate}
  \quad
  \twofortyeight{1/3/2, 3/3/2, 3/2/2, 1/1/2, 3/1/4, 1/0/4, 3/0/8}
  {\draw[ double distance=1.5pt
      , arrows = {-Latex[]}
  ]  (23.center) node[above] {\clap{\scriptsize merge left\qquad}} -- (13.center);}
  \quad
  \explain\mapsto{perform \\[-5pt] merge}
  \quad
  \twofortyeight{3/2/4, 1/1/4, 3/1/4, 1/0/4, 3/0/8}
  {}
  \quad
  \explain\Rotation{rotate \\[-5pt] back}
  \quad
  \twofortyeight{1/3/4, 2/1/4, 2/3/4, 3/1/4, 3/3/8}
  {}
\]

Another common example of symmetry is that of binary
search trees, in which elements smaller than the root go to the left
subtree, while larger elements go to the right. Operations such as
search, insertion, or deletion behave completely
symmetrically. One way of avoiding repetition by parametrisation is to
store both subtrees in an array of length two, and then use the
offset~$1 - i$ to refer to the tree opposite to the one at offset~$i$.
In practice, though, the additional complexity is often not worth
losing the high-level abstraction, and programmers prefer to duplicate
code.

But we can again avoid this repetition with the same approach and
implement the code only for one case. For example, when inserting an
element into the right subtree, we can flip the tree (and its ordering
relation!), insert the element into the left subtree, and then flip
the resulting tree back:
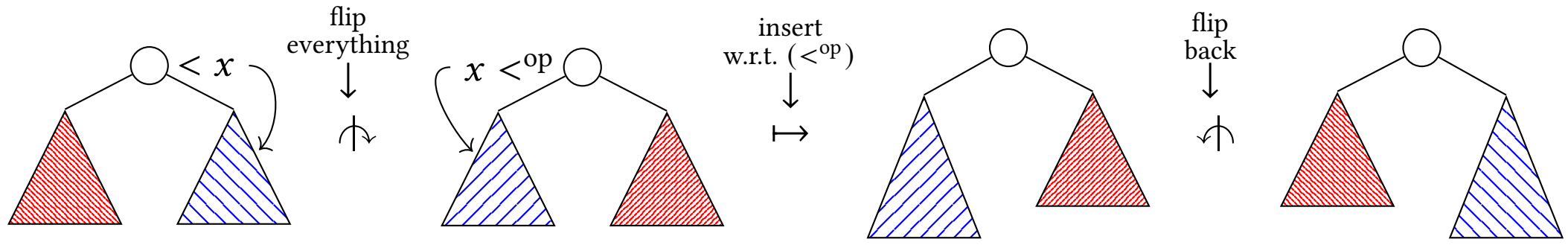
\usetikzlibrary{shapes.geometric, patterns, patterns.meta}
\[
  \newcommand{\insertion}[1]{\vcenter{\hbox{
        \begin{tikzpicture}[
            scale=0.75,
            inner/.style={circle, draw, minimum width=2.5mm},
            tree/.style={isosceles triangle, isosceles triangle stretches=true, draw, shape border rotate=90,minimum width=10mm},
            h/.style={minimum height=10mm},
            h1/.style={minimum height=12.5mm, yshift=-2.5mm},
            level 1/.style={sibling distance=20mm},
            transform/.style={bend left, dashed, shorten >=2.5mm, shorten <=2.5mm, -{stealth}}
          ] #1
  \end{tikzpicture}}}}
  \insertion{
    \node[inner] (t3) at (0, -4) {}
    [child anchor=north]
    child {node[pattern={Lines[angle=-45, distance=1pt]}, pattern color=red, tree, h] {}}
    child {node[pattern={Lines[angle=-45]}, pattern color=blue, tree, h] {}};
    \node (x') at (0.6, -4) {${} < x$};
    \path (x') edge[->, out=0, in=45] (1.3, -5);
  }
  \;\;\:\,
  \explain{\curvearrowright\hskip-10pt|}{flip\\[-5pt]everything}
  \;\;\:\,
  \insertion{\node[inner] (t1) at (0, 0) {}
    [child anchor=north]
    child {node[pattern={Lines[angle=45]}, pattern color=blue, tree, h] {}}
    child {node[pattern={Lines[angle=45, distance=1pt]}, pattern color=red, tree, h] {}};
    \node (x) at (-0.8, 0) {$x <^{\mathrm{op}} {}$};
    \path (x) edge[->, out=180, in=135] (-1.3, -1);
  }
  \;\;\:\,
  \explain\mapsto{insert\\[-5pt]w.r.t.~$(<^{\mathrm{op}})$}
  \;\;\:\,
  \insertion{
    \node[inner] (t2) at (4, 0) {}
    [child anchor=north]
    child {node[pattern={Lines[angle=45]}, pattern color=blue, tree, h1] {}}
    child {node[pattern={Lines[angle=45, distance=1pt]}, pattern color=red, tree, h] {}};
  }
  \;\;\:\,
  \explain{\curvearrowleft\hskip-8.5pt|}{flip\\[-5pt]back}
  \;\;\:\,
  \insertion{
    \node[inner] (t4) at (4, -4) {}
    [child anchor=north]
    child {node[pattern={Lines[angle=-45, distance=1pt]}, pattern color=red, tree, h] {}}
    child {node[pattern={Lines[angle=-45]}, pattern color=blue, tree, h1] {}};
  }
\]
A reader may justifiably worry about the cost of flipping the whole
tree each time we descend right, but a more efficient implementation
fits the same pattern.

Our final example comes from mathematics,
where one often invokes a \emph{without loss of generality} (or \WLOG)
argument when we treat a canonical case, and derive the remaining
cases by symmetry. Consider the pigeonhole principle given as an
example in the Wikipedia entry for \WLOG{}~\cite{wiki:wlog}:
\begin{quote}
If three objects are each painted either red or blue, then there must
be at least two objects of the same color.

\textbf{Proof.} Assume, without loss of generality, that the first
object is red. If either of the other two objects is red, then we are
finished; if not, then the other two objects must both be blue and we
are also finished.
\end{quote}
On paper, \WLOG appeals to an intuitive understanding that nothing has
been lost by considering a special case. But if the symmetry is complex
or if we want to convince a proof assistant, we need
to explicate the reduction of the general case
to the special one.

We can generalise the special case as a parametric proof that covers
all cases. Above, we could take the colour of the first object as a
parameter and then instantiate it to both red and blue. But as
\citeauthor{DBLP:conf/tphol/Harrison09} argues in his seminal study of
\WLOG in the context of the HOL proof assistant---and we agree---this
formulation is not faithful to the informal
proof~\cite{DBLP:conf/tphol/Harrison09}.

Modern proof assistants offer special tactics to treat \WLOG
arguments, but approach them from various directions. Some tactics or
theorems named \texttt{wlog} work in a very restricted setting, such
as symmetric relations or negation, and some cut in arbitrary
additional statements that do not involve symmetries.
These divergent approaches point towards a missing
mathematical foundation. Furthermore, none treat the
underlying symmetry as a first-class citizen, and as
\citeauthor{dijkstra:wlog} argues---and we agree---``omissions [of the
underlying symmetry] make most `\WLOG{} proofs' rather
unsatisfactory''~\cite{dijkstra:wlog}.

We can explicate this symmetry by following the same
pattern we used for programming. In the excluded case where the first object is blue, we can
exchange the colours to make it red, use the specialised argument, and
then revert the colours back to the original. For example, one of
the three cases where there is another blue object in addition to the
first one proceeds as:
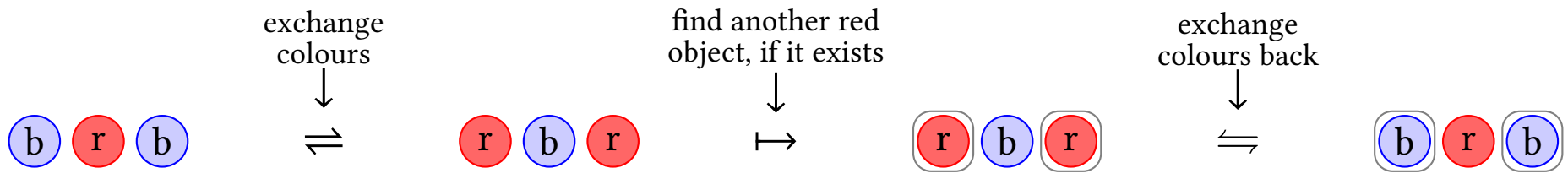
\[
  \newcommand{\balls}[2]{\vcenter{\hbox{
    \begin{tikzpicture}[
      scale=0.5,
      obj/.style={circle, draw, inner sep=0pt, minimum size=4mm},
      rdot/.style={obj, fill=red!60, draw=red},
      bdot/.style={obj, fill=blue!20, draw=blue},
      pair/.style={rounded corners, thin, draw=gray},
      >=stealth, font=\small
    ]
      \pgfmathsetmacro\dx{-0.03}
      \foreach \col/\f in {#1}
        \node[\f] at (\col+0.5, 0.5) {\ifthenelse{\equal{\f}{bdot}}{b}{r}};
      #2
    \end{tikzpicture}
  }}}
  \balls{0/bdot, 1/rdot, 2/bdot}{}
  \qquad
  \explain{\rightleftharpoons}{exchange\\[-5pt]colours}
  \qquad
  \balls{0/rdot, 1/bdot, 2/rdot}{}
  \qquad
  \explain{\mapsto}{find another red\\[-5pt]object, if it exists}
  \qquad
  \balls{0/rdot, 1/bdot, 2/rdot}{
    \draw[pair] (0-\dx,0-\dx) rectangle (1+\dx,1+\dx);
    \draw[pair] (2-\dx,0-\dx) rectangle (3+\dx,1+\dx);
  }
  \qquad
  \explain{\leftrightharpoons}{exchange\\[-5pt]colours back}
  \qquad
  \balls{0/bdot, 1/rdot, 2/bdot}{
    \draw[pair] (0-\dx,0-\dx) rectangle (1+\dx,1+\dx);
    \draw[pair] (2-\dx,0-\dx) rectangle (3+\dx,1+\dx);
  }
\]
while if both remaining objects are red, we similarly proceed as:
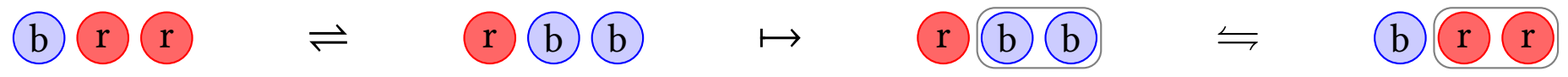
\[
  \newcommand{\balls}[2]{\vcenter{\hbox{
    \begin{tikzpicture}[
      scale=0.5,
      obj/.style={circle, draw, inner sep=0pt, minimum size=4mm},
      rdot/.style={obj, fill=red!60, draw=red},
      bdot/.style={obj, fill=blue!20, draw=blue},
      pair/.style={rounded corners, thin, draw=gray},
      >=stealth, font=\small
    ]
      \pgfmathsetmacro\dx{-0.03}
      \foreach \col/\f in {#1}
        \node[\f] at (\col+0.5, 0.5) {\ifthenelse{\equal{\f}{bdot}}{b}{r}};
      #2
    \end{tikzpicture}
  }}}
  \balls{0/bdot, 1/rdot, 2/rdot}{}
  \qquad
  \rightleftharpoons
  \qquad
  \balls{0/rdot, 1/bdot, 2/bdot}{}
  \qquad
  \mapsto
  \qquad
  \balls{0/rdot, 1/bdot, 2/bdot}{
    \draw[pair] (1-\dx,0-\dx) rectangle (3+\dx,1+\dx);
  }
  \qquad
  \leftrightharpoons
  \qquad
  \balls{0/bdot, 1/rdot, 2/rdot}{
    \draw[pair] (1-\dx,0-\dx) rectangle (3+\dx,1+\dx);
  }
\]

In this paper, we develop this pattern into a uniform construction, and
use it to eliminate symmetric cases in definitions of functions and in reasoning.
On the surface, this pattern is conceptually simple, and so one may,
just like us, be quickly tempted to implement it, either in a
programming language or in a proof assistant. We
experimented in simply- and dependently-typed languages, and
unsurprisingly, each obstacle pushed us back to pen-and-paper redesigns and
forced us to enrich the underlying mathematical foundation.

For this reason, while still keeping computational applications in
mind, this work focuses on these mathematical foundations.
We postpone a bona fide implementation to future work. Such future
implementations will allow us to explore symmetric programming on more
involved examples, in a machine-checked environment, design the best
interface for exposing it to the programmer, to build a comprehensive
library for symmetric programming and mechanised reasoning, and
revisit existing \WLOG formalisations systematically. While we keep these
downstream goals in mind, we pursue none of them in the
present work, and remain solely mathematical.

Our contributions:
\begin{itemize}
  \item In \secref{wlog construction}, we identify three ingredients:
    \emph{group actions} capturing the symmetries of the input and
    output, a \emph{canoniser} map that assigns a canonising group
    element to each input, and a \emph{core function} that defines the
    construction on canonical inputs. We then characterise our symmetric \WLOG{}
    construction as the unique symmetry-preserving
    (i.e. equivariant) extension of the core function to all the inputs.
  \item In \secref{dependent types}, we extend this theory to the
    setting of \emph{dependent types}. We show how via
    the Curry-Howard correspondence, symmetry-invariant propositions
    correspond to types equipped with a group action. We show how proofs
    via the symmetric \WLOG{} argument amount to dependent functions obtained by
    the above construction.
  \item In \secref{algebraic data types}, we further extend the theory
    to (simply- and dependently-typed) \emph{algebraic data types} and
    their induction principles. The main construct we introduce is
    \emph{symmetry strength}, which defines a group action in terms of
    data constructors, paving the way for symmetric reasoning on data types,
    and efficient partial evaluation.
\end{itemize}
We conclude with related and future work in \secref{conclusion}.
\ifthenelse{\boolean{WithSupplementary}}{%
  The Appendix has further details and proofs.%
}{%
  The accompanying Technical Report~\cite{kammar-pretnar:symmetric-programming-and-reasoning-tech-report}
  is an otherwise identical copy of this manuscript, apart from
  this sentence and detailed appendices supporting the claims we make, which we cross-reference from here.%
}

\begin{remark}
  Our development has been informed by category theoretic
  structure. To keep exposition accessible, we do not rely on the
  language of and concepts from category theory in the lion's share of this
  work.  The notable exception is our account of symmetric data types
  (\secref{algebraic data types}). There, generic accounts of the
  computational structure are close to their underlying categorical
  structure, and consequently we explicitly employ the categorical
  language. Readers who wish to reconstruct the category theoretic
  picture can consult \figref{indexed-datatypes-summary} in
  \subsecref{indexed data types} alongside the development.
\end{remark}

\section{Without loss of generality}
\seclabel{wlog construction}

Having observed the common pattern in the preceding examples, let us
detail the ingredients:

\begin{enumerate}
  \item Capture the inherent symmetries of our problem using \textbf{group actions}.
  \item Obtain a \textbf{canonising} symmetry for each input.
  \item Give the \textbf{core function}, defined only on canonical elements.
\end{enumerate}
The \textbf{symmetric \WLOG construction} combines these ingredients,
covering all inputs:
\[
  \begin{gathered}\text{a general}\\\text{input}
  \end{gathered}
  \quad
  \explain\mapsto{canonise input\\[-5pt] using a symmetry}
  \quad
  \begin{gathered}\text{a canonical}\\\text{input}
  \end{gathered}
  \qquad
  \explain\mapsto{apply the\\[-5pt]core function}
  \qquad
  \begin{gathered}\text{an output for the}\\\text{canonical input}
  \end{gathered}
  \quad
  \explain\mapsto{transform output back\\[-5pt] using the inverse symmetry}
  \quad
  \begin{gathered}\text{the desired output}\\\text{for the general input}
  \end{gathered}
\]
Our previous examples fit this pattern as:
\medskip

{\small
\noindent
\begin{tabularx}{\textwidth}{rp{2.7cm}XX}
  \toprule
  & \textbf{2048 merge} & \textbf{tree insertion} & \textbf{pigeonhole principle} \\
  \midrule
  \textbf{general input} & board \& direction & element \& tree & arbitrary colouring \\[0.5em]
  \textbf{desired output} & merged board & extended tree & general proof \\[0.5em]
  \textbf{symmetry} & rotation & flipping & colour exchange \\[0.5em]
  \textbf{canonical input} & direction is left & inserted key $\leq$ root key& first object is red\\[0.5em]
  \textbf{core function} & merge left & insert left as needed & special-case proof\\[0.5em]
  \bottomrule
\end{tabularx}}

\subsection{Groups and actions}
\subseclabel{groups-and-actions}

Before turning to symmetric \WLOG{} abstractions, let us start with a
short group theory tutorial. A \emph{concrete symmetry} on a set is a
bijection from the set to itself. When we compare different sets, we
need to relate their symmetries. A \emph{group} gives these symmetries
a name, and its \emph{actions} allow us to use the same abstract
symmetry to refer to concrete symmetries on different sets.

\paragraph{Groups}

Recall that a \emph{group} $\G = \seq*{\carrier G, \unit, (\mul),
\ginv*-}$  is a tuple
consisting of:
a \emph{carrier} set $\carrier \G$, whose elements we call \emph{symmetries};
a \emph{neutral} element $\unit \in \carrier \G$;
a \emph{composition} $(\mul) : \carrier \G \to \carrier \G \to \carrier \G$; and
an \emph{inverse} map $\ginv*{-} : \carrier \G \to \carrier \G$;
such that:
\begin{itemize}
  \item composition is associative: $g_1 \mul (g_2 \mul g_3) = (g_1 \mul g_2) \mul g_3$ for all $g_1, g_2, g_3 \in \carrier\G$;
  \item the neutral element is neutral w.r.t.~composition: $\unit \mul g = g \mul \unit = g$ for all $g \in \carrier\G$; and
  \item the inverse map picks inverses: $g \mul \inv{g} = \inv{g} \mul g = \unit$ for all $g \in \carrier\G$.
\end{itemize}

\noindent{}
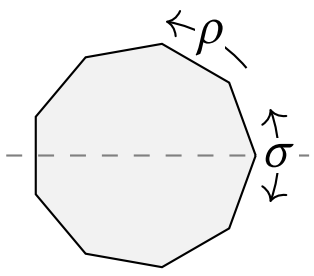
\begin{wrapfigure}[5]{l}*
  \begin{tikzpicture}[scale=0.75]
    \filldraw[fill=gray!10] (0:1) -- (40:1) -- (80:1) -- (120:1) -- (160:1) -- (200:1)-- (240:1)-- (280:1) -- (320:1) -- (0:1);
    \draw[dashed, gray, overlay] (-1.2,0) -- (1.5,0);
    \draw[<->, overlay] (20:1.2) arc[start angle=20, end angle=-20, radius=1.2] node[midway,fill=white,inner sep=1pt] {$\sigma$};
    \draw[<-, overlay] (80:1.2) arc[start angle=80, end angle=40, radius=1.2] node[midway,fill=white,inner sep=0] {$\rho$};
  \end{tikzpicture}
\end{wrapfigure}
Two famous examples of groups are the \emph{symmetric group}~$\Sym n$
of all permutations of the set $\Fin n \definedby \set{0, \dots,
n - 1}$, and the \emph{dihedral group}~$\Dih n$ of all
symmetries of a regular $n$-gon, i.e. $n$~rotations and
$n$~reflections. The dihedral group is \emph{generated} by a $2 \pi /
n$ rotation~$\rho$ and a single reflection~$\sigma$,
meaning every other symmetry can be obtained by a suitable composition
of $\rho$ and $\sigma$.\par\WFclear
\begin{wrapfigure}{r}*
  \newcommand{\twofortyeight}[1]{
    \begin{tikzpicture}[scale=0.4, tile/.style={rounded corners=0.05cm, minimum size=0.35cm, draw=white, thick, font=\sffamily}
      ]
      \foreach \x in {0,1,2,3} {
        \foreach \y in {0,1,2,3} {
          \node[tile, fill=gray!10] (\x\y) at (\x, \y) {};
        }
      }
      \foreach \row/\col/\num in {#1}
      \node[tile, fill=gray!10!orange!\num0] at (\col, 3 - \row) {};
      \draw[->,overlay] (4,2) arc[start angle=0, end angle=90, radius=2];
  \end{tikzpicture}}
  \twofortyeight{0/1/2, 0/3/2, 1/3/2, 2/1/2, 2/3/4, 3/1/4, 3/3/8}
\end{wrapfigure}%
\noindent{}%
In our examples, both the mirroring symmetry of trees and exchange of
colours are described by the symmetric group~$\Sym 2 =
\set{\stet,\swap}$, where $\stet$ is the identity permutation and
$\swap$ flips the two elements. The 2048 board is symmetric with
respect to the dihedral group~$\Dih 4$, though for our use, it is
sufficient to consider the \emph{cyclic group}~$\Cyc{4}$, generated by
the quarter-turn rotation marked here as~$\Rotation$.

\paragraph{Actions}

Groups capture symmetries abstractly. Symmetries between elements of a
particular set are captured by \emph{group actions}. Given a group
$\G$, recall that a \emph{$\G$-action} $\A = \seq*{\carrier A,
(\act)}$ consists of a \emph{carrier} set $\carrier \A$; and a map
$(\act) : \carrier \G \to \carrier \A \to \carrier \A$, writing
$(\act[\A])$ when ambiguous, such that:
\begin{itemize}
  \item every element is symmetric to itself via the neutral symmetry:
    $\unit \act x = x$ for $x \in \carrier\A$; and
  \item symmetry is transitive via composition:
    $(g_1 \mul g_2) \act x = g_1 \act (g_2 \act x)$ for
    $g_1, g_2 \in \carrier G$, $x \in \carrier A$.
\end{itemize}

\begin{example}
  \examplelabel{symmetric-group-action}
  The symmetric group $\Sym n$ acts on the finite sets $\Fin n = \set{0, 1,
    \dots, n - 1}$ by $\perm \act i \definedby \perm i$,
  and on vectors $\Vct n X \definedby X^n$ by
  \(
    \perm \act \seq{x_0, \dots, x_{n-1}} \definedby \seq{x_{\inv\perm 0}, \dots, x_{\inv\perm (n-1)}}
  \)
  sending the $i$-th element to the position  $\perm i$.
\end{example}

\begin{example}
  \examplelabel{2048 action}
  We define the sets of valid $n \times n$ boards of 2048 by:
  \[
    \Tile \definedby \set0\union \set{2^i \suchthat i \in \pnaturals}
    \qquad
    \Row n = \Vct n \Tile
    \qquad
    \Board n = \Vct n (\Row n)
  \]
  And the action $( \act ) : \Cyc 4 \to \Board n \to \Board n$ is
  given by rotating the board entries as:
  \[
    \Rotation \act
    \begin{psmallmatrix}
      x_{1,1} & \cdots & x_{1,n} \\
      \vdots & \ddots & \vdots \\[1ex]
      x_{n,1} & \cdots & x_{n,n}
    \end{psmallmatrix}
    =
    \begin{psmallmatrix}
      x_{1,n} & \cdots & x_{n,n} \\
      \vdots & \ddots & \vdots \\[1ex]
      x_{1,1} & \cdots & x_{n,1}
    \end{psmallmatrix}
  \]
  Since $\Rotation$ generates $\Cyc 4$, any other rotation is of the
  form $\Rotation^k$ for some $k$, and so the action of other
  rotations is uniquely determined by the group action laws to be
  ${\Rotation^k \act b = \Rotation \act (\Rotation^{k - 1} \act b)}$.

  Note, however, that not all definitions on generators extend to the
  whole group, as they must also respect the equalities between group
  elements. For example, since $\Rotation^4 = \unit$, we must also
  have that $\Rotation^4 \act b = b$ for any board~$b$, and the above
  definition respects that since rotating the entries four times gives
  us back the original matrix.
\end{example}

\paragraph{Equivariance}

The game 2048 exhibits symmetry not only by permitting rotations of
the board, but also by the fact that merging respects it: if we
rotate the board and then merge in the rotated direction, we get the
same result as first merging and then rotating the resulting board:
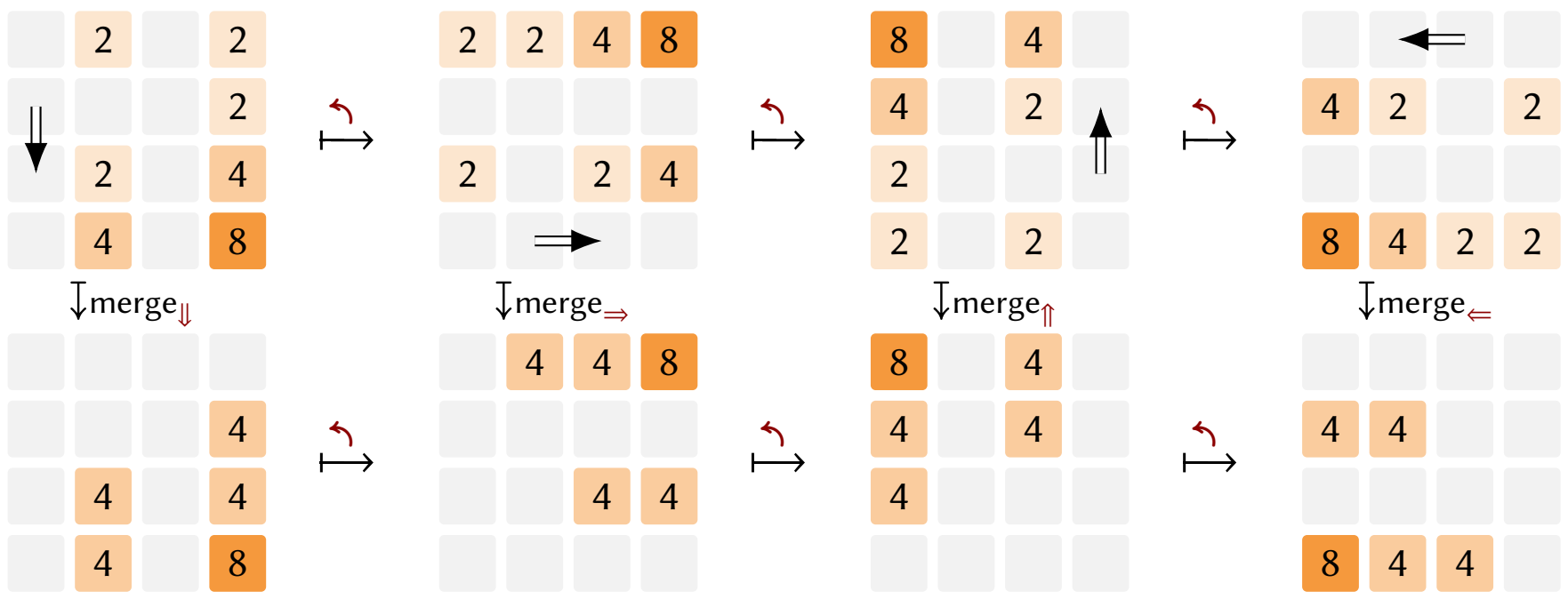
\[
  \newcommand{\twofortyeight}[2]{\vcenter{\hbox{
        \begin{tikzpicture}[scale=0.5, tile/.style={rounded corners=0.05cm, minimum size=0.45cm, draw=white, thick, font=\sffamily}
          ]
          \foreach \x in {0,1,2,3} {
            \foreach \y in {0,1,2,3} {
              \node[tile, fill=gray!10] (\x\y) at (\x, \y) {};
            }
          }
          \foreach \row/\col/\num in {#1}
          \node[tile, fill=gray!10!orange!\num0] at (\col, 3 - \row) {\small\num};
          #2
  \end{tikzpicture}}}}
  \begin{array}{ccccccc}
    \twofortyeight{0/1/2, 0/3/2, 1/3/2, 2/1/2, 2/3/4, 3/1/4, 3/3/8}
    {\draw[overlay, double distance=1.5pt
        , arrows = {-Latex[]}
    ]  (02.center) -- (01.center);}
    &\xmapsto{\Rotation}
    &\twofortyeight{2/0/2, 0/0/2, 0/1/2, 2/2/2, 0/2/4, 2/3/4, 0/3/8}
    {\draw[overlay, double distance=1.5pt
        , arrows = {-Latex[]}
    ]  (10.center) -- (20.center);}
    &\xmapsto{\Rotation}
    &\twofortyeight{3/2/2, 3/0/2, 2/0/2, 1/2/2, 1/0/4, 0/2/4, 0/0/8}
    {\draw[overlay, double distance=1.5pt
        , arrows = {-Latex[]}
    ]  (31.center) -- (32.center);}
    &\xmapsto{\Rotation}
    &\twofortyeight{1/3/2, 3/3/2, 3/2/2, 1/1/2, 3/1/4, 1/0/4, 3/0/8}
    {\draw[overlay, double distance=1.5pt
        , arrows = {-Latex[]}
    ]  (23.center) -- (13.center);}
    \\
    \downmapsto \scriptstyle\merge_{\Dn}
    &
    &\downmapsto \scriptstyle\merge_{\Rt}
    &
    &\downmapsto \scriptstyle\merge_{\Up}
    &
    &\downmapsto \scriptstyle\merge_{\Lt}
    \\
    \twofortyeight{1/3/4, 2/1/4, 2/3/4, 3/1/4, 3/3/8}{}
    &\xmapsto{\Rotation}
    &\twofortyeight{0/1/4, 2/2/4, 0/2/4, 2/3/4, 0/3/8}{}
    &\xmapsto{\Rotation}
    &\twofortyeight{2/0/4, 1/2/4, 1/0/4, 0/2/4, 0/0/8}{}
    &\xmapsto{\Rotation}
    &\twofortyeight{3/2/4, 1/1/4, 3/1/4, 1/0/4, 3/0/8}{}
  \end{array}
\]
\emph{Equivariance} captures symmetry preservation.
Given $\G$-actions $\A$ and $\B$, recall that an
\emph{equivariant} map
$\phi : \A \equito \B$
is a
map $\phi : \carrier
\A \to \carrier \B$ satisfying
\(
  \phi (g \act[\A] a) = g \act[\B] \phi a
\)
for all $g \in \carrier \G$ and $a \in \carrier \A$.

\begin{example}\examplelabel{action intro: permutation}
  Take any $f : X \to Y$. Its pointwise application
  $\vecmap f : \Vct n X \to \Vct n Y$
  is equivariant with respect to the permuting action of $\Sym n$ on
  vectors since:
  \begin{align*}
    \vecmap f (\perm \act \seq{x_0, \dots, x_{n-1}})
    &=
    \vecmap f (\seq{x_{\inv\perm 0}, \dots, x_{\inv\perm (n-1)}})
    =
    \seq{f x_{\inv\perm 0}, \dots, f x_{\inv\perm (n-1)}} \\
    &=
    \perm \act \seq{f x_0, \dots, f x_{n-1}}
    =
    \perm \act (\vecmap f \seq{x_0, \dots, x_{n-1}})
    \tag*{\qedhere}
  \end{align*}
\end{example}

Before the next example, recall that a \emph{product} $\A \times \B$
of $\G$-actions $\A$ and $\B$ is given on
\(
  \carrier{\A \times \B} \definedby {\carrier \A} \times {\carrier \B}
\)
component-wise by:
\(
  g \act[\A \times \B] \seq{a, b} \definedby \seq{g \act[\A] a, g \act[\B] b}
\).
Furthermore, the \emph{exponential action} $\A \to \B$ is given on
the set of all functions, not only the equivariant ones,
\(
  \carrier{\A \to \B}\definedby \carrier\A \to \carrier\B
\)
by \emph{conjugation}:
\(
  (g \act[\A \to \B] f) \definedby \fun a g \act[\B]  f(\inv g\act[\A] a)
\).
Finally, the \emph{trivial} action $\Triv \X$ on a set $\X$ is defined by
\(
  \carrier{\Triv \X} \definedby \X
\) and \(
  g \act x \definedby x \). An \emph{invariant} map is an equivariant
map $f : \A \equito \Triv \X$.

\begin{example}
  \examplelabel{tree example}
  We can equip the set $\Tree$ of binary trees with values in
  $\naturals$ (fixed for simplicity) with the recursive $\Sym2$-action
  $\swap \act[\Tree] {\Leaf} = \Leaf$ and ${\swap} \act[\Tree] \Node(\ell, n, r) =
  \Node(\swap \act r,n,\swap \act \ell)$. Furthermore, let $\TreePath$
  be the set of lists over step directions $\TreeLabel = \set{\Lft,\Rgt}$. We can
  equip $\TreePath$ with an $\Sym2$-action by mapping the
  $\Sym2$-action $\swap \act \Lft = \Rgt$ and $\swap \act \Rgt = \Lft$.

  Take $\lookup : \Tree \to \TreePath \to \Maybe\naturals$, which
  takes a path through a tree (a list of steps left or right) and
  returns the resulting value, if it exists. This function is
  invariant under symmetry: if we flip both the tree and the direction
  of all the steps, we should get back the same value. More precisely,
  $\lookup$ is equivariant in its curried form $\Tree \equito
  (\TreePath \to \Triv(\Maybe\naturals))$ with respect to the
  exponential action, while the uncurried form $\Tree \times \TreePath
  \equito \Triv(\Maybe\naturals)$ is invariant.
\end{example}

\subsection{The \WLOG construction}
\subseclabel{wlog-construction}

\paragraph{Canonisers}

The first step in our construction is to map each input to a
\emph{canonical} representative.  Crucially, we canonise via
an explicit symmetry in the group, rather than through an arbitrary
function, as we need to apply the inverse symmetry to reconstruct the
output for the original input.

For example, we canonise search trees by flipping the tree when the
inserted key is strictly larger than the root key. Thus, the remaining
canonical inputs are those that: insert into an empty tree, insert the
redundant key equal to the one in the root; and insert to the left
subtree. For the pigeonhole principle example, the canonising symmetry
is the identity if the first object is red, and $\swap$ that exchanges
the colours if the first object is blue. Thus, the canonical elements
are triples in which the first object is red.

\begin{definition}
  Let $\A$ be a $\G$-action. A \emph{canoniser} for $\A$ is a function
  $\canoniser : \carrier A \to \carrier G$. For each $a \in \carrier
  A$, we call the symmetry $\canoniser a \in \carrier G$ the
  \emph{$\canoniser$-canonising symmetry} of $a$, and the element
  $\canoniser a \act a \in \carrier A$ the \emph{canonical form}
  of $a$.
  We define the set of \emph{canonical} elements to be
  \(
    \Canon[\A]\canoniser \definedby \set*{\canoniser a \act a
    \suchthat* a \in \carrier \A}
  \).
\end{definition}

Canonisers need not respect the structure of the
action on $A$, as many of the interesting examples indeed do not.
However, there are still some examples that respect it, and we study
them and their impact on the equivariance of the \WLOG{} construction
in \subsecref{wlog-equivariance}.

\begin{example}
  \examplelabel{sorting canoniser}
  Let $P$ be a \emph{strict partial order}, i.e., a set equipped with
  an irreflexive and transitive relation~$<_P$. We say that an
  $n$-tuple $\vec p \in \Vct n P$ is \emph{sorted} when $p_{i} <
  p_{j}$ implies $i < j$ for all indices $i, j \in \Fin n$. Then, a
  \emph{sorting function} is a function $\sort : \Vct n P \to
  \carrier{\Sym n}$ that maps a tuple to a permutation that sorts it.
  Every such function is a canoniser for the permuting action of $\Sym
  n$ on $\Vct n P$ from \exampleref{symmetric-group-action}.

  Given an $n$-tuple $\vec p$, the sorted tuple $(\sort \vec p) \act
  \vec p$ is the canonical form of $\vec p$, and every $p \in \Canon\sort$
  is sorted. If the sorting function returns the identity
  permutation for sorted tuples---for example when the sorting is
  stable---the set $\Canon\sort$ is exactly the set of sorted tuples.
\end{example}

\begin{example}
  \examplelabel{2048 canoniser}
  Continuing \exampleref{2048 action}, the input for the 2048's merge
  is the board together with the merge direction. Define $\Direction
  \definedby \set{\Rt, \Up, \Lt, \Dn}$ and its action $(\act) : \Cyc 4
  \to \Direction \to \Direction$ by:
  \[
    \Rotation \act \Rt \definedby \Up
    \quad
    \Rotation \act \Up \definedby \Lt
    \quad
    \Rotation \act \Lt \definedby \Dn
    \quad
    \Rotation \act \Dn \definedby \Rt
  \]
  Define $\mkLt[] : \Direction \to \carrier{\Cyc 4}$ to pick the
  unique rotation turning its input left, i.e. $(\mkLt d) \act d = \Lt
  $:
  \[
    \mkLt \Rt \definedby \Rotation^2 = \twiceRotation
    \qquad
    \mkLt \Up \definedby \Rotation
    \qquad
    \mkLt \Lt \definedby \unit = \nonRotation
    \qquad
    \mkLt \Dn \definedby \inv\Rotation = \unRotation
  \]
  Thus the only canonical direction is left
  $\Canon{\mkLt[]} = \set{\Lt}$. We can extend the
  canoniser to input pairs $\Direction \times \Board n$ by
  $\mkLt \compose \projection_1$.
\end{example}

We frequently encounter similar canonisers that depend only on a part of the
input.

\begin{lemma}\lemmalabel{projection canoniser}
  Let $\A$ and $\B$ be $\G$-actions. Every $\A$-canoniser $\canoniser
  : \carrier\A \to \carrier\G$ induces an
  $(\A\times\B)$-canoniser $\canoniser \compose \projection_1 :
  \carrier {\A \times\B} \to \carrier \G$. We then have
  $\Canon{\parent{\canoniser \compose \projection_1}} =
  \Canon\canoniser \times \carrier\B$.
\end{lemma}

\paragraph{Core functions}

Given $\G$-actions $\A$ and $\B$, we want to define a
function $\A \to \B$ while avoiding symmetric cases. Once we have an
$\A$-canoniser~$\canoniser$ and its set of canonical
elements~$\Canon\canoniser$ in place, the only thing missing is the
\emph{core function} $f : \Canon\canoniser \to \carrier\B$:
merging of tiles in 2048, insertion into
a subtree, or the special case in the pigeonhole principle.
Typical symmetric \WLOG arguments only explicate canonical elements and core functions,
but gloss over groups, actions, and canonisers.

\begin{example}
  \examplelabel{2048 core function}
  Continuing \exampleref{2048 canoniser}, let us define the core
  function merging the tiles in 2048. We choose left merges as the
  canonical direction since vertical merges go against the structure
  of vectors of rows, while rightward merges have edge cases
  coalescing empty tiles. First, we define $\mergeLeft[\Row{}] : \Row
  n \to \Row n$ using pattern matching as all the relevant tiles are
  on the left:
  \begin{gather*}
    \mergeLeft[\Row{0}] \seq{}
    \definedby \seq{}
    \quad
    \mergeLeft[\Row{1}] (x :: \seq{})
    \definedby \seq{x}
    \quad
    \mergeLeft[\Row{1+n}] (0 :: r)
    \definedby \mergeLeft[\Row{n}](r) \++ \seq{0}
    \\
    \mergeLeft[\Row{2+n}] (x :: y :: r)
    \definedby
    \begin{cases}
      2 x :: \mergeLeft[\Row{n}](r) \++ \seq{0} & (x = y) \\
      \mergeLeft[\Row{1+n}](x :: r) \++ \seq{0} & (y = 0) \\
      x :: \mergeLeft[\Row{1+n}](y :: r) & (x \ne y \ne 0)
    \end{cases}
  \end{gather*}
  We merge the whole board simply by merging each row:
  \[
    \mergeLeft : \Board n \to \Board n
    \qquad
    \mergeLeft \definedby \vecmap \mergeLeft[\Row{n}]
    \tag*{\qedhere}
  \]
\end{example}

\paragraph{The construction}

We now formally define the \WLOG{} construction we sketched a number of
times, extending a core function $f : \Canon\canoniser \to \carrier
\B$ to the whole carrier $\carrier\A$.  Take any $a \in \carrier\A$
and compute its canonical form $\canoniser a \act a$. By
definition, the canonical form lies in $\Canon\canoniser$, so we can
apply the core function to obtain $f (\canoniser a \act a)$. Even
though this result has the expected type~$\B$, we still need to apply
the inverse symmetry $\ginv*{\canoniser a}$.
Indeed, making a downward merge in 2048 would result in a
rotation, followed by a leftward merge, followed by the inverse rotation.

\begin{definition}
  Let $\A$ and $\B$ be $\G$-actions, $\canoniser : \carrier\A \to
  \carrier\G$ a canoniser, and $f : \Canon\canoniser \to \carrier\B$
  a core function. Then, the \emph{\WLOG construction}~$\wlg(c,f) :
  \carrier\A \to \carrier\B$ is defined as
  \[
  \myfigbox{$
    \wlg(\canoniser,f) \definedby
    \fun{a} \ginv*{\canoniser a}\act[\B] f(\canoniser a \act[\A] a)
  $}
  \]
\end{definition}
Using \exampleref{2048 core function}, we define $\merge : \Direction \times \Board n
\to \Board n$ as $\wlg(\mkLt \compose \projection_1, \mergeLeft \compose \projection_2)$.

\subsection{Equivariance of the \WLOG construction}
\subseclabel{wlog-equivariance}

In addition to extending the core function $f : \Canon\canoniser \to
\carrier\B$ to the whole domain, we often want the extension $\phi :
\carrier\A \to \carrier\B$ to be equivariant. This requirement implies that the
core function $f$ must already behave equivariantly where defined. To
see this, consider a canonical element $x \in \Canon\canoniser$ that
also has a symmetric canonical element $g \act x \in
\Canon\canoniser$. Then:
\[
\myfigbox{
  $
  f(g \act x)
  \explain={$\phi$ extends $f$}
  \phi(g \act x)
  \explain={$\phi$ is equivariant}
  g \act \phi(x)
  \explain={$\phi$ extends $f$}
  g \act f x
  $}
\]
We generalise the notion of equivariance to domains that may
not be closed under the action:

\begin{definition}
  Take $\G$-actions $\A$, $\B$, a subset $X \subseteq \carrier\A$,
  and a function $f : X \to \carrier\B$. We call $f$ \emph{equivariant},
  writing $f : (X \subseteq A)
  \equito \B$, when
  \(
    g \act x \in X \implies f (g \act x) = g \act f x
  \)
  for $g \in \carrier\G$, $x \in X$.
\end{definition}

\begin{example}
  Every core function for the 2048 canoniser from \exampleref{2048
  canoniser} is vacuously equivariant: only $\Lt$ is
  canonical, and any $g \in \Cyc{4}\setminus\set\unit$ changes the
  direction to a non-canonical input. Therefore, if $x, g \act x$ are
  canonical, then $g = \unit$ and $g \act f\,x = f\,x = f(g \act x)$.

  Insertion into binary search trees features non-trivial corner cases
  of core function equivariance. Flipping the empty tree yields back the
  empty tree, which is again a canonical element. Thus inserting a key
  into a ``flipped'' empty tree needs to be the same as first inserting
  the key into the empty tree and then flipping the result. However, the
  resulting tree is a single root with two empty subtrees, which is
  invariant under flipping, thus the core function is equivariant.
  A similar case occurs when inserting a key that is equal to the root value.
\end{example}

Equivariance of core functions is easily overlooked when canonical
elements are treated in isolation, especially since the overlap
usually occurs on trivial cases. On a more positive note, the
condition is not only necessary, but also sufficient for the existence
of an equivariant extension. Furthermore, \WLOG{} gives the unique
such extension:

\begin{theorem}
  \theoremlabel{wlog-construction}
  Let $\A$ and $\B$ be $\G$-actions and $\canoniser : \carrier\A \to
  \carrier\G$ a canoniser. If $f : (\Canon\canoniser \subseteq \A)
  \equito \B$ is an equivariant core function, then $\wlg(c,f) : \A
  \equito \B$ is the unique equivariant function extending $f$ to the
  whole $\carrier\A$, along the inclusion $\Canon\canoniser \subset \carrier\A$.
\end{theorem}

\begin{example}
  Recall the $2048$ function
  $\merge = \wlg(\mkLt, \mergeLeft) : \Direction \times \Board n \to
  \Board n$.
  It indeed behaves as expected: for leftward merges, it
  extends $\mergeLeft$, while for other directions, it is uniquely
  determined by the equivariant behaviour.
\end{example}

In concrete cases, equivariance of core functions is often easily
discharged, but we can further relax it when the
canoniser~$\canoniser$ has additional properties. A natural property
is the equivariance of the canoniser. Since $\parent{\canoniser a \mul
\inv g} \act (g \act a) = \canoniser a \act a$ is canonical, it is
natural to ask whether $\canoniser a \mul \inv g$ is the
$\canoniser$-canonising symmetry of $g \act a$, i.e. whether
$\canoniser(g \act a) = \canoniser a \mul \inv g$. More succinctly,
this property means $\canoniser$ is an equivariant map $\canoniser :
\A \equito \inv\G$ where $\inv\G$ is the action of $\G$ on itself,
defined by $g \act[\inv\G] h \definedby h \mul \inv g$. It validates
the axioms since:
\begin{gather*}
  \unit \act[\inv\G] s = s \mul \inv\unit = s
  \qquad
  g \act[\inv\G] (h \act[\inv\G] s)
  =
  (s \mul \inv h)\mul \inv g = s \mul \inv*{g \mul h} = (g \mul h)\act[\inv\G] s
\end{gather*}

\begin{definition}
  A canoniser $\canoniser : \carrier\A \to \carrier\G$ is
  \emph{strictly equivariant} when $\canoniser\parent{g \act a} =
  \canoniser a \mul \inv g$, i.e. when it is an equivariant map
  $\canoniser : \A \equito \inv\G$.
\end{definition}

As we shall soon see, strict equivariance is a powerful assumption,
but also a very strong restriction, thus we reserve the name
\emph{equivariant canonisers} for a weaker and more natural
notion that we introduce in \subsecref{equivariant canonisers}, after
we discuss dependent types.

\begin{example}
  \examplelabel{strictly equivariant canonisers}
  The canoniser $\mkLt[]$ from \exampleref{2048 canoniser} is strictly
  equivariant since $\mkLt d$ is the unique rotation that makes $d$
  canonical. On the other hand, if in \exampleref{sorting canoniser},
  we take a non-total ordering $\P$ and a stable sorting algorithm,
  the canoniser $\sort : \Vct n P \to \carrier{\Sym n}$ is not
  strictly equivariant since for every two incomparable elements $p, q$,
  we have:
  \begin{align*}
    \mspace{-10mu}\sort\parent{(0 \leftrightarrow 1)\act \seq{p,q}}
    &= \sort\seq{q,p} = \id
    \neq (0 \leftrightarrow 1) = \id \mul \inv{(0 \leftrightarrow 1)} = (0 \leftrightarrow 1) \act \sort\seq{p,q}
    \mspace{-10mu}
    \tag*{\qedhere}
  \end{align*}
\end{example}

Strict equivariance guarantees equivariance of arbitrary core functions,
and thus also of the \WLOG{} construction. The converse also holds,
though we do not need it for our applications.

\begin{theorem}
  \theoremlabel{all core functions are partially equivariant}
  Let $c : \carrier\A \to \carrier\G$ be a canoniser. Then,
  $\canoniser$ is strictly equivariant iff for every action $\B$,
  every core function $f : \Canon\canoniser \to \carrier\B$ is
  equivariant.
\end{theorem}

Strict equivariance is too strong, as it restricts the action to
be \emph{semi-regular}, meaning that only the identity can fix an
element: $g \act a = a \implies g = \unit$.

\begin{lemma}[semi-regularity]\lemmalabel{semi-regularity}
  If $\A$ has a strictly equivariant canoniser, then $\A$ is
  semi-regular.
\end{lemma}

For example, the $\Cyc 4$ action on $\Board n$ is not semi-regular,
since all rotations preserve the empty 2048 board. It is only the
addition of the direction, which all non-trivial symmetries affect,
that makes the action on $\Direction \times \Board n$ semi-regular.

Strictly equivariant canonisers satisfy $\canoniser(\canoniser a
\act a) = \canoniser a \mul \inv*{\canoniser a} = \unit$, meaning that
canonisation is a one-step process and that further canonisation
symmetries are identities. This observation leads us to a weaker natural property
of canonisers, and one that all examples in this paper enjoy:

\begin{definition}
  A canoniser $\canoniser$ is \emph{immediate} when for all $a \in
  \carrier\A$, we have $\canoniser(\canoniser a \act a) = \unit$.
\end{definition}

\begin{proposition}\propositionlabel{strictly equivariant canonisers are immediate}
  A strictly equivariant canoniser $\canoniser : \A \equito \inv\G$ is immediate.
\end{proposition}

The converse does not hold: as we saw in \exampleref{strictly
equivariant canonisers}, a stable sorting function is not a strictly
equivariant canoniser, but it is immediate since the canonical
elements are exactly the sorted tuples, and the sorting permutation
for sorted tuples is the identity.

We will often derive a canoniser for $\A\times\B$ from a canoniser for
$\A$ through projection. For example, we
derive the canoniser for $\Direction\times\Board{n}$ from that of $\Direction$.

\begin{lemma}\lemmalabel{equivariance of projection canonisers}
  For $G$-actions $\A, \B$, if $\carrier \A \xto{\canoniser} \carrier
  \G$ is strictly equivariant/immediate, then so is
  $\carrier \A \times \carrier \B \xto{\canoniser
  \compose \projection_1} \carrier
  \G$.
\end{lemma}

To apply the \WLOG construction, we need to provide both the
canoniser~$\canoniser$ and the core function~$f$. If we have
$\canoniser a = \unit$ for some $a \in \carrier\A$, then $a =
\canoniser a \act a \in \Canon\canoniser$, and so we need to define
the value $f a$. In the ideal case, we want to consider each element
of $\carrier\A$ only once and define the core function only on
elements with trivial canonising symmetries. More precisely, we want
the sets $\set{a \in \carrier A \suchthat \canoniser a \ne \unit}$ and
$\Canon\canoniser$ to be disjoint. But in general, the two sets may
overlap. In the extreme case, if
\(
  \canoniser : \Direction \to \carrier{\Cyc4}
\)
is the constant function
\(
  \fun{\_} \Rotation
\),
they are both equal to the whole domain $\Direction$. However,
immediate canonisers are exactly those for which the domain
$\carrier\A$ cleanly splits into two:

\begin{proposition}
  \propositionlabel{immediate canonical is kernel}
  A canoniser $\canoniser : \carrier\A \to \carrier\G$ is immediate
  iff $\Canon\canoniser = \inv{\canoniser}[\unit] \definedby \set{a
  \in \carrier A \suchthat \canoniser a = \unit}$.
\end{proposition}

\section{Symmetric dependent types}
\seclabel{dependent types}

To precisely show how our construction captures traditional \WLOG
arguments, we need to move to a more expressive setting. For that, we
use dependent types, which also allow us to impose richer invariants
on programming examples such as binary search trees.

We start by recalling the framework of \emph{families}
(\subsecref{families}), which gives a mathematical interpretation of
familiar dependent-type notions. Afterwards, we generalise our
constructs and results to a dependent setting (\subsecref{indexed
actions}) and show how via the Curry-Howard correspondence, they
capture proofs that use \WLOG arguments (\subsecref{propositional
reasoning}). Finally, we use the additional power of dependent types
to define a more refined notion of canoniser equivariance
(\subsecref{equivariant canonisers}).

\subsection{Families}
\subseclabel{families}

A \emph{family}~$\Ctx \types X$ is a pair consisting of a set $\Ctx$
and an assignment of a set $X \ctx$ to each element $\ctx \in \Ctx$.
We call $\Ctx$ the \emph{indexing} set, its elements $\ctx \in \Ctx$
\emph{indices}, and each set $X \ctx$ the \emph{fibre} over the index
$\ctx$. A \emph{term}~$\Ctx \types t : X$ in a family $\Ctx \types X$
is an assignment of an element $t \ctx \in X \ctx$ to each $\ctx \in
\Ctx$. A family $\Ctx \types X$ is a \emph{subfamily} of $\Ctx \types
Y$, which we write as $\Ctx \types X \subseteq Y$ if $X \ctx \subseteq
Y \ctx$ for all $\ctx \in \Ctx$.

\begin{example}
  \examplelabel{families example}
  To eliminate the necessity for the $\Maybe$ type in the tree lookup
  from \exampleref{tree example}, let us define the family $\Tree
  \types \NodeInName$ of \emph{node paths} that contains all paths
  ending at an internal node of the index tree. More precisely, the
  fibre $\NodeIn \btree$ is empty if $\btree$ is empty, and is
  otherwise the set of lists over $\TreeLabel = \set{\Lft,\Rgt}$ which are either
  $\Nil$, representing the root of the index tree, or $d \:: p$ that
  start with a step direction $d$ followed by a path $p$ in that
  subtree. This set can be given inductively as:
  \begin{gather*}
    \begin{array}[b]{c}
    \inferrule{
      ~
    }{
      \Nil \in \NodeIn \Big( \NodeV{\ell}x{r}\Big)
    }\quad
    \inferrule{
      p \in \NodeIn \ell
    }{
      (\Lft \:: p) \in \NodeIn \Big(\NodeV{\ell}x{r}\Big)
    }\quad
    \inferrule{
      p \in \NodeIn r
    }{
      (\Rgt \:: p) \in \NodeIn \Big(\NodeV{\ell}x{r}\Big)
    }
    \end{array}
    \tag*{\qedhere}
  \end{gather*}
\end{example}

When specifying fibres or terms as assignments of the form $\fun\mctx
X$ or $\fun\mctx t$, which explicitly use a generic index $\mctx
\in \Ctx$, we write them as $\mctx : \Ctx \types X$ and $\mctx : \Ctx
\types t : X$.
We will continue to distinguish these uses, denoting generic indices
for the purpose of defining a family by $\mctx : \Ctx$, and use $\mctx
\in \Ctx$ when we refer to elements of an indexing set $\Ctx$ for any other
purpose. Note that despite using a convenient type-theoretic notation,
we continue to keep our development purely mathematical, constructing
objects without any recourse to a syntactic type system.  Were we to
present such a syntactic type system and differentiate between
syntactic contexts $\Ctx$ and their interpretations as indexing sets
$\sem\Ctx$, then we would distinguish between $\ctx : \Ctx$ and $\ctx
\in \sem\Ctx$.

Continuing this approach, we allow fibres and terms to depend on a
context of variables. For the empty context, we take the indexing set
$\terminal \definedby \set\unitval$ and identify families $\terminal
\types X$ with their only fibre~$X \unitval$. Then, we write fibres
over $\terminal$ simply as $\types X$, and similarly for terms.

For context extension, we employ the \emph{Grothendieck construction}.
Recall that for a family $\Ctx \types X$, it is defined as the set
$\coprod_{\Ctx}X \definedby \coprod_{\ctx \in \Ctx} X \ctx$,
consisting of pairs $\pair\ctx x$ where $\ctx \in \Ctx$ and $x \in X
\ctx$, together with a projection function $\projection_1
\definedby \parent{\fun{\pair\ctx x} \ctx}: \coprod_{\Ctx}X \to \Ctx$.
We write $\Ctx, \mx : X \types Y$ and $\Ctx, \mx : X \types t : Y$ for
fibres and terms indexed by $\coprod_{\Ctx}X$ and use $\mx$ to refer
to the second component of the index pair from $\coprod_{\Ctx}X$ in $Y$
and $t$.

Given two families $\Ctx \types \mX$ and $\Ctx, \x : \mX \types \mY$,
we have the family of \emph{dependent pairs} $\Ctx \types
\DPair\x\mX\mY$ whose fibre at $\ctx \in \Ctx$ is the Grothendieck
construction $\coprod_{\mx \in \mX \ctx}\mY \pair\ctx\mx$, and the
family of \emph{dependent functions} $\Ctx \types \DFun\x\mX\mY$ whose
fibre at $\ctx$ is the product $\prod_{\mx \in \mX \ctx}\mY
\pair\ctx\mx$. As usual, we write $\Fun\mX\mY$ when the codomain
family $\mY$ does not depend on the argument of the function. To
distinguish between indices and arguments of dependent functions $\Ctx
\types f : \DFun\x\mX\mY$, we apply them as $f_{\ctx} \mx$
for $\ctx \in \Ctx$ and $\mx \in \mX \ctx$, or even as $f \mx$
when $\ctx$ can be inferred.

\begin{example}
  \examplelabel{node lookup}
  With dependent types, we can refine the lookup function $\Tree \to
  \TreePath \to \Maybe\naturals$ from \exampleref{tree example} to a
  dependent function $\types \lookup : \DFun{\BoundV t}{\Tree}
  \Fun{\NodeIn{\BoundV t}} \naturals$, as the path now must point to a node
  in the tree. We can define it as:
  \begin{gather*}
    \lookup \Big({\NodeV \ell n r}\Big) \Nil \definedby n
    \qquad
    \lookup \Big(\NodeV {t_{\Lft}}n {t_{\Rgt}}\Big) (d \:: p) \definedby \lookup \, t_{d} \, p
  \end{gather*}
  Note that since we have two trees $t_{\Lft}, t_{\Rgt}$ rather than a
  function $\TreeLabel \to \Tree$, the parametrisation $t_d$ that
  we use in the second case is just repetition in disguise. Later, we
  will show how to properly avoid this repetition by using the \WLOG
  construction.
\end{example}

\subsection{Indexed actions}
\subseclabel{indexed actions}

Just as in \exampleref{tree example}, the function $\lookup$ from
\exampleref{node lookup} is invariant with respect to flipping:
flipping both the tree and the path returns the same value. But note
that once we flip the tree $t$, the flipped path no longer lies in
$\NodeIn{t}$, but in $\NodeIn{(\swap \act t)}$. For example:
\[
\swap \act \Bigg(\Lft \:: \Rgt \:: \Nil \in \NodeIn   \bigg(
  \begin{tree diagram}
    \& \& 3 \\
    \& 2 \& \\
    1 \& \& 4
    \treeLink1322
    \treeLink2231
    \treeLink2233
\end{tree diagram}\bigg)\Bigg)
=
\Rgt \:: \Lft \:: \Nil \in \NodeIn   \bigg(
  \begin{tree diagram}
    3 \& \& \\
    \& 2 \& \\
    4 \& \& 1
    \treeLink1122
    \treeLink2231
    \treeLink2233
\end{tree diagram}\bigg)
\tag*{\qedhere}
\]
In general, an action over a family $\Ctx \types A$ should thus
consist of an action over the index set $\Ctx$, and an action over
each fibre $A \ctx$, which tracks the indexing action: if $a \in A
\ctx$, then $g \act a \in A (g \act \ctx)$.
\begin{definition}
  Given a $\G$-action $\Ctx$, an \emph{indexed $\G$-action $\Ctx
  \types \A$} consists of a \emph{carrier} family $\carrier\Ctx \types
  \carrier\A$ and a dependent function $(\iact{-}) : \DFun {\BoundV
  g}{\carrier \G} \DFun {\BoundV \ctx}{\carrier \Ctx}
  \Fun{\carrier\A\ctx} \carrier \A (g \act[\Ctx] \ctx)$, which we
  apply as $g \iact{\ctx} \ma$, such that
  \(
    \unit \iact{\ctx} \ma = \ma
  \) and \(
    (g \mul h) \iact{\ctx} \ma = g \iact{h \act \ctx} (h \iact{\ctx} \ma)
  \).
\end{definition}
When $\A$ is ambiguous, we add a superscript and write
$\iact[\A]{\ctx}$, and when  we can infer the index $\ctx$, we elide
it and write $g \act \ma$ instead of $g \iact{\ctx} \ma$.

\begin{example}
  The action of the cyclic group $\Cyc 4$ on 2048 boards can be
  generalised to an indexed action on the family of arbitrary matrices
  $\naturals \times \naturals \types \TConst{Mat}$. First, take $\Ctx$
  to be the action of $\Cyc 4$ on dimensions $\carrier\Ctx \definedby
  \naturals \times \naturals$, defined as:
  \(
    \Rotation\act\pair mn \definedby \pair nm
  \).
  We then have a $\Ctx$-indexed action defined as in \exampleref{2048
  action}:
  \[
    \Rotation
    \iact{\pair mn}
    \Bigg(
      \bigg(
    \begin{smallmatrix}
      x_{1,1} & \cdots & x_{1,n} \\
      \vdots & \ddots & \vdots \\[1ex]
      x_{m,1} & \cdots & x_{m,n}
    \end{smallmatrix}
    \bigg)
    \in
    \Mat mn\Bigg)
    =
    \begin{psmallmatrix}
      x_{1,n} & \cdots & x_{m,n} \\
      \vdots & \ddots & \vdots \\[1ex]
      x_{1,1} & \cdots & x_{m,1}
    \end{psmallmatrix}
    \in \Mat nm
    \tag*{\qedhere}
  \]
\end{example}

\begin{example}
  Take indexed $\G$-actions $\Ctx \types \A$ and $\Ctx, \x : \A \types
  \B$. Then, we can define the \emph{dependent pair} action $\Ctx
  \types \DPair\x\A\B$ as:
  \[
    \carrier{\Ctx \types \DPair\x\A\B} \definedby \carrier\Ctx \types \DPair\x{\carrier\A}{\carrier\B}
    \qquad
    \g \iact[\DPair\x\A\B]{\ctx}{\pair \ma\mb} \definedby
    \pair{\g\iact[\A]\ctx\ma}{\g\iact[\B]{\pair\ctx\ma}\mb}
  \]
  Furthermore, we can define the \emph{dependent function} action
  $\Ctx \types \DFun\x\A\B$ as:
  \begin{gather*}
    \carrier{\Ctx \types \DFun\x\A\B} \definedby \carrier\Ctx \types \DFun\x{\carrier\A}{\carrier\B}
    \\
    g\iact[\DFun\x\A\B]\ctx f \definedby
    \fun{(a : \carrier\A(g \act \ctx))}
      g \iact[\B]{\pair\ctx{\inv g \act\ma}} f_{\ctx}(\inv g \iact[\A]{g \act \ctx}\ma)
  \end{gather*}
  It is straightforward to check that both definitions satisfy the
  indexed action laws.
\end{example}

\begin{definition}
  Given a $\G$-action $\Ctx$ and an indexed $\G$-action $\Ctx \types
  \A$, a term $\carrier\Ctx \types t : \carrier\A$ is
  \emph{equivariant} when its components are: $t (g \act[\Ctx] \ctx) =
  g \iact{\ctx} t \ctx$ for all $\ctx \in \carrier\Ctx$, $g \in
  \carrier\G$.
\end{definition}

Equivariance of terms is preserved by reindexing and substitution.
Furthermore, an equivariant term $\Ctx, \x : \A \types t : \B$ induces
an equivariant dependent function $\Ctx \types (\fun{\x:\A}t) :
\DFun\x\A\B$, and two equivariant terms $\Ctx \types t : \A$ and $\ctx
: \Ctx \types s : \B \pair\ctx{t \ctx}$ induce an equivariant term
given by the dependent pair in each fibre
$\Ctx \types \pair ts : \DPair\x\A\B$. For more details and proofs,
see~\appref[C]{dependent types}.

The ease of compositionality should come as no surprise given that the
definition of an indexed action is in fact just a reformulation of the
notion of a family internally to the topos of $\G$-actions:

\begin{proposition}\propositionlabel{Grothendieck action}
  Let $\Ctx$ be a $\G$-action. Then, a $\Ctx$-indexed action $\A$
  amounts to an action $\coprod_{\Ctx}\A$ over the Grothendieck
  construction for which the projection is equivariant:
  \(
    \carrier{\coprod_{\Ctx}\A} = \coprod_{\carrier\Ctx}\carrier\A
  \) and \(
    \projection_1 :  \coprod_{\Ctx}\A \equito \Ctx
  \).
\end{proposition}

We have used this connection above to reformulate standard notions
from dependent types to the symmetric setting, and continue to do so
with canonisers, core functions, and the \WLOG construction. Ignoring
the additional indices, we recover the simply-typed notions from
\secref{wlog construction}.
\begin{definition}
  Let $\Ctx \vdash \A$ be an indexed $\G$-action. An \emph{indexed
  canoniser} for $\A$ is a $\carrier\Ctx$-indexed term $\carrier\Ctx \types \canoniser :
  \carrier\A \to \carrier\G$. For each $\ma \in \carrier\A\ctx$, we
  call the symmetry $\canoniser_{\ctx} \ma \in \carrier\G$
  the \emph{$\canoniser$-canonising symmetry} of $\ma$, and the
  element $\canoniser \ma\iact{\ctx} \ma$ the \emph{canonical form} of
  $\ma$. We define the family of \emph{canonical} elements
  $\carrier\Ctx \types \Canon\canoniser$ with the fibres
  \(
    (\Canon\canoniser) \ctx \definedby \set{
      \canoniser \ma\iact{\ctx'} \ma
      \suchthat
      \ctx' \in \Ctx,
      \ma \in \carrier\A\ctx',
      \canoniser \ma\act\ctx' = \ctx
    }
  \)
  for each
  $\ctx \in \carrier\Ctx$.
\end{definition}

\begin{example}
  To capture the symmetry of the lookup function from \exampleref{node
  lookup}, we define a canoniser for the indexed $\Sym 2$-action on
  $\mtree : \Tree \types \NodeIn\mtree$:
  \[
    \canoniser_{\Node \triple\ell n r}\Nil \definedby \stet \qquad
    \canoniser_{\Node \triple\ell n r}(\Lft \:: p) \definedby \stet \qquad
    \canoniser_{\Node \triple\ell n r}(\Rgt \:: p) \definedby \swap
  \]
  The canonical elements in $\NodeIn\btree$ are \( \Nil \) and \(
  (\Lft \:: p) \in \NodeIn(\Node\triple\ell n r) \) where $p \in
  \NodeIn\ell$ can be an arbitrary path, even one that moves right on
  any of its steps.
\end{example}

\begin{definition}
  Take indexed $\G$-actions $\Ctx \types \A$ and $\Ctx, x : \A \types
  \B$, a subfamily $\carrier\Ctx \types X \subseteq \carrier\A$, and
  a function $\carrier\Ctx \types f : \DFun{\x}{X}\carrier\B$.
  We write $\Ctx \types f : \EFun{\x}{X \subseteq \A}\B$ and say that
  $f$ is \emph{equivariant} when
  \(
    g \iact[\A]{\ctx} x \in X (g \act \ctx)
    \implies
    f_{g\act \ctx}(g\iact[\A]{\ctx}x) = g\iact[\B]{\pair\ctx x}f_{\ctx}x
  \)
  for all $g \in \carrier\G$, $\ctx \in \carrier\Ctx$, and $x \in X\ctx$.
\end{definition}
We can now define the dependently-typed version of the \WLOG{} construction:
\begin{definition}
  Let $\Ctx \types \A$ and $\Ctx, \x : \A \types \B$ be indexed
  $\G$-actions, $\Ctx \types \canoniser : \carrier\A \to \carrier\G$
  an indexed canoniser, and $\Ctx \types f :
  \DFun{\x}{\Canon\canoniser} \carrier\B$ a dependent core function.
  Then, the \emph{indexed \WLOG construction}~$\wlg(c,f) :
  \DFun{\x}{\carrier\A} \carrier\B$ is defined as:
  \[
  \myfigbox{$
    \wlg(\canoniser,f)_{\ctx} \definedby \fun{(\ma : \carrier\A\ctx)}
    \inv*{\canoniser_{\ctx}\ma}\iact[\B]{\pair{\canoniser_{\ctx} a \act[\Ctx] \ctx}{\canoniser_{\ctx} a \iact[\A]{\ctx} a}}
  f_{\canoniser_{\ctx}\ma \act[\Ctx] \ctx}(\canoniser_{\ctx} \ma\iact[\A]{\ctx} \ma)
  $}
  \]
\end{definition}

With all the indices precisely spelt out, the definition can look a
bit intimidating, but if we elide them we get
\(
  \wlg(\canoniser,f)
    = \fun{\ma} \inv*{\canoniser\ma}\iact[\B]{} f(\canoniser \ma\iact[\A]{} \ma)
\)
as in the simply-typed case. And just as before in
\theoremref{wlog-construction}, the \WLOG construction gives the
unique equivariant extension of a core function.

\begin{theorem}[indexed representation]
  \theoremlabel{indexed-wlog-construction}
  Let $\Ctx \types \A$ and $\Ctx, \x : \A \types \B$ be indexed
  $\G$-actions and $\Ctx \types \canoniser : \carrier\A \to
  \carrier\G$ an indexed canoniser for $\A$. If $\Ctx \types f :
  \EFun{\x}{\Canon\canoniser \subseteq \A} \B$ is an equivariant
  dependent core function, then  $\wlg(c,f) : \EFun{\x}{\A} \B$ is
  the unique equivariant function extending~$f$ to the whole
  $\carrier\Ctx \types \carrier\A$.
\end{theorem}

\begin{example}
  \examplelabel{pigeonhole principle}
  Let us revisit the pigeonhole principle~\cite{wiki:wlog} from the
  introduction and extend it in two ways. First, instead of a proof,
  let us construct a function
  \(
    \mathsf{findSame} : \DFun{\BoundV{xyz}}{\set{\RED, \BLU}^3} \Type{Same}{\BoundV{xyz}}
  \)
  where
  \(
    \Type{Same}{\BoundV{xyz}} \definedby \set{
      (i,j) : (\Fin 3)^2 \suchthat i \neq j \land \projection_i \BoundV{xyz} = \projection_j \BoundV{xyz}
    }
  \)
  is the family of distinct positions of the same colour, indexed by
  the colouring $\BoundV{xyz}$.

  And second, we shall see that in the special case of the proof,
  there is another opportunity to use the \WLOG{} argument.
  Informally, our proof goes as follows:
  \begin{quote}
  \wLOG{}, the first object is red. The other two objects either have
  the same colour or different colours. If they have the same colour,
  choose $(1,2)$. If they have different colours, \WLOG{} the second
  object is red, so choose $(0,1)$.
  \end{quote}
  Such nested uses of \WLOG{} may be suspicious and confusing, which
  is exactly why it is important to make all aspects of \WLOG{}
  arguments explicit. In this case, this use is valid. For example,
  when the second object is not red and the third is red, the
  inner argument exchanges their positions:
  \[
    \newcommand{\balls}[2]{\vcenter
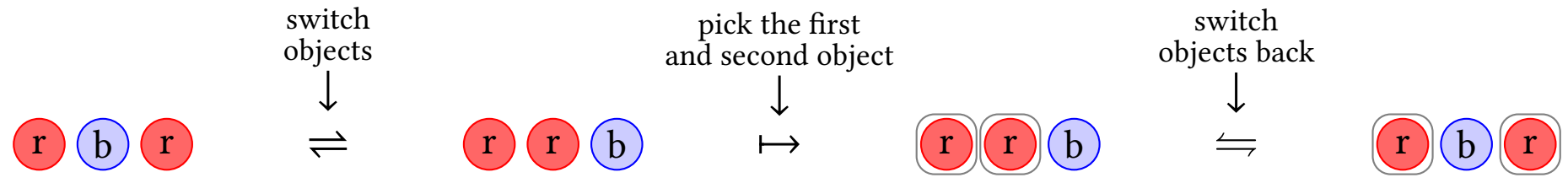
{\hbox{
      \begin{tikzpicture}[
        scale=0.5,
        obj/.style={circle, draw, inner sep=0pt, minimum size=4mm},
        rdot/.style={obj, fill=red!60, draw=red},
        bdot/.style={obj, fill=blue!20, draw=blue},
        pair/.style={rounded corners, thin, draw=gray},
        >=stealth, font=\small
      ]
        \pgfmathsetmacro\dx{-0.03}
        \foreach \col/\f in {#1}
          \node[\f] at (\col+0.5, 0.5) {\ifthenelse{\equal{\f}{bdot}}{b}{r}};
        #2
      \end{tikzpicture}
    }}}
    \balls{0/rdot, 1/bdot, 2/rdot}{}
    \qquad
    \explain{\rightleftharpoons}{switch\\[-5pt]objects}
    \qquad
    \balls{0/rdot, 1/rdot, 2/bdot}{}
    \qquad
    \explain{\mapsto}{pick the first\\[-5pt]and second object}
    \qquad
    \balls{0/rdot, 1/rdot, 2/bdot}{
      \draw[pair] (0-\dx,0-\dx) rectangle (1+\dx,1+\dx);
      \draw[pair] (1-\dx,0-\dx) rectangle (2+\dx,1+\dx);
    }
    \qquad
    \explain{\leftrightharpoons}{switch\\[-5pt]objects back}
    \qquad
    \balls{0/rdot, 1/bdot, 2/rdot}{
      \draw[pair] (0-\dx,0-\dx) rectangle (1+\dx,1+\dx);
      \draw[pair] (2-\dx,0-\dx) rectangle (3+\dx,1+\dx);
    }
  \]

  In the outer level, $\Sym 2$ acts on $\set{\RED,\BLU}$ by flipping
  the two colours, and on the domain $\set{\RED,\BLU}^3$ with the
  component-wise product action. On the codomain $\set{\RED,\BLU}^3
  \types \Type{Same}$, we lift the trivial action to preserve the
  elements of the fibre, but change the index. This is a sub-action
  (i.e. an action that preserves a subset) since if objects at two
  positions are of the same colour, they remain so after we exchange
  the colours. For example, $\smash{\swap\iact{(\RED,\RED,\BLU)}(0,1)}
  = (0,1) \in \Type{Same}{(\BLU, \BLU,\RED)}$

  By swapping colours, we make the first object red, thus the outer
  canoniser is:
  \[
    \mathsf{makeFirstRed}(\RED,\mb,\mc) \definedby \stet
    \qquad
    \mathsf{makeFirstRed}(\BLU,\mb,\mc) \definedby \swap
  \]
  and using it, we can define
  \[
  \mathsf{findSame} \definedby \wlg\Bigg(
    \mathsf{makeFirstRed},
    \fun{(\RED, \mb, \mc)}
      \begin{cases}
        (1,2) &(\mb = \mc) \\
        \mathsf{findOtherRed}(\mb,\mc) &(\mb \neq \mc)
      \end{cases}
  \Bigg)
  \]
  where we still need to define
  \[
  \mathsf{findOtherRed} : \DFun{(\y,\z)}{\set{(\mb,\mc) \in \set{\RED,\BLU}^2\suchthat \mb \ne \mc}}
  \Type{Same}{(\RED,\y,\z)}
  \]
  Again, we employ symmetry, but now, the $\Sym2$-action on the domain
  swaps the two components of the pair, but not their colour, i.e.,
  $\swap\act(\mb,\mc) \definedby (\mc, \mb)$. For the codomain, define
  an action on $\Fin 3$ that swaps $1$ and $2$, and extend it
  component-wise to the product $(\Fin 3)^2$. Since $\mb$ and $\mc$
  are of different colours, it is easy to check that this action again
  preserves the defining property of $\Type{Same}$ and thus lifts to a
  sub-action on the codomain. For example,
  $\smash{\swap\iact{(\RED,\BLU)}(0,1) = (0,2) \in \Type{Same}{(\RED, \BLU,
  \RED)}}$.

  The inner canoniser now makes the second object red:
  \[
  \mathsf{makeSecondRed}(\RED, \BLU) \definedby \stet
  \qquad
  \mathsf{makeSecondRed}(\BLU, \RED) \definedby \swap
  \]
  and so
  \(
  \mathsf{findOtherRed}\definedby \wlg(\mathsf{makeSecondRed},
  \fun{(\RED,\BLU)}(0,1))
  \).
  Equivariance is immediate since both core functions are vacuously
  equivariant.
\end{example}

\subsection{Propositional reasoning}
\subseclabel{propositional reasoning}

So far we have used the \WLOG{} construction to define functions. To
formalise mathematical proofs, we apply the Curry-Howard
correspondence and identify propositions with types, and their proofs
with terms. We could have also approached this differently and e.g.,
transferred the \WLOG{} construction to higher-order logic or
first-order logic, but this would require us to duplicate the
definitions for concepts such as group actions or equivariance for
proofs.

For concreteness, let us first borrow a running example of Schur's
inequality from \citet{DBLP:conf/tphol/Harrison09}, which states that
for all $x, y, z, t \ge 0$, we have
\[
  x^{t}(x - y)(x - z)+y^{t}(y - z)(y - x)+z^{t}(z - x)(z - y) \ge 0
\]
Since the inequality is symmetric in the three variables, the standard
proof assumes that \WLOG, $x \le y \le z$ and rearranges the
expression to an evidently non-negative form.

Let $\Prop \definedby \set{\Tru,\Fls}$ be the set of logical values.
Each \emph{proposition} $\Ctx \types \phi : \Prop$ determines a family
$\Ctx \types \Type{Proof} \phi$ such that for each $\ctx \in \Ctx$,
the fibre $(\Type{Proof} \phi) \ctx = \terminal = \set{\unitval}$ when
$\phi\ctx = \Tru$ and $(\Type{Proof} \phi) \ctx = \emptyset$ when
$\phi\ctx = \Fls$. As is conventional, we overload the notation and
identify propositions with their proofs, writing $\Ctx \types
\Type{Proof} \phi$ simply as $\Ctx \types \phi$. Then, any proof term
$\Ctx \types p : \phi$ is an explicit witness that $\phi$ holds at
every $\ctx \in \Ctx$.

\begin{example}
  \examplelabel{schur invariance}
  Schur's inequality is the proposition $(\x, \y, \z) :
  \preals^3 \types \Type{Schur}(x, y, z) : \Prop$ where
  \[
    \Type{Schur}(x, y,z) \definedby \forall t \ge 0.\
    x^{t}(x-y)(x-z)+y^{t}(y-z)(y-x)+z^{t}(z-x)(z-y) \ge 0
  \]
  Recall from \exampleref{symmetric-group-action} that $\Sym 3$ acts
  on $\preals^3$ by permutation. The proposition $\Type{Schur}$ is
  invariant under those permutations.
\end{example}

\begin{definition}
  A proposition $\Ctx \types \phi : \Prop$ is \emph{invariant} when it
  is equivariant as a term with respect to the trivial $\G$-action on
  $\Prop$, i.e. $\phi (\g \act \ctx) = \phi \ctx$ for all $\ctx \in
  \Ctx$.
\end{definition}

In addition to equivariant substitutions, like all equivariant terms,
invariant propositions are also preserved by logical connectives and
quantifiers. Each invariant proposition $\Ctx \types \phi : \Prop$
induces a $\Ctx$-indexed action on its proof family $\Ctx \types
\phi$. Moreover, given a $\Ctx$-indexed action $\A$, each invariant
proposition $\Ctx, \x : \A \types \phi : \Prop$ induces a
$\Ctx$-indexed action on the subfamily $\Ctx \types \set{\x : \A
\suchthat \phi} \subseteq \A$. For more details and proofs, again
see~\appref[C]{dependent types}.

\begin{example}
  Coming back to Schur's inequality, we can see that the \WLOG
  argument again fits the now familiar pattern:
  \[
    x_0, x_1, x_2 \in \preals
    \quad
    \explain\mapsto{sort using a\\[-5pt]permutation $\perm^{-1}$}
    \quad
    x_{\perm0} \le x_{\perm1} \le x_{\perm2}
    \quad
    \explain\mapsto{prove using \\[-5pt]additional assumption}
    \quad
    \Type{Schur}(x_{\perm0}, x_{\perm1}, x_{\perm2})
    \quad
    \explain\mapsto{$\Type{Schur}$ is invariant\\[-5pt]under permutations}
    \quad
    \Type{Schur}(x_0, x_1, x_2)
  \]
  Let us use the \WLOG construction to obtain a proof, i.e. a
  function $\DFun{(\x, \y, \z)}{\preals^3} \Type{Schur}(x, y, z)$.
  From \exampleref{schur invariance} we already have an $\Sym
  3$-action on $\preals^3$, and since $\Type{Schur}$ is an invariant
  proposition, we also get an indexed $\Sym 3$-action on $\preals^3
  \types \Type{Schur}$. For the canoniser, we take a stable sorting
  function $\sort : \preals^3 \to \Sym 3$ from \exampleref{sorting
  canoniser} so that $\Canon\sort$ is exactly the set of sorted
  triples. The last thing we need is the core function $f :
  \set{(x, y, z) : \preals^3 \suchthat x \le y \le z} \to \Schur(x, y,
  z)$, which we can define using the rearrangement in the standard proof.
\end{example}

The resulting function $\wlg(\sort, f)$ is equivariant since in the
case of propositions, core functions are always equivariant due to
proof irrelevance:

\begin{lemma}\lemmalabel{core functions to propositions invariance}
  Let $\Ctx, \x : \A \types \phi : \Prop$ be an invariant proposition,
  and $\Ctx \types \canoniser : \carrier\A \to \carrier\G$ a
  canoniser.  Every core function $\Ctx \types f : \Canon\canoniser
  \to \phi$ is equivariant.
\end{lemma}

Harrison suggests proving Schur's inequality using the following
tactic\footnotemark{}, which states that it
is sufficient to prove permutation invariant propositions only for
sorted triples~\cite[page 44]{DBLP:conf/tphol/Harrison09}:
\begin{align*}
  \inferrule{
    \forall \x,\y,\z .\ P(\x,\y,\z) \Rightarrow P(\y,\x,\z) \land P(\x,\z,\y)\\
    \forall \x,\y,\z .\ (\x \le \y) \land (\y \le \z) \Rightarrow P(\x,\y,\z)
  }{
    \forall \x,\y,\z .\ P(\x,\y,\z)
  }
\end{align*}
\footnotetext{Available here: \url{https://github.com/jrh13/hol-light/blob/e5236a5/int.ml\#L612}.}%
Let us see how this bespoke tactic relates to our general \WLOG{}
construction. In contrast to our approach, the symmetry group $\Sym 3$
is left implicit, as is the canoniser. Instead, only its set of
canonical elements $\set{(x, y, z) : \preals^3 \suchthat x \le y \le
z}$ is spelt out. Furthermore, the first assumption corresponds to an
indexed $\Sym 3$-action on the proposition $P$, and the second
assumption is exactly the core function.

The latter is evident, while for the former, first recall from
\exampleref{2048 action} that a group action is already uniquely
determined by the action of any set of generators. The symmetry
group $\Sym 3$ is generated by the two transpositions $(0
\leftrightarrow 1)$ and $(1 \leftrightarrow 2)$, hence the action of
$\Sym 3$ is determined by maps $(0 \leftrightarrow 1) \act {-}$ and
$(1 \leftrightarrow 2) \act {-}$. Spelling out the indices more
precisely, these two maps are:
\begin{align*}
  (0 \leftrightarrow 1) \iact{(x, y, z)} {-} &: P(x, y, z) \to P((0 \leftrightarrow 1) \act (x, y, z))
  &\text{equivalently}\quad
  &P(x, y, z) \to P(y, x, z)
  \\
  (1 \leftrightarrow 2) \iact{(x, y, z)} {-} &: P(x, y, z) \to P((1 \leftrightarrow 2) \act (x, y, z))
  &\text{equivalently}\quad
  &P(x, y, z) \to P(x, z, y)
\end{align*}
And this is exactly the contents of the first assumption.

As in \exampleref{2048 action}, we must ensure that such a definition
by generator cases respects all the equalities of the group. For
example in any $\Sym n$-action, $(0 \leftrightarrow 1) \act {-}$ must
be its own inverse (amongst other relations) since $(0 \leftrightarrow
1) \mul (0 \leftrightarrow 1) = \unit$. However, in the case of
propositions, once again proof irrelevance implies this condition.

Exposing the symmetric structure in proofs also allows us to use
existing results in group theory. Consider Harrison's proof in the
tactic, which explicitly lists all six
permutations~\cite{DBLP:conf/tphol/Harrison09}. For $\Sym 3$, this
number may still be manageable, but in general, the group~$\Sym n$ has
$n!$ elements, so when $n$ increases, the size of the proof script
quickly becomes prohibitively large. Also, such an approach precludes
a proof of a general $\wlg_n$ principle if $n$ is a variable, as one
cannot explicitly list all the $n!$ cases in the script. In contrast,
we can lean on the general group-theoretic result, whose proof is
independent of the concrete value of $n$, that $\Sym n$ is generated,
for example, by a single cycle $(0 \, 1 \, 2 \, \cdots \, n - 1)$ and
a single transposition of adjacent elements such as $(0
\leftrightarrow 1)$. So it suffices to prove two cases
to define any propositional $\Sym n$-action, yielding a single
parametric principle $\wlg_n$.

\subsection{Equivariant canonisers}
\subseclabel{equivariant canonisers}

In \subsecref{wlog-equivariance}, we saw that all \WLOG{} constructions
of a canoniser are equivariant iff the canoniser is strictly
equivariant (\theoremref{all core functions are partially
  equivariant}). This result is significant because it allows us to do
away completely with the equivariance of the core function, which
involves comparing every pair of symmetric canonical elements.
However, only the semi-regular actions can
have such canonisers~(\lemmaref{semi-regularity}). We now investigate a weaker
notion of equivariance. It is also characterised through equivariant
extension, but of \emph{stable} core functions. A core function $f$ is
stable when every \emph{stabilising} symmetry~$g$ of an
element---$g\act a = a$---also stabilises the image $f\,a$. This
necessary condition is natural and often easy to check.  Furthermore,
if we assume the axiom of choice, all actions admit equivariant
canonisers, making them a viable---albeit theoretical---class of canonisers.

Both $\canoniser (g \act
a)$ and $\canoniser a \mul \inv g$ canonise $g \act a$. Strictly
equivariant canonisers equate these two symmetries:
$\canoniser (g \act a) = \canoniser a \mul \inv g$.
Equivariant canonisers merely relate
them: $\canoniser (g \act a) \stabeq{} \canoniser a \mul \inv g$.

\begin{definition}
  \definitionlabel{indistinguishability}
  Take a $\G$-action $\A$. Then, for each $a \in \carrier\A$, we equip
  $\carrier\G$ with an \emph{indistinguishability} relation $g \stabeq
  a h \iff g \act a = h \act a$. Let $[g]_{\stabeq a}$ be the
  equivalence class of $g$ in $\carrier\G / \stabeq a$. Define an
  indexed $\G$-action $\A \types \G / \A$ with the fibre $(\G / \A) \, a
  \definedby \carrier\G / \stabeq a$ and with $\act[\G / \A]$ defined
  as:
  $
  g \iact[\G / \A] a{[h]_{\stabeq a}} \definedby [h \mul \inv g]_{\stabeq {g \act a}}
  $.
\end{definition}

The action is well-defined since $h_1 \stabeq a h_2$ implies $h_1 \mul
\inv g \stabeq{g \act a} h_2 \mul \inv g$. The relation $\stabeq a$ is
closely related to the \emph{stabiliser} $\stabilise a \definedby
\set{g \in \carrier\G \suchthat g \act a = a} = [\unit]_{\stabeq a}$, i.e. a subgroup of all
symmetries that preserve $a$. In particular, $g \stabeq a h$ is
equivalent to $(\inv h \mul g) \act a = a$, i.e. $\inv h \mul g \in
\stabilise a$.

\begin{remark}
The equivalence classes of the indistinguishability relation are
precisely the cosets of the stabiliser group. Therefore, we can recast the
celebrated Orbit-Stabiliser theorem as an \emph{equivariant} isomorphism between
the actions on orbits and stabiliser cosets.

Given a $\G$-action $\A$, the fibres of the dependent family of
\emph{orbits} contain all the symmetric elements of an index, i.e. $a
: \carrier\A \types \OrbitOf[\G]a \definedby \set{h \act a \suchthat h
\in \carrier\G}$. On orbits, we can define an indexed action as:
\[
  g \iact a (h \act a \in \OrbitOf a)
  \definedby
  h \act a
  =
  (h \mul \inv g) \act (g \act a)
  \in
  \OrbitOf{g \act a}
\]
Even though the action does not change the fibres, it is not
trivial, since it changes the indices.
\end{remark}

\begin{theorem}[Orbit-Stabiliser]
  \theoremlabel{orbit stabiliser}
  Let $\A$ be a $\G$-action. We have an equivariant isomorphism:
  \[
  a : \A \types \phi_a : (\G / \A) a \overset{\isomorphic}\equito \OrbitOf{a}
  \qquad
  \phi_a [h]_{\stabeq a} \definedby h \act a
  \qquad
  \inv \phi_a(h \act a) \definedby [h]_{\stabeq a}
  \]
\end{theorem}

The indexed action $\G / \A$ generalises $\inv\G$ and allows us to
consider a weaker property in which the equality $\canoniser\parent{g
\act[\A] a} = g \act[\inv\G] \canoniser a$ of strictly equivariant
canonisers is replaced by indistinguishability:

\begin{definition}
  A canoniser $\canoniser : \carrier\A \to \carrier\G$ is
  \emph{equivariant} when post-composing it with the quotient map
  yields an equivariant map $(\fun a[\canoniser\,a]_a) : \EFun
  a\A{(\carrier\G / \A) a}$, i.e. when $\canoniser (g \act[\A]
  a) \stabeq{g \act a} g \act[\inv\G] \canoniser a$.
\end{definition}

\begin{example}
  For a totally ordered set $P$, we can define the $\Sym 2$-action on
  $P \times P$ by $\swap\act\pair pq \definedby \pair qp$ and a
  canoniser $\canoniser : P \times P \to \carrier{\Sym2}$ such that
  $\canoniser \pair pq = \stet$ if $p \le q$, and $\canoniser \pair pq
  = \swap$ if $p > q$. Then $\canoniser$ is not strictly equivariant
  since for any $p \in P$, we have:
  \[
  \canoniser\parent{\swap \act \pair pp} = \canoniser\pair pp = \stet \neq \swap
  = \stet \mul {\inv\swap} = \swap\iact{\pair pp}\canoniser\pair pp
  \]
  It is, however, equivariant, since
  \(
  \canoniser\parent{\swap \act \pair pp} =
  \stet \stabeq{\pair pp} \swap =
  \swap\iact{\pair pp}\canoniser\pair pp
  \).
\end{example}

Note that for every equivariant function $f$ and every stabilising
element we have:
\[
\begin{tabular}[c]{@{}c@{}}
\myfigbox{
  $
  g \in \stabilise a
  \implies
  g \act f a \explain={equivariance} f(g \act a) \explain{=}{stabiliser} f a
  \implies
  g \in \stabilise {f \, a}
  $
}
\end{tabular}
\qquad\text{i.e.:}\qquad
\myfigbox{$
  \stabilise a \subseteq \stabilise {f \, a}
$}
\]
We thus arrive at the following condition on core functions:
\begin{definition}
  Take indexed $\G$-actions $\Ctx \types \A$ and $\Ctx, x : \A \types
  \B$, a subfamily $\carrier\Ctx \types X \subseteq \carrier\A$. A
  function $\carrier\Ctx \types f : \DFun{\x}{X}\carrier\B$
  is \emph{stable} when it preserves the stabiliser: $\stabilise x
  \subset \stabilise {f\,x}$ for all $x \in X\ctx$.
\end{definition}

\begin{example}
  Just as in \exampleref{strictly equivariant canonisers}, consider a
  non-total ordering $P$ and a stable sorting function, and the
  permutation sub-action on all tuples with distinct elements
  $\set{\vec p \in P^n \suchthat \forall i,j \in \Fin n. p_i \neq
  p_j}$. Then, we can define $\minimal : \carrier P^n \to \Fin n$ that
  finds the index of a minimal element by $\minimal \definedby \wlg(\sort,
  (\fun{\_}0))$.

  Now, for any two incomparable $p, q$, both $\pair pq$ and $\pair qp$
  are sorted, thus canonical. However:
  \[
    (0 \leftrightarrow 1) \act ((\fun{\_}0) \seq{p,q})
    = (0 \leftrightarrow 1) \act 0
    = 1 \neq 0 = (\fun{\_}0) \seq{q,p}
    = (\fun{\_}0) \parent{(0 \leftrightarrow 1) \act \seq{p,q}}
  \]
  and so $(\fun{\_}0)$ is not equivariant on $\Canon\sort$. It is,
  however, vacuously stable. Indeed, take a canonical $\vec p$ and a
  $\vec p$-stabilising permutation $\perm \in \stabilise{\vec p}$.
  Since all the elements in $\vec p$ are distinct, yet $\perm\act \vec
  p = \vec p$, we have that $\perm = \id$. Therefore $\perm \in
  \stabilise{f\vec p} = \stabilise 0$.
\end{example}

So, in general, stability is strictly weaker than equivariance.
Nonetheless, in the presence of an equivariant canoniser, stability
and equivariance coincide.

\begin{theorem}\theoremlabel{all stable core functions are equivariant}
  Let $\Ctx \types \A$ be an indexed $G$-action and let $\carrier\Ctx \types
  \canoniser : \carrier\A \to \carrier\G$ be an indexed canoniser.
  Then, $\canoniser$ is equivariant iff for every indexed $G$-action
  $\Ctx, x : \A \types \B$, every stable $\Ctx$-indexed core function
  $\Ctx \types f : (x : \Canon\canoniser) \to \carrier\B$
  is equivariant.
\end{theorem}

Unlike before, where only semi-regular actions admitted strictly
equivariant canonisers, here every action admits an equivariant
canoniser, although we may need to invoke the axiom of choice.

\begin{theorem}\theoremlabel{equivariant canoniser choice}
  Using the axiom of choice, every indexed $\G$-action has an equivariant canoniser.
\end{theorem}

To us, this use of the axiom of choice indicates that identifying
symmetry is a creative activity, rather than a mechanical one.

\section{Symmetric algebraic data types}
\seclabel{algebraic data types}

Let us return to the tree lookup function $\lookup : \DFun{\BoundV
t}{\Tree} \Fun{\NodeIn{\BoundV t}} \naturals$ from \exampleref{node
lookup}. Consider the symmetry in the two recursive cases
\[
  \lookup \Big(\NodeV \ell n r\Big) (\Lft \:: p) \definedby \lookup \, \ell \, p
  \qquad
  \lookup \Big(\NodeV \ell n r\Big) (\Rgt \:: p) \definedby \lookup \, r \, p
\]
and define the canoniser\label{goLeft canoniser first occurrence} on paths by $\goLeft \, \Nil =
\goLeft (\Lft \:: p) = \stet$ and $\goLeft (\Rgt \::
p) = \swap$. Then, we can define $\lookup$ as follows, with dependent
types ruling out inaccessible cases:
\[
\lookup \definedby \fun{\Big(\NodeV \ell n r\Big)} \wlg\Bigg(
  \goLeft,
    \begin{cases}
      \Nil &\mapsto n \\
      \Lft \:: p' &\mapsto \lookup \, \ell \, p'
    \end{cases}
\Bigg)
\]
However, note that $\lookup$ and the core function are mutually
recursive: $\lookup$ is defined in terms of the core function, which
then recursively calls $\lookup$ on descent. This recursion is
structural, but via interdependence through a higher-order function.

Languages such as Agda and Idris have an implicit structural recursion
and a termination checker that tries to detect size changes.  This
type of implicit mutual recursion, called through the canoniser and
group action, is typically beyond the termination checker's ability to
detect a size change. Some languages provide eliminators explicitly,
such as Rocq and Lean, generating an appropriate
recursion or induction principle, which the user can use to:
(1) define the action on the data type; (2) compose the recursion
principle with \WLOG{} reasoning; and (3) develop induction principles
for equivariant reasoning. In this section we will isolate an
additional piece of data in addition to the data type, in the form of
an action of the group on the data type's \emph{constructors}. From
this structure we derive (1)--(3). Furthermore, an implementation that
naively defines $\lookup$ as above is inefficient. It flips the tree
each time the path changes the step direction. Our theory suggests
implementation techniques that avoid this inefficiency.

We proceed as follows. We use the well-established framework of \emph{initial
algebras}. After a brief recall of functors and their initial
algebras~(\subsecref{initial algebras}), we extend them to the
symmetric setting~(\subsecref{initial equivariant algebras}) through
\emph{symmetry strengths} that describe actions in terms of data type
constructors. They package folds of equivariant algebras as
equivariant traversals. Next, we show how to reuse canonisers for
traversals over the same constructors (\subsecref{wlog-traversals}),
and how symmetry strengths avoid unnecessary data type traversals by
progressively acting only on a single layer of
constructors~(\subsecref{partial evaluation}). Finally, we combine
these results with the previous section (\secref{dependent types}),
adapting data types to the
indexed setting~(\subsecref{indexed data types}).

\subsection{Initial algebras}
\subseclabel{initial algebras}

In \emph{initial algebra semantics}, each data type signature is
specified by a \emph{structure functor}~$F : \Set \to \Set$. For
example, trees over natural numbers and paths in them from
\exampleref{tree example} are given by the following two functors,
which closely correspond to their usual data type definitions:
\begin{align*}
  \mathbf{data}\, \Tree &=
    \Leaf \Or \Node \, \Tree \, \naturals \, \Tree
  &
  \mathbf{data}\, \TreePath &=
    \Nil \Or \TreeLabel \:: \TreePath \\
  \struct[\Tree] X &\definedby \terminal \amalg X \times \naturals \times X
  &
  \struct[\TreePath] X &\definedby \terminal \amalg \TreeLabel \times X
  \intertext{
    Each functor $F$ has an evident component-wise map $F : (X \to Y) \to (F X
    \to F Y)$, for example:
  }
  \fmap[\Tree]\, f\, (\injection_1\unitval) &\definedby \injection_1\unitval &
  \fmap[\TreePath]\, f\, (\injection_1\unitval) &\definedby \injection_1\unitval &
   \\
  \fmap[\Tree]\, f\, (\injection_2\triple{\ell}n{r}) &\definedby \injection_2\triple{f\,\ell}n{f\,r} &
  \fmap[\TreePath]\, f\, (\injection_2\pair{d}p) &\definedby \injection_2\pair{d}{f\,p}
\end{align*}

An \emph{algebra}~$\pair\X f$ for a functor $F$ is a set $\X$ together
with a function $f : F \X \to \X$, which describes the evaluation of
constructors. For example, the node-counting algebra is the
$\struct[\Tree]$-algebra $\pair\naturals {c : (\terminal \amalg \naturals
\times \naturals \times \naturals) \to \naturals}$, given by
\(
  c (\injection_1\unitval) \definedby 0
\) and \(
  c (\injection_2\triple{m_1}\_{m_2}) \definedby 1 + m_1 + m_2 \). A
\emph{homomorphism} of $F$-algebras $h : \pair\X f \to \pair\Y g$ is a
function $h : \X \to \Y$ such that $h \circ f = g \circ \fmap h$.

We interpret data types by \emph{initial} algebras, denoted by
$\pair{\fix F}{(\roll : F \fix F \to \fix F)}$, whose defining
universal property is that for every other $F$-algebra $\pair\X f$,
there is a unique homomorphism $\fold f : \pair{\fix F}{\roll} \to
\pair\X f$. This homomorphism interprets structurally recursive
functions. For example, the initial $\struct[\Tree]$-algebra, which we
label by $\pair{\Tree}{\roll[\Tree]}$, satisfies:
\begin{align*}
  \roll[\Tree] \injection_1\unitval = \Leaf
  \qquad
  \roll[\Tree] \injection_2\triple{\ell}{n}{r} = \Node\,{\ell}\,{n}\,{r}
\end{align*}
When readability is more important than precision, we are going
to elide $\roll$, and identify e.g. $\injection_1\unitval$ with
$\Leaf$, and $\injection_2\triple{\ell}{n}{r}$ with
$\Node\,{\ell}\,{n}\,{r}$.

For a given $F$-algebra $\pair\X f$, the universal property of the
initial algebra becomes the recurrence:
\[
  \fold f\, \Leaf = f(\injection_1\unitval)
  \qquad
  \fold f\, (\Node\,{\ell}\,{n}\,{r}) = f(\injection_2\triple{\fold f\,\ell}n{\fold f\,r})
\]
Through the algebra $\pair{\naturals}{c}$, we can compute the number of nodes in a tree $t$ by $\fold c\, t$ since
\[
  \fold c\, \Leaf = 0
  \qquad
  \fold c\, (\Node\,{\ell}\,{n}\,{r}) = 1 + \fold c\, \ell + \fold c\, r
\]

\subsection{Initial equivariant algebras}
\subseclabel{initial equivariant algebras}

To transfer our framework to the context of data types, it is not
sufficient to equip the initial $F$-algebra with a $\G$-action $(\act)
: \G \to \fix F \to \fix F$, as this does not tell us what to do for
other $F$-algebras. Instead, we take a more refined construct that
describes the interaction of the group with the data type
constructors:

\begin{definition}
  Let $\G$ be a group. A \emph{$\G$-symmetry-strong functor}~$F =
  \pair{\carrier F}{\gstrength[\G][F]}$, or a \emph{$\G$-strong
  functor} for short, is a functor $\carrier F : \Set \to \Set$
  together with a \emph{$\G$-symmetry strength}~$\gstrength[\G][F]$,
  or \emph{$\G$-strength} for short, which is a natural transformation
  \(
    \gstrength[\G][F]_{\X} : (\carrier\G \times \carrier F X) \to \carrier F(\carrier\G \times X)
  \)
  satisfying
  \begin{align*}
    \gstrength_{\X}(\unit, u) &= \carrier F(\fun x\pair-{\unit}x)\, u \\
    \gstrength_{\X}(g\mul h, u) &=
    \carrier F (\fun{\triple{g'}{h'}{x}}\pair{g'\mul h'}x)
    (\gstrength_{G\times\X}(g,
    \gstrength_{\X}(h, u)))
  \end{align*}
  for all $u \in \carrier F\X$ and for all $g,h \in \carrier\G$.
\end{definition}

Note that unlike group actions, symmetry strengths also feature the
group in the codomain, which allows us to specify how to act on
particular sub-terms.

\begin{example}
  \examplelabel{tree strengths}
  We can imagine two different $\Sym 2$-actions on trees, one that
  recursively flips the whole tree, and one that flips only the top
  node, with corresponding strengths:
  \begin{align*}
    \swap \act
    \parent{
      \begin{tree diagram}
        \& \& 3 \\
        \& 2 \& \\
        1 \& \& 4
        \treeLink1322
        \treeLink2231
        \treeLink2233
    \end{tree diagram}}
    &=
    \parent{
      \begin{tree diagram}
        3 \& \& \\
        \& 2 \& \\
        4 \& \& 1
        \treeLink1122
        \treeLink2231
        \treeLink2233
    \end{tree diagram}}
    &
    \swap {\act}'
    \parent{
      \begin{tree diagram}
        \& \& 3 \\
        \& 2 \& \\
        1 \& \& 4
        \treeLink1322
        \treeLink2231
        \treeLink2233
    \end{tree diagram}}
    &=
    \parent{
      \begin{tree diagram}
        3 \& \& \\
        \& 2 \& \\
        1 \& \& 4
        \treeLink1122
        \treeLink2231
        \treeLink2233
    \end{tree diagram}}
    \\
    \gstrength\pair\swap{\Node\, \ell\, n\, r} &\definedby
    \Node\, {\pair \swap r}\,n\,{\pair \swap \ell}
    &
    {\gstrength}'\pair\swap{\Node\, \ell\, n\, r}
    &\definedby \Node\,{\pair \stet r}\,n\,{\pair \stet\ell}
  \end{align*}
  The first strength propagates $\swap$ to the subtrees,
  while the second strength propagates $\stet$.

  We only need to define the strength for $\swap$ since the first axiom
  forces it at $\pair\stet u$ to be $\fmap(\fun x \pair\stet
  x) u$, i.e.,
  \(
    \gstrength\pair\stet{\Node\, \ell\, n\, r} \definedby
    {\gstrength}'\pair\stet{\Node\, \ell\, n\, r}
    \definedby
    \Node {\pair\stet\ell} n {\pair\stet r}
  \).
  For leaves, we define both strengths to act trivially
  \(
    \gstrength\pair g \Leaf \definedby
    {\gstrength}'\pair g \Leaf
    \definedby \Leaf
  \).
\end{example}

\begin{remark}\remarklabel{cartesian vs symmetry strengths}
  The notion of a symmetry strength is related to, but different from the
  usual Cartesian or monoidal strength. Recall that a \emph{Cartesian
  strength} for a functor $F : \Set \to \Set$ is a natural
  transformation
  \(
    \strength[F]_{\Ctx, \X}  : \Ctx \times F\X \to F(\Ctx \times \X)
  \)
  satisfying
  \(
    \strength_{\terminal,
      \X}(\unitval, u) = \fmap(\fun x\pair-{\unitval}x)u
  \)
  for all $u \in F\X$, and
  \(
    \strength_{\Ctx\times\Dtx, \X}((\ctx, \dtx), u) =
    \fmap(\fun{\triple {\ctx'}{\dtx'}x}\pair{\pair {\ctx'}{\dtx'}}x)
    \,\parent{\strength_{\Ctx, \Dtx\times\X}(
    \ctx, \strength_{\Dtx, \X}(\dtx, u))}
    \) for all $\ctx \in \Ctx$, and $\dtx \in \Dtx$.
    (Cf.~\applemmaref[D.1]{Cartesian strengths are symmetric strengths} in the Appendix
    for the diagrams.)
  Every Cartesian strength is also a $\G$-strength, which leaves
  constructors as they are and propagates the same group element into
  the sub-terms. For example, for trees, it acts as:
  \[
    \strength_{\carrier\G, \X} \pair g \Leaf \definedby \Leaf
    \qquad
    \strength_{\carrier\G, \X} \pair g {\Node\,\ell\,n\,r}
    \definedby \Node\,{\pair g\ell}\,n\,{\pair g r}
  \]
  Symmetry strengths are an instance of the more general
  \emph{distributive laws} between a
  monad and a functor~\cite{beck:distributive-laws}. We use this
  perspective pervasively in our proofs (cf.~\appref[F]{cat
    dev}). Here, the monad is the \emph{free $\G$-action} monad, a
  special case of the \emph{Writer monad}, sometimes known as the \emph{complexity
    monad}~\cite{cenciarelli-moggi:a-syntactic-approach-to-modularity-in-denotational-semantics}.
  We use this monadic perspective in \subsecref{partial evaluation} and
  \subsecref{indexed data types}.
\end{remark}

We use $\G$-symmetry strengths to lift any action on $\X$ to an action
on $F\X$:
\begin{lemma}
  \lemmalabel{strength lifts actions}
  Let $\G$ be a group and $F$ a $\G$-strong functor. Then, for every
  $\G$-action $\A$, we have a $\G$-action $F\A$ given by:
  \[
  \carrier{F\A} \definedby \carrier F\,\carrier\A
  \qquad\smash{
    (\act[{F\A}]) : \carrier\G \times \carrier F\carrier\A \xto{\gstrength_{\carrier\A}}
    \carrier F(\carrier\G \times \carrier\A) \xto{\fmap (\act[\A])} \carrier F\carrier\A}
  \]
  Furthermore, if $h : \A \equito \B$ is equivariant, so is $\fmap h : F \A \equito F \B$.
\end{lemma}

\begin{example}
  The two tree symmetry strengths from \exampleref{tree strengths}
  induce the two expected actions: action induced by the first strength
  flips the node and propagates symmetry, while for the second strength, the action
  only flips the top node and does not propagate the symmetry further.
  Cartesian strengths induce the pointwise lifting of actions.
\end{example}

Symmetry-strong functors not only lift actions, but as we will see
shortly, also uniquely determine actions on their initial algebras.
Even more, the resulting algebra structure and the action are
compatible in the following sense:

\begin{definition}%
  \definitionlabel{equivariant F-algebra}
  Let $F$ be a $\G$-strong functor. An \emph{equivariant $F$-algebra}
  $\pair \A f$ consists of a $\G$-action $\A$ and an equivariant map
  $f : F \A \equito \A$, i.e., an $\carrier F$-algebra $\pair{\carrier\A}f$
  such that:
      \diagram{22}
  An \emph{equivariant algebra homomorphism} $h : \pair\A f \equito \pair \B
  g$ is simultaneously an $\carrier F$-homomorphism $h : \pair{\carrier\A}f \to
  \pair{\carrier\B}g$ that is also equivariant $h : \A \equito \B$.
\end{definition}

Recall that initial algebra semantics supports a modular account of
data types~\cite{swierstra:data-types-a-la-carte}: coproducts $F_1 \amalg
F_2$ of functors are given pointwise by $(F_1 \amalg F_2) X = F_1 X \amalg F_2
X$, and an algebra for this coproduct functor is given by an algebra
structure for each functor over the same carrier set. This modularity
extends to symmetry-strong functors and their equivariant algebras:
\begin{proposition}[modularity]\propositionlabel{modularity}
  Let $F_1, F_2 : \Set \to \Set$ be $\G$-strong functors with symmetry strengths
  $\gstrength_1, \gstrength_2$. The cotupled map
  \(
    \coseq{\strength_1, \strength_2}\pair g{\injection_ix}
    \definedby
    \injection_i\strength_i\pair g{x}
  \)
  is a $\G$-symmetry strength for the coproduct functor $\carrier F_1 \amalg \carrier F_2 : \Set \to \Set$.
  Moreover, given an action $\A$, an equivariant algebra $f : (F_1
  \amalg F_2) \A \equito \A$ for this strength amounts to a pair of
  equivariant algebra structures $f_i : F_i\A \equito \A$.
\end{proposition}

Using (cf.~\appref[D]{algebraic data types}) an
appropriate notion of parameterised initiality originally proposed by
\citeauthor{bird-paterson:generalised-folds-for-nested-datatypes}~\cite{%
  bird-paterson:generalised-folds-for-nested-datatypes},
the symmetry strength lets us equip the initial algebra with an action
that makes it the \emph{initial equivariant algebra}. Then, to define
an equivariant homomorphism $h : \fix F \equito \A$, we need to define
an equivariant $F$-algebra $g : F\A \equito \A$, possibly with
the \WLOG construction.

\begin{theorem}[equivariant initiality]
  \theoremlabel{representation for algebras}
  Let $F$ be a $\G$-strong functor and $\roll : \carrier F\fix
  \carrier F \to \fix \carrier F$ the initial algebra of $\carrier F$ in $\Set$.
  Then, there is a unique action $\fix F = \pair{\fix \carrier F}{(\act[\fix F])}$ satisfying
      \diagram{50}
  making $\pair{\fix F}\roll$ into an initial equivariant $F$-algebra, i.e., the initial $F$-algebra on $\G$-actions:
  for any equivariant $F$-algebra ${f : F\A \equito A}$, the unique
  $\carrier F$-homomorphism is equivariant: $\fold f : \fix F \equito
  \A$.
\end{theorem}

\begin{example}\examplelabel{simply typed tree path}
  \examplelabel{tree path}
  To see all the modular building blocks in action, define the path lookup function
  from \exampleref{tree example}. The mirroring symmetry of the type
  $\TreePath$ is captured by the strength:
  \begin{align*}
    \strength[\TreePath]\pair g\Nil &\definedby \Nil
    &
    \strength[\TreePath]\pair g{d \:: x} &\definedby (g \act d) \:: \pair g x
  \end{align*}
  and we can compute that the action given by
  \theoremref{representation for algebras}
  satisfies:
  \begin{align*}
    g \act (d \:: p) &=
    \big((\act) \compose (\id \times \roll)\big) \pair g{\injection_2\pair{d}{p}}
    \explain*={equivariant algebra}
    \big(\roll\compose {\struct[\TreePath](\act)} \compose \strength \big)\pair g{\injection_2\pair{d}{p}} \\&
    \explain+={strength def}
    \big(\roll\compose {\struct[\TreePath](\act)}\big)(\injection_2\pair{g \act d}{\pair g p}) =
    \roll (\injection_2\pair{g \act d}{g \act p}) =
    (g \act d) \:: g \act p
    \end{align*}
  A similar calculation for $\Nil$ shows we get the expected action
  for $\TreePath$, and a similar set of calculations shows we get the
  expected recursive action for $\Tree$ from the propagating strength
  from \exampleref{tree strengths}.
  We now construct $\lookup$ and simultaneously show it is
  equivariant as a map \(
    \TreePath \equito (\Tree \to \Maybe\naturals)
  \).
  By \theoremref{representation for
  algebras}, we can instead define an equivariant algebra:
  \[
    \lookupElim : \struct[\TreePath](\Tree \to \Maybe\naturals) \equito
    (\Tree \to \Maybe\naturals)
  \]
  Thanks to the Modularity \propositionref{modularity}, it suffices to define two
  equivariant algebra structures for each case of summand of
  $\struct[\TreePath] = \terminal \amalg \TreeLabel \times {-}$.

  The $\Nil$ summand's algebra structure
  \(
    \lookupElim_{\Nil} : \terminal \equito (\Tree \to \Maybe\naturals)
  \):
  \[
    \lookupElim_{\Nil}\unitval \, \Leaf \definedby \Nothing
    \qquad
    \lookupElim_{\Nil} \unitval \, {(\Node\, \ell\, n\, r)} \definedby \Just n
  \]
  To show that $\lookupElim_{\Nil}$ is equivariant, we refer to the fact that
  currying preserves equivariance, and calculate the behaviour of the uncurried
  version on the only interesting case:
  \[
    \lookupElim_{\Nil}\!\parent{\swap \act \pair\unitval{\Node\,\ell\,{n}\,r}}
    =
    \lookupElim_{\Nil}\!\pair\unitval{\Node\,\parent{\swap\act r}\,\!{n}\,\!\parent{\swap\act\ell}}
    =
    \Just n
    =
    \swap\act \Just n
    \]
  For $(\::)$, define the algebra
  \(
  \lookupElim_{(\::)} :
  \text{$\TreeLabel$} \times
    (\Tree \to \Maybe\naturals) \equito (\Tree \to
    \Maybe\naturals)
  \).
  Here, we use \WLOG reasoning, as stepping right is the same as
  stepping left on the flipped tree with the rest of the path flipped.
  Define the $\TreeLabel$-canoniser
  \(
  \canoniser' : \TreeLabel \equito \inv-{\Sym 2}
  \) that ensures canonical steps are left steps, i.e.
  \(
    \canoniser' \Lft = \stet
    \) and \(
    \canoniser' \Rgt = \swap
  \).
  Then, extend it to a canoniser~$\canoniser$ for the whole domain
  through projection via \lemmaref{projection canoniser}. Its
  canonical elements are the tuples $\seq{\Lft, f}$,
  where $f$ is the recursive call we make when traversing (necessarily left).
  Thus, we eliminated the symmetric cases in the following
  core function
  \(
  \lookLeft : \Canon\canoniser \to (\Tree \to \Maybe\naturals)
  \):
  \[
    \lookLeft\seq{\Lft, f} \definedby \begin{cases}
      \Leaf &\mapsto \Nothing \\
      \Node \, \ell \, n \, r &\mapsto  f \ell
    \end{cases}
  \]
  Now, we only need to assemble the pieces. As $\canoniser'$ is
  strictly equivariant, so is $\canoniser$ by \lemmaref{equivariance
  of projection canonisers}, and by \theoremref{all core functions
  are partially equivariant}, all core functions are equivariant,
  including $\lookLeft$. Thus $\lookupElim_{(\::)}$ is equivariant by
  \theoremref{wlog-construction}. By \propositionref{modularity}, so
  is $\lookupElim$, and by \theoremref{representation for
  algebras} so is $\lookup$.

  To see that $\lookup$ indeed matches the expected behaviour, let us
  compute its behaviour on the symmetric case, recalling that the
  exponential action conjugates, $(g \act f)(a) = g \act f(\inv g \act
  a)$:
  \begin{align*}
    &\lookup \, (\Rgt \:: p) \, (\Node \, \ell \, n \, r)
    =
    \big(\fold(\lookupElim) \, (\Rgt \:: p)\big) (\Node \, \ell \, n \, r) \\
    = {}
    &\big(\lookupElim \, (\injection_2\pair\Rgt{\lookup\, p})\big) (\Node \, \ell \, n \, r)
    =
    \big(\swap \act \lookLeft\pair\Lft{\swap \act \lookup\, p}\big) (\Node \, \ell \, n \, r) \\
    = {}
    &\swap \act \Big(\lookLeft\pair\Lft{\swap \act \lookup\, p}
        \big(\Node \, (\swap \act r) \, n \, (\swap \act \ell)\big)\Big) \\
    = {}
    &\swap \act \big( (\swap \act \lookup\, p)\,(\swap \act r) \big)
    =
    \swap \act \big( \swap \act (\lookup\, p \, (\swap \act \swap \act r)) \big)
    =
    \lookup \, p \, r
    \tag*{\qedhere}
  \end{align*}
\end{example}

\subsection{Data type traversals \WLOG{}}
\subseclabel{wlog-traversals}
The steps in \exampleref{simply typed tree path} are common, and we often
traverse a data type in different ways but with the same canoniser.
In this section, we present reusable abstractions that streamline these steps.

\begin{definition}
  Let $F$ be a $\G$-strong functor. A \emph{canoniser} for $F$ is a
  natural transformation $\canoniser_{\X} : \carrier F\X \to
  \carrier\G$. Each such canoniser induces the \emph{canonical
  elements functor}:
  \[
  \Canon\canoniser : \Action\G \to \Set
  \quad
  (\Canon\canoniser)\A \definedby \Canon[F\A]{\canoniser_{\carrier\A}} \definedby
  \set{
    (\canoniser_{\carrier\A}u)\act[F\A] u \suchthat u \in \carrier F\carrier\A
  }\quad
  (\Canon\canoniser)h : \fun u (\carrier Fh)u
  \]
\end{definition}

A canoniser $\canoniser$ thus induces a natural assignment of
canonisers $\canoniser_{\carrier\A} : \carrier F\carrier\A \to
\carrier \G$ for each lifted $\G$-action $F\A$.
The naturality condition means canonisers are determined by their
component at the singleton $\terminal$, i.e., by a canoniser for
$\carrier F\terminal$. The unique extension of such a canoniser
$\canoniser_{\terminal} : \carrier F\terminal \to \carrier\G$ to a
canoniser for $F$ is given by the following composition
$\canoniser_{\X} : \carrier F\X \xto{\carrier F\seq{}} \carrier
F\terminal \xto{\canoniser_{\terminal}} \carrier\G$.

\begin{example}\examplelabel{goLeft canoniser}
  The $\Sym 2$-strong functor $F_{\TreePath}$ has the
  canoniser (cf.~p.~\pageref{goLeft canoniser first occurrence})
  and canonical elements:
  \[
  \goLeft\,\Nil \definedby \stet \defines \goLeft (\Lft\:: p)
  \quad
  \goLeft (\Rgt\:: p) \definedby \swap
  \qquad
  \Canon[\A]\goLeft \definedby
  \set{\Nil, \Lft\:: a \suchthat a \in \carrier\A}
  \tag*{\qedhere}
  \]
\end{example}

We can now streamline the data we need to specify to traverse a
symmetric data type using an algebra, or to reason about such
traversals using symmetry:
\begin{definition}
  Let $F$ be a $\G$-strong functor and $\canoniser : \carrier F \to
  \carrier\G$ a canoniser for it. A \emph{$\canoniser$-core
  $F$-algebra} $\pair\A f$ consists of a $\G$-action $\A$ equipped
  with a function $f : (\Canon\canoniser)\A \to \carrier\A$. It is
  \emph{equivariant} when $f$ is equivariant $f :
  \parent{(\Canon\canoniser)\A \subset F\A} \equito \A$. A
  \emph{homomorphism} of equivariant $\canoniser$-core $F$-algebras $h
  : \pair\A f \to \pair \B g$ is an equivariant map $h : \A \equito
  \B$ satisfying: $h (f\, u) = g\, ((\Canon\canoniser) h\,u)$.
\end{definition}

We can rearrange the constructs in \exampleref{simply typed tree path}
using these concepts:
\begin{example}\examplelabel{simply typed lookup as core algebra}
  We have the following equivariant $\goLeft$-core $F_{\TreePath}$-algebra:
  \begin{gather*}
  \lookupCore : (\Canon\goLeft)(\Tree \to \Triv{\Maybe\naturals}) \to
  \carrier{\Tree \to \Triv{\Maybe\naturals}} = (\Tree\to\Maybe\naturals)
  \\
  \Nil \mapsto \begin{cases}
    \Leaf            &\mapsto \Nothing \\
    \Node \ell\,n\,r &\mapsto \Just n
  \end{cases}
  \qquad\quad
  \parent{
    \Lft\:: (\Tree\xto{\mathrm{rec} }\Maybe\naturals)
  } \mapsto \begin{cases}
    \Leaf &\mapsto \Nothing \\
    \Node\,\ell\,n\,r &\mapsto \mathrm{rec}\,\ell
  \end{cases}
  \tag*{\qedhere}
  \end{gather*}
\end{example}

These ingredients streamline the steps of \exampleref{simply typed tree path},
using the portmanteau \underline{f}old w\underline{log}:
\[
\myfigbox{$
\parent{\Canon[F\A]\canoniser_{\carrier\A} = (\Canon\canoniser)\A \xto f \carrier\A}
\mapsto
\parent{F\carrier \A \xto{\wlg(\canoniser_{\carrier\A}, f)} \carrier\A}
\mapsto
\parent{\flg[\A](\canoniser, f) :
  \fix\carrier F \xto{\fold \wlg(\canoniser_{\carrier\A}, f)} \carrier \A}
$}
\]
We chain \theoremref{wlog-construction} and the Equivariant Initiality \theoremref{representation for
  algebras}:
\begin{theorem}[core initiality]
  \theoremlabel{representation for core algebras}
  Let $F$ be a $\G$-strong functor, $\roll : \carrier F\fix
  \carrier F \to \fix \carrier F$ an initial algebra of $\carrier F$,
  and $\canoniser$ a canoniser for $F$. Then restricting $\roll$ to canonical elements
  makes $\fix F$ into the initial $\canoniser$-core $F$-algebra
  $\roll : (\Canon\canoniser)\fix F \equito \fix F$:
  for any equivariant $\canoniser$-core $F$-algebra ${f :
    (\Canon\canoniser)\A \equito A}$, the unique equivariant
  $\canoniser$-core $F$-algebra homomorphism
  into $f$ is:\/ $\flg(\canoniser, f) : \fix F \equito \A$.
\end{theorem}

This theorem implies, for example, that the $\lookup$ function from
\exampleref{simply typed lookup as core algebra} is the unique
equivariant $\goLeft$-core $F_{\TreePath}$-homomorphism given by
$\flg$.

\begin{example}\examplelabel{subtree projection}
  We reuse the canoniser to define the \emph{subtree projection} function
  \begin{gather*}
  \subtree \definedby \flg(\goLeft, \subtreeCore): \TreePath \equito (\Tree \to \Maybe\Tree)
    \\
  \subtreeCore : (\Canon\goLeft)(\Tree \to \Maybe\Tree) \equito (\Tree \to \Maybe\Tree)
  \\\!\!
  \Nil\mapsto (t \mapsto \Just\,t)
  \qquad
  (\Lft\:: \Tree\xto{\mathrm{rec}}\Maybe\Tree) \mapsto \begin{cases}
    \Leaf & \mapsto \Nothing \\
    \Node\,\ell\,n\,r & \mapsto \mathrm{rec}\,\ell
  \end{cases}
  \tag*{\qedhere}
  \end{gather*}
\end{example}

\subsection{Partially evaluating actions}
\subseclabel{partial evaluation}

Both $\lookup$ from \exampleref{simply typed tree path} and $\subtree$
from \exampleref{subtree projection} give us the result we want. But
implementations that naively follow our definitions
will repeatedly flip their arguments, and the
$\subtree$ implementation will also flip its result, leading to high time and space
complexity. Faithful to our motivation, we only demonstrate which
mathematical structures underlie a more efficient implementation, and
leave a systematic account of performance engineering to future work.

We use a general principle from modern partial evaluation folklore:
instead of performing an expensive operation repeatedly, axiomatise
its properties and design an efficient implementation for its free
algebra. Then evaluate the free algebra back into the desired
structure after partial evaluation. In our case, we shall build on top
of the \emph{free action} $\gast \X$ on a set $\X$, given by
$\carrier{\gast \X} = \carrier \G \times \X$ and $g \act[\gast \X]
\pair h x = \pair {g \mul h} x$. The free action lets us
accumulate the symmetries while leaving the input unchanged, and a
pair $\pair h x \in \carrier{\gast \X}$ captures a delayed evaluation
of $h \act[\A] x$ for some $\G$-action $\A$ with $\carrier{\A} = \X$.

In particular, consider $\gast \fix F$, the free $\G$-action on
the initial $F$-algebra. When at some point in a definition we want to
inspect $\pair g t$, we should not simply unroll $t$, as that ignores the
accumulated symmetries. We also do not want to compute $g \act t$
in implementations, since that is inefficient. What we do instead is use the symmetry
strength to act only on the first layer of the data type and
propagate the action into a delayed evaluation of a subtree
represented by the free-action monad:
\[
  \carrier{\gast \fix F}
  =
  \carrier \G \times \fix F
  \xto{\carrier\G \times \unroll}
  \carrier \G \times F \fix F
  \xto{\gstrength[F]_{\fix F}}
  F (\carrier \G \times \fix F)
  =
  F (\carrier{\gast \fix F})
\]
From \theoremref{representation for algebras}, we get $g \act t =
\fmap(\act)(\strength \pair g t)$ (ignoring unrolling), meaning that
the symmetry strength both gives us a symmetry-respecting pattern
matching of a data type, and represents a symmetry-accumulating implementation that
avoids inefficient repeated
transformations.

To see the construction in action, we can equivalently compute $\subtree \, p\, t$
with the free action:
\[
\begin{array}{@{}l@{}l@{}}
\subtree \, &p\, t =
\\&\mathrm{let}\, \pair gs = \subtreeAux\,p \pair
\stet t
\\&
\mathrel{\mathrm{in}} g\act s
\end{array}
\quad
\begin{array}{@{}l@{}}
\text{where: }\\
\quad\subtreeAux : \TreePath
\equito (\gast[\Sym 2] \Tree \to \gast[\Sym 2]\Maybe\Tree)\\
\quad \subtreeAux = \flg(\goLeft, \subtreeAuxCore)
\end{array}
\]
and $\subtreeAuxCore$
is an equivariant $\goLeft$-core
$F_{\TreePath}$-algebra over $\gast[\Sym 2] \Tree \to \gast[\Sym
2]\Maybe\Tree$:
\begin{gather*}
\subtreeAuxCore : \Canon\goLeft\parent{\gast[\Sym 2]\Tree \to \gast[\Sym 2]\Maybe\Tree}
\to \parent{\gast[\Sym 2]\Tree \to \gast[\Sym 2]\Maybe\Tree}\\
\begin{aligned}
\Nil &\mapsto \parent{\fun{\pair gt} \pair g{\Just t}} \\
\parent{
    \Lft\::  \parent{\begin{array}{@{}l@{}}
    \gast[\Sym 2] \Tree \xto{\mathrm{rec}}\\
    \qquad\gast[\Sym 2]\Maybe\Tree
  \end{array}}
}&\mapsto \parent{\fun{\pair gt} \begin{cases}
  \pair g \Nothing
  & (\strength \pair gt = \Leaf) \\
  \mathrm{rec}\pair{g_\ell}\ell  & (\strength \pair gt = \Node\, {\pair{g_\ell}\ell} \, n \, \pair{g_r}r)
  \end{cases}}
\end{aligned}
\end{gather*}
Whether the tree in the last case is flipped or not, applying the
strength ensures the correct sub-structure resides in the left
sub-tree and $g_\ell$ appropriately propagates the symmetry. More
importantly, an implementation of this design never flips the tree. For example, we have:
\begin{align*}
\subtreeAux&(\Rgt\::\Nil)\pair\stet{\Node\,\ell\,n\,r} \\&=
\swap\act\subtreeAuxCore(\Lft\::\subtreeAux (\swap \act \Nil))(\swap \act \pair\stet{\Node\,\ell\,n\,r})\\&=
\swap\act\subtreeAuxCore(\Lft\::\subtreeAux \Nil)\pair\swap{\Node\,\ell\,n\,r}\\&=
\swap\act\subtreeAux \Nil\pair\swap r=
\swap\act\pair\swap{\Just r} =
\pair{\stet}{\Just r}
\end{align*}
where in the reduction of $\subtreeAuxCore$, we used that $\strength
\pair\swap{\Node\,\ell\,n\,r} = \Node\,\pair\swap r\,n\,\pair\swap
\ell$. As we can see, the only evaluated actions were the free ones,
and the subtree $r$, for example, remained unflipped throughout the
argument.

There are further opportunities for optimisation. For example, instead
of stacking output symmetries $g \act -$ (in contrast to the input
ones that get absorbed by the free action) we could combine them into
a single symmetry as $g \act (h \act a) = (g \mul h) \act a$. In fact,
in this specific example, we can go even further and ignore the output
symmetries altogether. By the Core Initiality
\theoremref{representation for core algebras}, we
have $(\act[\Maybe\Tree])
\compose \subtreeAux\,p\pair gt = \subtree\,p(g \act[\Tree]t)$.
Furthermore, we can show that
$\projection_1\parent{\subtreeAux\,p\,\pair gt} = g$, hence
$\subtree\,p\,t = \projection_2\parent{\subtreeAux\,p\pair \stet t}$
and the output symmetry is never used, but is merely a reasoning
device for the correctness of the implementation.

\subsection{Indexed data types}
\subseclabel{indexed data types}

In \exampleref{simply typed tree path}, we fell back to simple types,
but thanks to the generality of the initial algebra semantics
approach, a dependently-typed development follows the same lines and
recovers the finer version from \exampleref{node lookup}.
Moreover, it uncovers nuances that degenerate in the simply-typed case. For
example, the notion of a symmetry strength must in fact be phrased in
terms of the \emph{free action} $\gast \X$ over a family. The carrier
of the free action
coincides with the Cartesian product $\carrier\G \times \X$ in the
simply-typed case, but differs in the indexed case.
As we pointed out in \remarkref{cartesian vs symmetry strengths},
viewing this product as
the free-action monad also emphasises the role of the symmetry strength
as a monad-functor distributive law.
Once we explicate
this structure, much of the development amounts to repeating the
simply-typed development with the adjective `indexed' peppered
throughout. \appref[A]{indexed adts} repeats the development, and we
describe the changes here. \figref{indexed-datatypes-summary}
summarises the difference between the simply typed theory and its
dependently typed counterpart as follows. Where appropriate, we also
describe the categorical concept.

    \begin{figure}%
      \begin{center}%
        \begingroup
\rowcolors{2}{white}{gray!10}
\setlength\tabcolsep{0pt}
{\small
\begin{tabular}{@{}@{}p{.26\textwidth}<{\centering}@{}p{.39\textwidth}<{\centering}@{}p{.35\textwidth}<{\centering}@{}@{}}\toprule
  \textbf{simply-typed concept} & \textbf{dependently-typed concept} & \textbf{categorical concept} \\\midrule
$\Set$ & $\Fam_{\carrier\Ctx}$ & category $\Cat{C}$\\
functor $F : \Set \to \Set$
       & functor $F : \Fam_{\carrier\Ctx} \to \Fam_{\carrier\Ctx}$
                            & functor $F : \Cat C \to \Cat C$\\
free $\G$-action monad
$\carrier PX \definedby \carrier\G\times X$
       & free $\Ctx$-indexed $\G$-action monad
         $\carrier P(\carrier\Ctx \types X) \definedby \carrier\Ctx \types \gast X$
                            & monad structure $P$ over $\carrier P : \Cat C \to \Cat C$
\\
{\centering $\G$-action\par
$\carrier\G \times \carrier\A
  \xto{\pair ga \mapsto g \act a} \carrier\A
  \vphantom{\xto{\iact{\inv g}}}
  $
}
       &{\centering$\Ctx$-indexed $\G$-action\par
         $\gast \carrier\A \xto{\pair ga \mapsto g \iact{\inv g \act \mctx} a}
           \carrier\A$
         }
                            & {\centering$P$-algebra\par
                              $P\carrier\A \to\vphantom{\xto{\iact{\inv g}}} \carrier\A$
                              }
                            \\
$\G$-strength $\strength : \carrier\G\times FX \to F(\carrier G\times X)$
       & $\Ctx$-indexed $\G$-strength
           $\strength : \gast (FX) \to F(\gast X)$
                            & monad-functor distributive law
                              $\strength : PF \to FP$
\\
functor $S\carrier\X \definedby \carrier\G \to \X$
       & functor $S(\carrier\Ctx \types \X) \definedby
                    \carrier\Ctx\types \lollybox\X$
                            & right-adjoint $\carrier P \leftadjointto E$
                            (cf.~\apptheoremref[F.4]{generalised parameterised initiality})
\\
parameterised traversal
$\carrier\G\times\fix F \xto{\pfold\parent{FX \xto f X}} X$
      & indexed parameterised traversal
$\gast\fix F \xto{\pfold\parent{FX \xto f X}} X$
                            & parameterised fold
                               $T\fix F \xto{\pfold\parent{FX \xto f X}} X$
\\
equivariant $F$-algebra
$f : F\A \equito \A$
     & indexed equivariant $F$-algebra
       $f : F\A \equito \A$
                            & $F$-algebra for lifted functor
                              $F : \Cat C^P \to \Cat C^P$
\\
canoniser $\canoniser : F \to \carrier\G$
      & indexed canoniser $\canoniser : F \to \carrier\G$
                            & 
\\
canonical elements
  $\Canon\canoniser : \Action\G \to \Set$
      & canonical elements
          $\Canon\canoniser : \Action\G_{\Ctx} \to \Fam_{\carrier\Ctx}$
          & canonical elements
     $\Canon\canoniser : \Cat{C}^P \to \Cat C$
\\\bottomrule
\end{tabular}}
\endgroup
        \caption{Summary of symmetric data type theory}%
        \figlabel{indexed-datatypes-summary}%
      \end{center}%
    \end{figure}%

Although the categorical theory does not capture all
concepts, e.g.~canonisers, it suffices to assign to almost every concept its
type. The theory includes indexed versions of the Modularity
\propositionref{modularity}, the Equivariant Initiality
\theoremref{representation for algebras} and Core Initiality
\theoremref{representation for core algebras}.

We generalise from the category $\Set$ to $\Fam_{\carrier\Ctx}$, the
category of $\carrier\Ctx$-indexed sets and vertical maps.  The main
technical device we use to generalise from $\Set$ to
$\Fam_{\carrier\Ctx}$ is the free $\Ctx$-indexed $\G$-action
adjunction between $\Fam_{\carrier\Ctx}$ and $\Action\G_{\Ctx}$---the
category of $\Ctx$-indexed $\G$-sets and vertical equivariant maps.
Its monadicity means we can capture its structure categorically, and
we use the existence of a right adjoint $S$ to the resulting monad $T$
over $\Fam_{\carrier\Ctx}$ in one of the proofs (cf.~\apptheoremref[F.4]{generalised parameterised initiality}).

For $\G$-action $\Ctx$, we define, for each
$\carrier\Ctx$-family $\X$, the \emph{free $\Ctx$-indexed $\G$-action over} $\X$:
\begin{gather*}
\carrier{\DPair{\mg}\G{\mg\act\X}}_{\ctx} \definedby
\DPair{\mg}{\carrier\G}{\X_{\inv \mg \act \ctx}}
\qquad
g\iact{\ctx}\pair{h}{x : \X_{\inv h\act \ctx}} \definedby
\pair{g\mul h}{x : \X_{\inv h \act \inv g \act g \act \ctx}}
\end{gather*}
We also define $\DFun{\mg}\G{\inv\mg\act\X}$,
the \emph{cofree $\Ctx$-indexed $\G$-action over} $\X$.
Its fibres are the dependent-function fibres
\(
\carrier{\DFun{\mg}\G{\inv\mg\act\X}}_{\ctx} \definedby
\DFun{\mg}{\carrier\G}{\X_{\mg \act \ctx}}
\), defined by:
\[
g\iact{\ctx}\parent{f : \DFun{\g[3]}{\carrier\G}\X_{\g[3]\act\ctx}}
\definedby\fun h f(h\mul g) \in \DFun{\mg[2]}{\carrier\G}\X_{\mg[2]\act g\act\ctx}
\]
They are the two adjoints to
the forgetful functor from $\Ctx$-indexed actions to $\Ctx$-indexed families:
\begin{theorem}\theoremlabel{descent}
  For a $\G$-action $\Ctx$ and family $\carrier\Ctx \types \X$, the
  map $\Ctx \types \fun x\pair\unit x : \X \to
  \carrier{\DPair\mg\G(\mg\act\X)}$ is the universal $\Ctx$-indexed
  $\G$-action over $\X$.  The map
  $\carrier\Ctx\types\eval\pair{-}\unit :
  \carrier{\DFun\mg\G{\inv\mg\act\X}} \to \X$ is the universal
  $\Ctx$-indexed $\G$-action under $\X$. The resulting two adjunctions
  are (co)monadic.
\end{theorem}
This theorem generalises the non-indexed case, where
$\carrier\G\times\X$ has the structure of the free $\G$-action over
$\X$.  The existence of the adjoints and their (co)monadicity follow
from abstract reasons. Let: $\oint F$ be the Grothendieck
construction for a presheaf $F : \opposite{\mathbb C} \to \Set$;
$\abs{\mathcal C}$ be the discrete category underlying a category
$\mathcal C$, and $\hat{\mathbb C}$ the presheaf category of $\mathbb
C$. Consider a $\G$-action~$\Ctx$ as a presheaf over the category with one
object and morphisms the $\G$-elements. We then have the following two equivalences. First, $\Ctx$-indexed actions
are equivalent to presheaves over $\oint\Ctx$
\(
\Action\G_{\Ctx}\simeq \widehat{\oint\Ctx}
\). Forgetting the functorial action then recovers the category of
$\carrier\Ctx$-indexed families, a presheaf category itself:
\(
\Fam_{\carrier\Ctx} \simeq \widehat{\oint{\carrier\Ctx}} = \mspace{-5mu}\widehat{\ {\abs{\oint\Ctx}}\ }
\).
The forgetful functor $\Action\G_{\Ctx} \to \Fam_{\carrier\Ctx}$
arises as precomposition by $J: {\oint\Ctx} \from \abs{{\oint\Ctx}} $.
Consequently, the two adjoints are given by (co)end
formulae, and their monadicity follows by surjectivity of this
functor~\cite[Prop.~1.1]{fiore-menni:reflective-kleisli-subcategories}.
This situation is exploited by
\citeauthor{fiore-szamozvancev:soas-theory} to account for
mathematical and computational representations of abstract syntax with
binding and substitution~\cite{fiore-szamozvancev:formal-soas,%
  fiore-szamozvancev:soas-theory,%
  szamozvancev:thesis}.
One can use the coend formula to calculate the explicit descriptions
we gave the two adjoints. We show directly
(cf.~\appref[E]{indexed data types}) that these descriptions are the
adjoints, and for completeness, we also show monadicity and comonadicity using
Beck's theorem. We then have the situation in \figref{marcelo situation}.
Therefore, $P$ is a monad, $E$ is a comonad, and $P \leftadjointto E$.
Moreover, $P^2 \definedby P\compose P \leftadjointto E^2$.
\begin{figure}
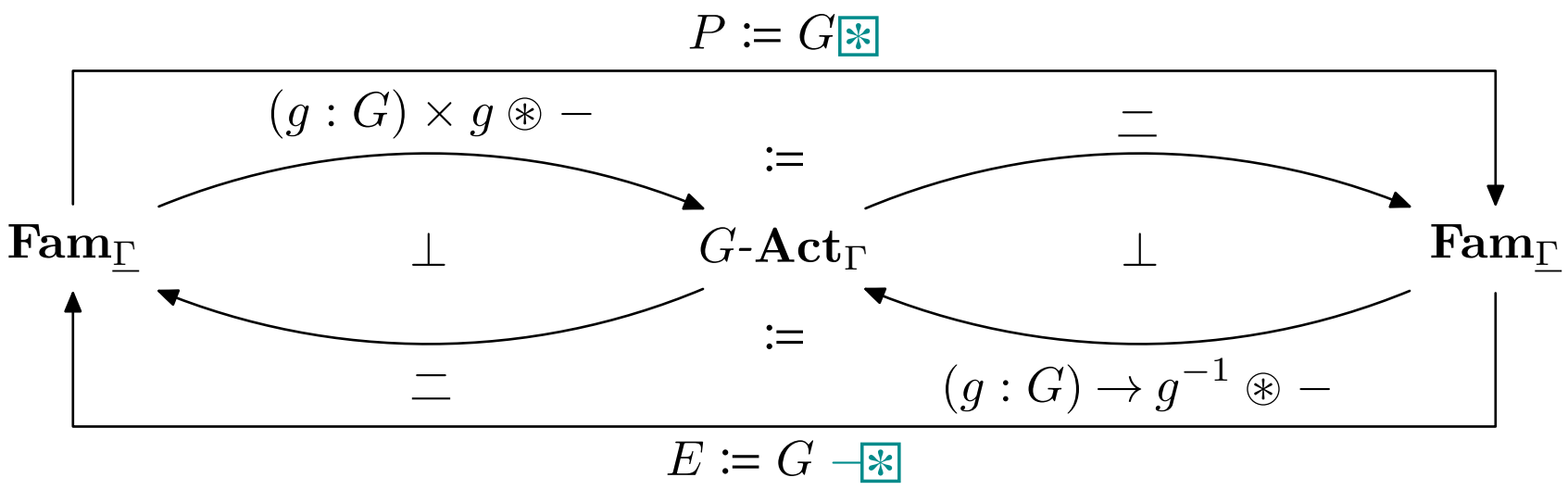

  \diagram{30}
  \caption{Categorical situation for indexed $\G$-action}\figlabel{marcelo situation}
\end{figure}

\section{Conclusion}
\seclabel{conclusion}

\subsection{Related work}
\subseclabel{related work}

\paragraph{\wLOG in theorem provers}

Given that \WLOG is a common phrase in mathematical proofs, it comes
as no surprise that several major theorem provers feature a similarly
named concept. What is more compelling is that the treatment is far
from uniform, pointing toward a missing uniform foundation. Our
work provides a uniform foundation for symmetric uses of \WLOG{}.

For example, the Rocq library, SSReflect~\cite{ssreflect}, Isabelle/HOL's Archive of Formal Proofs~\cite{unruh:wlog-archive-of-formal-proofs} and the
Lean~4 Mathlib~\cite{lean-mathlib}, define tactics named
$\mathtt{wlog}$, which cut in an additional statement~$\psi$ when proving
$\phi$. Rocq's tactic\footnote{%
  Rocq Reference Manual:
  \tiny\url{https://rocq-prover.org/doc/V9.2.0/refman/proof-engine/ssreflect-proof-language.html\#rocq:tacn.wlog}%
}
creates an additional subgoal $(\psi \Rightarrow \phi) \Rightarrow
\phi$.  Isabelle/HOL's tactic similarly adds assumptions \WLOG{}, and
further accounts for generalisation of variables, explicitly
documenting symmetric arguments as the motivation.
Lean's tactic\footnote{%
  Mathlib4 documentation:
  \tiny\url{https://leanprover-community.github.io/mathlib4_docs/Mathlib/Tactic/WLOG.html}
}
additionally assumes $\lnot \psi$, adding a subgoal $(\psi \Rightarrow
\phi) \Rightarrow \lnot \psi \Rightarrow \phi$. In all three tactics, the
additional subgoal (e.g., $\psi$ in Rocq and Lean) can be arbitrary and not tied to any
particular symmetry, so it could be worth investigating how many of
its uses correspond to the assumption of canonicity in our framework.

Agda Stdlib defines\footnote{%
  Agda Stdlib:
  \tiny\url{https://github.com/agda/agda-stdlib/blob/c2abce1b8e72ccae49f73791d0996eaf23674357/src/Relation/Binary/Consequences.agda\#L197}
}
$\mathtt{wlog} : (\forall (a\, b : A). a \, R \, b \Rightarrow Q \, a
\, b) \Rightarrow (\forall a\, b. Q \, a \, b )$ for any total
relation~$R$ and a symmetric predicate~$Q$. Here, we have a swapping
$\Sym 2$-action on the domain $A \times A$, while $Q$ provides an
appropriate indexed action on the codomain. The
totality of $R$
encodes the canoniser
implicitly, and the assumption exactly
corresponds to our core function.
\citet{dijkstra:wlog} uses a similar setting to criticise \WLOG arguments
for not explicating the underlying symmetry.

A richer collection of symmetries can be found in
\citepos{DBLP:conf/tphol/Harrison09} seminal paper on \WLOG tactics in
HOL. In addition to the tactic discussed in \subsecref{propositional
reasoning}, Harrison develops four additional tactics for \WLOG{}
reasoning in geometry, using a bank of results about invariance and
equivariance of various operations under translations and orthonormal
transformations. Three tactics are immediate special cases of our
principle, while the fourth one is a reduction from a predicate on
points of the form $(\x,\y,0) \in \reals^3$ to a predicate on
$\reals^2$. \citet{schmoetten-palmer-fleuriot:formalising-geometric-axioms-for-minkowski-spacetime-and-without-loss-of-generality-theorems,%
schmoetten-palmer-fleuriot:towards-formalising-schutz-axioms-for-minkowski-spacetime-in-isabelle-hol%
} use Harrison's approach and prove further \WLOG{} lemmata when
formalising Schutz's axioms for Minkowski
spacetime~\cite{schutz:independent-axioms-for-Minkowski-space-time}.

In summary, our work does not account for all
previous principles named $\mathtt{wlog}$. Instead, we offer a crisp
criterion by which one can distinguish symmetric \WLOG{} arguments.

\paragraph{Nominal techniques}

Nominal techniques are a close topic in programming languages, an
established approach to treating bound variables, in which
$\alpha$-equivalence is defined with groups of permutations of atoms acting on terms built from such
atoms~\cite{DBLP:journals/iandc/Pitts03,pitts:nominal-sets-book,DBLP:journals/siglog/Pitts16}.
Our approach borrows from, builds upon, and is analogous to these
nominal techniques. Like nominal techniques, we make pervasive use of, and
rely on, equivariance and related concepts~\cite{pitts:nominal-sets-book}.
Our use of symmetry strengths is similar to \citepos{blanchette-et-al:bindings-as-bounded-natural-functors} use of bounded natural functors.
The proof obligations we
require to establish the existence of recursively defined functions
and their compatibility with the relevant operators is also similar to
the recursors and induction principles in nominal techniques~\cite{%
  gabbay-pitts:a-new-approach-to-abstract-syntax-with-variable-binding,%
  norrish:recursive-function-definition-for-types-with-binders,%
  pitts:alpha-structural-recursion-and-induction,%
  urban-berghofer:a-recursion-combinator-for-nominal-datatypes-implemented-in-isabelle-hol,%
  blanchette-et-al:bindings-as-bounded-natural-functors,%
  popescu:nominal-recursors-as-epi-recursors,%
  popescu-gunter:recursion-principles-for-syntax-with-bindings-and-substitution%
}, and so on. Nominal techniques are typically used
with the group of finitely-supported permutations of names, but not always. The
finite-support requirement is sometimes relaxed~\cite{blanchette-et-al:bindings-as-bounded-natural-functors,
dowek-gabbay-mulligan:permissive-nominal-terms-and-their-unification,
cheney:nominal-logic-programming}.
The names we permute may have additional structure~\cite{bojanczyk-klin-lasota:automata-with-group-actions}.
Nominal metatheory typically employs \WLOG{} reasoning pervasively, cf., e.g., \citepos{cheney:nominal-logic-programming} thesis.

Unlike nominal techniques, our recursion principles do not solely
account for binders and $\alpha$-equivalence. They do require the
additional structure of canonisers. Whereas in nominal techniques one
stays equivariant throughout, in symmetric programming there is no
such requirement and outside the definition of a \WLOG{} construction,
code or reasoning may violate equivariance. We want to revisit this
comparison after we develop symmetric programming and reasoning
further. An even more interesting direction is to combine symmetric
programming and nominal techniques, pervasively maintaining nominal
equivariance, but potentially violating equivariance for non-nominal
symmetries.  Mechanically representing and reasoning about nominal data
is a wide area of research~\cite{%
  poplmark1,%
  poplmark2,%
  schafer-et-al:autosubst,%
  aydemir-et-al:lngen,%
  urban:nominal-isabelle/hol,%
  forster-stark:coq-a-la-carte,%
  kathrin-stark:thesis,%
  stark-et-al:autosubst-2,%
  pientka:beluga,%
  gacek:thesis,%
  gacek-et-al:abella-journal,%
  sewell-et-al:ott,%
  allais-et-al:type-and-scope-safe-syntax-journal,%
  gabbay-pitts:a-new-approach-to-abstract-syntax-with-variable-binding,%
  chlipala:parametric-hoas,%
  chargueraud:locally-nameless,%
  weirich-et-al:binders-unbound,%
  poulsen-et-al:definitional-interpreters,%
  cheney:towards-names,%
  ghani-et-al:cyclic-structures,%
  hamana:initial-semantics-for-cyclic-sharing,%
  polonowski:automatically-generated-syntaxes,%
  keuchel-et-al:needle-and-knot,%
  keuchel-jeuring:generic-conversions,%
  keuchel:msc-thesis,%
  benton-et-al:strongly-typed-terms,%
  lee-et-al:gmeta,%
  erdi:well-scoped-syntax,%
  copello-et-al:variable-convention,%
  morris-et-al:regular-tree-types,%
  hofmann:semantical-analysis-of-ho-abstract-syntax,%
  pfenning-elliott:hoas,%
  altenkirch-reus,%
  blanchette-et-al:bindings-as-bounded-natural-functors}.
While there are first steps towards nominal
type
theories~\cite{aydemir-bohannon-weirich:nominal-reasoning-techniques-in-coq,pitts-matthiesen-derikx:a-dep-type-theory-with-abstractable-names},
we still lack an implementation of a nominal dependently-typed system.

\paragraph{Relational parametricity.}
If we view each symmetry $g \in \G$ as a relation $a \equiv_g b$ given
by $g\act a = b$, then equivariance becomes relational preservation:
$a \equiv_g b \implies f\,a \equiv_g f\,b$. This perspective suggests
some connections with logical
relations~\cite{plotkin:lambda-definability-and-logical-relations} and
with relational
parametricity~\cite{reynolds:types-abstraciton-and-parametric-polymorphism}.
A possible connection is the work of \citet{DBLP:conf/popl/Atkey14} on
using parametricity and Noether's theorem to derive invariants of
physical systems. A shared theme is identifying the group actions that
describe a system's symmetries.  However, applying Noether's theorem
requires a notion of differentiation, which is not necessarily present in our
general setting. One direction for future work is to investigate which
of our action combinators can be translated to smooth-manifold
settings, and whether they can serve as compositional operators for
describing physical systems. It may be fruitful to explore other
connections. For example, there might be some proofs for free
theorems~\cite{wadler:theorems-for-free} that appeal to parametricity
through relations defined via symmetry. There might be symmetric
invariants for restricted programming languages (e.g., with nominal
features) provable via logical relations.

\paragraph{Symmetry in programming languages and type theory}
A more recent development on symmetries in programming is a Hoare
logic for symmetry properties by~\citet{10.1145/3763080}. It features
triples that relate the symmetries of preconditions to those of
postconditions. In particular, a triple of the form $(\G,\A) C
(\H,\B)_\phi$ uses a $\G$-action $\A$ to capture symmetries of input
states and an $\H$-action $\B$ to capture symmetries of output states
after executing the program $C$. The program $C$ is required to be
equivariant with respect to a group homomorphism $\phi : \G \to \H$.
More precisely:
\(
  \forall g \in \G, a \in \carrier\A. C(g \act[\A] a) = \phi(g) \act[\B] C(a)
\).
Beyond general action constructions such as (non-dependent) products
and pre-/post-composition, the paper develops a forward-reasoning
system that simultaneously derives an action on the output and proves
the equivariance of $C$. As we move from mathematical foundations
toward more practical symmetric programming, we
may adapt these rules to ease the construction of the group
actions required by the \WLOG construction.

\paragraph{Symmetry elsewhere in computer science}

Computer science is vast, and group theory is pervasive, so there are
many points of contact between the two. Rather than attempting a
comprehensive review, we highlight a few representative areas and
point to recent surveys for further reading.

In computer algebra and computational group theory, groups are both
the object of study and a computational resource, shaping algorithm
design and complexity analysis in areas such as permutation-group
algorithms and constructive
recognition~\citep{Seress_2003,Holt2005HandbookOC}. The
concept of \emph{canonical labelling} from
nauty/Traces~\cite{MCKAY201494}, which finds the vertex permutation
mapping a graph into its representative element w.r.t. graph
isomorphism, exactly matches our canonisers. It is worth investigating
how many algorithms building on it fit into our \WLOG pattern.

In constraint programming and satisfiability, symmetries have long
been a central concern. A large body of work studies symmetry-breaking
techniques, which add constraints that rule out symmetric solutions
while preserving
satisfiability~\citep{DBLP:reference/fai/GentPP06,DBLP:conf/aaai/Walsh12,janota2025breakingsymmetriesquantifiedgraph}.
In machine learning, symmetry appears both as an
architectural prior (equivariance/invariance) and as an analysis
tool. In computer vision, it supports tasks such as
recognition, reconstruction, and compression of geometric
data~\citep{basheer2025symmetry,zhao2025symmetryneuralnetworkparameter,cornish2025stochasticneuralnetworksymmetrisation}.

Similarly, model checking techniques and tools exploit symmetry to
mitigate the state-space explosion problem~\cite{%
  clarke-jha:symmetry-and-induction-model-checking,%
  donaldson:thesis,%
  clarke2018handbook,%
  miller-donaldson-calder:symmetry-in-temporal-logic-model-checking,%
  sistla-godefroid:symmetry-and-reduced-symmetry-in-model-checking%
}. Model checkers typically try to compute the relevant symmetries~\citep[e.g.][]{
  wahl-d'silva:a-lazy-approach-to-symmetry-reduction,%
  chu-jaffar:a-complete-method-for-symmetry-reduction-in-safety-verification,%
  leuschel-et-al:symmetry-reduction-for-B-by-permutation-flooding,%
  wahl-damm-hermanns:adaptive-symmetry-reduction,%
  bosnacki-dams-holenderski:symmetric-spin%
},
and in the general case this computational task is equivalent to the
graph isomorphism problem~\cite{jha:thesis}, whose precise complexity
class is still open.

\subsection{Future work}

On the foundational side, it may be interesting to explore richer data type constructions, such
as: co-data and co-recursion via terminal co-algebras; and universes.
The theory of distributive laws also applies to $F$-coalgebras,
suggesting there is a similar theory of coinductive
data-types. Universes offer various directions, depending on the
symmetries one may want to equip the inhabitants of the universe.
One option is to equip a universe of types with the trivial action,
as we did with the universe of propositions~(\subsecref{propositional reasoning}).

A richer direction involves a universe $\mathcal U_G$ of $\G$-actions, for some or
each group $\G$.  Here each value of the universe is a code for the
underlying carrier and the $\G$-action. This universe supports the
various type constructors using the usual type-interpretation in the
category of $\G$-actions.  While we can equip this universe with a
trivial $\G$-action, we can also equip it with other actions.

For example, given a symmetry $g \in \underline G$ and an inhabitant  $A \in \mathcal U_G$ with
$\sem{A} = (\underline A, \circledast)$, define $g
\circledast_{\mathcal U_G} A$ as the action with
the carrier unchanged $\underline{g\circledast A} \definedby \underline A$
which acts through conjugation $h \circledast_{g \circledast A} x \definedby
(g^{-1} \ast h \ast g) \circledast x$.  These definitions give
a map $\carrier\G \times \mathcal U_G \to \mathcal U_G$, and moreover this
map is a $G$-action over $\mathcal U_G$.  So in this case there is
some room to explore symmetries on a universe of $G$-actions and
potential to employ symmetric programming/reasoning here. In another direction the
equivariant initial algebra theorem may internalise to an equivariant
function on codes.

In another kind of universe, each inhabitant describes a group $\G$
and a code in a universe $\mathcal U_{[\![G]\!]}$ of $\G$-actions. One
can then similarly look for useful actions and symmetric
programming/reasoning there. Another direction in this area would be
to explore codes for homomorphisms between codes for $G$-actions and
actions and corresponding transport functions between their decoded
$G$-actions and actions. This direction may open up
questions regarding coherence of the resulting type theory.

That said, our development does not need a universe, it is merely
compatible with these various choices, and we plan to explore them in
future work.

On the practical side, we want to refine the developed modular mathematical
abstractions into programming constructs,
guided by more involved symmetric programming and reasoning
examples. In the first instance, we want to complete the full set of
operations on self-balancing binary search
trees~\cite{mcbride:neighbours-in-order}.  Another promising example
is \citepos{escardo:signed-digit-numerals} representation of exact
reals using infinite streams of signed-digit numerals. Operations on
these streams and proofs about these operations contain multiple
symmetric cases, including swapping the roles of symmetric variables,
and reflecting the digits around $0$.  Computable representations of
the constructive Dedekind reals~\cite{kaucher:interval-analysis-in-the-extended-interval-space-IR,bauer-taylor:dedekind-reals-in-abstract-stone-duality,bauer:efficient-computation-with-dedekind-reals} employ symmetric operations on
the cut's lower and upper sets, with symmetric programming potential.

It may be worth exploring
symmetric programming and reasoning for model checking. Since
symmetric programming exposes symmetric program paths explicitly, it
may alleviate the need to automatically detect symmetry when checking
symmetric programs. A more interesting direction is whether a
symmetry-detecting model checker can suggest refactoring opportunities
that use symmetry more explicitly. Another direction involves using
symmetric programming when implementing model checkers or constraint
solvers.

In addition to being interesting in their own right, such examples should guide
us in developing abstractions for performance engineering, as well as on how to most
ergonomically present groups, actions, canonisers, core functions, and
equivariance to the programmer. For example, immediate canonisers
(see~\propositionref{immediate canonical is kernel}) could admit an
interface similar to \emph{lazy
constructors}~\cite{DBLP:journals/pacmpl/LorenzenLSL25}, where we
extend a data type definition with non-canonical constructors together
with their non-trivial canonising symmetries, but list only canonical
cases in pattern matching, since the non-canonical ones are covered by
the \WLOG construction.

We already explored the landscape with two experiments
in Idris and Lean. Having a clean mathematical resolution
of issues that surfaced, such as termination and efficiency, we can
now return to both the programming language and the
proof assistant: not to increase confidence in the
results' correctness, but to develop a library of reusable
results and lay a foundation for future implementation and mechanisation.

\vfill{}

\pagebreak
\begin{acks}
  Supported by a Royal Society University Research Fellowship, an Advanced
  Research and Invention Agency (ARIA) Safeguarded AI grant
  Qbs4Safety, and an Alan Turing Institute Seed Funding grant.
  This material is partly based upon work supported by the Air Force
  Office of Scientific Research under award number FA9550-21-1-0024.

  We are grateful to the anonymous referees and
  the following individuals for useful comments,
  suggestions, and pointers:
Guillaume X.~Allais,
Andrej Bauer,
Greg Brown,
James Cheney,
Rob Cornish,
Pierre-Evariste Dagand,
Marcelo P.~Fiore,
Nicola Gambino,
Jeremy Gibbons,
Justin Hsu,
Martin E.~Hyland,
Andr\'as Kov\'acs,
Neel Krishnaswami,
Sam Lindley,
Justus Matthiesen,
Amaury Mazoyer,
James McKinna,
Georg Moser,
Sean K.~Moss,
Bruno Oliveira,
Andrew M.~Pitts,
Greg Restall,
Tom Schrijvers,
Alex Simpson,
Sam Staton,
Tina Šfiligoj,
Armin Walch,
Robert Wright,
Nick Wu,
Jeremy D.~Yallop, and
Nejc Zajc.

The second author
used Anthropic's Claude models to obtain feedback on the prose and to
proofread drafts for errors.
\paragraph{Copyright addendum.}
  For the purpose of open access, the authors have applied a Creative
  Commons attribution (CC BY) licence to any Author Accepted
  Manuscript version arising.
\end{acks}

\bibliographystyle{ACM-Reference-Format}

\bibliography{ref}


\begin{thebibliography}{98}


\ifx \showCODEN    \undefined \def \showCODEN     #1{\unskip}     \fi
\ifx \showDOI      \undefined \def \showDOI       #1{#1}\fi
\ifx \showISBNx    \undefined \def \showISBNx     #1{\unskip}     \fi
\ifx \showISBNxiii \undefined \def \showISBNxiii  #1{\unskip}     \fi
\ifx \showISSN     \undefined \def \showISSN      #1{\unskip}     \fi
\ifx \showLCCN     \undefined \def \showLCCN      #1{\unskip}     \fi
\ifx \shownote     \undefined \def \shownote      #1{#1}          \fi
\ifx \showarticletitle \undefined \def \showarticletitle #1{#1}   \fi
\ifx \showURL      \undefined \def \showURL       {\relax}        \fi
\providecommand\bibfield[2]{#2}
\providecommand\bibinfo[2]{#2}
\providecommand\natexlab[1]{#1}
\providecommand\showeprint[2][]{arXiv:#2}

\bibitem[Abel et~al\mbox{.}(2019)]%
        {poplmark2}
\bibfield{author}{\bibinfo{person}{Andreas Abel}, \bibinfo{person}{Guillaume
  Allais}, \bibinfo{person}{Aliya Hameer}, \bibinfo{person}{Brigitte Pientka},
  \bibinfo{person}{Alberto Momigliano}, \bibinfo{person}{Steven Schäfer},
  {and} \bibinfo{person}{Kathrin Stark}.} \bibinfo{year}{2019}\natexlab{}.
\newblock \showarticletitle{POPLMark reloaded: Mechanizing proofs by logical
  relations}.
\newblock \bibinfo{journal}{\emph{Journal of Functional Programming}}
  \bibinfo{volume}{29} (\bibinfo{year}{2019}), \bibinfo{pages}{e19}.
\newblock
\urldef\tempurl%
\url{https://doi.org/10.1017/S0956796819000170}
\showDOI{\tempurl}


\bibitem[Allais et~al\mbox{.}(2021)]%
        {allais-et-al:type-and-scope-safe-syntax-journal}
\bibfield{author}{\bibinfo{person}{Guillaume Allais}, \bibinfo{person}{Robert
  Atkey}, \bibinfo{person}{James Chapman}, \bibinfo{person}{Conor {McBride}},
  {and} \bibinfo{person}{James McKinna}.} \bibinfo{year}{2021}\natexlab{}.
\newblock \showarticletitle{A type- and scope-safe universe of syntaxes with
  binding: their semantics and proofs}.
\newblock \bibinfo{journal}{\emph{Journal of Functional Programming}}
  \bibinfo{volume}{31} (\bibinfo{year}{2021}).
\newblock
\urldef\tempurl%
\url{https://doi.org/10.1017/S0956796820000076}
\showDOI{\tempurl}


\bibitem[Altenkirch and Reus(1999)]%
        {altenkirch-reus}
\bibfield{author}{\bibinfo{person}{Thorsten Altenkirch} {and}
  \bibinfo{person}{Bernhard Reus}.} \bibinfo{year}{1999}\natexlab{}.
\newblock \showarticletitle{Monadic Presentations of Lambda Terms Using
  Generalized Inductive Types}. In \bibinfo{booktitle}{\emph{Computer Science
  Logic, 13th International Workshop, {CSL} '99, 8th Annual Conference of the
  EACSL, Madrid, Spain, September 20-25, 1999, Proceedings}}
  \emph{(\bibinfo{series}{Lecture Notes in Computer Science},
  Vol.~\bibinfo{volume}{1683})}, \bibfield{editor}{\bibinfo{person}{J{\"{o}}rg
  Flum} {and} \bibinfo{person}{Mario Rodr{\'{\i}}guez{-}Artalejo}} (Eds.).
  \bibinfo{publisher}{Springer}, \bibinfo{pages}{453--468}.
\newblock
\urldef\tempurl%
\url{https://doi.org/10.1007/3-540-48168-0\_32}
\showDOI{\tempurl}


\bibitem[Atkey(2014)]%
        {DBLP:conf/popl/Atkey14}
\bibfield{author}{\bibinfo{person}{Robert Atkey}.}
  \bibinfo{year}{2014}\natexlab{}.
\newblock \showarticletitle{From parametricity to conservation laws, via
  Noether's theorem}. In \bibinfo{booktitle}{\emph{The 41st Annual {ACM}
  {SIGPLAN-SIGACT} Symposium on Principles of Programming Languages, {POPL}
  '14, San Diego, CA, USA, January 20-21, 2014}},
  \bibfield{editor}{\bibinfo{person}{Suresh Jagannathan} {and}
  \bibinfo{person}{Peter Sewell}} (Eds.). \bibinfo{publisher}{{ACM}},
  \bibinfo{pages}{491--502}.
\newblock
\urldef\tempurl%
\url{https://doi.org/10.1145/2535838.2535867}
\showDOI{\tempurl}


\bibitem[Aydemir et~al\mbox{.}(2007)]%
        {aydemir-bohannon-weirich:nominal-reasoning-techniques-in-coq}
\bibfield{author}{\bibinfo{person}{Brian Aydemir}, \bibinfo{person}{Aaron
  Bohannon}, {and} \bibinfo{person}{Stephanie Weirich}.}
  \bibinfo{year}{2007}\natexlab{}.
\newblock \showarticletitle{Nominal Reasoning Techniques in Coq (Extended
  Abstract)}.
\newblock \bibinfo{journal}{\emph{Electronic Notes in Theoretical Computer
  Science}} \bibinfo{volume}{174}, \bibinfo{number}{5} (\bibinfo{year}{2007}),
  \bibinfo{pages}{69--77}.
\newblock
\showISSN{1571-0661}
\urldef\tempurl%
\url{https://doi.org/10.1016/j.entcs.2007.01.028}
\showDOI{\tempurl}
\newblock
\shownote{Proceedings of the First International Workshop on Logical Frameworks
  and Meta-Languages: Theory and Practice (LFMTP 2006)}.


\bibitem[Aydemir et~al\mbox{.}(2008)]%
        {aydemir-et-al:lngen}
\bibfield{author}{\bibinfo{person}{Brian Aydemir}, \bibinfo{person}{Arthur
  Chargu\'{e}raud}, \bibinfo{person}{Benjamin~C. Pierce},
  \bibinfo{person}{Randy Pollack}, {and} \bibinfo{person}{Stephanie Weirich}.}
  \bibinfo{year}{2008}\natexlab{}.
\newblock \showarticletitle{Engineering formal metatheory}. In
  \bibinfo{booktitle}{\emph{Proceedings of the 35th Annual ACM SIGPLAN-SIGACT
  Symposium on Principles of Programming Languages}} (San Francisco,
  California, USA) \emph{(\bibinfo{series}{POPL '08})}.
  \bibinfo{publisher}{Association for Computing Machinery},
  \bibinfo{address}{New York, NY, USA}, \bibinfo{pages}{3–15}.
\newblock
\showISBNx{9781595936899}
\urldef\tempurl%
\url{https://doi.org/10.1145/1328438.1328443}
\showDOI{\tempurl}


\bibitem[Aydemir et~al\mbox{.}(2005)]%
        {poplmark1}
\bibfield{author}{\bibinfo{person}{Brian~E. Aydemir}, \bibinfo{person}{Aaron
  Bohannon}, \bibinfo{person}{Matthew Fairbairn}, \bibinfo{person}{J.~Nathan
  Foster}, \bibinfo{person}{Benjamin~C. Pierce}, \bibinfo{person}{Peter
  Sewell}, \bibinfo{person}{Dimitrios Vytiniotis}, \bibinfo{person}{Geoffrey
  Washburn}, \bibinfo{person}{Stephanie Weirich}, {and} \bibinfo{person}{Steve
  Zdancewic}.} \bibinfo{year}{2005}\natexlab{}.
\newblock \showarticletitle{Mechanized Metatheory for the Masses: The PoplMark
  Challenge}. In \bibinfo{booktitle}{\emph{Theorem Proving in Higher Order
  Logics}}, \bibfield{editor}{\bibinfo{person}{Joe Hurd} {and}
  \bibinfo{person}{Tom Melham}} (Eds.). \bibinfo{publisher}{Springer Berlin
  Heidelberg}, \bibinfo{address}{Berlin, Heidelberg}, \bibinfo{pages}{50--65}.
\newblock
\showISBNx{978-3-540-31820-0}


\bibitem[Basheer and Mishra(2025)]%
        {basheer2025symmetry}
\bibfield{author}{\bibinfo{person}{Ramzan Basheer} {and}
  \bibinfo{person}{Deepak Mishra}.} \bibinfo{year}{2025}\natexlab{}.
\newblock \showarticletitle{Symmetry group equivariant convolutions for
  representation learning: a survey}.
\newblock \bibinfo{journal}{\emph{International Journal of Data Science and
  Analytics}}  \bibinfo{volume}{22} (\bibinfo{date}{12} \bibinfo{year}{2025}).
\newblock
\urldef\tempurl%
\url{https://doi.org/10.1007/s41060-025-00994-7}
\showDOI{\tempurl}


\bibitem[Bauer(2008)]%
        {bauer:efficient-computation-with-dedekind-reals}
\bibfield{author}{\bibinfo{person}{Andrej Bauer}.}
  \bibinfo{year}{2008}\natexlab{}.
\newblock \showarticletitle{Efficient computation with {D}edekind reals}.
\newblock  (\bibinfo{year}{2008}).
\newblock
\urldef\tempurl%
\url{https://math.andrej.com/wp-content/uploads/2008/08/abstract-cca2008.pdf}
\showURL{%
\tempurl}
\newblock
\shownote{Fifth International Conference on Computability and Complexity in
  Analysis (CCA 2008)}.


\bibitem[Bauer and Taylor(2009)]%
        {bauer-taylor:dedekind-reals-in-abstract-stone-duality}
\bibfield{author}{\bibinfo{person}{Andrej Bauer} {and} \bibinfo{person}{Paul
  Taylor}.} \bibinfo{year}{2009}\natexlab{}.
\newblock \showarticletitle{The {D}edekind Reals in Abstract {S}tone Duality}.
\newblock \bibinfo{journal}{\emph{Mathematical Structures in Computer Science}}
   \bibinfo{volume}{19} (\bibinfo{year}{2009}), \bibinfo{pages}{757--838}.
\newblock
\urldef\tempurl%
\url{https://doi.org/10.1017/S0960129509007695}
\showDOI{\tempurl}


\bibitem[Beck(1969)]%
        {beck:distributive-laws}
\bibfield{author}{\bibinfo{person}{Jon Beck}.} \bibinfo{year}{1969}\natexlab{}.
\newblock \showarticletitle{Distributive laws}. In
  \bibinfo{booktitle}{\emph{Seminar on Triples and Categorical Homology
  Theory}}, \bibfield{editor}{\bibinfo{person}{B.~Eckmann}} (Ed.).
  \bibinfo{publisher}{Springer Berlin Heidelberg}, \bibinfo{address}{Berlin,
  Heidelberg}, \bibinfo{pages}{119--140}.
\newblock
\showISBNx{978-3-540-36091-9}


\bibitem[Benton et~al\mbox{.}(2012)]%
        {benton-et-al:strongly-typed-terms}
\bibfield{author}{\bibinfo{person}{Nick Benton}, \bibinfo{person}{Chung-Kil
  Hur}, \bibinfo{person}{Andrew~J. Kennedy}, {and} \bibinfo{person}{Conor
  McBride}.} \bibinfo{year}{2012}\natexlab{}.
\newblock \showarticletitle{Strongly Typed Term Representations in Coq}.
\newblock \bibinfo{journal}{\emph{J. Autom. Reason.}} \bibinfo{volume}{49},
  \bibinfo{number}{2} (\bibinfo{date}{Aug.} \bibinfo{year}{2012}),
  \bibinfo{pages}{141–159}.
\newblock
\showISSN{0168-7433}
\urldef\tempurl%
\url{https://doi.org/10.1007/s10817-011-9219-0}
\showDOI{\tempurl}


\bibitem[Bird and Paterson(1999)]%
        {bird-paterson:generalised-folds-for-nested-datatypes}
\bibfield{author}{\bibinfo{person}{Richard Bird} {and} \bibinfo{person}{Ross
  Paterson}.} \bibinfo{year}{1999}\natexlab{}.
\newblock \showarticletitle{Generalised folds for nested datatypes}.
\newblock \bibinfo{journal}{\emph{Form. Asp. Comput.}} \bibinfo{volume}{11},
  \bibinfo{number}{2} (\bibinfo{date}{Sept.} \bibinfo{year}{1999}),
  \bibinfo{pages}{200–222}.
\newblock
\showISSN{0934-5043}
\urldef\tempurl%
\url{https://doi.org/10.1007/s001650050047}
\showDOI{\tempurl}


\bibitem[Blanchette et~al\mbox{.}(2019)]%
        {blanchette-et-al:bindings-as-bounded-natural-functors}
\bibfield{author}{\bibinfo{person}{Jasmin~Christian Blanchette},
  \bibinfo{person}{Lorenzo Gheri}, \bibinfo{person}{Andrei Popescu}, {and}
  \bibinfo{person}{Dmitriy Traytel}.} \bibinfo{year}{2019}\natexlab{}.
\newblock \showarticletitle{Bindings as bounded natural functors}.
\newblock \bibinfo{journal}{\emph{Proc. ACM Program. Lang.}}
  \bibinfo{volume}{3}, \bibinfo{number}{POPL}, Article \bibinfo{articleno}{22}
  (\bibinfo{date}{Jan.} \bibinfo{year}{2019}), \bibinfo{numpages}{34}~pages.
\newblock
\urldef\tempurl%
\url{https://doi.org/10.1145/3290335}
\showDOI{\tempurl}


\bibitem[Bojanczyk et~al\mbox{.}(2011)]%
        {bojanczyk-klin-lasota:automata-with-group-actions}
\bibfield{author}{\bibinfo{person}{Mikolaj Bojanczyk}, \bibinfo{person}{Bartek
  Klin}, {and} \bibinfo{person}{Slawomir Lasota}.}
  \bibinfo{year}{2011}\natexlab{}.
\newblock \showarticletitle{Automata with Group Actions}. In
  \bibinfo{booktitle}{\emph{2011 IEEE 26th Annual Symposium on Logic in
  Computer Science}}. \bibinfo{pages}{355--364}.
\newblock
\urldef\tempurl%
\url{https://doi.org/10.1109/LICS.2011.48}
\showDOI{\tempurl}


\bibitem[Bo{\v{s}}na{\v{c}}ki et~al\mbox{.}(2002)]%
        {bosnacki-dams-holenderski:symmetric-spin}
\bibfield{author}{\bibinfo{person}{Dragan Bo{\v{s}}na{\v{c}}ki},
  \bibinfo{person}{Dennis Dams}, {and} \bibinfo{person}{Leszek Holenderski}.}
  \bibinfo{year}{2002}\natexlab{}.
\newblock \showarticletitle{Symmetric Spin}.
\newblock \bibinfo{journal}{\emph{International Journal on Software Tools for
  Technology Transfer}} \bibinfo{volume}{4}, \bibinfo{number}{1}
  (\bibinfo{date}{01 Oct} \bibinfo{year}{2002}), \bibinfo{pages}{92--106}.
\newblock
\showISSN{1433-2779}
\urldef\tempurl%
\url{https://doi.org/10.1007/s100090200074}
\showDOI{\tempurl}


\bibitem[Cenciarelli and Moggi(1993)]%
  {cenciarelli-moggi:a-syntactic-approach-to-modularity-in-denotational-semantics}
\bibfield{author}{\bibinfo{person}{Pietro Cenciarelli} {and}
  \bibinfo{person}{Eugenio Moggi}.} \bibinfo{year}{1993}\natexlab{}.
\newblock \showarticletitle{A syntactic approach to modularity in denotational
  semantics}.
\newblock \bibinfo{journal}{\emph{Proc. 5th CTCS, CWI Technical report}}
  (\bibinfo{year}{1993}).
\newblock


\bibitem[Chargu{\'e}raud(2012)]%
        {chargueraud:locally-nameless}
\bibfield{author}{\bibinfo{person}{Arthur Chargu{\'e}raud}.}
  \bibinfo{year}{2012}\natexlab{}.
\newblock \showarticletitle{The locally nameless representation}.
\newblock \bibinfo{journal}{\emph{Journal of automated reasoning}}
  \bibinfo{volume}{49} (\bibinfo{year}{2012}), \bibinfo{pages}{363--408}.
\newblock


\bibitem[Cheney(2004)]%
        {cheney:nominal-logic-programming}
\bibfield{author}{\bibinfo{person}{James Cheney}.}
  \bibinfo{year}{2004}\natexlab{}.
\newblock \emph{\bibinfo{title}{Nominal Logic Programming}}.
\newblock \bibinfo{thesistype}{Ph.\,D. Dissertation}. \bibinfo{school}{Cornell
  University, {USA}}.
\newblock


\bibitem[Cheney(2005)]%
        {cheney:towards-names}
\bibfield{author}{\bibinfo{person}{James Cheney}.}
  \bibinfo{year}{2005}\natexlab{}.
\newblock \showarticletitle{Toward a general theory of names: binding and
  scope}. In \bibinfo{booktitle}{\emph{Proceedings of the 3rd ACM SIGPLAN
  Workshop on Mechanized Reasoning about Languages with Variable Binding}}
  (Tallinn, Estonia) \emph{(\bibinfo{series}{MERLIN '05})}.
  \bibinfo{publisher}{Association for Computing Machinery},
  \bibinfo{address}{New York, NY, USA}, \bibinfo{pages}{33–40}.
\newblock
\showISBNx{1595930728}
\urldef\tempurl%
\url{https://doi.org/10.1145/1088454.1088459}
\showDOI{\tempurl}


\bibitem[Chlipala(2008)]%
        {chlipala:parametric-hoas}
\bibfield{author}{\bibinfo{person}{Adam Chlipala}.}
  \bibinfo{year}{2008}\natexlab{}.
\newblock \showarticletitle{Parametric higher-order abstract syntax for
  mechanized semantics}. In \bibinfo{booktitle}{\emph{Proceedings of the 13th
  ACM SIGPLAN International Conference on Functional Programming}} (Victoria,
  BC, Canada) \emph{(\bibinfo{series}{ICFP '08})}.
  \bibinfo{publisher}{Association for Computing Machinery},
  \bibinfo{address}{New York, NY, USA}, \bibinfo{pages}{143–156}.
\newblock
\showISBNx{9781595939197}
\urldef\tempurl%
\url{https://doi.org/10.1145/1411204.1411226}
\showDOI{\tempurl}


\bibitem[Chu and Jaffar(2012)]%
  {chu-jaffar:a-complete-method-for-symmetry-reduction-in-safety-verification}
\bibfield{author}{\bibinfo{person}{Duc-Hiep Chu} {and} \bibinfo{person}{Joxan
  Jaffar}.} \bibinfo{year}{2012}\natexlab{}.
\newblock \showarticletitle{A Complete Method for Symmetry Reduction in Safety
  Verification}. In \bibinfo{booktitle}{\emph{Computer Aided Verification}},
  \bibfield{editor}{\bibinfo{person}{P.~Madhusudan} {and}
  \bibinfo{person}{Sanjit~A. Seshia}} (Eds.). \bibinfo{publisher}{Springer
  Berlin Heidelberg}, \bibinfo{address}{Berlin, Heidelberg},
  \bibinfo{pages}{616--633}.
\newblock
\showISBNx{978-3-642-31424-7}


\bibitem[Clarke et~al\mbox{.}(2018)]%
        {clarke2018handbook}
\bibfield{author}{\bibinfo{person}{Edmund~M Clarke}, \bibinfo{person}{Thomas~A
  Henzinger}, \bibinfo{person}{Helmut Veith}, \bibinfo{person}{Roderick Bloem},
  {et~al\mbox{.}}} \bibinfo{year}{2018}\natexlab{}.
\newblock \bibinfo{booktitle}{\emph{Handbook of model checking}}.
  Vol.~\bibinfo{volume}{10}.
\newblock \bibinfo{publisher}{Springer}.
\newblock


\bibitem[Clarke and Jha(1995)]%
        {clarke-jha:symmetry-and-induction-model-checking}
\bibfield{author}{\bibinfo{person}{Edmund~M. Clarke} {and}
  \bibinfo{person}{Somesh Jha}.} \bibinfo{year}{1995}\natexlab{}.
\newblock \bibinfo{booktitle}{\emph{Symmetry and induction in model checking}}.
\newblock \bibinfo{publisher}{Springer Berlin Heidelberg},
  \bibinfo{address}{Berlin, Heidelberg}, \bibinfo{pages}{455--470}.
\newblock
\showISBNx{978-3-540-49435-5}
\urldef\tempurl%
\url{https://doi.org/10.1007/BFb0015260}
\showDOI{\tempurl}


\bibitem[Copello et~al\mbox{.}(2021)]%
        {copello-et-al:variable-convention}
\bibfield{author}{\bibinfo{person}{Ernesto Copello}, \bibinfo{person}{Nora
  Szasz}, {and} \bibinfo{person}{{\'A}lvaro Tasistro}.}
  \bibinfo{year}{2021}\natexlab{}.
\newblock \showarticletitle{Formalization of metatheory of the Lambda Calculus
  in constructive type theory using the Barendregt variable convention}.
\newblock \bibinfo{journal}{\emph{Mathematical Structures in Computer Science}}
  \bibinfo{volume}{31}, \bibinfo{number}{3} (\bibinfo{year}{2021}),
  \bibinfo{pages}{341–360}.
\newblock
\urldef\tempurl%
\url{https://doi.org/10.1017/S0960129521000335}
\showDOI{\tempurl}


\bibitem[Cornish(2025)]%
        {cornish2025stochasticneuralnetworksymmetrisation}
\bibfield{author}{\bibinfo{person}{Rob Cornish}.}
  \bibinfo{year}{2025}\natexlab{}.
\newblock \bibinfo{title}{Stochastic Neural Network Symmetrisation in Markov
  Categories}.
\newblock
\showeprint[arxiv]{2406.11814}~[stat.ML]
\urldef\tempurl%
\url{https://arxiv.org/abs/2406.11814}
\showURL{%
\tempurl}


\bibitem[Dijkstra(1997)]%
        {dijkstra:wlog}
\bibfield{author}{\bibinfo{person}{Edsger~W. Dijkstra}.}
  \bibinfo{year}{1997}\natexlab{}.
\newblock \showarticletitle{WLOG, or the misery of the unordered pair
  (EWD1223)}. In \bibinfo{booktitle}{\emph{Mathematical Methods in Program
  Development}}, \bibfield{editor}{\bibinfo{person}{Manfred Broy} {and}
  \bibinfo{person}{Birgit Schieder}} (Eds.). \bibinfo{publisher}{Springer
  Berlin Heidelberg}, \bibinfo{address}{Berlin, Heidelberg},
  \bibinfo{pages}{33--34}.
\newblock
\showISBNx{978-3-642-60858-2}


\bibitem[Donaldson(2007)]%
        {donaldson:thesis}
\bibfield{author}{\bibinfo{person}{Alastair~F. Donaldson}.}
  \bibinfo{year}{2007}\natexlab{}.
\newblock \emph{\bibinfo{title}{Automatic techniques for detecting and
  exploiting symmetry in model checking}}.
\newblock \bibinfo{thesistype}{Ph.\,D. Dissertation}.
  \bibinfo{school}{University of Glasgow, {UK}}.
\newblock
\urldef\tempurl%
\url{http://theses.gla.ac.uk/1336/}
\showURL{%
\tempurl}


\bibitem[Dowek et~al\mbox{.}(2010)]%
        {dowek-gabbay-mulligan:permissive-nominal-terms-and-their-unification}
\bibfield{author}{\bibinfo{person}{Gilles Dowek}, \bibinfo{person}{Murdoch~J.
  Gabbay}, {and} \bibinfo{person}{Dominic~P. Mulligan}.}
  \bibinfo{year}{2010}\natexlab{}.
\newblock \showarticletitle{Permissive nominal terms and their unification: an
  infinite, co-infinite approach to nominal techniques}.
\newblock \bibinfo{journal}{\emph{Logic Journal of the IGPL}}
  \bibinfo{volume}{18}, \bibinfo{number}{6} (\bibinfo{date}{12}
  \bibinfo{year}{2010}), \bibinfo{pages}{769--822}.
\newblock
\showISSN{1367-0751}
\urldef\tempurl%
\url{https://doi.org/10.1093/jigpal/jzq006}
\showDOI{\tempurl}
\showeprint{https://academic.oup.com/jigpal/article-pdf/18/6/769/17505709/jzq006.pdf}


\bibitem[{\'E}rdi(2018)]%
        {erdi:well-scoped-syntax}
\bibfield{author}{\bibinfo{person}{Gerg{{o}} {\'E}rdi}.}
  \bibinfo{year}{2018}\natexlab{}.
\newblock \bibinfo{title}{Generic Description of Well-Scoped, Well-Typed
  Syntaxes}.
\newblock
\showeprint[arxiv]{1804.00119}~[cs.PL]
\urldef\tempurl%
\url{https://arxiv.org/abs/1804.00119}
\showURL{%
\tempurl}


\bibitem[Escard{ó}(1998)]%
        {escardo:signed-digit-numerals}
\bibfield{author}{\bibinfo{person}{Martín~H{ö}tzel Escard{ó}}.}
  \bibinfo{year}{1998}\natexlab{}.
\newblock \showarticletitle{Effective and sequential definition by cases on the
  reals via infinite signed-digit numerals}.
\newblock \bibinfo{journal}{\emph{Electronic Notes in Theoretical Computer
  Science}}  \bibinfo{volume}{13} (\bibinfo{year}{1998}),
  \bibinfo{pages}{53--68}.
\newblock
\showISSN{1571-0661}
\urldef\tempurl%
\url{https://doi.org/10.1016/S1571-0661(05)80214-2}
\showDOI{\tempurl}
\newblock
\shownote{Comprox III, Third Workshop on Computation and Approximation}.


\bibitem[Fiore and Menni(2005)]%
        {fiore-menni:reflective-kleisli-subcategories}
\bibfield{author}{\bibinfo{person}{Marcelo Fiore} {and}
  \bibinfo{person}{Mat{\'\i}as Menni}.} \bibinfo{year}{2005}\natexlab{}.
\newblock \showarticletitle{Reflective Kleisli subcategories of the category of
  Eilenberg-Moore algebras for factorization monads.}
\newblock \bibinfo{journal}{\emph{Theory and Applications of Categories
  [electronic only]}}  \bibinfo{volume}{15} (\bibinfo{year}{2005}),
  \bibinfo{pages}{40--65}.
\newblock
\urldef\tempurl%
\url{http://eudml.org/doc/125160}
\showURL{%
\tempurl}


\bibitem[Fiore and Szamozvancev(2022)]%
        {fiore-szamozvancev:formal-soas}
\bibfield{author}{\bibinfo{person}{Marcelo Fiore} {and}
  \bibinfo{person}{Dmitrij Szamozvancev}.} \bibinfo{year}{2022}\natexlab{}.
\newblock \showarticletitle{Formal metatheory of second-order abstract syntax}.
\newblock \bibinfo{journal}{\emph{Proc. ACM Program. Lang.}}
  \bibinfo{volume}{6}, \bibinfo{number}{POPL}, Article \bibinfo{articleno}{53}
  (\bibinfo{date}{Jan.} \bibinfo{year}{2022}), \bibinfo{numpages}{29}~pages.
\newblock
\urldef\tempurl%
\url{https://doi.org/10.1145/3498715}
\showDOI{\tempurl}


\bibitem[Fiore and Szamozvancev(2025)]%
        {fiore-szamozvancev:soas-theory}
\bibfield{author}{\bibinfo{person}{Marcelo Fiore} {and}
  \bibinfo{person}{Dmitrij Szamozvancev}.} \bibinfo{year}{2025}\natexlab{}.
\newblock \bibinfo{title}{Familial Model of Second-Order Abstract Syntax}.
  (\bibinfo{year}{2025}).
\newblock
\newblock
\shownote{unpublished}.


\bibitem[Fiore et~al\mbox{.}(1999)]%
        {fiore-plotkin-turi:ast}
\bibfield{author}{\bibinfo{person}{Marcelo~P. Fiore},
  \bibinfo{person}{Gordon~D. Plotkin}, {and} \bibinfo{person}{Daniele Turi}.}
  \bibinfo{year}{1999}\natexlab{}.
\newblock \showarticletitle{Abstract Syntax and Variable Binding}. In
  \bibinfo{booktitle}{\emph{14th Annual {IEEE} Symposium on Logic in Computer
  Science, Trento, Italy, July 2-5, 1999}}. \bibinfo{publisher}{{IEEE} Computer
  Society}, \bibinfo{pages}{193--202}.
\newblock
\urldef\tempurl%
\url{https://doi.org/10.1109/LICS.1999.782615}
\showDOI{\tempurl}


\bibitem[Fiore and Ranchod(2025)]%
        {fiore-ranchod:substructural-ast}
\bibfield{author}{\bibinfo{person}{Marcelo~P. Fiore} {and}
  \bibinfo{person}{Sanjiv Ranchod}.} \bibinfo{year}{2025}\natexlab{}.
\newblock \showarticletitle{Substructural Abstract Syntax with Variable Binding
  and Single-Variable Substitution}. In \bibinfo{booktitle}{\emph{2025 40th
  Annual ACM/IEEE Symposium on Logic in Computer Science (LICS)}}.
  \bibinfo{pages}{196--208}.
\newblock
\urldef\tempurl%
\url{https://doi.org/10.1109/LICS65433.2025.00022}
\showDOI{\tempurl}


\bibitem[Forster and Stark(2020)]%
        {forster-stark:coq-a-la-carte}
\bibfield{author}{\bibinfo{person}{Yannick Forster} {and}
  \bibinfo{person}{Kathrin Stark}.} \bibinfo{year}{2020}\natexlab{}.
\newblock \showarticletitle{Coq {\`a} la Carte {---} A Practical Approach to
  Modular Syntax with Binders}. In \bibinfo{booktitle}{\emph{9th ACM SIGPLAN
  International Conference on Certified Programs and Proofs, CPP 2020, New
  Orleans, USA}}. \bibinfo{publisher}{ACM}.
\newblock


\bibitem[Gabbay and Pitts(2002)]%
        {gabbay-pitts:a-new-approach-to-abstract-syntax-with-variable-binding}
\bibfield{author}{\bibinfo{person}{Murdoch~J. Gabbay} {and}
  \bibinfo{person}{Andrew~M. Pitts}.} \bibinfo{year}{2002}\natexlab{}.
\newblock \showarticletitle{A New Approach to Abstract Syntax with Variable
  Binding}.
\newblock \bibinfo{journal}{\emph{Formal Aspects Comput.}}
  \bibinfo{volume}{13}, \bibinfo{number}{3-5} (\bibinfo{year}{2002}),
  \bibinfo{pages}{341--363}.
\newblock
\urldef\tempurl%
\url{https://doi.org/10.1007/S001650200016}
\showDOI{\tempurl}


\bibitem[Gacek(2009)]%
        {gacek:thesis}
\bibfield{author}{\bibinfo{person}{Andrew Gacek}.}
  \bibinfo{year}{2009}\natexlab{}.
\newblock \emph{\bibinfo{title}{A Framework for Specifying, Prototyping, and
  Reasoning about Computational Systems}}.
\newblock \bibinfo{thesistype}{Ph.\,D. Dissertation}.
  \bibinfo{school}{University of Minnesota}.
\newblock


\bibitem[Gacek et~al\mbox{.}(2012)]%
        {gacek-et-al:abella-journal}
\bibfield{author}{\bibinfo{person}{Andrew Gacek}, \bibinfo{person}{Dale
  Miller}, {and} \bibinfo{person}{Gopalan Nadathur}.}
  \bibinfo{year}{2012}\natexlab{}.
\newblock \showarticletitle{A two-level logic approach to reasoning about
  computations}.
\newblock \bibinfo{journal}{\emph{Journal of Automated Reasoning}}
  \bibinfo{volume}{49}, \bibinfo{number}{2} (\bibinfo{year}{2012}),
  \bibinfo{pages}{241--273}.
\newblock


\bibitem[Gent et~al\mbox{.}(2006)]%
        {DBLP:reference/fai/GentPP06}
\bibfield{author}{\bibinfo{person}{Ian~P. Gent}, \bibinfo{person}{Karen~E.
  Petrie}, {and} \bibinfo{person}{Jean{-}Fran{\c{c}}ois Puget}.}
  \bibinfo{year}{2006}\natexlab{}.
\newblock \showarticletitle{Symmetry in Constraint Programming}.
\newblock In \bibinfo{booktitle}{\emph{Handbook of Constraint Programming}},
  \bibfield{editor}{\bibinfo{person}{Francesca Rossi}, \bibinfo{person}{Peter
  van Beek}, {and} \bibinfo{person}{Toby Walsh}} (Eds.).
  \bibinfo{series}{Foundations of Artificial Intelligence},
  Vol.~\bibinfo{volume}{2}. \bibinfo{publisher}{Elsevier},
  \bibinfo{pages}{329--376}.
\newblock
\urldef\tempurl%
\url{https://doi.org/10.1016/S1574-6526(06)80014-3}
\showDOI{\tempurl}


\bibitem[Ghani et~al\mbox{.}(2006)]%
        {ghani-et-al:cyclic-structures}
\bibfield{author}{\bibinfo{person}{Neil Ghani}, \bibinfo{person}{Makoto
  Hamana}, \bibinfo{person}{Tarmo Uustalu}, {and} \bibinfo{person}{Varmo
  Vene}.} \bibinfo{year}{2006}\natexlab{}.
\newblock \showarticletitle{Representing Cyclic Structures as Nested
  Datatypes}.
\newblock  (\bibinfo{date}{April} \bibinfo{year}{2006}),
  \bibinfo{pages}{173--188}.
\newblock


\bibitem[Gonthier and Mahboubi(2010)]%
        {ssreflect}
\bibfield{author}{\bibinfo{person}{Georges Gonthier} {and}
  \bibinfo{person}{Assia Mahboubi}.} \bibinfo{year}{2010}\natexlab{}.
\newblock \showarticletitle{An introduction to small scale reflection in Coq}.
\newblock \bibinfo{journal}{\emph{Journal of Formalized Reasoning}}
  \bibinfo{volume}{3}, \bibinfo{number}{2} (\bibinfo{date}{Jan.}
  \bibinfo{year}{2010}), \bibinfo{pages}{95–152}.
\newblock
\urldef\tempurl%
\url{https://doi.org/10.6092/issn.1972-5787/1979}
\showDOI{\tempurl}


\bibitem[Hamana(2009)]%
        {hamana:initial-semantics-for-cyclic-sharing}
\bibfield{author}{\bibinfo{person}{Makoto Hamana}.}
  \bibinfo{year}{2009}\natexlab{}.
\newblock \showarticletitle{Initial Algebra Semantics for Cyclic Sharing
  Structures}. In \bibinfo{booktitle}{\emph{Typed Lambda Calculi and
  Applications}}, \bibfield{editor}{\bibinfo{person}{Pierre-Louis Curien}}
  (Ed.). \bibinfo{publisher}{Springer Berlin Heidelberg},
  \bibinfo{address}{Berlin, Heidelberg}, \bibinfo{pages}{127--141}.
\newblock
\showISBNx{978-3-642-02273-9}


\bibitem[Harrison(2009)]%
        {DBLP:conf/tphol/Harrison09}
\bibfield{author}{\bibinfo{person}{John Harrison}.}
  \bibinfo{year}{2009}\natexlab{}.
\newblock \showarticletitle{Without Loss of Generality}. In
  \bibinfo{booktitle}{\emph{Theorem Proving in Higher Order Logics, 22nd
  International Conference, TPHOLs 2009, Munich, Germany, August 17-20, 2009.
  Proceedings}} \emph{(\bibinfo{series}{Lecture Notes in Computer Science},
  Vol.~\bibinfo{volume}{5674})}, \bibfield{editor}{\bibinfo{person}{Stefan
  Berghofer}, \bibinfo{person}{Tobias Nipkow}, \bibinfo{person}{Christian
  Urban}, {and} \bibinfo{person}{Makarius Wenzel}} (Eds.).
  \bibinfo{publisher}{Springer}, \bibinfo{pages}{43--59}.
\newblock
\urldef\tempurl%
\url{https://doi.org/10.1007/978-3-642-03359-9\_3}
\showDOI{\tempurl}


\bibitem[Hofmann(1999)]%
        {hofmann:semantical-analysis-of-ho-abstract-syntax}
\bibfield{author}{\bibinfo{person}{Martin Hofmann}.}
  \bibinfo{year}{1999}\natexlab{}.
\newblock \showarticletitle{Semantical analysis of higher-order abstract
  syntax}. In \bibinfo{booktitle}{\emph{Proceedings. 14th Symposium on Logic in
  Computer Science (Cat. No. PR00158)}}. \bibinfo{pages}{204--213}.
\newblock
\urldef\tempurl%
\url{https://doi.org/10.1109/LICS.1999.782616}
\showDOI{\tempurl}


\bibitem[Holt et~al\mbox{.}(2005)]%
        {Holt2005HandbookOC}
\bibfield{author}{\bibinfo{person}{Derek~F. Holt}, \bibinfo{person}{Bettina
  Eick}, {and} \bibinfo{person}{Eamonn~A. O'Brien}.}
  \bibinfo{year}{2005}\natexlab{}.
\newblock \showarticletitle{Handbook of Computational Group Theory}.
\newblock
\urldef\tempurl%
\url{https://api.semanticscholar.org/CorpusID:116937731}
\showURL{%
\tempurl}


\bibitem[Janota et~al\mbox{.}(2025)]%
        {janota2025breakingsymmetriesquantifiedgraph}
\bibfield{author}{\bibinfo{person}{Mikoláš Janota}, \bibinfo{person}{Markus
  Kirchweger}, \bibinfo{person}{Tomáš Peitl}, {and} \bibinfo{person}{Stefan
  Szeider}.} \bibinfo{year}{2025}\natexlab{}.
\newblock \bibinfo{title}{Breaking Symmetries in Quantified Graph Search: A
  Comparative Study}.
\newblock
\showeprint[arxiv]{2502.15078}~[cs.LO]
\urldef\tempurl%
\url{https://arxiv.org/abs/2502.15078}
\showURL{%
\tempurl}


\bibitem[Jha(1996)]%
        {jha:thesis}
\bibfield{author}{\bibinfo{person}{Somesh Jha}.}
  \bibinfo{year}{1996}\natexlab{}.
\newblock \emph{\bibinfo{title}{Symmetry and Induction in Model Checking}}.
\newblock \bibinfo{thesistype}{Ph.\,D. Dissertation}. \bibinfo{school}{Carnegie
  Mellon University}.
\newblock


\bibitem[Kaucher(1980)]%
        {kaucher:interval-analysis-in-the-extended-interval-space-IR}
\bibfield{author}{\bibinfo{person}{Edgar Kaucher}.}
  \bibinfo{year}{1980}\natexlab{}.
\newblock \bibinfo{booktitle}{\emph{Interval Analysis in the Extended Interval
  Space IR}}.
\newblock \bibinfo{publisher}{Springer Vienna}, \bibinfo{address}{Vienna},
  \bibinfo{pages}{33--49}.
\newblock
\showISBNx{978-3-7091-8577-3}
\urldef\tempurl%
\url{https://doi.org/10.1007/978-3-7091-8577-3_3}
\showDOI{\tempurl}


\bibitem[Keuchel(2011)]%
        {keuchel:msc-thesis}
\bibfield{author}{\bibinfo{person}{Steven Keuchel}.}
  \bibinfo{year}{2011}\natexlab{}.
\newblock \bibinfo{title}{Generic Programming With Binders and Scope}.
\newblock


\bibitem[Keuchel and Jeuring(2012)]%
        {keuchel-jeuring:generic-conversions}
\bibfield{author}{\bibinfo{person}{Steven Keuchel} {and}
  \bibinfo{person}{Johan~T. Jeuring}.} \bibinfo{year}{2012}\natexlab{}.
\newblock \showarticletitle{Generic conversions of abstract syntax
  representations}. In \bibinfo{booktitle}{\emph{Proceedings of the 8th ACM
  SIGPLAN Workshop on Generic Programming}} (Copenhagen, Denmark)
  \emph{(\bibinfo{series}{WGP '12})}. \bibinfo{publisher}{Association for
  Computing Machinery}, \bibinfo{address}{New York, NY, USA},
  \bibinfo{pages}{57–68}.
\newblock
\showISBNx{9781450315760}
\urldef\tempurl%
\url{https://doi.org/10.1145/2364394.2364403}
\showDOI{\tempurl}


\bibitem[Keuchel et~al\mbox{.}(2016)]%
        {keuchel-et-al:needle-and-knot}
\bibfield{author}{\bibinfo{person}{Steven Keuchel}, \bibinfo{person}{Stephanie
  Weirich}, {and} \bibinfo{person}{Tom Schrijvers}.}
  \bibinfo{year}{2016}\natexlab{}.
\newblock \showarticletitle{Needle \& Knot: Binder Boilerplate Tied Up}. In
  \bibinfo{booktitle}{\emph{Proceedings of the 25th European Symposium on
  Programming Languages and Systems - Volume 9632}}.
  \bibinfo{publisher}{Springer-Verlag}, \bibinfo{address}{Berlin, Heidelberg},
  \bibinfo{pages}{419–445}.
\newblock
\showISBNx{9783662494974}
\urldef\tempurl%
\url{https://doi.org/10.1007/978-3-662-49498-1_17}
\showDOI{\tempurl}


\bibitem[Lee et~al\mbox{.}(2012)]%
        {lee-et-al:gmeta}
\bibfield{author}{\bibinfo{person}{Gyesik Lee}, \bibinfo{person}{Bruno C. D.~S.
  Oliveira}, \bibinfo{person}{Sungkeun Cho}, {and} \bibinfo{person}{Kwangkeun
  Yi}.} \bibinfo{year}{2012}\natexlab{}.
\newblock \showarticletitle{GMeta: A Generic Formal Metatheory Framework for
  First-Order Representations}. In \bibinfo{booktitle}{\emph{Programming
  Languages and Systems}}, \bibfield{editor}{\bibinfo{person}{Helmut Seidl}}
  (Ed.). \bibinfo{publisher}{Springer Berlin Heidelberg},
  \bibinfo{address}{Berlin, Heidelberg}, \bibinfo{pages}{436--455}.
\newblock
\showISBNx{978-3-642-28869-2}


\bibitem[Leuschel et~al\mbox{.}(2006)]%
        {leuschel-et-al:symmetry-reduction-for-B-by-permutation-flooding}
\bibfield{author}{\bibinfo{person}{Michael Leuschel}, \bibinfo{person}{Michael
  Butler}, \bibinfo{person}{Corinna Spermann}, {and} \bibinfo{person}{Edd
  Turner}.} \bibinfo{year}{2006}\natexlab{}.
\newblock \showarticletitle{Symmetry Reduction for B by Permutation Flooding}.
  In \bibinfo{booktitle}{\emph{B 2007: Formal Specification and Development in
  B}}, \bibfield{editor}{\bibinfo{person}{Jacques Julliand} {and}
  \bibinfo{person}{Olga Kouchnarenko}} (Eds.). \bibinfo{publisher}{Springer
  Berlin Heidelberg}, \bibinfo{address}{Berlin, Heidelberg},
  \bibinfo{pages}{79--93}.
\newblock
\showISBNx{978-3-540-68761-0}


\bibitem[Lorenzen et~al\mbox{.}(2025)]%
        {DBLP:journals/pacmpl/LorenzenLSL25}
\bibfield{author}{\bibinfo{person}{Anton Lorenzen}, \bibinfo{person}{Daan
  Leijen}, \bibinfo{person}{Wouter Swierstra}, {and} \bibinfo{person}{Sam
  Lindley}.} \bibinfo{year}{2025}\natexlab{}.
\newblock \showarticletitle{First-Order Laziness}.
\newblock \bibinfo{journal}{\emph{Proc. {ACM} Program. Lang.}}
  \bibinfo{volume}{9}, \bibinfo{number}{{ICFP}} (\bibinfo{year}{2025}),
  \bibinfo{pages}{734--762}.
\newblock
\urldef\tempurl%
\url{https://doi.org/10.1145/3747530}
\showDOI{\tempurl}


\bibitem[Mac~Lane(1978)]%
        {mac-lane:cwm}
\bibfield{author}{\bibinfo{person}{Saunders Mac~Lane}.}
  \bibinfo{year}{1978}\natexlab{}.
\newblock \bibinfo{booktitle}{\emph{Categories for the Working Mathematician}
  (\bibinfo{edition}{2nd} ed.)}.
\newblock \bibinfo{publisher}{Springer New York, NY}.
\newblock
\showISBNx{978-0-387-98403-2}
\urldef\tempurl%
\url{https://doi.org/10.1007/978-1-4757-4721-8}
\showDOI{\tempurl}


\bibitem[McBride(2014)]%
        {mcbride:neighbours-in-order}
\bibfield{author}{\bibinfo{person}{Conor~Thomas McBride}.}
  \bibinfo{year}{2014}\natexlab{}.
\newblock \showarticletitle{How to keep your neighbours in order}. In
  \bibinfo{booktitle}{\emph{Proceedings of the 19th ACM SIGPLAN International
  Conference on Functional Programming}} (Gothenburg, Sweden)
  \emph{(\bibinfo{series}{ICFP '14})}. \bibinfo{publisher}{Association for
  Computing Machinery}, \bibinfo{address}{New York, NY, USA},
  \bibinfo{pages}{297–309}.
\newblock
\showISBNx{9781450328739}
\urldef\tempurl%
\url{https://doi.org/10.1145/2628136.2628163}
\showDOI{\tempurl}


\bibitem[McKay and Piperno(2014)]%
        {MCKAY201494}
\bibfield{author}{\bibinfo{person}{Brendan~D. McKay} {and}
  \bibinfo{person}{Adolfo Piperno}.} \bibinfo{year}{2014}\natexlab{}.
\newblock \showarticletitle{Practical graph isomorphism, II}.
\newblock \bibinfo{journal}{\emph{Journal of Symbolic Computation}}
  \bibinfo{volume}{60} (\bibinfo{year}{2014}), \bibinfo{pages}{94--112}.
\newblock
\showISSN{0747-7171}
\urldef\tempurl%
\url{https://doi.org/10.1016/j.jsc.2013.09.003}
\showDOI{\tempurl}


\bibitem[Mehta and Hsu(2025)]%
        {10.1145/3763080}
\bibfield{author}{\bibinfo{person}{Vaibhav Mehta} {and} \bibinfo{person}{Justin
  Hsu}.} \bibinfo{year}{2025}\natexlab{}.
\newblock \showarticletitle{A Hoare Logic for Symmetry Properties}.
\newblock \bibinfo{journal}{\emph{Proc. ACM Program. Lang.}}
  \bibinfo{volume}{9}, \bibinfo{number}{OOPSLA2}, Article
  \bibinfo{articleno}{302} (\bibinfo{date}{Oct.} \bibinfo{year}{2025}),
  \bibinfo{numpages}{25}~pages.
\newblock
\urldef\tempurl%
\url{https://doi.org/10.1145/3763080}
\showDOI{\tempurl}


\bibitem[Miller et~al\mbox{.}(2006)]%
        {miller-donaldson-calder:symmetry-in-temporal-logic-model-checking}
\bibfield{author}{\bibinfo{person}{Alice Miller}, \bibinfo{person}{Alastair
  Donaldson}, {and} \bibinfo{person}{Muffy Calder}.}
  \bibinfo{year}{2006}\natexlab{}.
\newblock \showarticletitle{Symmetry in temporal logic model checking}.
\newblock \bibinfo{journal}{\emph{ACM Comput. Surv.}} \bibinfo{volume}{38},
  \bibinfo{number}{3} (\bibinfo{date}{Sept.} \bibinfo{year}{2006}),
  \bibinfo{pages}{8–es}.
\newblock
\showISSN{0360-0300}
\urldef\tempurl%
\url{https://doi.org/10.1145/1132960.1132962}
\showDOI{\tempurl}


\bibitem[Morris et~al\mbox{.}(2006)]%
        {morris-et-al:regular-tree-types}
\bibfield{author}{\bibinfo{person}{Peter Morris}, \bibinfo{person}{Thorsten
  Altenkirch}, {and} \bibinfo{person}{Conor {McBride}}.}
  \bibinfo{year}{2006}\natexlab{}.
\newblock \showarticletitle{Exploring the Regular Tree Types}. In
  \bibinfo{booktitle}{\emph{Types for Proofs and Programs}},
  \bibfield{editor}{\bibinfo{person}{Jean-Christophe Filli{\^a}tre},
  \bibinfo{person}{Christine Paulin-Mohring}, {and} \bibinfo{person}{Benjamin
  Werner}} (Eds.). \bibinfo{publisher}{Springer Berlin Heidelberg},
  \bibinfo{address}{Berlin, Heidelberg}, \bibinfo{pages}{252--267}.
\newblock
\showISBNx{978-3-540-31429-5}


\bibitem[Norrish(2004)]%
        {norrish:recursive-function-definition-for-types-with-binders}
\bibfield{author}{\bibinfo{person}{Michael Norrish}.}
  \bibinfo{year}{2004}\natexlab{}.
\newblock \showarticletitle{Recursive Function Definition for Types with
  Binders}. In \bibinfo{booktitle}{\emph{Theorem Proving in Higher Order
  Logics}}, \bibfield{editor}{\bibinfo{person}{Konrad Slind},
  \bibinfo{person}{Annette Bunker}, {and} \bibinfo{person}{Ganesh
  Gopalakrishnan}} (Eds.). \bibinfo{publisher}{Springer Berlin Heidelberg},
  \bibinfo{address}{Berlin, Heidelberg}, \bibinfo{pages}{241--256}.
\newblock
\showISBNx{978-3-540-30142-4}


\bibitem[Pfenning and Elliott(1988)]%
        {pfenning-elliott:hoas}
\bibfield{author}{\bibinfo{person}{F. Pfenning} {and} \bibinfo{person}{C.
  Elliott}.} \bibinfo{year}{1988}\natexlab{}.
\newblock \showarticletitle{Higher-order abstract syntax}. In
  \bibinfo{booktitle}{\emph{Proceedings of the ACM SIGPLAN 1988 Conference on
  Programming Language Design and Implementation}} (Atlanta, Georgia, USA)
  \emph{(\bibinfo{series}{PLDI '88})}. \bibinfo{publisher}{Association for
  Computing Machinery}, \bibinfo{address}{New York, NY, USA},
  \bibinfo{pages}{199–208}.
\newblock
\showISBNx{0897912691}
\urldef\tempurl%
\url{https://doi.org/10.1145/53990.54010}
\showDOI{\tempurl}


\bibitem[Pientka(2010)]%
        {pientka:beluga}
\bibfield{author}{\bibinfo{person}{Brigitte Pientka}.}
  \bibinfo{year}{2010}\natexlab{}.
\newblock \showarticletitle{Beluga: Programming with Dependent Types,
  Contextual Data, and Contexts}. In \bibinfo{booktitle}{\emph{Functional and
  Logic Programming}}, \bibfield{editor}{\bibinfo{person}{Matthias Blume},
  \bibinfo{person}{Naoki Kobayashi}, {and} \bibinfo{person}{Germ{\'a}n Vidal}}
  (Eds.). \bibinfo{publisher}{Springer Berlin Heidelberg},
  \bibinfo{address}{Berlin, Heidelberg}, \bibinfo{pages}{1--12}.
\newblock
\showISBNx{978-3-642-12251-4}


\bibitem[Pitts(2003)]%
        {DBLP:journals/iandc/Pitts03}
\bibfield{author}{\bibinfo{person}{Andrew~M. Pitts}.}
  \bibinfo{year}{2003}\natexlab{}.
\newblock \showarticletitle{Nominal logic, a first order theory of names and
  binding}.
\newblock \bibinfo{journal}{\emph{Inf. Comput.}} \bibinfo{volume}{186},
  \bibinfo{number}{2} (\bibinfo{year}{2003}), \bibinfo{pages}{165--193}.
\newblock
\urldef\tempurl%
\url{https://doi.org/10.1016/S0890-5401(03)00138-X}
\showDOI{\tempurl}


\bibitem[Pitts(2006)]%
        {pitts:alpha-structural-recursion-and-induction}
\bibfield{author}{\bibinfo{person}{Andrew~M. Pitts}.}
  \bibinfo{year}{2006}\natexlab{}.
\newblock \showarticletitle{Alpha-structural recursion and induction}.
\newblock \bibinfo{journal}{\emph{J. ACM}} \bibinfo{volume}{53},
  \bibinfo{number}{3} (\bibinfo{date}{May} \bibinfo{year}{2006}),
  \bibinfo{pages}{459–506}.
\newblock
\showISSN{0004-5411}
\urldef\tempurl%
\url{https://doi.org/10.1145/1147954.1147961}
\showDOI{\tempurl}


\bibitem[Pitts(2013)]%
        {pitts:nominal-sets-book}
\bibfield{author}{\bibinfo{person}{Andrew~M. Pitts}.}
  \bibinfo{year}{2013}\natexlab{}.
\newblock \bibinfo{booktitle}{\emph{Nominal Sets: Names and Symmetry in
  Computer Science}}.
\newblock \bibinfo{publisher}{Cambridge University Press}.
\newblock


\bibitem[Pitts(2016)]%
        {DBLP:journals/siglog/Pitts16}
\bibfield{author}{\bibinfo{person}{Andrew~M. Pitts}.}
  \bibinfo{year}{2016}\natexlab{}.
\newblock \showarticletitle{Nominal techniques}.
\newblock \bibinfo{journal}{\emph{{ACM} {SIGLOG} News}} \bibinfo{volume}{3},
  \bibinfo{number}{1} (\bibinfo{year}{2016}), \bibinfo{pages}{57--72}.
\newblock
\urldef\tempurl%
\url{https://doi.org/10.1145/2893582.2893594}
\showDOI{\tempurl}


\bibitem[Pitts et~al\mbox{.}(2015)]%
        {pitts-matthiesen-derikx:a-dep-type-theory-with-abstractable-names}
\bibfield{author}{\bibinfo{person}{Andrew~M. Pitts}, \bibinfo{person}{Justus
  Matthiesen}, {and} \bibinfo{person}{Jasper Derikx}.}
  \bibinfo{year}{2015}\natexlab{}.
\newblock \showarticletitle{A Dependent Type Theory with Abstractable Names}.
\newblock \bibinfo{journal}{\emph{Electronic Notes in Theoretical Computer
  Science}}  \bibinfo{volume}{312} (\bibinfo{year}{2015}),
  \bibinfo{pages}{19--50}.
\newblock
\showISSN{1571-0661}
\urldef\tempurl%
\url{https://doi.org/10.1016/j.entcs.2015.04.003}
\showDOI{\tempurl}
\newblock
\shownote{Ninth Workshop on Logical and Semantic Frameworks, with Applications
  (LSFA 2014)}.


\bibitem[Plotkin(1973)]%
        {plotkin:lambda-definability-and-logical-relations}
\bibfield{author}{\bibinfo{person}{G.~D. Plotkin}.}
  \bibinfo{year}{1973}\natexlab{}.
\newblock \bibinfo{booktitle}{\emph{Lambda {Definability} and {Logical}
  {Relations}}}.
\newblock \bibinfo{type}{Memorandum} SAI-RM-4. \bibinfo{institution}{University
  of Edinburgh}.
\newblock
\urldef\tempurl%
\url{https://homepages.inf.ed.ac.uk/gdp/publications/logical_relations_1973.pdf}
\showURL{%
\tempurl}


\bibitem[Polonowski(2013)]%
        {polonowski:automatically-generated-syntaxes}
\bibfield{author}{\bibinfo{person}{Emmanuel Polonowski}.}
  \bibinfo{year}{2013}\natexlab{}.
\newblock \showarticletitle{Automatically Generated Infrastructure for De
  Bruijn Syntaxes}. In \bibinfo{booktitle}{\emph{Interactive Theorem Proving}},
  \bibfield{editor}{\bibinfo{person}{Sandrine Blazy},
  \bibinfo{person}{Christine Paulin-Mohring}, {and} \bibinfo{person}{David
  Pichardie}} (Eds.). \bibinfo{publisher}{Springer Berlin Heidelberg},
  \bibinfo{address}{Berlin, Heidelberg}, \bibinfo{pages}{402--417}.
\newblock
\showISBNx{978-3-642-39634-2}


\bibitem[Popescu(2024)]%
        {popescu:nominal-recursors-as-epi-recursors}
\bibfield{author}{\bibinfo{person}{Andrei Popescu}.}
  \bibinfo{year}{2024}\natexlab{}.
\newblock \showarticletitle{Nominal Recursors as Epi-Recursors}.
\newblock \bibinfo{journal}{\emph{Proc. ACM Program. Lang.}}
  \bibinfo{volume}{8}, \bibinfo{number}{POPL}, Article \bibinfo{articleno}{15}
  (\bibinfo{date}{Jan.} \bibinfo{year}{2024}), \bibinfo{numpages}{32}~pages.
\newblock
\urldef\tempurl%
\url{https://doi.org/10.1145/3632857}
\showDOI{\tempurl}


\bibitem[Popescu and Gunter(2011)]%
  {popescu-gunter:recursion-principles-for-syntax-with-bindings-and-substitution}
\bibfield{author}{\bibinfo{person}{Andrei Popescu} {and}
  \bibinfo{person}{Elsa~L. Gunter}.} \bibinfo{year}{2011}\natexlab{}.
\newblock \showarticletitle{Recursion principles for syntax with bindings and
  substitution}. In \bibinfo{booktitle}{\emph{Proceedings of the 16th ACM
  SIGPLAN International Conference on Functional Programming}} (Tokyo, Japan)
  \emph{(\bibinfo{series}{ICFP '11})}. \bibinfo{publisher}{Association for
  Computing Machinery}, \bibinfo{address}{New York, NY, USA},
  \bibinfo{pages}{346–358}.
\newblock
\showISBNx{9781450308656}
\urldef\tempurl%
\url{https://doi.org/10.1145/2034773.2034819}
\showDOI{\tempurl}


\bibitem[Poulsen et~al\mbox{.}(2018)]%
        {poulsen-et-al:definitional-interpreters}
\bibfield{author}{\bibinfo{person}{{Casper Bach} Poulsen},
  \bibinfo{person}{Arjen Rouvoet}, \bibinfo{person}{Andrew Tolmach},
  \bibinfo{person}{Robbert Krebbers}, {and} \bibinfo{person}{Eelco Visser}.}
  \bibinfo{year}{2018}\natexlab{}.
\newblock \showarticletitle{Intrinsically-Typed Definitional Interpreters for
  Imperative Languages}.
\newblock \bibinfo{journal}{\emph{Proceedings of the ACM on Programming
  Languages}} \bibinfo{volume}{2}, \bibinfo{number}{POPL} (\bibinfo{date}{10
  Jan.} \bibinfo{year}{2018}), \bibinfo{pages}{1--34}.
\newblock
\showISSN{2475-1421}
\urldef\tempurl%
\url{https://doi.org/10.1145/3158104}
\showDOI{\tempurl}


\bibitem[Reynolds(1983)]%
        {reynolds:types-abstraciton-and-parametric-polymorphism}
\bibfield{author}{\bibinfo{person}{John~C. Reynolds}.}
  \bibinfo{year}{1983}\natexlab{}.
\newblock \showarticletitle{Types, Abstraction and Parametric Polymorphism}. In
  \bibinfo{booktitle}{\emph{Information Processing 83, Proceedings of the
  {IFIP} 9th World Computer Congress, Paris, France, September 19-23, 1983}},
  \bibfield{editor}{\bibinfo{person}{R.~E.~A. Mason}} (Ed.).
  \bibinfo{publisher}{North-Holland/IFIP}, \bibinfo{pages}{513--523}.
\newblock


\bibitem[Sch{\"a}fer et~al\mbox{.}(2015)]%
        {schafer-et-al:autosubst}
\bibfield{author}{\bibinfo{person}{Steven Sch{\"a}fer}, \bibinfo{person}{Gert
  Smolka}, {and} \bibinfo{person}{Tobias Tebbi}.}
  \bibinfo{year}{2015}\natexlab{}.
\newblock \showarticletitle{Completeness and Decidability of de Bruijn
  Substitution Algebra in Coq}. In \bibinfo{booktitle}{\emph{Proceedings of the
  2015 Conference on Certified Programs and Proofs, CPP 2015, Mumbai, India,
  January 15-17, 2015}}. \bibinfo{publisher}{ACM}, \bibinfo{pages}{67--73}.
\newblock


\bibitem[Schmoetten et~al\mbox{.}(2021)]%
  {schmoetten-palmer-fleuriot:formalising-geometric-axioms-for-minkowski-spacetime-and-without-loss-of-generality-theorems}
\bibfield{author}{\bibinfo{person}{Richard Schmoetten},
  \bibinfo{person}{Jake~E. Palmer}, {and} \bibinfo{person}{Jacques~D.
  Fleuriot}.} \bibinfo{year}{2021}\natexlab{}.
\newblock \showarticletitle{Formalising Geometric Axioms for Minkowski
  Spacetime and Without-Loss-of-Generality Theorems}. In
  \bibinfo{booktitle}{\emph{Proceedings of the 13th International Conference on
  Automated Deduction in Geometry, {ADG} 2021, Hagenberg, Austria/virtual,
  September 15-17, 2021}} \emph{(\bibinfo{series}{{EPTCS}},
  Vol.~\bibinfo{volume}{352})}, \bibfield{editor}{\bibinfo{person}{Predrag
  Janicic} {and} \bibinfo{person}{Zolt{\'{a}}n Kov{\'{a}}cs}} (Eds.).
  \bibinfo{pages}{116--128}.
\newblock
\urldef\tempurl%
\url{https://doi.org/10.4204/EPTCS.352.13}
\showDOI{\tempurl}


\bibitem[Schmoetten et~al\mbox{.}(2022)]%
  {schmoetten-palmer-fleuriot:towards-formalising-schutz-axioms-for-minkowski-spacetime-in-isabelle-hol}
\bibfield{author}{\bibinfo{person}{Richard Schmoetten},
  \bibinfo{person}{Jake~E. Palmer}, {and} \bibinfo{person}{Jacques~D.
  Fleuriot}.} \bibinfo{year}{2022}\natexlab{}.
\newblock \showarticletitle{Towards Formalising Schutz' Axioms for Minkowski
  Spacetime in Isabelle/HOL}.
\newblock \bibinfo{journal}{\emph{J. Autom. Reason.}} \bibinfo{volume}{66},
  \bibinfo{number}{4} (\bibinfo{year}{2022}), \bibinfo{pages}{953--988}.
\newblock
\urldef\tempurl%
\url{https://doi.org/10.1007/S10817-022-09643-1}
\showDOI{\tempurl}


\bibitem[Schutz(1997)]%
        {schutz:independent-axioms-for-Minkowski-space-time}
\bibfield{author}{\bibinfo{person}{John~W Schutz}.}
  \bibinfo{year}{1997}\natexlab{}.
\newblock \bibinfo{booktitle}{\emph{Independent axioms for Minkowski
  space-time}}. Vol.~\bibinfo{volume}{373}.
\newblock \bibinfo{publisher}{CRC Press}.
\newblock


\bibitem[Seress(2003)]%
        {Seress_2003}
\bibfield{author}{\bibinfo{person}{Ákos Seress}.}
  \bibinfo{year}{2003}\natexlab{}.
\newblock \bibinfo{booktitle}{\emph{Permutation Group Algorithms}}.
\newblock \bibinfo{publisher}{Cambridge University Press}.
\newblock


\bibitem[Sewell et~al\mbox{.}(2010)]%
        {sewell-et-al:ott}
\bibfield{author}{\bibinfo{person}{Peter Sewell},
  \bibinfo{person}{Francesco~Zappa Nardelli}, \bibinfo{person}{Scott Owens},
  \bibinfo{person}{Gilles Peskine}, \bibinfo{person}{Thomas Ridge},
  \bibinfo{person}{Susmit Sarkar}, {and} \bibinfo{person}{Rok Strni{\v s}a}.}
  \bibinfo{year}{2010}\natexlab{}.
\newblock \showarticletitle{Ott: Effective tool support for the working
  semanticist}.
\newblock \bibinfo{journal}{\emph{Journal of Functional Programming}}
  \bibinfo{volume}{20}, \bibinfo{number}{1} (\bibinfo{year}{2010}),
  \bibinfo{pages}{71–122}.
\newblock
\urldef\tempurl%
\url{https://doi.org/10.1017/S0956796809990293}
\showDOI{\tempurl}


\bibitem[Sistla and Godefroid(2004)]%
        {sistla-godefroid:symmetry-and-reduced-symmetry-in-model-checking}
\bibfield{author}{\bibinfo{person}{A.~Prasad Sistla} {and}
  \bibinfo{person}{Patrice Godefroid}.} \bibinfo{year}{2004}\natexlab{}.
\newblock \showarticletitle{Symmetry and reduced symmetry in model checking}.
\newblock \bibinfo{journal}{\emph{ACM Trans. Program. Lang. Syst.}}
  \bibinfo{volume}{26}, \bibinfo{number}{4} (\bibinfo{date}{July}
  \bibinfo{year}{2004}), \bibinfo{pages}{702–734}.
\newblock
\showISSN{0164-0925}
\urldef\tempurl%
\url{https://doi.org/10.1145/1011508.1011511}
\showDOI{\tempurl}


\bibitem[Stark(2020)]%
        {kathrin-stark:thesis}
\bibfield{author}{\bibinfo{person}{Kathrin Stark}.}
  \bibinfo{year}{2020}\natexlab{}.
\newblock \emph{\bibinfo{title}{Mechanising Syntax with Binders in Coq}}.
\newblock \bibinfo{thesistype}{Ph.\,D. Dissertation}. \bibinfo{school}{Saarland
  University}.
\newblock


\bibitem[Stark et~al\mbox{.}(2019)]%
        {stark-et-al:autosubst-2}
\bibfield{author}{\bibinfo{person}{Kathrin Stark}, \bibinfo{person}{Steven
  Sch{\"a}fer}, {and} \bibinfo{person}{Jonas Kaiser}.}
  \bibinfo{year}{2019}\natexlab{}.
\newblock \showarticletitle{Autosubst 2: Reasoning with Multi-Sorted de Bruijn
  Terms and Vector Substitutions}.
\newblock \bibinfo{journal}{\emph{8th ACM SIGPLAN International Conference on
  Certified Programs and Proofs, CPP 2019, Cascais, Portugal, January 14-15,
  2019}} (\bibinfo{year}{2019}).
\newblock


\bibitem[Swierstra(2008)]%
        {swierstra:data-types-a-la-carte}
\bibfield{author}{\bibinfo{person}{Wouter Swierstra}.}
  \bibinfo{year}{2008}\natexlab{}.
\newblock \showarticletitle{Data types {\`{a}} la carte}.
\newblock \bibinfo{journal}{\emph{J. Funct. Program.}} \bibinfo{volume}{18},
  \bibinfo{number}{4} (\bibinfo{year}{2008}), \bibinfo{pages}{423--436}.
\newblock
\urldef\tempurl%
\url{https://doi.org/10.1017/S0956796808006758}
\showDOI{\tempurl}


\bibitem[Szamozvancev(2026)]%
        {szamozvancev:thesis}
\bibfield{author}{\bibinfo{person}{Dmitrij Szamozvancev}.}
  \bibinfo{year}{2026}\natexlab{}.
\newblock \emph{\bibinfo{title}{{Categorical models of second-order abstract
  syntax}}}.
\newblock \bibinfo{thesistype}{Ph.\,D. Dissertation}.
  \bibinfo{school}{{University of Cambridge}}.
\newblock


\bibitem[{The mathlib Community}(2020)]%
        {lean-mathlib}
\bibfield{author}{\bibinfo{person}{{The mathlib Community}}.}
  \bibinfo{year}{2020}\natexlab{}.
\newblock \showarticletitle{The {L}ean {M}athematical {L}ibrary}. In
  \bibinfo{booktitle}{\emph{Proceedings of the 9th {ACM} {SIGPLAN}
  International Conference on Certified Programs and Proofs}}
  \emph{(\bibinfo{series}{CPP 2020})}. \bibinfo{publisher}{ACM},
  \bibinfo{address}{New Orleans, LA, USA}.
\newblock
\urldef\tempurl%
\url{https://doi.org/10.1145/3372885.3373824}
\showDOI{\tempurl}


\bibitem[Unruh(2024)]%
        {unruh:wlog-archive-of-formal-proofs}
\bibfield{author}{\bibinfo{person}{Dominique Unruh}.}
  \bibinfo{year}{2024}\natexlab{}.
\newblock \showarticletitle{Without Loss of Generality}.
\newblock \bibinfo{journal}{\emph{Archive of Formal Proofs}}
  (\bibinfo{date}{August} \bibinfo{year}{2024}).
\newblock
\showISSN{2150-914x}
\newblock
\shownote{\url{https://isa-afp.org/entries/Wlog.html}, Formal proof
  development}.


\bibitem[Urban(2008)]%
        {urban:nominal-isabelle/hol}
\bibfield{author}{\bibinfo{person}{Christian Urban}.}
  \bibinfo{year}{2008}\natexlab{}.
\newblock \showarticletitle{Nominal Techniques in {Isabelle/HOL}}.
\newblock \bibinfo{journal}{\emph{J. Autom. Reason.}} \bibinfo{volume}{40},
  \bibinfo{number}{4} (\bibinfo{year}{2008}), \bibinfo{pages}{327--356}.
\newblock
\urldef\tempurl%
\url{https://doi.org/10.1007/S10817-008-9097-2}
\showDOI{\tempurl}


\bibitem[Urban and Berghofer(2006)]%
  {urban-berghofer:a-recursion-combinator-for-nominal-datatypes-implemented-in-isabelle-hol}
\bibfield{author}{\bibinfo{person}{Christian Urban} {and}
  \bibinfo{person}{Stefan Berghofer}.} \bibinfo{year}{2006}\natexlab{}.
\newblock \showarticletitle{A Recursion Combinator for Nominal Datatypes
  Implemented in Isabelle/HOL}. In \bibinfo{booktitle}{\emph{Automated
  Reasoning}}, \bibfield{editor}{\bibinfo{person}{Ulrich Furbach} {and}
  \bibinfo{person}{Natarajan Shankar}} (Eds.). \bibinfo{publisher}{Springer
  Berlin Heidelberg}, \bibinfo{address}{Berlin, Heidelberg},
  \bibinfo{pages}{498--512}.
\newblock
\showISBNx{978-3-540-37188-5}


\bibitem[Wadler(1989)]%
        {wadler:theorems-for-free}
\bibfield{author}{\bibinfo{person}{Philip Wadler}.}
  \bibinfo{year}{1989}\natexlab{}.
\newblock \showarticletitle{Theorems for Free!}. In
  \bibinfo{booktitle}{\emph{Proceedings of the fourth international conference
  on Functional programming languages and computer architecture, {FPCA} 1989,
  London, UK, September 11-13, 1989}},
  \bibfield{editor}{\bibinfo{person}{Joseph~E. Stoy}} (Ed.).
  \bibinfo{publisher}{{ACM}}, \bibinfo{pages}{347--359}.
\newblock
\urldef\tempurl%
\url{https://doi.org/10.1145/99370.99404}
\showDOI{\tempurl}


\bibitem[Wahl(2007)]%
        {wahl-damm-hermanns:adaptive-symmetry-reduction}
\bibfield{author}{\bibinfo{person}{Thomas Wahl}.}
  \bibinfo{year}{2007}\natexlab{}.
\newblock \showarticletitle{Adaptive Symmetry Reduction}. In
  \bibinfo{booktitle}{\emph{Computer Aided Verification}},
  \bibfield{editor}{\bibinfo{person}{Werner Damm} {and} \bibinfo{person}{Holger
  Hermanns}} (Eds.). \bibinfo{publisher}{Springer Berlin Heidelberg},
  \bibinfo{address}{Berlin, Heidelberg}, \bibinfo{pages}{393--405}.
\newblock
\showISBNx{978-3-540-73368-3}


\bibitem[Wahl and D'Silva(2010)]%
        {wahl-d'silva:a-lazy-approach-to-symmetry-reduction}
\bibfield{author}{\bibinfo{person}{Thomas Wahl} {and}
  \bibinfo{person}{Vijay~Victor D'Silva}.} \bibinfo{year}{2010}\natexlab{}.
\newblock \showarticletitle{A lazy approach to symmetry reduction}.
\newblock \bibinfo{journal}{\emph{Formal Aspects Comput.}}
  \bibinfo{volume}{22}, \bibinfo{number}{6} (\bibinfo{year}{2010}),
  \bibinfo{pages}{713--733}.
\newblock
\urldef\tempurl%
\url{https://doi.org/10.1007/S00165-009-0131-X}
\showDOI{\tempurl}


\bibitem[Walsh(2012)]%
        {DBLP:conf/aaai/Walsh12}
\bibfield{author}{\bibinfo{person}{Toby Walsh}.}
  \bibinfo{year}{2012}\natexlab{}.
\newblock \showarticletitle{Symmetry Breaking Constraints: Recent Results}. In
  \bibinfo{booktitle}{\emph{Proceedings of the Twenty-Sixth {AAAI} Conference
  on Artificial Intelligence, July 22-26, 2012, Toronto, Ontario, Canada}},
  \bibfield{editor}{\bibinfo{person}{J{\"{o}}rg Hoffmann} {and}
  \bibinfo{person}{Bart Selman}} (Eds.). \bibinfo{publisher}{{AAAI} Press},
  \bibinfo{pages}{2192--2198}.
\newblock
\urldef\tempurl%
\url{https://doi.org/10.1609/AAAI.V26I1.8437}
\showDOI{\tempurl}


\bibitem[Weirich et~al\mbox{.}(2011)]%
        {weirich-et-al:binders-unbound}
\bibfield{author}{\bibinfo{person}{Stephanie Weirich},
  \bibinfo{person}{Brent~A. Yorgey}, {and} \bibinfo{person}{Tim Sheard}.}
  \bibinfo{year}{2011}\natexlab{}.
\newblock \showarticletitle{Binders unbound}.
\newblock \bibinfo{journal}{\emph{SIGPLAN Not.}} \bibinfo{volume}{46},
  \bibinfo{number}{9} (\bibinfo{date}{Sept.} \bibinfo{year}{2011}),
  \bibinfo{pages}{333–345}.
\newblock
\showISSN{0362-1340}
\urldef\tempurl%
\url{https://doi.org/10.1145/2034574.2034818}
\showDOI{\tempurl}


\bibitem[{Wikipedia contributors}(2026)]%
        {wiki:wlog}
\bibfield{author}{\bibinfo{person}{{Wikipedia contributors}}.}
  \bibinfo{year}{2026}\natexlab{}.
\newblock \bibinfo{title}{Without loss of generality --- {Wikipedia}{,} The
  Free Encyclopedia}.
\newblock
\urldef\tempurl%
\url{https://en.wikipedia.org/w/index.php?title=Without_loss_of_generality&oldid=1336885984}
\showURL{%
\tempurl}
\newblock
\shownote{[Online; accessed 8-March-2026]}.


\bibitem[Zhao et~al\mbox{.}(2025)]%
        {zhao2025symmetryneuralnetworkparameter}
\bibfield{author}{\bibinfo{person}{Bo Zhao}, \bibinfo{person}{Robin Walters},
  {and} \bibinfo{person}{Rose Yu}.} \bibinfo{year}{2025}\natexlab{}.
\newblock \bibinfo{title}{Symmetry in Neural Network Parameter Spaces}.
\newblock
\showeprint[arxiv]{2506.13018}~[cs.LG]
\urldef\tempurl%
\url{https://arxiv.org/abs/2506.13018}
\showURL{%
\tempurl}


\end{thebibliography}

\newpage
\ifthenelse{\boolean{WithSupplementary}}{%
\appendix
\section*{Supplementary material}

We proceed as follows:
\begin{itemize}
\item \appref{indexed adts} presents the theory of symmetric indexed
  data types, as it follows almost identical structure to its
  simply-typed counterpart.
\item \appref{simply-typed wlog details} contains further details
  about the simply-typed \WLOG{} construction (\secref{wlog construction}).
\item \appref{dependent types} contains further details
  about the dependently-typed \WLOG{} construction (\secref{dependent
    types}).
\item \appref{algebraic data types} contains further details and
  proofs about symmetric simply typed data types (\subsecref{initial
    algebras}--\subsecref{partial evaluation}).  This material is a
  special case of its dependently-typed counterparts. Much of it can
  also be done more abstractly and categorically. Here we prove much
  of it more concretely and with fewer abstractions to help make the
  material more accessible.
\item \appref{indexed data types} contains further details and proofs
  about symmetric dependently typed data types (\subsecref{indexed
    data types} and \appref{indexed adts}). Here we give concrete
  descriptions of additional auxiliary results, and omitted details
  from concrete examples.
\item \appref{cat dev} contains a more abstract category theoretic
  development that unifies parts of both theories of data types
  (cf.~summary figure in \subsecref{indexed data types}).
\end{itemize}

\section{Indexed algebraic data types}
\applabel{indexed adts}

In \subsecref{indexed data types} we summarised the changes to the
simply-typed development of symmetric data types. In this appendix, we
spell the theory out in full detail, for completeness.

To extend the theory to indexed data types $\carrier\Ctx \types
\TConst D$, we use endofunctors $\struct[\TConst D] :
\Fam_{\carrier\Ctx} \to \Fam_{\Ctx}$ over the category
$\Fam_{\carrier\Ctx}$ of families indexed by a set $\carrier\Ctx$ and
vertical maps between them. The data type $\TConst D$ denotes an
initial $\struct[\TConst D]$-algebra $\carrier\Ctx \types \roll :
\struct[\TConst D] \TConst D \to \TConst D$. As in the non-indexed
case, we equip those endofunctors with a symmetry strength relative to
an ambient $\G$-action $\Ctx = \pair{\carrier\Ctx}{(\act)}$ on the set
of indices. This strength will be a certain natural transformation
$\strength : \gast \struct[\TConst D] \X \to \struct[\TConst D](\gast
\X)$, where $(\gast-)$ is the functor part of the \emph{free
$\Ctx$-indexed $\G$-action} monad.  Initial algebras for such
endofunctors enjoy an indexed version of the parameterised initiality
and representation theorems, and stating and establishing those is our
main goal.

We'll use the indexed family $\mt : \Tree \types \NodeIn \mt$ as our
running example, but unlike \exampleref{node lookup}, we avoid the
redundant cases through symmetric programming. As our endofunctor, we
take:
\begin{gather*}
(\struct[\NodeIn]\X)_{\Leaf} \definedby \emptyset
\qquad
(\struct[\NodeIn]\X)_{\Node\,\ell\, n\,r} \definedby
  \Nil \Or \Lft\:: \X_{\ell} \Or \Rgt\:: \X_{r}
\\
(\fmap f)_{u}\Nil \definedby \Nil
\qquad
(\fmap f)_{u}(\Lft\:: x) \definedby \Lft\:: f_{\ell}x
\qquad
(\fmap f)_{u}(\Rgt\:: y) \definedby \Rgt\:: f_{r}y
\tag{where: $\mt : \Tree \types f : \X \to \Y$, $u = \Node\, \ell\, n\, r$}
\end{gather*}
The initial algebra has as carrier $\mt :
\Tree \types \NodeIn\mt$ from \exampleref{families example}.

We denote the multiplication
of the free $\Ctx$-indexed $\G$-action monad $P$ by:
\[
(\muast) : \gast(\gast \X) \to \gast\X
\qquad
\pair g{\pair h{x : \X_{\inv h \act \inv g \act \ctx}}} \definedby
\pair{g\mul h}{x : \X_{\inv*{g\mul h}\act\ctx}}
\]
It is the functor $(\gast)$ that generalises the notion of strength to
the indexed setting:
\begin{definition}\definitionlabel{indexed symmetry strength}
  Let $\Ctx$ be a $\G$-action and $F : \Fam_{\carrier\Ctx} \to
  \Fam_{\carrier\Ctx}$ be an endofunctor. A \emph{$\Ctx$-indexed
  $\G$-symmetry strength} is a natural transformation $\strength :
  \gast F\X \to F(\gast \X)$, satisfying:
  \begin{gather*}
  \strength_{\ctx}(\unit, u) = \fmap(\fun x\pair\unit x) u
  \qquad
  \fmap(\muast)
  \compose
    \strength(g, \strength(h, x)) =
  \strength(g\mul h, x)
  \end{gather*}
\end{definition}

\begin{example}%
  \examplelabel{path strength}
  We define an $\Sym2$-symmetry strength for $\mt : \Tree \types \NodeIn\mt$.
  Only the $\swap$ case is interesting, as the first strength axiom forces the
  definition for $\stet$:
  \begin{gather*}
  \strength[\NodeIn]_{\parent{\NodeV \ell n r}}
  \pair{\text{$\swap$}}{\Nil : \NodeIn {\NodeV{\swap\act r}n{\swap\act \ell}}}
  \definedby \Nil : \NodeIn \parent{\NodeV{\ell}n{r}}
  \\
  \strength[\NodeIn]_{\parent{\NodeV \ell n r}}\pair\swap{\Lft \:: (x : \X_{\swap\act r})} \definedby
  \Rgt \:: \pair\swap{x : \X_{\swap \act r}} \in (\struct[\NodeIn](\gast\X))_{\parent{\NodeV \ell nr}}
  \\
  \strength[\NodeIn]_{\parent{\NodeV \ell n r}}\pair\swap{\Rgt \:: (x : \X_{\swap\act \ell})} \definedby
  \Lft \:: \pair\swap{x : \X_{\swap \act \ell}} \in (\struct[\NodeIn](\gast\X))_{\parent{\NodeV \ell nr}}
  \end{gather*}
  To prove the second strength axiom, only the case $g \definedby h
  \definedby \swap$ is interesting though straightforward:
  \begin{align*}
    \fmap(\muast)&\compose\strength(\swap, \strength(\swap, \Nil))
    = \fmap(\muast)\compose\strength(\swap, \Nil)
    = \fmap(\muast)\Nil
    = \Nil
    \\&= \strength(\swap\mul\swap, \Nil)
    \\
    \fmap(\muast)&\compose\strength(\swap, \strength(\swap, \Lft\:: x))
    = \fmap(\muast)\compose\strength(\swap, \Rgt\::\pair\swap x)
    \\&= \fmap(\muast)(\Lft\::\pair{\swap\mul\swap} x)
    = \Lft \:: \pair\stet x
    = \strength(\swap\mul\swap, \Lft \::x)
  \end{align*}
  and symmetrically for $\Rgt::x$. We cannot avoid this
  symmetric argument using wlog as we are setting up the symmetric
  action by defining the strength. Later uses of $\wlg{}$
  will exploit this work.
\end{example}

The appropriate category-theoretic concept that abstracts this notion
of strength is a \emph{monad-functor distributive law}, and we will
use this perspective in our category theoretic development in
\appref{cat dev}. As may be clear, when $\terminal \types X$ is a
single-fibred family, $\terminal \types \gast X$ is the single-fibred
family $\carrier G \times X$, subsuming the simply-typed case.

The following lemma justifies checking only the interesting cases, as
we did in \exampleref{path strength}:
\begin{lemma}\lemmalabel{indexed check only non-trivial}
  Let $\strength$ be the structure of an indexed $\G$-symmetric
  strength, satisfying the first axiom. Then the second axiom holds
  when either $g = \unit$ or $h = \unit$.
\end{lemma}

Our main initiality theorem uses an indexed notion of equivariant
algebras (cf.~\definitionref{equivariant F-algebra}):
\begin{definition}
  Let $\Ctx$ be a $\G$-action, $F : \Fam_{\carrier\Ctx} \to
  \Fam_{\carrier\Ctx}$ an endofunctor equipped with a symmetry
  strength. For every indexed action $\A$, the \emph{lifted action}
  is given by:
  \[
  \carrier{F_{\strength}\A} \definedby F\carrier\A
  \qquad
  g \iact\ctx u \definedby \fmap (\act) \strength_{g\act\ctx}\pair gt
  \]
  An \emph{equivariant $F$-algebra} $\pair\A f$ for an endofunctor
  equipped with a symmetry strength, consists of an indexed action
  $\A$ equipped with an equivariant algebra structure $\Ctx \types f :
  F_{\strength}\A \equito \A$. An equivariant $F$-algebra homomorphism is
  then a family map between the carrier families of two such algebras
  that is simultaneously equivariant and $F$-algebra homomorphic.
\end{definition}

We generalise the Equivariant Initiality \theoremref{representation for algebras}
to indexed families:
\begin{theorem}[indexed equivariant initiality]%
  \theoremlabel{indexed representation}
  Let $\roll : F\fix F \to \fix F$ be an initial algebra for a
  functor $F$ over $\Ctx$-indexed families with a $\G$-symmetry strength
  $\strength : \gast
  F\X \to F(\gast\X)$. There is a unique map $(\act) : \gast \fix F \to \fix F$
  satisfying \(
  g \act (\roll u)
  =
  \roll (\fmap(\act)(\strength(g, u)))
  \).
  It makes $\pair{\fix f}{\roll}$
  into an initial equivariant $F$-algebra: for every
  equivariant $F$-algebra $\pair\A{f : F_{\strength}\A \equito \A}$,
  the
  $F$-homomorphism into $\carrier\A$
   is equivariant: $\fold f :
  \fix F \equito \A$.
\end{theorem}

Like its simply-typed variant, the proof of this theorem uses an
appropriate notion of parameterised initiality originally proposed by
\citeauthor{bird-paterson:generalised-folds-for-nested-datatypes}~\cite{%
  bird-paterson:generalised-folds-for-nested-datatypes}.

\begin{example}
  The action induced by the Indexed Equivariant Initiality
  \theoremref{indexed representation} satisfies:
  \begin{align*}
  \swap\act(\Lft \:: \Rgt \:: \Nil) &=
  \roll\compose\fmap(\act)\strength(\swap, \Lft \:: (\Rgt \:: \Nil)) =
  \roll\compose\fmap(\act)(\Rgt \:: \pair\swap{\Rgt \:: \Nil})
  \\&=
  \Rgt\::(\swap \act (\Rgt \:: \Nil))=
  \Rgt\::\Lft \:: (\swap \act \Nil)=
  \Rgt\::\Lft \:: \Nil
  \end{align*}
  as we may expect from the example at the beginning of \subsecref{indexed actions}.
\end{example}

\begin{example}%
  \examplelabel{wlog node lookup}%
  Let's implement the dependently-typed node lookup function and
  simultaneously show its equivariance $\lookup :
  \DPair\mt{\Tree}{\NodeIn\mt} \equito \naturals$. We'll
  construct an indexed equivariant map $\mt : \Tree \types
  \lookup_{\mt} : \NodeIn\mt \equito \naturals$. By the Indexed
  Equivariant Initiality \theoremref{indexed representation}, it
  suffices to equip $\naturals$ with an equivariant
  $\struct[\NodeIn]$-algebra structure $\lookupElim : \struct[\NodeIn]
  \naturals \equito \naturals$. Informally, \WLOG{}, the path to the
  node is either empty or starts with the $\Lft$eft subtree,
  eliminating one symmetric case. Formally, define a
  canoniser by
  \(
  \canoniser\Nil \definedby \canoniser(\Lft\:: n) \definedby \stet
  \) and \(
  \canoniser(\Rgt\:: n) \definedby \swap
  \).
  Its canonical elements are $\Nil$ and $\Lft\::n$.
  This canoniser is equivariant, but not strictly:
  \begin{align*}
  \canoniser(\swap \act \Nil) &=
  \canoniser\Nil = \stet \stabeq{\Nil} \swap = \stet\mul\inv\swap =
  \swap\act \canoniser\,\Nil \\
  \canoniser(\swap \act (\Lft \:: n)) &=
  \canoniser(\Rgt \:: n) =
  \swap = \stet \mul \inv\swap = \swap\act\canoniser(\Lft\:: n)
  \end{align*}
  and symmetrically for $\Rgt \:: n$. Again, we cannot avoid this
  symmetry with wlog since we are setting up the symmetric argument.

  Let's define the core function for
  the eliminator $\mt : \Tree \types \lookupCore :
  \Canon\canoniser \to \naturals$:
  \[
  \lookupCore_{\parent{\Node\, \ell\,n\,  r}} \Nil \definedby n
  \qquad
  \lookupCore_{\parent{\Node\,\ell\,n\,  r}} (\Lft \:: p) \definedby p
  \]
  and we do not need to consider the symmetric case.
  The stabiliser subgroup for $\naturals$ is the whole of $\Sym2$
  since its action is trivial. Consequently, every core function is
  stable, and we don't need to work out the stabilisers for canonical
  elements. For completeness, here they are.  The stabiliser subgroup
  for $\Nil$ is the whole of $\Sym2$, and the stabiliser subgroup for
  $\Lft\::p$ is the trivial subgroup $\set\stet$.

  We use the
  Indexed Representation \theoremref{indexed-wlog-construction} to
  show the extended function is equivariant: $\lookupElim \definedby
  \wlg(c,f) : \struct[\NodeIn]\naturals \equito \naturals$.
  \end{example}

As in the simply-typed case (cf.~\subsecref{wlog-traversals}), we
reuse the same canoniser for multiple traversals of the same symmetric
indexed data type, and streamline the steps in \exampleref{wlog node lookup}.

\begin{definition}
  Let $\Ctx$ be a $\G$-action, and $F$ be a $\Ctx$-indexed
  $\G$-strong.  An \emph{indexed canoniser} for $F$ is a natural
  transformation $\carrier\Ctx \types \canoniser : F \to
  \carrier\G$. Each such canoniser induces the \emph{canonical
  elements} functor $\Canon\canoniser : \Action\G_{\Ctx} \to
  \Fam_{\carrier\Ctx}$ given by:
  \[
  ((\Canon\canoniser)\A)_{\ctx} \definedby
  (\Canon[F\A]\canoniser_{\carrier\A})_{\ctx} \definedby
  \set{
    g \iact[F\A]\dtx u \suchthat
    \begin{array}{@{}l@{}}
      g = \canoniser_{\carrier\A; \dtx}u, g \act \dtx = \ctx,\\
      u \in (\carrier F\carrier\A)_{\dtx}
    \end{array}
  }
  \qquad
  (\Canon\canoniser)h : \fun u (\carrier Fh)u
  \]
\end{definition}

An indexed canoniser is uniquely determined by its component at the
terminal family $\carrier\Ctx \types \terminal$.

\begin{example}
  The indexed $\Sym 2$-strong functor $\struct[\NodeIn]$ has the
  canoniser (cf.\exampleref{goLeft canoniser})
  \[
  \goLeft_{\Node n\,\ell\,r}\,\Nil \definedby \stet \defines \goLeft_{\Node \ell\,n\,r} (\Lft\:: p)
  \quad
  \goLeft_{\Node \ell\,n\,r} (\Rgt\:: p) \definedby \swap
  \]
  The canonical elements are:
  \[
  (\Canon[\A]\goLeft)_{\Leaf} \definedby \emptyset
  \qquad
  (\Canon[\A]\goLeft)_{\Node \ell\,n\,r} \definedby
  \set{\Nil, \Lft\:: a \suchthat a \in \carrier\A_{\Node \ell\,n\,r}}
  \tag*{\qedhere}
  \]
\end{example}

Continuing in tandem with the development of \subsecref{indexed data types},
we define the data we use to traverse an indexed symmetric data type:

\begin{definition}\definitionlabel{indexed core algebras}
  Let $F$ be a $\Ctx$-indexed $\G$-strong functor and $\canoniser : \carrier F \to
  \carrier\G$ a canoniser for it. An \emph{indexed $\canoniser$-core
  $F$-algebra} $\pair\A f$ consists of a $\Ctx$-indexed $\G$-action $\A$ equipped
  with a vertical map $\carrier\Ctx \types f : (\Canon\canoniser)\A \to \carrier\A$. It is
  \emph{equivariant} when $f$ is equivariant $\Ctx \types f :
  \parent{(\Canon\canoniser)\A \subset F\A} \equito \A$. A
  \emph{homomorphism} of equivariant indexed $\canoniser$-core $F$-algebras $h
  : \pair\A f \to \pair \B g$ is an equivariant map $\Ctx \types h : \A \equito
  \B$ satisfying: $h (f\, u) = g\, ((\Canon\canoniser) h\,u)$.
\end{definition}

\begin{example}\examplelabel{dependently typed lookup as core algebra}
  We have the following equivariant $\goLeft$-core $\struct[{\NodeIn}]$-algebra:
  \begin{gather*}
  \lookupCore : (\Canon\goLeft)(\Triv{\naturals}) \to
  \carrier{\Triv{\naturals}} = \naturals
  \\
  \lookupCore_{\Node \ell\,n\,r}
  \Nil \definedby n
  \qquad
  \lookupCore_{\Node \ell\,n\,r}
  \parent{
    \Lft\:: (k \in \naturals)
  } \definedby k
  \tag*{\qedhere}
  \end{gather*}
\end{example}

Next, chain together the traversal and \WLOG{} constructions:
\[
\carrier\Ctx\types
\parent{\Canon[F\A]\canoniser_{\carrier\A} = (\Canon\canoniser)\A \xto f \carrier\A}
\mapsto
\parent{F\carrier \A \xto{\wlg(\canoniser_{\carrier\A}, f)} \carrier\A}
\mapsto
\parent{\flg[\A](\canoniser, f) :
  \fix\carrier F \xto{\fold \wlg(\canoniser_{\carrier\A}, f)} \carrier \A}
\]
And similarly chain \theoremref{indexed-wlog-construction} and the
Indexed Equivariant Initiality \theoremref{indexed representation}:

\begin{theorem}[indexed core initiality]
  \theoremlabel{indexed representation for core algebras}
  Let $F$ be a $\Ctx$-indexed $\G$-strong functor, $\carrier\Ctx \types
  \roll : \carrier F\fix
  \carrier F \to \fix \carrier F$ an initial algebra of $\carrier F$,
  and $\canoniser$ a canoniser for $F$. Then restricting $\roll$ to canonical elements
  makes $\fix F$ into the initial indexed $\canoniser$-core $F$-algebra
  $\roll : (\Canon\canoniser)\fix F \equito \fix F$:
  for any equivariant indexed $\canoniser$-core $F$-algebra ${f :
    (\Canon\canoniser)\A \equito A}$, the unique equivariant
  indexed $\canoniser$-core $F$-algebra homomorphism
  into $f$ is:\/ $\flg(\canoniser, f) : \fix F \equito \A$.
\end{theorem}

This theorem implies, by \exampleref{dependently typed lookup as core
  algebra}, that we have an equivariant function:
\[
t : \Tree \types \lookup \definedby
\flg(\goLeft, \lookupCore) : \text{$\NodeIn t \equito \Triv \naturals$}
\mathbf{}
\]

\section{Simply-typed \WLOG: further details}
\applabel{simply-typed wlog details}

In this Appendix, we provide further details to \secref{wlog construction}.
Our first proof obligation is a comment we immediately before \exampleref{tree example}:

\begin{lemma}[well-known]\lemmalabel{action exponential}
  Let $\A$ and $\B$ be $\G$-actions. The exponential $\A \to \B$ in $\Action\G$ is given by the set of functions between their carriers:
  \begin{gather*}
  \carrier{\A \to \B}\definedby \carrier\A \to \carrier\B
  \qquad
  (g \act[\A \to \B] f) \definedby \fun a g \act[\B] f(\inv g\act[\A]a)
  \\
  \eval : (\A \to \B) \times \A \equito \B
  \qquad
  \eval \pair fa \definedby f\,a
  \end{gather*}
\end{lemma}
\begin{proof}
  We indeed have an action. Unit preservation:
  \[
  (\unit[\G] \act[\A \to \B] f)a
  = \unit[\G]\act[\B] f(\inv{\unit[\G]}\act a) = f\,a
  \implies
  \unit[\G] \act[\A \to \B] f = f
  \]
  Action preservation:
  \begin{align*}
  (g \act[\A \to \B] h \act[\A\to\B] f)a
  &=
  g \act[\B] (h \act[\A\to\B] f)(\inv g \act[\A] a)
  =
  g \act[\B] (h \act[\B] f(\inv h \act[\A] (\inv g \act[\A] a)))
  \\&=
  (g \mul[\G] h) \act[\B] f\parent{\inv*{\smash{g \mul[\G] h}}\act[\A] a}
  =
  \parent{(g \mul[\G] h) \act[\A \to \B] f} a
  \\&\implies
  (g \act[\A \to \B] h \act[\A\to\B] f) = (g \mul[\G] h) \act[\A \to \B] f
  \end{align*}

  Evaluation is equivariant. Calculate:
  \begin{align*}
  \eval(g \act[(\A\to\B)\times \A] \pair fa)
  &=
  \eval\pair{g\act[\A\to\B] f}{g\act[\A] a}
  =
  (g\act[\A\to\B]f)\,(g\act[\A] a)
  \\&=
  g \act[\B] (f(\inv g \act[A] (g \act[\A] a)))
  =
  g \act[\B] (f\,a)
  =
  g \act[\B] \eval\pair fa
  \tag*{\qedhere}
  \end{align*}

  Given any equivariant map $h : \C \times \A \equito \B$, its curried
  map is also equivariant:
  \begin{align*}
    (g \act[\A \To \B] \curry h c)a
    &=
    g \act[\B] \curry h (c; \inv g \act[\A] a)
    =
    g \act[\B] h (c, \inv g \act[\A] a)
    \explain={equivariance}
    h (g \act[\C] c, g \act \inv g \act[\A] a)
    \\&=
    h (g \act[\C] c, a)
    =
    (\curry h (g \act[\C]c) a
    \\&\implies\quad
    g \act[\A \to \B] \curry h c = \curry h (g \act[\C] c)
  \end{align*}
  and so $\curry h : \C \equito (\A \to \B)$. The curried function
  lifts $h$ along evaluation. Now if we have an equivariant map $g :
  \C \equito (\A \to \B)$ that lifts $h$ along evaluation, then by the
  set-theoretic argument, it coincides with $\curry h$.
\end{proof}

Next, we characterise the canonical elements of
canonisers of the form $\carrier \A \times \carrier \B \xto{\projection_1}\A \xto\canoniser \carrier\G$:

\begin{proof}[of \lemmaref{projection canoniser}]%
  Take an arbitrary canonical element $(\canoniser \compose \projection_1) \seq{a, b}
  \act \seq{a, b} \in \Canon{\parent{\canoniser \compose
  \projection_1}}$. We then have
  \[
    (\canoniser \compose \projection_1) \seq{a, b} \act \seq{a, b}
    = \canoniser a \act \seq{a, b}
    = \seq{\canoniser a \act a, \canoniser a \act b}
    \in \Canon\canoniser \times \carrier\B
  \]
  Conversely, given an arbitrary pair $\seq{\canoniser a \act a, b} \in
  \Canon\canoniser \times \carrier\B$, let $b' = \inv*{\canoniser a} \act b$.
  We then have:
  \[
    \seq{\canoniser a \act a, b}
    = \seq{\canoniser a \act a, (\canoniser a \mul \inv*{\canoniser a}) \act b}
    = \canoniser a \act \seq{a, b'}
    = ((\canoniser \compose \projection_1) \seq{a, b'}) \act \seq{a, b'}
    \in \Canon{\parent{\canoniser \compose \projection_1}}
  \]
  Therefore $\Canon{\parent{\canoniser \compose
  \projection_1}} = \Canon\canoniser \times \carrier\B$.
\end{proof}

Next, we proof the main equivariance theorem for the \WLOG{} construction:
\begin{proof}[of \theoremref{wlog-construction}]%
  Let us first show that $\wlg(c, f)$ is equivariant. Take arbitrary
  $g \in \carrier\G$ and $a \in \carrier\A$. By definition and by the
  group and action axioms, we can rewrite:
  \begin{align*}
    \wlg(c, f)(g \act a)
    &=
    \inv{c (g \act a)} \act f(c (g \act a) \act (g \act a)) \\
    &=
    \inv{c (g \act a)} \act f(c (g \act a) \act (g \act \inv{(c a)} \act c a \act a)) \\
    &=
    \inv{c (g \act a)} \act f\big((c (g \act a) \mul g \mul \inv{(c a)}) \act (c a \act a)\big) \\
    \intertext{
      Since both $c a \act a$ and $c (g \act a) \act (g \act a)$ belong to $\Canon{\canoniser}$, equivariance of $f$ allows us to rewrite:
    }
    &=
    \inv{c (g \act a)} \act \big((c (g \act a) \mul g \mul \inv{(c a)}\big) \act f(c a \act a)) \\
    &=
    g \act (\inv{(c a)} \act f(c a \act a)) \\
    &=
    g \act \wlg(c, f)\,a
  \end{align*}
  as desired.

  Next, let us show that $\wlg(c, f)$ extends $f$. Take an arbitrary
  canonical element $a \in \Canon\canoniser$. Since $c a \act a$
  also lies in $\in \Canon\canoniser$, the action axioms imply:
  \[
    \wlg(c, f)\,a
    =
    \inv{(c a)} \act f(c a \act a)
    \explain={$f$ is equivariant}
    \inv{(c a)} \act c a \act f a
    =
    f a
  \]

  Finally, take an arbitrary $\phi : \A \equito \B$ that extends $f$
  along $\Canon\canoniser\subset\carrier\A$. Then, for an arbitrary
  $a \in \carrier\A$, we have
  \[
    \phi a
    =
    \phi (\inv{(c a)} \act c a \act a)
    \explain={$\phi$ equivariant}
    \inv{(c a)} \act \phi (c a \act a)
    \explain={$\phi$ extends $f$}
    \inv{(c a)} \act f (c a \act a)
    \explain={by definition}
    \wlg(c, f)\,a
  \]
  as desired.
\end{proof}

The order of the next few results
differs from the order in which they appear in the main text.
They concern the characterisation of
strictly equivariant and immediate canonisers from
\subsecref{wlog-construction}.
Strict equivariance seem to be a more natural concept to the
audience when we present this work, so the main text starts there.
Here we first characterise immediate canonisers, and use this
characterisation as part of the proof. This order make it easier to
see that we avoid circular reasoning.

We start with immediate canonisers as every strictly equivariant
canoniser is also immediate:
\begin{proof}[of \propositionref{strictly equivariant canonisers are immediate}]%
  Calculate:
  \[
  \canoniser(\canoniser a \mul a)
  \explain={equivariance}
  \canoniser a \mul \inv*{\canoniser a}
  =
  \unit
  \]
  as desired.
\end{proof}

We characterise immediate canonisers
via the condition $\Canon\canoniser =
\inv\canoniser[\unit]$:

\begin{proof}[of \propositionref{immediate canonical is kernel}]%
  Take an arbitrary $\canoniser a \act a \in \Canon\canoniser$. Since
  $\canoniser$ is immediate, we have $\canoniser \parent{\canoniser a \act a}
  = \unit$, and so $\canoniser a \act a \in \inv{\canoniser}[\unit]$,
  thus $\Canon\canoniser \subseteq \inv{\canoniser}[\unit]$. The
  converse inclusion always holds. Indeed, if $a \in \inv{\canoniser}[\unit]$,
  then $\canoniser a = \unit$ and so $a = \canoniser a \act a \in
  \Canon\canoniser$.

  Conversely, let us show that $\canoniser$ is immediate and take an
  arbitrary $a \in \carrier\A$. We have $\canoniser a \act a \in
  \Canon\canoniser$ by definition, thus $\canoniser a \act a \in
  \inv{\canoniser}[\unit]$ by assumption, and so
  $\canoniser(\canoniser a \act a) = \unit$ as required.
\end{proof}

The next result characterises strict equivariance of a canoniser via the
equivariance of every core function. The proof uses the concept of
stabilisers that we introduce in \subsecref{equivariant canonisers}.
The proof is dependently-typed in nature, and it is best understood in
the context of \secref{dependent types} and \subsecref{equivariant
  canonisers} in particular. However, we make the proof self-contained
and re-introduce the concepts that are relevant for it.

\begin{proof}[of \theoremref{all core functions are partially equivariant}]%
  First, take a strictly equivariant canoniser $\canoniser :
  \carrier\A \to \carrier\G$, and let $\B$ be an arbitrary action. We
  need to show that every core function $f : \Canon\canoniser \to
  \carrier\B$ is equivariant. Take $a, g\act a \in
  \Canon\canoniser$. Since $\canoniser$ is strictly equivariant,
  by \propositionref{strictly equivariant canonisers are
  immediate}, it is an immediate canoniser. So by \propositionref{immediate canonical is kernel}, $\canoniser a = \unit =
  \canoniser (g \act a)$, therefore:
  \[
    \inv g = \canoniser a \mul \inv g
    \explain*={$\canoniser$ strictly equivariant}
    \canoniser\parent{g \act a}
    = \unit
  \]
  and equivariance holds vacuously.

  Conversely, assume that every core function is equivariant, and let us show that $\canoniser$ is strictly equivariant.
  Recall that for every $a \in \carrier\A$, we define the stabiliser sub-group of $a$ as the set of symmetries that fix $a$:
  \[
    \stabilise a \definedby \set{s \in \carrier\G \suchthat s \act a = a}
  \]
  Take the following action over the
  the set of dependent pairs consisting of an
  element in $\carrier\A$ and one
  of its stabilising symmetries:
  \[
    \carrier\B \definedby (a : \carrier\A) \times \stabilise a
    \qquad
    g \act \pair as \definedby \pair{g \act a}{s \mul \inv g}
  \]
  Indeed, this definition is well-typed because:
  \[
    (s \mul \inv g)\mul (g \act a)
    =
    s \act (\inv g \mul g) \act a
    =
    s \act a
    =
    a
  \]
  It is also an action, since:
  \begin{gather*}
    \unit \act \pair as =
    \pair{\unit \act a}{s \act \inv\unit} =
    \pair as
    \\
    g \act h \act \pair as
    =
    \pair{g \act h \act a}{s \mul \inv h \mul \inv g}
    =
    \pair{(g \mul h) \act a}{s \mul \inv*{g \mul h }}
    =
    (g\mul h)\act \pair as
  \end{gather*}
  Define:
  \[
    f : \Canon\canoniser \to \carrier\B
    \qquad
    f a \definedby \pair a\unit
  \]
  By our assumption, its \WLOG extension is equivariant: $\wlg(c,f) :
  \A \equito \B$. Note that this map is given by:
  \[
    \wlg(f,c)a = \inv*{\canoniser a} \act f\parent{\canoniser a\mul a}
    = \inv*{\canoniser a} \act \pair{\canoniser a\mul a}\unit
    = \pair{a}{\inv*{\inv*{\canoniser a}}}
    = \pair{a}{\canoniser a}
  \]
  Now take any $a \in \carrier\A$ and $g \in
  \carrier\G$, and calculate:
  \begin{align*}
    \pair{g\act a}{\canoniser\parent{g\act a}}
    =
    \wlg(c,f)(g\act a)
    =
    g\act \wlg(c,f)a
    =
    g\act \pair{a}{\canoniser a}
    =
    \pair{g \act a}{\parent{\canoniser a}\mul \inv g}
  \end{align*}
  Comparing the two second components, we have
  \(
    \canoniser\parent{g \act a} = (\canoniser a)\mul \inv g
  \) and the canoniser is strictly equivariant.
\end{proof}

The next result shows that actions that admit strictly equivariant
canonisers must be semi-regular, which makes this condition too strong
for some actions:

\begin{proof}[of \lemmaref{semi-regularity}]%
  Assume $g \act a = a$. By equivariance of the canoniser, we thus have
  $\canoniser a = \canoniser\parent{g \act a} = \canoniser a \mul \inv g$ and by cancelling
  $\canoniser a$ on the left, we get $\unit = g$.
\end{proof}

The final result concerns inheriting strict equivariance and
immediateness from a canoniser:

\begin{proof}[of \lemmaref{equivariance of projection canonisers}]%
  To show strict equivariance, we have that the projection is an equivariant map $\projection_1 : \A \times \B
  \equito \A$, and equivariant maps compose.
  For immediateness, we calculate
    \begin{align*}
    \canoniser\compose\projection_1\parent{(\canoniser\compose\projection_1{\pair ab})\mul{\pair ab}}
    &=
    \canoniser\compose\projection_1\parent{\canoniser a\mul{\pair ab}}
    =
    \canoniser\compose\projection_1\parent{\pair {\canoniser a\mul a}{\canoniser a\mul b}}
    \\&=
    \canoniser\parent{\canoniser{a} \mul a}
    =
    \unit
    \end{align*}
    and so $\canoniser \compose \projection_1$ is immediate when $\canoniser$ is.
\end{proof}

\section{Dependent types: further details}
\applabel{dependent types}

We begin by establishing the correctness of the action over the
Grothendieck construction $\coprod_{\carrier\Ctx}\A$, as we will use
its properties pervasively, e.g., when extending contexts.

\begin{proof}[of \propositionref{Grothendieck action}]%
  Take a $\Ctx$-indexed action $\A$ and define an action on
  $\coprod_{\carrier\Ctx} \carrier\A$ by:
  \[
  g \act \pair{\ctx : \carrier\Ctx}{a : \carrier\A_{\ctx}}
  \definedby \pair{g \act \ctx}{g \iact{\ctx} a : \carrier\B (g \act \ctx)}
  \]
  Since $\projection_1(g \act\pair\ctx a) = g \act \ctx = \projection_1\pair\ctx a$,
  the projection is equivariant
  $\projection_1 : \coprod_{\Ctx}\A \equito \Ctx$.

  Conversely, given an action $\coprod_{\Ctx}\A$ as in the assumption,
  define $g \iact{\ctx} a \definedby \projection_2(g \act
  \pair{\ctx}{a})$. This definition is a well-typed dependent function $\ctx : \Ctx \types (\iact{}) : (g : \carrier\G) \to A\ctx \to A (g \act \ctx)$:
  \[
  g \iact\ctx a \in A\parent{\projection_1(g \act \pair \ctx a)} =
  A\parent{\g \act \ctx}
  \]
  It also satisfies the two conditions for an indexed action. The unit axiom:
  \[
  \unit \iact \ctx a = \projection_2 (\unit \act \pair \ctx a) = \projection_2\pair\ctx a = a
  \]
  The multiplication axiom:
  \begin{align*}
  h \iact{g \act \ctx} (g \iact \ctx a)
  &=
  \projection_2\parent{h \act
    \pair{g\act \ctx}
         {\projection_2\parent{g \act \pair \ctx a}}}
  \\&
  \explain={$\projection_1$ equivariant}
  \projection_2\parent{h \act
    \pair{\projection_1\parent{g\act \pair\ctx a}}
         {\projection_2\parent{g \act \pair \ctx a}}}
  \\&
  \explain={extensionality of pairs}
  \projection_2\parent{h \act \parent{g \act \pair \ctx a}}
  =
  \projection_2\parent{(h \mul g) \act \pair \ctx a}
  =
  (h \mul g) \iact\ctx a
  \end{align*}
  and so we have an indexed action. The two constructions are inverse
  to each other.
\end{proof}

The following example lets us work more formally with concrete
examples:

\begin{example}\examplelabel{equivariant Grothendieck}
The Grothendieck construction extends equivariantly to equivariant
terms and maps.  It sends equivariant terms $\Ctx \types t : \A$ to
equivariant sections \((\fun \ctx \pair{\ctx}{t_{\ctx}}) : \Ctx
\equito \coprod_{\Ctx}\A \). It sends a dependent equivariant function
$\Ctx \types f : \DFun\x\A \B$ to the following equivariant map
$\parent{\fun{\pair\ctx a}\pair{\ctx, \For x a}{f_{\ctx}a}} :
\coprod_{\Ctx}\A \equito \coprod_{\Ctx, \x : \A}\B$.
\end{example}

Next, we establish the indexed action axioms for the family of
dependent functions, and the indexed equivariance of related
operations.

\begin{lemma}\lemmalabel{dependent function action}
  The indexed action structure over the family of dependent functions
  is an action, evaluation is equivariant, and currying and uncurrying
  preserve equivariance.
\end{lemma}
\begin{proof}
  Indeed, the action axioms hold by calculating with the correct
  indices as in the ordinary case. Similarly for evaluation:
  \[
  \eval(g\iact{\ctx} f,\x)_{g\act \ctx, \For \x g\iact\ctx a}
  =
  g \iact{} f(\inv g \iact{} g \iact{} a)
  =
  g \iact{} fa
  =
  g\iact{}\eval(f, \x)_{\ctx, \For\x a}
  \]
  The bijective correspondence is given as in the set-theoretic
  case. If $t$ is equivariant, then:
  \begin{align*}
  (\fun{\x:\A}t)_{g \act \ctx}a &=
  g \iact{} t_{g\act \ctx, \For\x a}
  =
  g \iact{} t_{g\act \ctx, \For\x (g \act (\inv g \act a))}
  \\&\explain={equivariance}
  g \iact{} g\iact{} t_{\ctx, \For\x (\inv g \act a)}
  =
  (g \iact{} (\fun{\x : \A}t)_{\ctx})a
  \end{align*}
  Therefore $(\fun{\x:\A}t)_{g \act \ctx} = (g \iact{} (\fun{\x : \A}t)_{\ctx})$.

  Conversely, if $\fun{(\x:\A)}t$ is equivariant, then since
  $\eval\pair{-}\x$ is equivariant, $t$ is.
\end{proof}

We now turn to prove the indexed counterpart to the \WLOG{}
Representation \theoremref{wlog-construction}:

\begin{proof}[of \theoremref{indexed-wlog-construction}]%
  We can prove this theorem directly, but it may be more interesting
  to reduce it to the ordinary representation theorem. Applying the
  Grothendieck construction to the situation, we have:
  \begin{itemize}
  \item
    $\G$-actions $\coprod_{\Ctx}\A$ and $\coprod_{\Ctx, \x : \A}\B$;
  \item canoniser $\canoniser' \definedby \fun{\pair\ctx a}\canoniser_{\ctx}a :
    \carrier{\coprod_{\Ctx}\A} \to \carrier\G$;
  \item below we show that $\Canon{\canoniser'} = \set{\pair\ctx a \suchthat
    a \in \Canon{\canoniser, \ctx}} \defines \coprod_{\Ctx}\Canon{\canoniser}$;
  \item core function $f' \definedby
    \parent{\fun{\pair\ctx a}\pair{(\ctx, \For\x a)}{f_{\ctx}a}}
    : \Canon{\canoniser'} \to \carrier{\coprod_{\Ctx,\x:\A}\B}$;
  \item we have that $\wlg(\canoniser', f'; \pair\ctx a) =
    \pair{(\ctx, \For\x a)}{\wlg(\canoniser, f)_{\ctx}a}$;
  \item the core function $f'$ is equivariant iff $f$ is
    (by \exampleref{equivariant Grothendieck});
  \item below we show that $\wlg(\canoniser', f')$ is equivariant iff
    $\wlg(\canoniser, f)$ is equivariant;
  \item below we show that the function term $\carrier\Ctx \types g : \DFun\x{\carrier\A}{\carrier\B}$
    extends $f$ along $\Ctx \types \Canon{\canoniser} \subset \A$ iff
    the function $\fun {\pair\ctx a}\pair-{(\ctx, \For\x a)}{g_{\ctx}a}
    : \coprod_{\carrier\Ctx}{\carrier\A}
    \to \coprod_{\carrier\Ctx, \x : \carrier\A}\carrier\B$ extends
    $f'$ along $\Canon{\canoniser'}\subset
    \coprod_{\carrier\Ctx}\carrier\A$.
  \end{itemize}
  The theorem follows by unpacking and re-packing definitions in both directions. Let's discharge the outstanding proof obligations.

  Let $\pair\ctx a$ be canonical for $\canoniser'$, so $\pair\ctx a =
  (\canoniser_{\ctx}a') \act\pair{\ctx'}{a'}$. Then $\ctx' =
  \inv*{\canoniser a'}\act \ctx$ and $a' \in \A_{\inv*{\canoniser
      a'}\act \ctx}$, so $a \in (\Canon\canoniser)_{\ctx}$, and so
  $\pair\ctx a \in \coprod_{\Ctx}\Canon{\canoniser}$. These
  implications are invertible, proving the converse.

  Next, we show that the Grothendieck construction preserves the \WLOG
  extension:
  \begin{align*}
  \wlg(\canoniser', f'; \pair\ctx a) &=
  \inv*{\canoniser'\pair\ctx a} \act
    f'(\parent{\canoniser'\pair\ctx a}\act \pair\ctx a)
  \\&=
  \inv*{\canoniser_{\ctx} a} \act
    f'((\canoniser_{\ctx} a)\act \pair\ctx a)
  \\&=
  \inv*{\canoniser_{\ctx} a} \act
    f'\pair{(\canoniser_{\ctx} a)\act\ctx}{(\canoniser_{\ctx} a) \iact{} a}
  \\&=
  \inv*{\canoniser_{\ctx} a} \act
  \pair{\parent{(\canoniser_{\ctx} a)\act\ctx,
                  \For\x((\canoniser_{\ctx} a) \iact{} a)}}
       {f_{(\canoniser_{\ctx} a)\act\ctx}{(\canoniser_{\ctx} a) \iact{} a}}
  \\&=
  \pair{\parent{\ctx,
                  \For\x a}}
       {\inv*{\canoniser_{\ctx} a} \act
         f_{(\canoniser_{\ctx} a)\act\ctx}{(\canoniser_{\ctx} a) \iact{} a}}
  \\&=
  \pair{\parent{\ctx,
                  \For\x a}}
       {\wlg(\canoniser, f)_{\ctx}a}
  \end{align*}
  In addition, by \exampleref{equivariant Grothendieck}, one \WLOG
  extension is equivariant iff the other is equivariant.

  The final bullet point follows by functoriality of the Grothendieck
  construction.
\end{proof}

We now establish the bread-and-butter properties facilitating smooth
dependently-typed symmetric programming and symmetric propositional
reasoning. We start with dependent-function abstraction:

\begin{lemma}[abstraction]
  Function abstraction of an equivariant term $\Ctx, \x : \A \types t
  : \B$ is both:
  \begin{itemize}
  \item an equivariant term $\Ctx \types \parent{\fun{(\x:\A)}t} : \DFun\x\A\B$.
  \item an equivariant function $\Ctx \types \parent{\fun{(\x:\A)}t} : \EFun\x\A\B$.
  \end{itemize}
\end{lemma}
\begin{proof}
  Calculate, ignoring indices (as those type-check):
  \begin{align*}
    \parent{g\act(\fun{\x:\A}t)}_{\ctx}\ma
    &\explain{=}{action on\\dep. functions}
  g \act ((\fun{\x:\A}t)(\inv g \act \ma))
  \explain={abstraction def}
  g \act t_{\ctx, \For\x(\inv g \act \ma)}
  \explain={$t$ equivariant}
  t_{g \act \ctx, \For\x(g \act \inv g \act \ma)}
  \\&=
  t_{g \act \ctx, \For\x \ma}
  =
  (\fun{\x : \A}t)_{g\act\ctx} \ma
  \end{align*}
  showing abstraction is an equivariant term. It is also an
  equivariant function, for:
  \[
  (\fun{\x:\A}t)_{g\act\ctx}(g \act \ma)
  \explain={abstraction def}
  t_{g\act\ctx, \For \x{g\act\ma}}
  \explain={$t$ equivariant}
  g \act t_{\ctx, \For\x\ma}
  \]
  as we wanted.
\end{proof}

Substitution using an equivariant substitution preserves equivariance:

\begin{lemma}[reindexing and substitution]
Reindexing by an equivariant function $\msubst : \Ctx
\to \Dtx$ sends equivariant terms $\Dtx \types s : \B$ to equivariant
terms $\Ctx \types \subst s\msubst : \subst \B\msubst$, and moreover
every equivariant term $\Ctx \types t : \subst\B\msubst$ induces an
equivariant substitution $(\msubst, \For\y t) : \Ctx \equito (\Dtx, \y :
\B)$.
\end{lemma}
\begin{proof}
  Calculate:
  \[
    g\act \subst s\msubst_{\ctx}
    \explain={reindexing def}
    g\act s_{\msubst\ctx}
    \explain*={$s$ equivariant}
    s_{g\act \msubst\ctx}
    \explain={$\msubst$ equivariant}
    s_{\msubst(g\act \ctx)}
    =
    \subst s\msubst_{g\act \ctx}
  \]
  and so $\subst s\msubst$ is equivariant. Next, calculate:
  \begin{align*}
  (\msubst, \For\y t)(g \act \ctx)
  &\explain={substitution def}
  (\msubst(g\act\ctx), \For\y t_{g \act \ctx})
  \explain={$t$ equivariant}
  (\msubst(g\act\ctx), \For\y g\act t_{\ctx})
  \explain={action on context extension}
  g\act (\msubst\ctx, \For\y t_{\ctx})
  \\&=
  g\act((\msubst, \For\y t)\ctx)
  \end{align*}
  and so $(\msubst, \For\y t) : \Ctx \equito (\Dtx, \y : \B)$.
\end{proof}

Dependent pairing also preserves equivariant terms:

\begin{lemma}[dependent pairing]
  Two equivariant terms $\Ctx \types t : \A$ and $\Ctx \types s :
\subst\B{\For\x t}$ induce an equivariant term $\Ctx \types \dpair ts
: \DPair\x\A\B$.
\end{lemma}
\begin{proof}
  Calculate:
  \[
  (g \act (\dpair ts))_{\ctx}
  =
  (\dpair{g \act t_{\ctx}}{g \act s_{\ctx, \For\x t_{\ctx}}})
  \explain={$t$, $s$ equivariant}
  (\dpair{t_{g \act \ctx}}{s_{g\act\ctx, \For\x (g \act t_{\ctx})}})
  \]
  as we wanted.
\end{proof}

We now turn to propositions:

\begin{lemma}[subfamilies]
Each equivariant proposition $\Ctx, \x : \A \types \phi : \Prop$
induces an indexed action on the subfamily $\Ctx \types \set{\x : \A
  \suchthat \phi}$ that lifts $\A$ along the subset inclusions.
Equivariant terms $\Ctx \types t : \A$ satisfying $\phi$ lift to
equivariant terms inhabiting the associated subspace i.e., $\Ctx
\types \lift t : \set{\x : \A \suchthat \phi}$.
\end{lemma}
\begin{proof}
  Equivariance means that if $a \in \carrier\A_{\ctx}$ satisfies
  $\phi$, then so does $g \act a$. Therefore $g \iact\ctx a \in
  \set{\x : \A \suchthat \phi}$, and the action lifts.  For
  equivariance of the lifted term, use injectivity of the coercion
  function $\Ctx \types \coerce \definedby \fun \ma \ma : \set{\x : \A
    \suchthat \phi} \equito \A$:
  \[
  \coerce_{g \act \ctx} (g \act (\lift t)_{\ctx}) \explain={$\coerce$ equivariance}
  g \act \coerce_{\ctx} \lift t_{\ctx} \explain={coercion}
  g \act t_{\ctx} \explain={$t$ equivariant}
  t_{g\act\ctx} =
  \coerce_{g\act \ctx} (\lift t)_{g\act\ctx}
  \]
  and so $g \act (\lift t)_{\ctx} = (\lift t)_{g\act\ctx}$.
\end{proof}

We use these facts implicitly when conducting reasoning:

\begin{lemma}
The logical connectives are equivariant $(\land), (\lor) : \Prop^2 \equito \Prop$,
$(\lnot) : \Prop \equito \Prop$, as well as the quantifiers $\forall,\exists :
(\A \to\Prop) \equito \Prop$.
\end{lemma}
\begin{proof}
  The logical connectives operate between trivial actions hence
  equivariant.  For the quantifiers, assume $\forall t = \Tru$, and
  so, for each $\ma \in \carrier\A_{\ctx}$, we have $t(\inv g \act
  \ma) = \Tru$, i.e.:
  \[
  \forall (g \act t) = \forall \x . t(\inv g \act
  \x) = \Tru
  \] Conversely, if $\forall (g \act t)$, then by
  instantiating with $g \act \ma$ we have that $t\ma = t(\inv g \act g
  \act \ma) = \Tru$. Therefore $\forall t = \forall g \act t$. A
  similar argument holds for $\exists$, or by use of de Morgan.
\end{proof}

We conclude our treatment of propositions by explicating how symmetric
reasoning \WLOG{} often manifests---via a property invariant under the
symmetry of interest:

\begin{proof}[of \lemmaref{core functions to propositions invariance}]%
  Take canonical $a$ and $g \iact\ctx a$. By type-checking, $g \iact{}
  f_{\ctx} a, f_{\ctx}(g \iact{} a)$ prove the same proposition,
  namely $\phi_{g\act\ctx, \For\x(g \iact{} a)}$.  Therefore, $g
  \iact{} f_{\ctx} a = f_{\ctx}(g \iact{} a)$.
\end{proof}

In some concrete cases, working with the entire group of permutations
incurs a combinatorial explosion. Group theorists often generate the
group from a smaller set of symmetries, and then only work with those.
As we briefly mentioned in the main text,
a group $\G$ is \emph{generated} by a subset $S \subset
\carrier\G$ when $\G$ is the smallest subgroup of $\G$ containing $S$, equivalently,
when every $\carrier\G$-element is the multiplication of
$\carrier\G$-elements or their inverses.
For example, the symmetry group $\Sym 3$ is generated by
the two transpositions $(1 \, 2)$ and $(2 \, 3)$.
We have an indexed variation of a well-known lemma:
\begin{lemma}\lemmalabel{equivariance by generators}
  Let $\Ctx \types \A$ and $\Ctx, \x : \A \types \B$ be indexed
  $\G$-actions.  If $S$ generates $\G$ and $\carrier\Ctx \types f :
  \DFun\x{\carrier\A}{\carrier\B}$ is equivariant w.r.t.~symmetries in
  $S$, then $f$ is equivariant $\Ctx \types f : \EFun\x{\A}{\B}$. When
  $\B = \Prop$ and $\inv{\bracks{S}} \subset S$,
  it suffices to show lax/oplax equivariance, i.e.:
  \[
  \parent{\forall g \in S, \forall \x. f(g\act \x) \implies f \x}
  \lor
  \parent{\forall g \in S, \forall \x. f(g\act \x) \impliedby f \x}
  \]
\end{lemma}
\begin{proof}[of \lemmaref{equivariance by generators}]
  Let $H \subset \carrier\G$ be the set of symmetries that $f$
  respects. Then $S \subset H$. Since $f (e \act a) = f a = e \act (f a)$, we have:
  $e \in H$. Take $h \in H$ we have:
  \[
  f(h \act a) = h \act (fa) \implies f (\inv h \act a) =
  \inv h \act f(h \act (\inv h \act a)) = \inv h \act (f a)
  \]
  so $\inv h \in H$.
  Take $g,h \in H$ and then:
  \[
  f ((g \mul h) \act a) = f (g \act (h \act a)) = g \act (f (h \act
  a)) = g \act (h \act (f a)) = (g \mul h)\act (f a)
  \]
  so $g \mul h \in H$.
  Thus $H \subset \carrier\G$ is a subgroup containing $S$, and so $H
  = \carrier\G$.

  When $\B = \Prop$ and $\inv{\bracks{S}} \subset S$, by substituting
  $\For\x{\inv g \act a}$ we have:
  \[
  f x \iff f (\inv g \act (g \act x))
  \explain\implies{$\inv{\bracks{S}} \subset S$}
  f (g \act x)
  \]
  For the left disjunct, $f$ is in fact equivariant for
  symmetries in $S$.  Similarly for the right disjunct.
\end{proof}

We now turn to study equivariant canonisers. It turns out that the
indexed setting is fruitful for this purpose. We first show that the
indexed action of \emph{distinguishable symmetries} $\G/\A$ from
\definitionref{indistinguishability} is well-defined:

\begin{lemma}\lemmalabel{distinguishable symmetry action well-defined}
  Let $\A$ be a $\G$-action. The action over distinguishable symmetries $\G/\A$
  is congruent w.r.t.~indistinguishability, hence is well-defined.
\end{lemma}
\begin{proof}
  Assume $s_1 \stabeq a s_2$. Calculate:
  \[
  (s_1 \mul \inv g)\act (g \act a) = s_1 \act a = s_2 \act a
  = (s_2 \mul \inv g)\act (g \act a)
  \]
  and so $s_1\mul \inv g \stabeq{g \act a} s_2\mul \inv g$, and the
  action is well-defined.
\end{proof}

In \subsecref{equivariant canonisers} we mentioned in passing the
stabiliser family alongside the indexed actions of the orbits and the
indistinguishable symmetries. The stabilisers also form an indexed
action:
\begin{example}
  The \emph{stabiliser} family has all the symmetries fixing an index
  $a \in \carrier\A$ as the fibre over it: $a : \carrier\A \types
  \stabilise[\G] a \definedby \set{s \in \carrier\G \suchthat s \act a
    = a}$.
  The stabiliser has an $\A$-indexed action:
  \[
  g \iact a{s} \definedby s \mul \inv g \in \stabilise{g \act a}\quad
  \text{since}\quad s \mul \inv g \act (g \act a) = s \act a = a
  \tag*{\qedhere}
  \]
\end{example}

\goodbreak

We now prove our reformulation of the famous Orbit-Stabiliser
\theoremref{orbit stabiliser}:
\begin{proof}[of \theoremref{orbit stabiliser}]%
  The bijectivity of $\phi$ is the well-known theorem. We show equivariance:
  \begin{align*}
  \phi_{g \act x}(g \iact x [h]_{\stabeq x})
  &\explain={action on $\G / \A$} \phi_{g \act x}([h \mul \inv g]_{\stabeq {g \act x}})
  = (h \mul \inv g) \act (g \act x)
  = h \act x
  \explain={action on $\OrbitOf{x}$} g \iact x (h \act x)
  \\&= g \iact x \phi_{x}[h]_{\stabeq x}
  \end{align*}
  and we conclude as inverses of equivariant functions are equivariant.
\end{proof}

Next, we characterise equivariant canonisers as those for which
stable and equivariant core functions coincide. First, an observation:

\begin{lemma}
  \lemmalabel{equivariant canoniser weakly immediate}
  Let $\canoniser : \carrier\A \to \carrier\G$ be an equivariant
  canoniser for a $\G$-action $\A$. For every canonical element $a \in
  \Canon\canoniser$, we have
  \(
  \canoniser a \stabeq a \unit
  \).
\end{lemma}
\begin{proof}
  Since $a$ is canonical, there is some $b \in \carrier\A$ such that
  $a = (\canoniser b)\act b$. Since $\canoniser$ is equivariant, we have:
  \[
  \canoniser a = \canoniser((\canoniser b) \act b)
  \explain*{\stabeq{a}}{equivariance\\$(\canoniser b)\act b = a$}
  (\canoniser b) \mul \inv*{\canoniser b}
  = \unit
  \]
  as we wanted.
\end{proof}

Now, for the proof:

\begin{proof}[\theoremref{all stable core functions are equivariant}]%
  First, assume $\canoniser : (a : \A) \equito (\G/\A)a$ is
  equivariant, and consider any stable core function $f :
  \Canon\canoniser \to \B$. We show that $f$ is equivariant.

  Consider any canonical $g \act a, a \in \Canon{\canoniser}$. Since
  $\canoniser$ is equivariant, by \lemmaref{equivariant canoniser
    weakly immediate}, we have $\canoniser a \stabeq a \unit$ and so
  $\canoniser a \in \stabilise a$.  By equivariance, we have:
  \[
  g \act[\G/\A] [\canoniser a]_{\stabeq a}
  =
  [\canoniser(g \act a)]_{\stabeq{g \act a}}
  \explain={\lemmaref{equivariant canoniser weakly immediate} for $\g \act a$}
  [\unit]_{\stabeq{\g \act a}}
  \implies
  \bracks{\canoniser a}_{\stabeq a}
  = \inv g \act \bracks\unit_{\stabeq{g \act a}}
  = \bracks g_{\stabeq a}
  \]
  and so $g \stabeq a \canoniser a \in \stabilise a$ and so $g \in
  \stabilise a \subset \stabilise {f a}$. Therefore:
  \[
  f(g \act a)
  \explain={$g \in \stabilise a$}
  f a
  \explain*={$g \in \stabilise{f a}$}
  g \act f a
  \]
  and so $f$ is equivariant.

  Conversely, assume that every stable core function is equivariant. We need
  to show the canoniser is equivariant.
  The proof adapts the argument in \theoremref{all core functions are partially equivariant}. Since we now have the indexed action
  vocabulary, the proof simplifies somewhat.

  Take the $\A$-indexed action $\B \definedby \G/\A$, and the following
  core function:
  \[
  f : \DFun\x{\Canon{\canoniser}} (\G/\A)\x
  \qquad
  f a \definedby [\unit]_{\stabeq a}
  \]
  It is stable. Indeed, take any canonical $a \in \Canon{\canoniser}$
  and a stabilising symmetry $g \in \stabilise a$. Since $\stabilise
  a$ is a sub-group, $\inv g \in \stabilise a$, and so $[\inv g]_{\stabeq a} =
  [\unit]_{\stabeq a}$. Calculate:
  \[
  g \iact[\G/\A] a (f a)
  =
  g \iact[\G/\A] a [\unit]_{\stabeq a}
  =
  \bracks{\unit\mul\inv g}_{\stabeq{g \act a}}
  \explain={$g \in \stabilise a$}
  \bracks{\inv g}_{\stabeq a}
  =
  [\unit]_{\stabeq a}
  =
  f a
  \]
  Therefore $g \in \stabilise{f a}$, and so $\stabilise a \subset
  \stabilise{f a}$, and so $f$ is stable.

  By assumption, $f$ is equivariant, and by the Indexed Representation
  \theoremref{indexed-wlog-construction} we have an equivariant
  $\wlg(\canoniser,f) : \EFun\x\A (\G/\A)\x$. We will show that
  $\wlg(\canoniser,f) = \parent{\fun a\bracks{\canoniser a}_{\stabeq a}}$ and therefore
  $\canoniser$ is equivariant.

  Take any $a \in \carrier\A$ and calculate:
  \[
  \wlg(\canoniser, f)a = \inv*{\canoniser a}\act f(\canoniser a\act a)
  =
  \inv*{\canoniser a}\act \bracks\unit_{\stabeq{\canoniser a\act a}}
  =
  \bracks{\inv*{\inv*{\canoniser a}}}_{\stabeq{(\inv*{\canoniser a}) \act \canoniser a\act a}}
  =
  \bracks{\canoniser a}_{\stabeq a}
  \]
  Therefore $\wlg(\canoniser,f) = \bracks{\canoniser a}_{\stabeq a}$, as we wanted.
\end{proof}

We conclude this section by proving that every action has an
equivariant canoniser. Recall that a \emph{transversal} of an action
is a (dependent) function $t : (a : \carrier \A) \to \G\act\bracks a$
that picks a representative for the orbit of its input: $t a = t b$
iff $\G\act\bracks a = \G\act\bracks b$.

\begin{proof}[\theoremref{equivariant canoniser choice}]%
  Consider the family of orbits $\x : \A \types \OrbitOf\x$, and use
  the axiom of choice to choose a transversal: $t : \DFun\x\A
  \OrbitOf\x$.  For each $a \in \carrier\A$, the elements $a$ and $t\,a \in
  \OrbitOf a$ are symmetric, and so there is some symmetry $g \in
  \carrier\G$ such that $g \act a = t\,a$. Use the axiom of choice to
  choose a canoniser $\canoniser : \A \to \carrier\G$ such that
  $\canoniser a \act a = t\,a$. We will show this canoniser is
  equivariant.

  Take any element $a \in \carrier\A$ and symmetry $g \in \carrier\G$.
  Then $g \act a$ and $a$ are in the same orbit, and therefore $t(g \act a) = t\,a$.
  Calculate:
  \[
  \parent{\canoniser(g \act a) \mul g}\act a
  =
  \canoniser(g \act a) \act (g\act a)
  =
  t(g \act a)
  =
  t\,a
  =
  \canoniser a \act a
  \]
  Therefore, $\canoniser(g \act a) \mul g \stabeq a \canoniser a$, and so
  $\bracks{\canoniser(g \act a) \mul g}_{\stabeq a} = \bracks{\canoniser a}_{\stabeq a}$.
  Calculate:
  \[
  \bracks{\canoniser(g\act a)}_{\stabeq{g \act a}}
  =
  \bracks{\canoniser(g\act a)\mul g \mul \inv g}_{\stabeq{g \act a}}
  =
  g \act \bracks{\canoniser(g\act a)\mul g}_{\stabeq a}
  =
  g \act \bracks{\canoniser a}_{\stabeq a}
  \]
  and so $\canoniser$ is equivariant.
\end{proof}

\section{Algebraic data types: further details}%
\applabel{algebraic data types}

In some sense, the theoretical development in this section is a
special case of the proceedings two sections. However, it requires
some additional abstractions that make the material less accessible:
indexed families (\appref{indexed data types}), or monadic adjunctions
and distributive laws (\appref{cat dev}). Therefore, we decided to
include more specialised proofs for the simply-typed case, that use
less categorical reasoning and mostly concern sets and functions.

We start by proving that every Cartesian strength is a $\G$-strength
(cf.~\remarkref{cartesian vs symmetry strengths}).

\begin{lemma}\lemmalabel{Cartesian strengths are symmetric strengths}
  Let $\strength : X \times FY \to F(X \times Y)$ be a Cartesian
  strength for a functor $F : \Set \to \Set$. Then
  $\strength_{\carrier\G,\X} : \carrier\G \times FX \to F(\carrier\G
  \times X)$ is a $\G$-strength.
\end{lemma}
\begin{proof}
  It's more natural to use diagrammatic reasoning here:
  
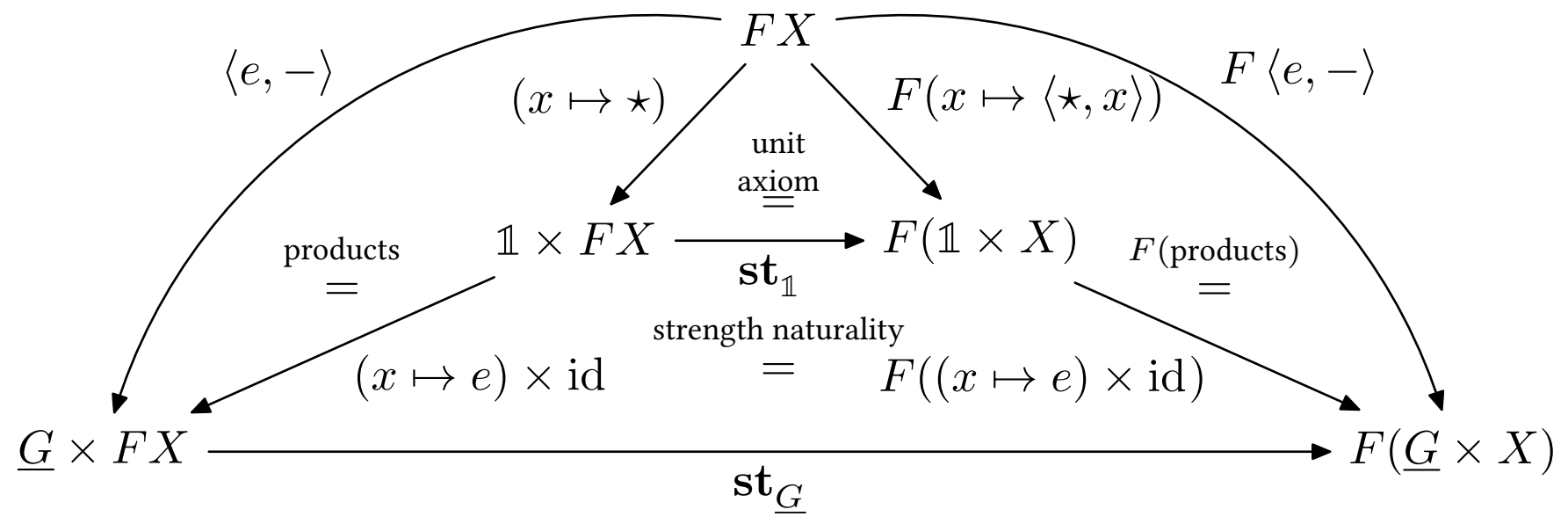
\diagram{15}
  
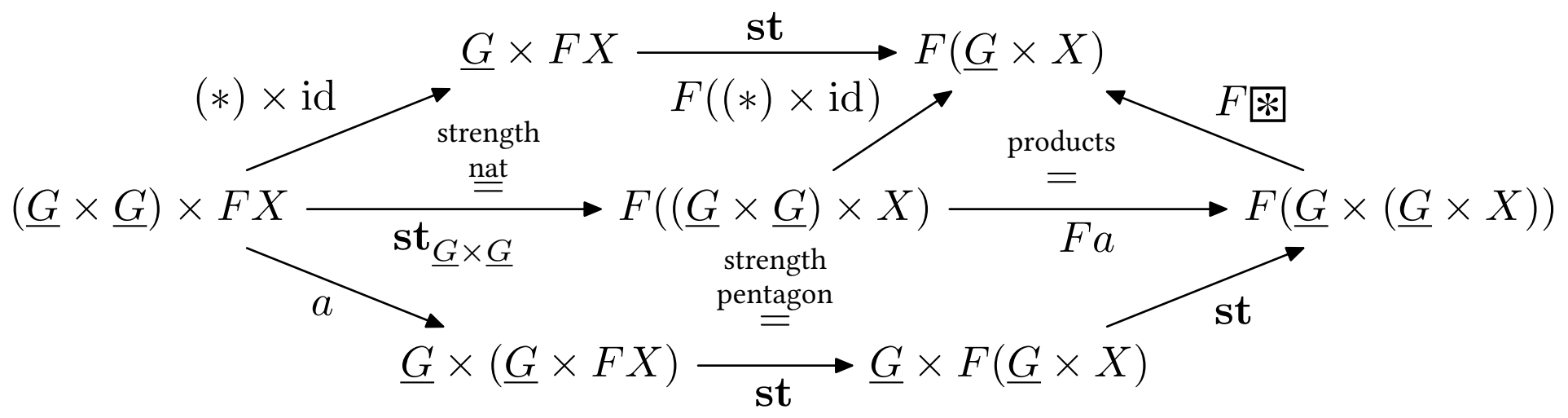
\diagram{16}
  as we wanted.
\end{proof}

Next, we prove that a $\G$-strength for $F$ and an $\G$-action $\A$
let us equip $F\carrier\A$ with a $\G$-action:

\begin{proof}[of \lemmaref{strength lifts actions}]%
  This definition is a $\G$-action by straightforward calculation:
  \[
    e \act t =
    \fmap (\act) (\strength(\unit, t))
    \explain={first axiom}
    \fmap (\act) (\fmap(\fun x \pair \unit x)t)
    \explain={map fusion}
    \fmap(\fun x \unit \act x)t
    =
    \fmap(\fun x \unit\act x)t
    =
    \fmap(\id)t
    =
    t
  \]
  \begin{align*}
    g \act (h \act t) &=
    \fmap (\act) (\strength(g, \fmap (\act) (\strength(h, t))))
    \\&\explain={strength nat.}
    \fmap (\act) (\fmap (\fun{\triple{g'}{h'}{x}}\pair{g'}{h'\act x})
      (\strength(g,
    (\strength(h, t)))))
    \\&\explain={map fusion}
    \fmap (\fun{\triple{g'}{h'}x}g'\act h'\act x)
    (\strength(g,
    (\strength(h, t))))
    \\&=
    \fmap (\fun{\triple{g'}{h'}x}(g'\mul h')\act x)
    (\strength(g,
    (\strength(h, t))))
    \\&\explain={map fusion}
    \fmap (\act) (\fmap (\fun{\triple{g'}{h'}x}\pair{g'\mul h'}x)
      (\strength(g,
    (\strength(h, t)))))
    \\&\explain={second axiom}
    \fmap (\act) (\strength(g \mul h, t))
    \defines
    (g\mul h) \act t
  \end{align*}
  and so we have an action.

  Let $h : \A \equito \B$ be equivariant. Calculate:
  \begin{align*}
    \fmap h (g \act t)
    &=
    \fmap h (\fmap (\act) (\gstrength (g, t)))
    \explain={map fusion}
    \fmap(\fun{\pair{g'}x}h(g' \act x))(\gstrength (g, t))
    \\&=
    \fmap(\fun{\pair{g'}x}g' \act h(x))(\gstrength (g, t))
    \explain={map fusion}
    \fmap(\act) \fmap(\id \times h)(\gstrength (g, t))
    \\&\explain={strength nat.}
    \fmap(\act) \gstrength(g, \fmap ht)
    =
    g \act \fmap h t
  \end{align*}
  and so we have an equivariant map $\fmap h : F\A \equito F\B$.
\end{proof}

Next, we prove that $\G$-strengths for each component functor induce a
$\G$-strength for the coproduct functor:

\begin{proof}[\propositionref{modularity}]%
  Follows by $\Set$'s distributivity: the natural
  isomorphism $\carrier\G \times (\X \amalg \Y) \isomorphic
  (\carrier\G \times \X) \amalg (\carrier\G \times \Y)$.
\end{proof}

The following technical result is the workhorse that allows us to both
derive the action on the data type from its strength, and to reason
about properties of parameterised maps. It is
prevalent in Fiore et al.'s work on abstract
syntax~\cite{fiore-szamozvancev:formal-soas,%
  fiore-szamozvancev:soas-theory,%
  fiore-ranchod:substructural-ast}, originally proposed by
\citeauthor{bird-paterson:generalised-folds-for-nested-datatypes}~\cite{%
  bird-paterson:generalised-folds-for-nested-datatypes}. Here we use
the following special case.

\begin{theorem}[parameterised initiality]%
  \theoremlabel{parameterised initiality}
  Let $A$ be a set, $F : \Set \to \Set$ a functor, $\roll : F\fix F
  \to \fix F$ an initial algebra, and $\strength : A\times FX \to F(A
  \times \X)$ a natural transformation in $\X$. For every $F$-algebra
  $f : F\X \to \X$ there is a unique \emph{$A$-parameterised}
  homomorphism $\pfold f : \pair{\fix F}{\roll} \to \pair{X}{f}$, i.e., a
  function $\pfold f : A \times \fix F \to \X$ satisfying:
  \[
  \pfold f\pair a{\roll t} = f(\fmap(\strength(a,t)))
  \]
\end{theorem}
\begin{proof}%
  Using both the given strength and the Cartesian-closed structure,
  define the following morphism:
  \diagram{17}
  Currying, we have $h \definedby \curry g : F(X^A) \to X^A$, and so
  by ordinary initiality, there is a unique homomorphism into $X^A$:
  \diagram{18}
  We define the required parameterised traversal as the uncurried function:
  \[
  \pfold f \definedby \uncurry k :
  A \times \fix F \xto{\id \times k} A \times X^A \xto\eval X
  \]
  We need to show that it satisfies the required equation, and that it
  is the unique such function. Calculate:
  
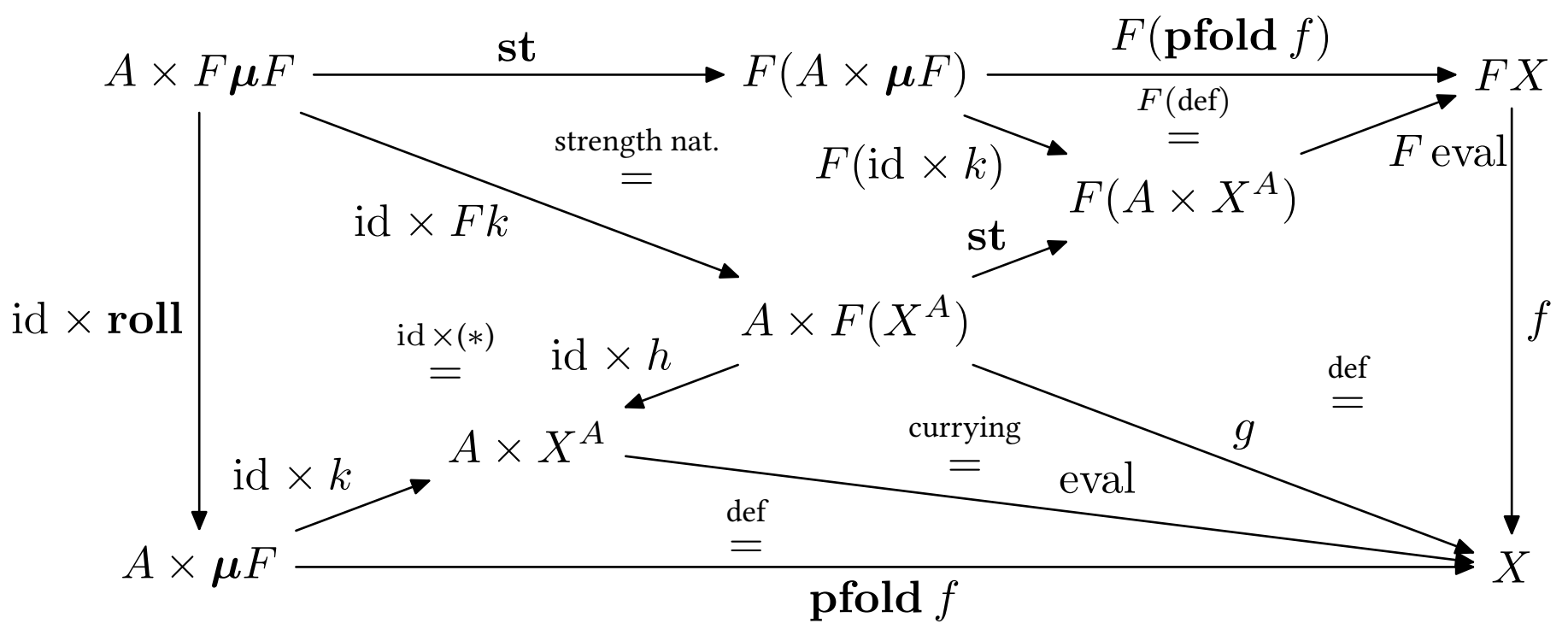
\diagram{19}
  Now, take any map $p : A\times \fix F \to \X$ that satisfies:
  
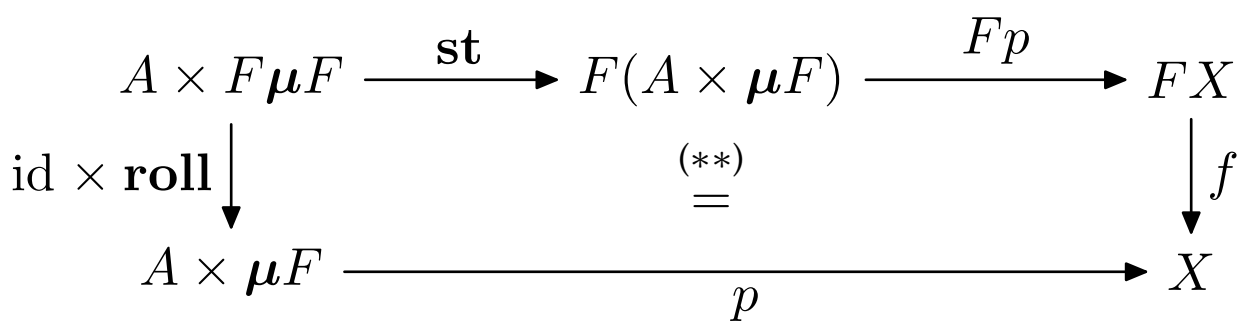
\diagram{20}
  Let $q \definedby \curry p : \fix F \to X^A$. Calculate:
  
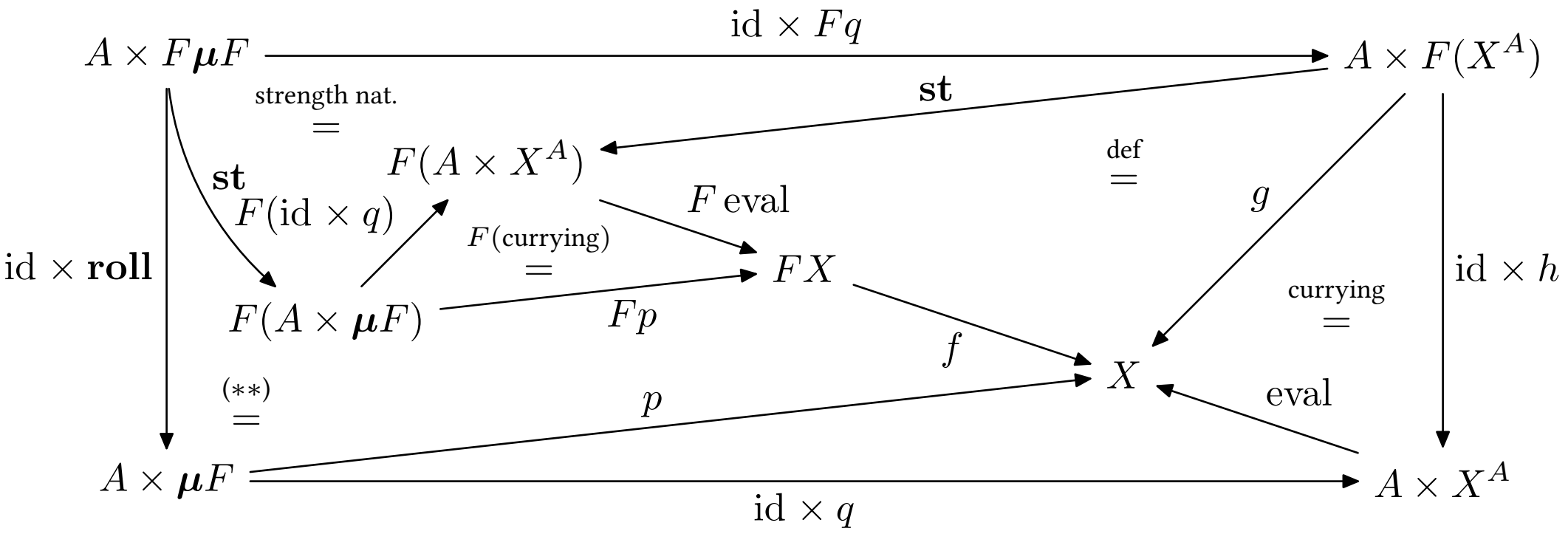
\diagram{21}
  and so, by the universal property of the exponential, $q$ satisfies
  ($*$), and so $q = k$, hence:
  \[
  p = \uncurry q = \uncurry k = \pfold f
  \]
  and we proved uniqueness.
\end{proof}

We now lay the capstone in the theory of data types, the Equivariant
Initiality \theoremref{representation for algebras}:

\begin{proof}[of \theoremref{representation for algebras}]%
  We have three things to do:
  \begin{itemize}
    \item show the unit axiom for $(\act)$.
    \item show the associativity axiom for $(\act)$.
    \item show that $\fold f$ is equivariant when $f$ is equivariant.
  \end{itemize}
  All three are straightforward calculations, appealing to initiality
  or parameterised initiality.

  For the unit axiom, appeal to initiality:
  
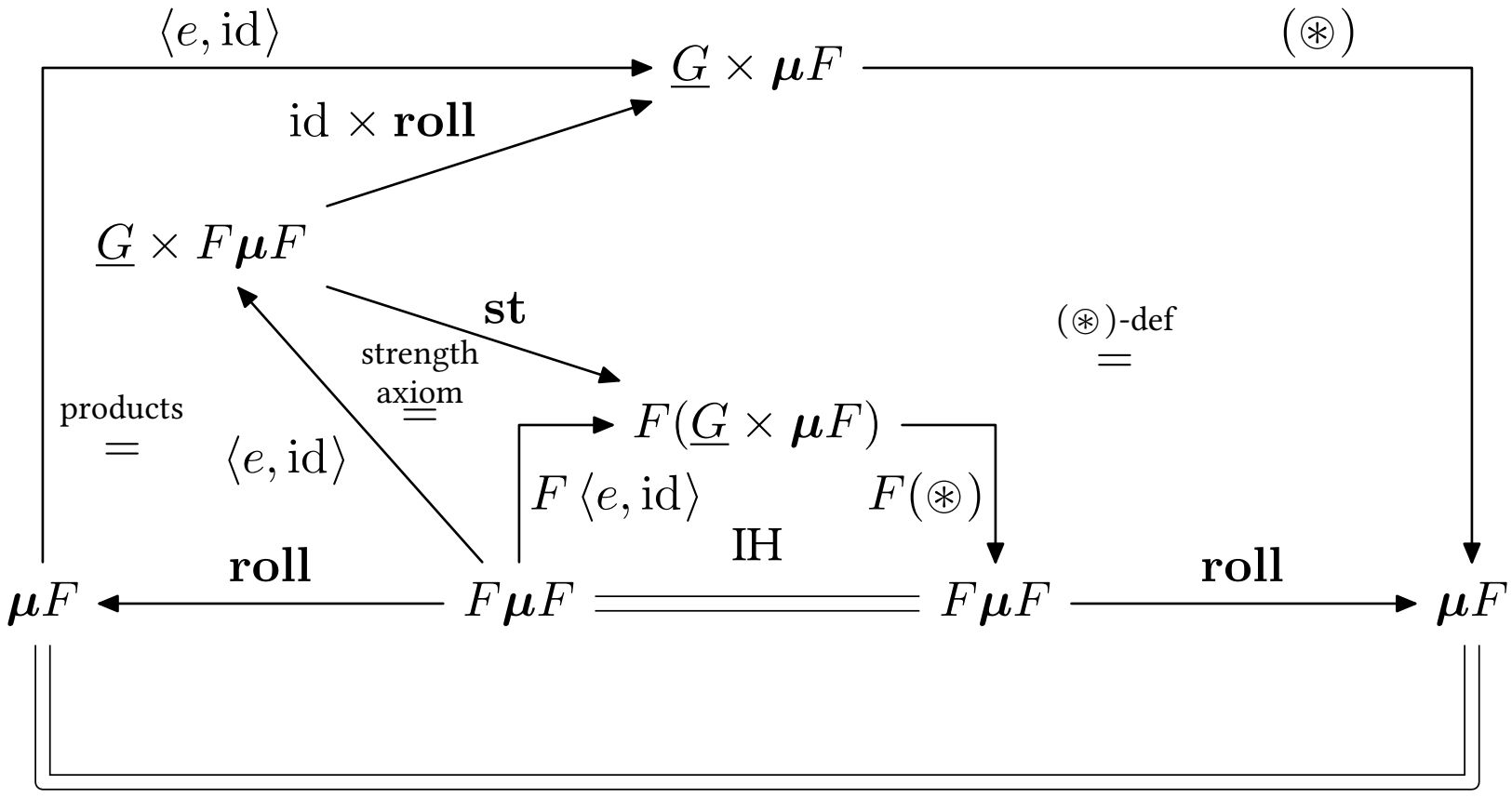
\diagram{23}
  and so, by initiality, $\unit \act t = t$.

  For associativity, we appeal to parameterised initiality, for the
  composite strength:
  
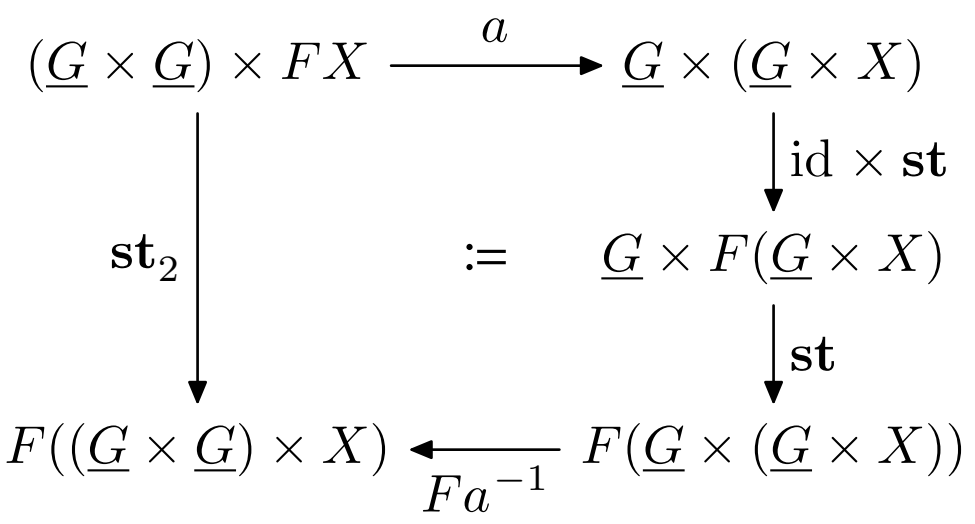
\diagram{24}
  via the  calculation in \figref{associativity of initial algebra action}.
  \begin{figure}
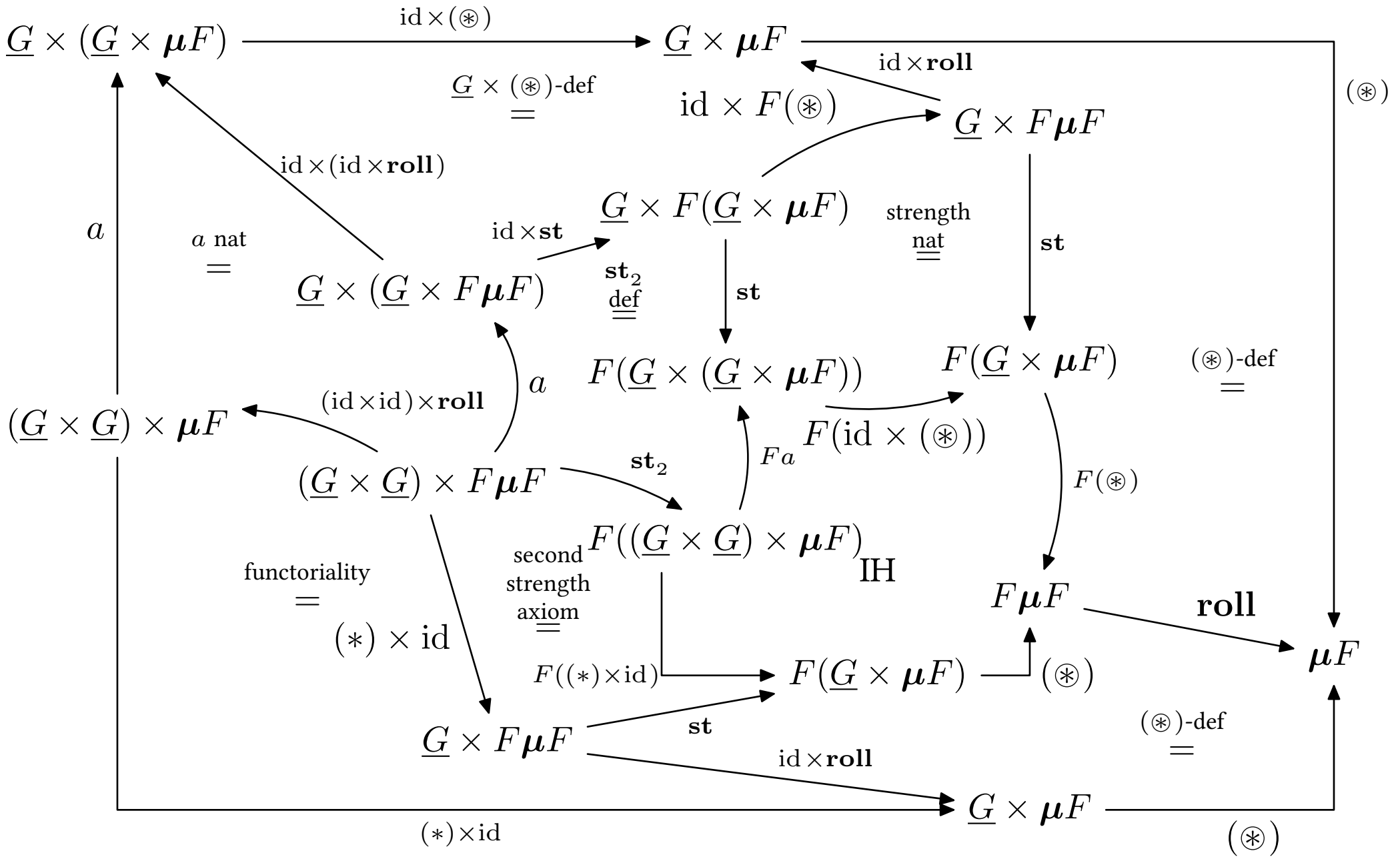

    \diagram{25}
    \caption{associativity of the action over the initial algebra}
    \figlabel{associativity of initial algebra action}
  \end{figure}
  Therefore, $(g \mul h) \act t = g \act (h \mul t)$, and we have
  an action. Note that the defining property of $(\act)$ makes the
  initial algebra structure equivariant $\roll : F\fix F \equito \fix F$.

  Finally, take any equivariant algebra $f : F\A \equito \A$, and
  consider the unique $F$-homomorphism $\fold f : \fix F \to
  \carrier\A$. We appeal to parametrised initiality w.r.t.~the
  $\G$-strength, and calculate as in \figref{equivariant initiality}.
  \begin{figure}
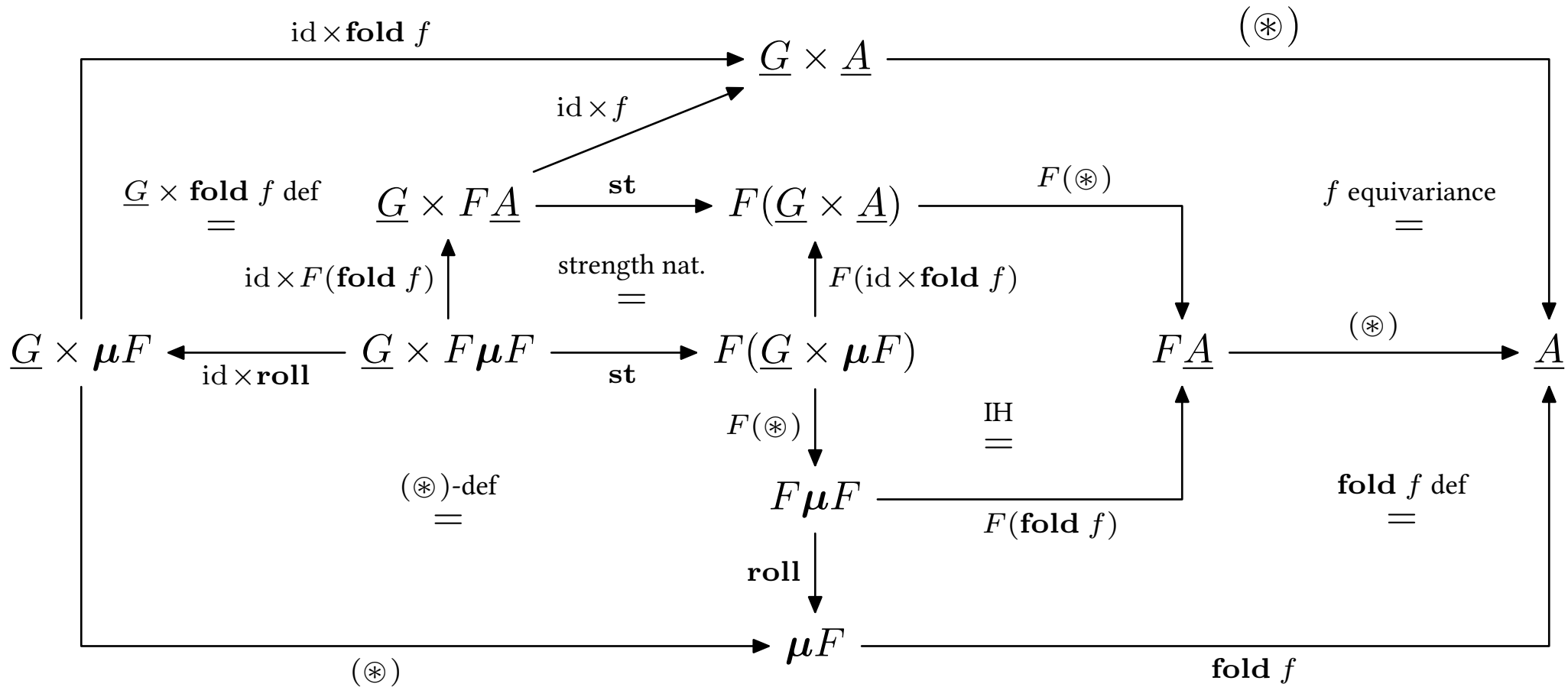

    \diagram{26}
    \caption{equivariance of the homomorphism into an equivariant algebra}
    \figlabel{equivariant initiality}
  \end{figure}
  Therefore $\fold f(g \act t) = g \act (\fold f)$, as we wanted.
\end{proof}

The proofs from \subsecref{wlog-traversals} are identical and special
cases to their indexed variants, and do not use abstract categorical
concepts. We therefore only prove them in the indexed case
(cf.~\appref{indexed data types}).

\section{Indexed data types}
\applabel{indexed data types}

The theory of indexed data types has a few more moving parts. So in
this case we do most of our proofs more abstractly in \appref{cat dev}.
In this appendix we establish the assumptions needed for that development, prove
a result about canonisers, who do not feature in the abstract account, and state
the indexed version of the Parameterised Initiality Theorem.

\begin{proof}[\theoremref{descent}]%
  Let's start with the left adjoint.
  It is an action by direct calculation:
  \[
  \unit \iact\ctx\pair gx \definedby \pair{\unit\mul g}x = \pair gx
  \qquad
  g \act h \act \pair sx \definedby \pair{g\mul(h\mul s)}x
  =
  \pair{(g\mul h)\mul s}x
  =
  (g\mul h)\act \pair sx
  \]
  Given any $\Ctx$-indexed $\G$-action $\A$ and a vertical map $h : \X
  \to \carrier\A$, define:
  \[
  h^{\dagger} : \DPair\g\G{\g\act\X}\equito \A
  \qquad
  h^\dagger_{\ctx}\pair g{x} \definedby \parent{g \iact{\inv g \act\ctx}x} \in \X_{\ctx}
  \]
  Indeed:
  \[
  h^\dagger_{g\act\ctx}(g \act \pair sx) =
  h^\dagger_{g\act\ctx}(\pair{g \mul s}x) =
  (g\mul s)\act (hx)
  =
  g\act (s\act (hx))
  =
  g\act h^\dagger_{\ctx}\pair sx
  \]
  Now, $h^{\dagger}$ extends $h$ along the unit, since:
  \[
  h^{\dagger}\pair\unit x = \unit\act hx = hx
  \]
  If $\overline h : \DPair\g\G{\g\act\X}\equito \A$ is any other such
  extension, we have:
  \[
  \overline h_{\ctx}\pair gx =
  \overline h_{\ctx}(g \act \pair\unit x)
  \explain={equivariance}
  g \act \overline h_{\inv g \act \ctx}\pair\unit x
  \explain*={$\overline h$ extends \\the unit}
  g \act h_{\inv g \act \ctx}x
  =
  h^\dagger_{\ctx}\pair gx
  \]
  and so $\overline h = h^\dagger$, as we wanted.

  The monadicity result follows by an application of Beck's
  monadicity theorem~\cite[Thm.~VI.7.1]{mac-lane:cwm}.
  Start with a split coequalier diagram, in which $\A$ and
  $\B$ are $\Ctx$-indexed actions, and $f,g : \A \equito \B$ are
  equivariant:
  
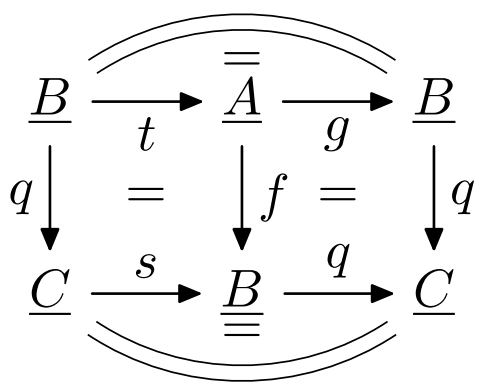
\diagram{28}
  First, equip $\carrier\C$ with an indexed action:
  \(
  h \iact\ctx c \definedby q_{g\act\ctx}(h\iact\ctx s_{\ctx} c)
  \).
  Indeed:
  \(
  \unit\iact\ctx c = q(\unit\iact\ctx s c)
  =
  q(s c) = c
  \)
  and:
  \[
  h \act k \act c = q(h\act s (q(k\act s c)))
  =q(h\act (k\iact\ctx s c))
  =q((h\mul k)\act s c)
  = (h\mul k)\act c
  \]
  The coequaliser $\carrier\Ctx \types q : \carrier\B \to \carrier\C$ in $\Fam_{\carrier\Ctx}$
  is in fact equivariant $\Ctx \types q: \B \equito \C$:
  \begin{align*}
  h\act(q\,b) &= q(h\act(s(q\,b)))= q(h\act(f(t\,b)))
  = q(f(h\act(t\,b)))
  = q(g(h\act (t\,b)))
  = q(h\act g(t\,b))
  \\&= q(h\act b)
  \end{align*}
  It is also a coequaliser in $\Action\G_{\Ctx}$. Given any
  coequalising equivariant map $\ell : \B \equito D$,
  the underlying family map $\ell : \carrier\B \to \carrier D$ is also
  coequalising. Therefore, the unique extension in $\Fam_{\carrier\Ctx}$ of $\ell$
  along $q$ is given by:
  \[
  r : \carrier \C \xto s \carrier\B \xto \ell \carrier D
  \]
  It is an equivariant map $r : \C \equito D$. Indeed:
  \begin{align*}
  r(h \act c) &= \ell(s(h \act c)) = \ell(s(q(h \act s\,c)) =
  \ell(f(t(h \act s\,c)) = \ell(g(t(h\act s\, c)))
  = \ell(h\act s\,c)
  \\&= h \act (\ell(s\,c)) = h \act (r\,c)
  \end{align*}
  Therefore $\carrier-$ lifts split coequalisers of parallel pairs to
  coequalisers.

  Now, if $\C$ already had an indexed $\A$-action structure and $q$
  was equivariant, then necessarily $h \act c = h\act q(s\,c) =
  q(h\act (s\,c))$, and so the action on $\carrier\C$ must be the one
  we defined. Therefore, $\carrier-$ reflects split coequalisers of
  parallel pairs as coequalisers.

  By Beck's theorem, the free indexed action adjunction is monadic.

  Turning to the right adjoint, we first prove it's an action:
  \[
  \unit\iact{\ctx}\phi
  \definedby
  \parent{\fun{h}\phi(h\mul\unit)}
  =
  \parent{\fun{h}\phi h} = \phi
  \] and:
  \begin{align*}
    (g \act k \act \phi)h =
    (k \act \phi)(h \mul g) =
    \phi((h\mul g) \mul k)
    =
    \phi(h\mul (g \mul k))
    =
    ((g\mul k)\act \phi)h
  \end{align*}
  and so \(
  g \act k \act \phi = (g\mul k)\act \phi
  \).

  Next, take any $\Ctx$-indexed $\G$-action $\A$, and a vertical map
  $\carrier\Ctx \types f : \carrier\A \to \X$. Define:
  \[
  \Ctx \types
  f^\dagger : \A \equito \parent{\DFun\g\G{\inv g \act \X}}
  \qquad
  f^\dagger_{\ctx} a \definedby \parent{\fun g f(g \act a)} \in
  \DFun\g{\carrier\G}\parent{\X_{\g \act \ctx}}
  \]
  It is equivariant:
  \[
  f^\dagger_{h\act \ctx}(h \act a) = \fun g f(g\act (h \act a))
  =
  \fun g f((g \mul h)\act a)
  =
  h \act (\fun g f (g \act a))
  =
  h \act f^\dagger_{\ctx} a
  \]

  It lifts $f$ along $\eval\pair{-}\unit$:
  \[
  \eval\pair{f^\dagger a}\unit = (f^\dagger a)\unit = f(\unit \act a) = f\,a
  \]
  Now if $\overline f : \A \equito \DFun\g\G{\inv\g\act \X}$ lifts $f$
  along $\eval\pair{-}\unit$, then:
  \[
  (\overline f a) g = (\overline f a)(\unit \mul g) =
  (g \act (\overline f a))(\unit) =
  \eval\pair{g \act (\overline f a)}\unit
  =
  \eval\pair{\overline f (g \act a)}\unit
  =
  f(g \act a)
  \]
  and so $\overline f = f^\dagger$, and we have a right adjoint.

  We use Beck's theorem in its dual formulation. Consider a split
  equaliser, with $\A$, $\B$ actions:
  
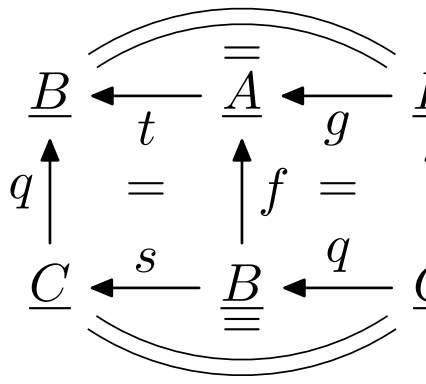
\diagram{29}
  Equip $\carrier\C$ with the following indexed $\G$-action structure:
  \(
  k \iact\ctx c \definedby s(k \iact\ctx (q\,c))
  \). It is indeed an action, for $\unit \act c = s(\unit \act (q\,c)) = s(q\,c) = c$
  and:
  \[
  k \act h \act c = s(k \act (q (s(h \act (q\,c))))) =
  s(k \act h \act (q\,c)) = s((k \mul h) \act (q\,c)) = (k \mul h)\act c
  \]
  We need to show the structure maps $q$ and $s$ (but not $t$!)
  are equivariant. For $q$:
  \[
  q(k \act c) = q(s(k \act (q\,c))) = k \act (q\,c)
  \]
  For $s$:
  \[
  k \act s b = s(k \act (q (s\,b))) = s(k\act b)
  \]
  Now, given an equalising equivariant map $\Ctx \types e : \X \equito \B$,
  by moving to families, we know the mediating map is $u : \X
  \xto{e} \B \xto{s} \C$, and therefore equivariant. Therefore
  $\carrier-$ lifts equalisers of split equalisers. Now, equip $\C$ with any
  action that makes $q : \C \to \B$ equivariant. We then have:
  \[
  k\act c = s(q(k\act c)) = s(k\act (q\,c))
  \]
  and so the action on $\C$ is the lifted one, making $q : \C \to \B$
  a coequaliser, i.e., $\carrier-$ reflects equalisers that split in
  $\carrier\Ctx$-families.
\end{proof}

Next, we justify not checking the second axiom of $\G$-strengths in
the cases $g = \unit$ or $h = \unit$:

\begin{proof}[\lemmaref{indexed check only non-trivial}]
  For $h = \unit$:
  \begin{align*}
    \fmap(\mapmul)
  (\gstrength(g,
   \gstrength(\unit, t)))
   &\explain={first axiom}
   \fmap (\mapmul)
  (\gstrength_{\X}(g,
   \fmap(\fun x \pair \unit x) t))
   \\&\explain={naturality}
   \fmap (\mapmul)
   (\fmap(\fun{\pair{g'}x}\triple{g'}\unit x)
     \gstrength(g, t))
   \\&\explain={map fusion}
   (\fmap(\fun{\pair{g'}x}\pair{g'\mul\unit}{x})
   \gstrength(g, t))
   \\&= \gstrength(g, t) = \gstrength(g\mul\unit, t)
  \end{align*}
  Next, for $g = \unit$:
  \begin{align*}
    \fmap(\mapmul)
  (\gstrength(\unit,
   \gstrength(h, t)))
   &\explain={first axiom}
   \fmap(\mapmul)(
   \fmap(\fun{\pair{h'}x}\triple\unit{h'}x)
   (\gstrength(h, t)))
   \\&\explain={map fusion}
   \fmap (\fun{\pair{h'}x}\pair{\unit\mul h'}x)
   (\gstrength(h, t))
   \\&= \gstrength(h, t) = \gstrength(\unit\mul h, t)
  \end{align*}
  as we wanted.
\end{proof}

Next, we state the indexed analogue to the Parameterised Initiality \theoremref{parameterised initiality} for the indexed
setting:
\begin{theorem}[indexed parameterised initiality]%
  \theoremlabel{indexed parameterised initiality}
  Let $\Ctx$ be a $\G$-action, $F : \Fam_{\carrier\Ctx} \to
  \Fam_{\carrier\Ctx}$ an endofunctor equipped with a symmetry
  strength, and $\carrier\Ctx \types \roll : F\fix F \to \fix F$ an
  initial $F$-algebra. For every $F$-algebra $f : F\X \to \X$,
  there is a unique \emph{parameterised homomorphism} into $\pair \X f$, i.e., a
  map
  $\pfold f : \gast \fix F \to \X$ satisfying:
  \(
  (\pfold f)(g, \roll u)
  =
  f\compose \fmap(\pfold f)\strength(g, u)
  \).
\end{theorem}

We will prove this theorem and the Indexed Equivariant Initiality
\theoremref{indexed representation} using the more abstract tools of
the next section.

The remainder of this section establishes the results concerning
canonisers for indexed $\G$-strong functors, which do not have a
general counterpart. By ignoring the indices, they also generalise the
corresponding results from \subsecref{wlog-traversals}.

\begin{lemma}\lemmalabel{canonical elements functor}
  Let $\Ctx$ be a $\G$-action, $F : \Fam_{\carrier\Ctx} \to
  \Fam_{\carrier\Ctx}$ be a functor equipped with a symmetry strength
  $\strength : \gast F \X \to F(\gast \X)$, and $\canoniser : F \to
  \carrier\G$ an indexed canoniser for $F$. The canonical elements extend to a functor
  $\Canon\canoniser : \Action\G_{\carrier\Ctx} \to
  \Fam_{\carrier\Ctx}$ given for $\Ctx \types h : \A \equito \B$ by:
  \[
  ({\Canon\canoniser})\,h\,({g \iact\dtx u}) \definedby
  g \iact\dtx  (F h u)
  \tag{$g = \canoniser_{\carrier\A; \dtx}u, g \act \dtx = \ctx, u \in (F\carrier\A)_{\dtx}$}
  \]
  Consequently, we have an equivariant function $\Ctx \types
  (\parent{\Canon\canoniser}h : (\Canon\canoniser)\A \equito
  (\Canon\canoniser)\A$. The inclusions $\Canon\canoniser \embed F$ are natural.
\end{lemma}
\begin{proof}
  First, since $h : \A \equito \B$ is equivariant, $Fh : F\A \equito F\B$ is
  equivariant. Therefore:
  \[
  (\Canon\canoniser)h(g \iact\dtx u) \definedby
  Fh(g \iact\dtx u) = g \iact\dtx (Fhu)
  \]
  and so the equation in the lemma statement defines
  a family map $k : (\Canon\canoniser)\A \to F\B$,
  independently of the representatives used.

  The crux of the proof is that, by $\canoniser$'s naturality,
  $\canoniser_{\carrier\B; \dtx}(\fmap h u) = \canoniser_{\carrier\A;
    \dtx}(u) = g$. Therefore, $g\iact\dtx(\fmap h u)$ is canonical,
  and $(\Canon\canoniser)h$ is well-defined by lifting $k :
  (\Canon\canoniser)\A \to F\B$ along the inclusion
  $(\Canon\canoniser)\B \embed F\B$. Functoriality follows by routine
  calculation. For equivariance, take:
  \begin{gather*}
    g \iact \dtx u \in ((\Canon\canoniser)\A)_{\ctx}
    \tag{$g = \canoniser_{\carrier\A; \dtx}u, g \act \dtx
      = \ctx, u \in (F\carrier\A)_{\dtx}$}
    \\
    p \iact \xtx v \in ((\Canon\canoniser)\A)_{\ctx}
    \tag{$p = \canoniser_{\carrier\A; \xtx}v, p \act \xtx
      = \ctx, v \in (F\carrier\A)_{\xtx}$}
  \end{gather*}
  and $r \in \stabilise \ctx \subset \carrier\G$ such that:
  \[
  r \iact\ctx g \iact \dtx u = p \iact\xtx v
  \]
  We then have:
  \begin{align*}
  \parent{\Canon\canoniser}h\, (p \iact\xtx v)
  &{}=
  p \iact\xtx (\fmap h\,v)
  \explain={$\fmap h : F\A \equito F\B$}
    \fmap h\,(p \iact\xtx v)
  =
  \fmap h\,(r\iact\ctx g \iact\dtx u)
  =
  r\iact\ctx g \iact\dtx (\fmap h\,u)
  \\
  &{}=
  r\iact\ctx(\parent{\Canon\canoniser}h\, (g \iact\dtx u))
  \end{align*}
  and so $\parent{\Canon\canoniser}h$ is equivariant.

  The naturality of the inclusions $\Canon\canoniser \embed F$ amounts
  to the calculation:
  \[
  (\Canon\canoniser)h(g\iact\dtx u) \definedby g\iact \dtx (Fh u)
  =
  Fh(g \iact \dtx u)
  \]
  which holds by equivariance of $Fh : F\A \equito F\B$.
\end{proof}

The following transformation lets us simultaneously work with
canonical elements in a symmetric fibre while carrying around their
canonising symmetry.

\begin{proposition}
  The transformation $\mctx : \carrier\Ctx \types
  \canonise[\canoniser] \definedby \fun {\BoundV u}
  \pair-{\inv*{\canoniser\BoundV u}}{\canoniser\BoundV u \iact\mctx
    \BoundV u}: F \To \gast (\Canon\canoniser)$ is natural.
\end{proposition}
\begin{proof}
  More generally, for every $g \in \carrier\G$, we have transformation:
  \[
  \ctx : \carrier\Ctx \types (g\iact\ctx\eta-) : \carrier\A \to \gast\carrier\A
  \qquad
  g\iact{\ctx}\eta a \definedby \pair-{g}{\inv g \iact\ctx a}
   = g \iact[\gast\carrier\A]{\inv g\act\ctx} {\pair\unit {\inv g \iact\ctx a}}
  \]
  Given an equivariant family map $\Ctx \types f : \A \equito \B$, we indeed have:
  \diagram{51}
  We now use the canoniser's naturality:
  \begin{align*}
  \canonise(\fmap f u) &{}\definedby
  \pair{\inv*{\canoniser (\fmap f u)}}{(\canoniser (\fmap f u))\act (\fmap f u)}
  \explain={$\canoniser : F \to \carrier\G$ natural}
  \pair{\inv*{\canoniser u}}{(\canoniser u)\act (\fmap f u)}
  \\&\explain={$\fmap f : F \A \equito F\B$}
  \pair{\inv*{\canoniser u}}{\fmap f \parent{(\canoniser u)\act(u)}}
  \explain={$(\inv*{\canoniser u}\act\eta-) : \carrier\C \to \gast\carrier\C$ natural}
  \fmap f \parent{\pair{\inv*{\canoniser u}}{(\canoniser u)\act u}}
  \\&\defines
  \fmap f (\canonise u)
  \end{align*}
  as we wanted.
\end{proof}

We can use this transformation to express the \WLOG{} construction:
\begin{lemma}
  We have:
  \[
  \carrier\Ctx \types
  \wlg(\canoniser, f) : F\carrier\A \xto{\canonise[\canoniser]}
  \carrier{\gast ((\Canon\canoniser)\A)}
  \xto{\gast f} \carrier{\gast\carrier\A}
  \xto{(\act)} \carrier\A
  \]
\end{lemma}
\begin{proof}
  Calculate:
  \begin{align*}
  u \xmapsto{\canonise[\canoniser]}
  \pair{\inv*{\canoniser u}}{(\canoniser u) \act u}
  \xmapsto{\gast f}
  \pair{\inv*{\canoniser u}}{f \parent{(\canoniser u) \act u}}
  \xmapsto{(\act)} {} &
  \inv*{\canoniser u}\act {f \parent{(\canoniser u) \act u}}
  \\&= \wlg(\canoniser, f)
  \end{align*}
  as we wanted.
\end{proof}

The next result establishes a universal property for the intermediate
step in the $\flg$ construction: the extension of a core algebra to an
$F$-algebra.

\begin{theorem}[indexed equivariant algebra representation]%
  \theoremlabel{indexed equivariant algebra representation}
  Let $\Ctx$ be a $\G$-action, $F$ a $\Ctx$-indexed $\G$-strong functor,
  and $\canoniser : F \To
\carrier\G$ an indexed canoniser for $F$, and consider an indexed
equivariant core-$F$ algebra $\Ctx \types f :
(\Canon\canoniser)\A \equito \A$. Then the \WLOG{} construction is the
unique equivariant $F$-algebra $\Ctx \types \wlg(\canoniser, f) :
F\A \equito \A$ extending $f$ along the inclusion
$\Canon\canoniser\embed F$.

Given two equivariant $F$-algebras $(\A, f)$ and $(\B, g)$
and an equivariant function $\Ctx \types h : \A \equito \B$, it is an
equivariant $F$-homomorphism $\Ctx \types h : (\A,f)
\equito (\B, g)$ iff it is an equivariant core-$\Canon\canoniser$
algebra homomorphism.
\end{theorem}
\begin{proof}
  Assume $f : (\Canon\canoniser)\A \equito \A$ is a core algebra.
  By the naturality of $\Canon\canoniser$, $f$ is a core
  function for the canoniser $\canoniser_{\A} : F\carrier A \to
  \carrier\G$. This function is equivariant w.r.t.~$\canoniser_{\A}$
  in the sense of the main paper iff it is equivariant w.r.t.~the
  canoniser $\canoniser$ for $F$ in the sense of \definitionref{indexed core algebras}.

  If $f$ is equivariant, then by the Indexed Representation
  \theoremref{indexed-wlog-construction}, $\wlg(\canoniser, f) = \wlg(\canoniser_{\carrier\A}, f)$ is
  equivariant extending $f$. Covnersely, if $h : F\A
  \equito \A$ is an equivariant algebra, then by the representation
  theorem, $h = \wlg(\canoniser_{\carrier\A}, f) = \wlg(\canoniser,
  f)$, as we wanted.

  Given another such equivariant algebra $g : F\B
  \equito \B$, and an equivariant map $h : \A \to \B$. The naturality
  and injectivity of the embedding $\Canon\canoniser \embed F$ show
  that $h$ is an $F$-algebra homomorphism iff it is a
  $\Canon\canoniser$-algebra homomorphism. Indeed, consider the
  composite square:
  \[
  \diagram*{52}
  \tag{$*$}
  \]
  Our goal is to show that the inner square $(**)$ commutes---$h$ is
  an $F$-algebra homomorphism---iff the outer square ($*$)
  commutes---$h$ is a $\Canon\canoniser$-algebra homomorphism. If
  ($**$) commutes, the whole diagram commutes. Assume ($*$) commutes.
  We have that $f$, $g$, and $h$ are equivariant, and the latter also
  implies that $Fh$ is equivariant. Therefore, the two
  sides of $(**)$ are equivariant extensions of the outer face ($*$)
  along the embedding $(\Canon\canoniser)\A\embed F\A$. By the
  indexed representation theorem, they coincide, i.e., ($**$) commutes.
\end{proof}

We conclude by proving the representation theorem for the indexed
$\flg$ construction:
\begin{proof}
  The proof assembles the previous results. First, by the indexed
  equivariant initiality theorem, the initial algebra is equivariant
  $\roll : F\fix F \equito \fix
  F$. By the algebra representation \theoremref{indexed
    equivariant algebra representation}, its restriction along the
  embedding $\Canon\canoniser \embed F$ is an equivariant algebra
  $\roll : (\Canon\canoniser)\fix F \to \fix
  F$.

  Now, given any equivariant core algebra $f : (\Canon\canoniser) \A
  \equito \A$, by the algebra representation \theoremref{indexed
    equivariant algebra representation}, its \WLOG{} construction is
  an equivariant algebra $g \definedby \wlg(\canoniser, f) :
  F\A \equito \A$.  By the equivariant initiality
  theorem, the unique $F$-homomorphism out of the initial algebra is
  equivariant $\flg f \definedby \fold g : \fix F \equito
  \A$. By the algebra representation \theoremref{indexed equivariant
    algebra representation}, it is also a
  $(\Canon\canoniser)$-homomorphism.

  Conversely, if $h : \fix F \equito \A$ is any
  equivariant core algebra homomorphism, then by \theoremref{indexed
    equivariant algebra representation}, it is an equivariant
  $F$-homomorphism. By the equivariant initiality
  theorem, $h = \fold g = \flg f$.
\end{proof}

\section{Category theoretic development}
\applabel{cat dev}

The amount of details in the theory of indexed symmetric data types is
high, but a few categorical abstractions keep the development
relatively clean. We start by recasting the parameterised initiality
theorems in this setting. Much of this is known but appears in
different places in the work of Fiore et
al.~\cite{fiore-plotkin-turi:ast}, and especially in the more recent
work with
Szamozvancev~\cite{fiore-szamozvancev:formal-soas,fiore-szamozvancev:soas-theory,szamozvancev:thesis}. We
provide references where available.
\begin{lemma}
  Let $P, F : \Cat C \to \Cat C$ be two endofunctors, and $\strength :
  PF \to FP$ a natural transformation. Then $F$ lifts along the
  forgetful functor $\fun{\pair xa}x$ to an endofunctor over
  $P$-algebras:
  \begin{gather*}
  \fun{\pair x{a : Px \to x}} \pair{Fx}{a_F : PFx \xto\strength FPx
    \xto{Fa} Fx}
  \\
  \fun{\parent{\pair xa \xto h \pair yb}}
  \parent{\pair{Fx}{a_F} \xto{Fh} \pair{Fy}{b_F}}
  \end{gather*}
\end{lemma}
The proof is by direct calculation, see Prop.~3.1.2 in
\citeauthor{szamozvancev:thesis}'s thesis. This type of result is
known from the early days of the concept, as in
\citeauthor{beck:distributive-laws}'s
work~\cite{beck:distributive-laws}.

When $P$ is a left adjoint $P \leftadjointto E$, we can swap
$P$-strengths for an endofunctor $F$ with $E$-costrengths for the same
endofunctor\footnote{We learned this fact from Marcelo
P.~Fiore in private communication.}:
\begin{lemma}[strength-costrength]
  Consider endofunctors $P, E, F : \Cat C \to \Cat C$. Let $P
  \leftadjointto E$ with unit $\eta : \Id \to EP$ and counit $\epsilon :
  PE \to \Id$. TFAE:
  \begin{itemize}
  \item $P$-strengths for $F$, i.e., natural transformations
    $\strength : PF \to FP$.
  \item $E$-costrengths for $F$, i.e., natural transformations
    $\costrength : FE \to EF$.
  \end{itemize}
  The equivalence is given by the following relationships:
  \begin{center}
    \diagram*{31} \qquad
    \diagram*{32}
  \end{center}
  and moreover satisfy:
  \begin{center}
    \diagram*{33}
    \qquad
    \diagram*{34}
  \end{center}
\end{lemma}
\begin{proof}
  By duality, we only need to show:
  \begin{itemize}
  \item That $\costrength$ is natural.
  \item That the strength associated with $\costrength$ is $\strength$.
  \item The diagram involving the units only ($\eta$ and $F\eta$).
  \end{itemize}

  The $\costrength$ is natural: it is a
  composition of natural transformations.

  We substitute the definition of $\costrength$ into the definition of
  $\strength$ and show it recovers $\strength$:
  \diagram{36}

  The diagram involving units only amounts to straightforward
  calculation:
  
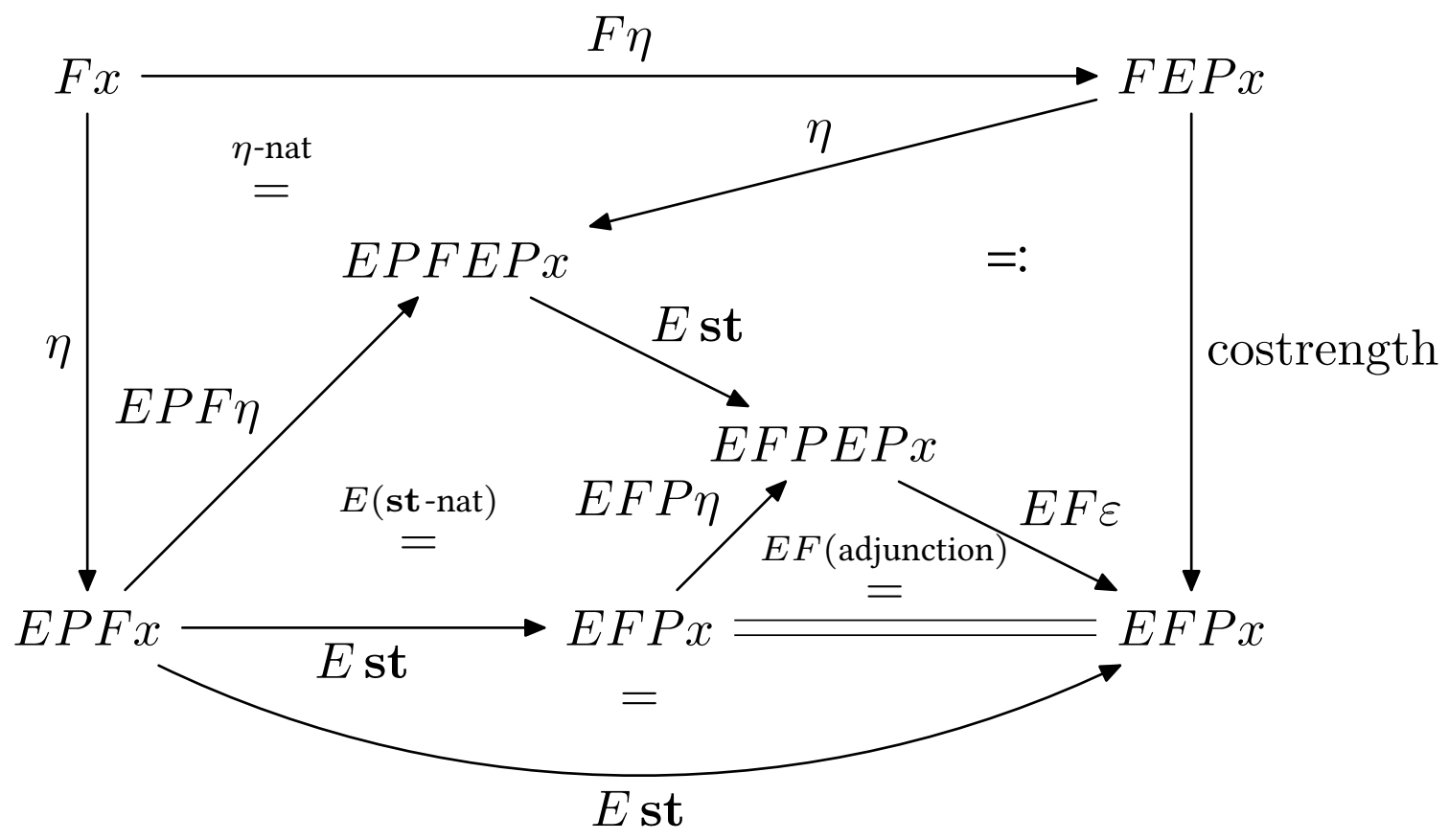
\diagram{35}
\end{proof}

We now prove a general version of parameterised initiality, already
prevalent in Fiore et al.'s work on abstract
syntax~\cite{fiore-szamozvancev:formal-soas,%
  fiore-szamozvancev:soas-theory,%
  fiore-ranchod:substructural-ast}, originally proposed by
\citeauthor{bird-paterson:generalised-folds-for-nested-datatypes}~\cite{%
  bird-paterson:generalised-folds-for-nested-datatypes}.

\begin{definition}
  Let $P, F : \Cat C \to \Cat C$ be endofunctors. An
  initial $F$-algebra $\roll : F\fix F \to \fix F$ is
  \emph{$P$-initial} when every $F$-algebra $f : Fx \to x$ has a
  unique morphism $\pfold f : P\fix F \to x$ satisfying: 
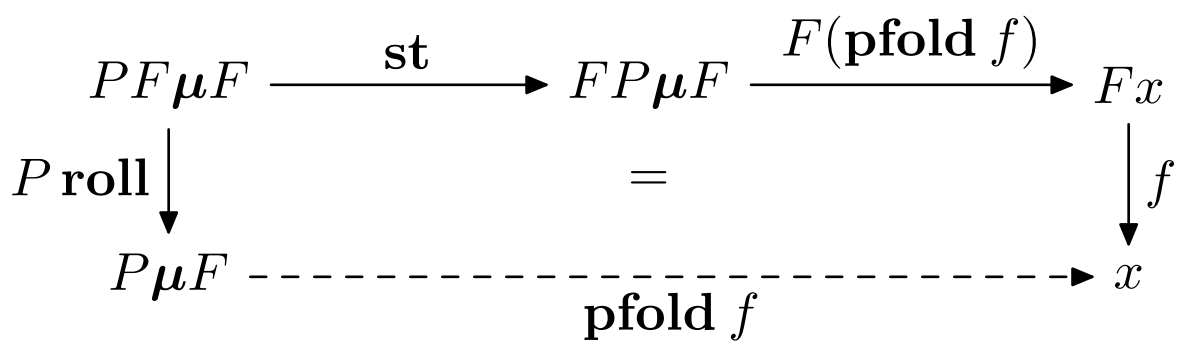
\diagram{37}
\end{definition}

\begin{theorem}[generalised parameterised initiality]%
  \theoremlabel{generalised parameterised initiality}
  Let $P \leftadjointto (E : \Cat C \to \Cat C)$ be an adjunction over
  a category $\Cat C$, with unit $\eta : \Id \to EP$ and counit
  $\epsilon : PE \to \Id$. Every initial algebra $\roll : F\fix F \to
  \fix F$, for an endofunctor $F : \Cat C \to \Cat C$ equipped with a
  natural transformation $\strength : PF \to FP$, is $P$-initial.
\end{theorem}

For completeness, we include the proof of this result in this
formulation.

\begin{proof}
  Let $f : Fx \to x$ be any $F$-algebra.  We equip $Ex$ with an
  $F$-algebra structure. It is more direct to work with the costrength:
  \[
  g : FEx \xto\costrength EFx \xto{Ef} Ex
  \]
  and so, by initiality, we have a unique morphism $h \definedby \fold
  g : \fix F \to Ex$ satisfying:
  
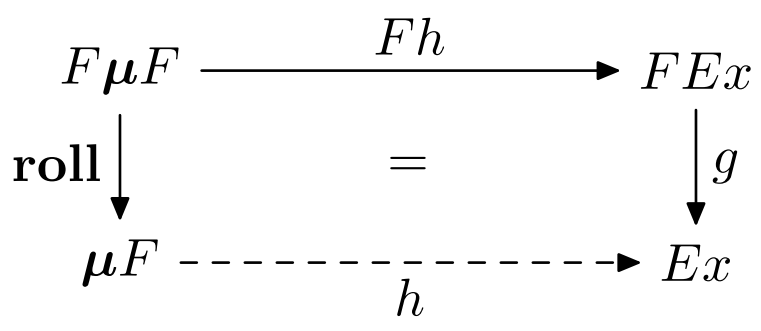
\diagram{38}
  Now take the mate of this morphism:
  \[
  \pfold f :P\fix F \xto{Ph} PEx \xto{\epsilon} x
  \]
  This morphism satisfies the required property:
  
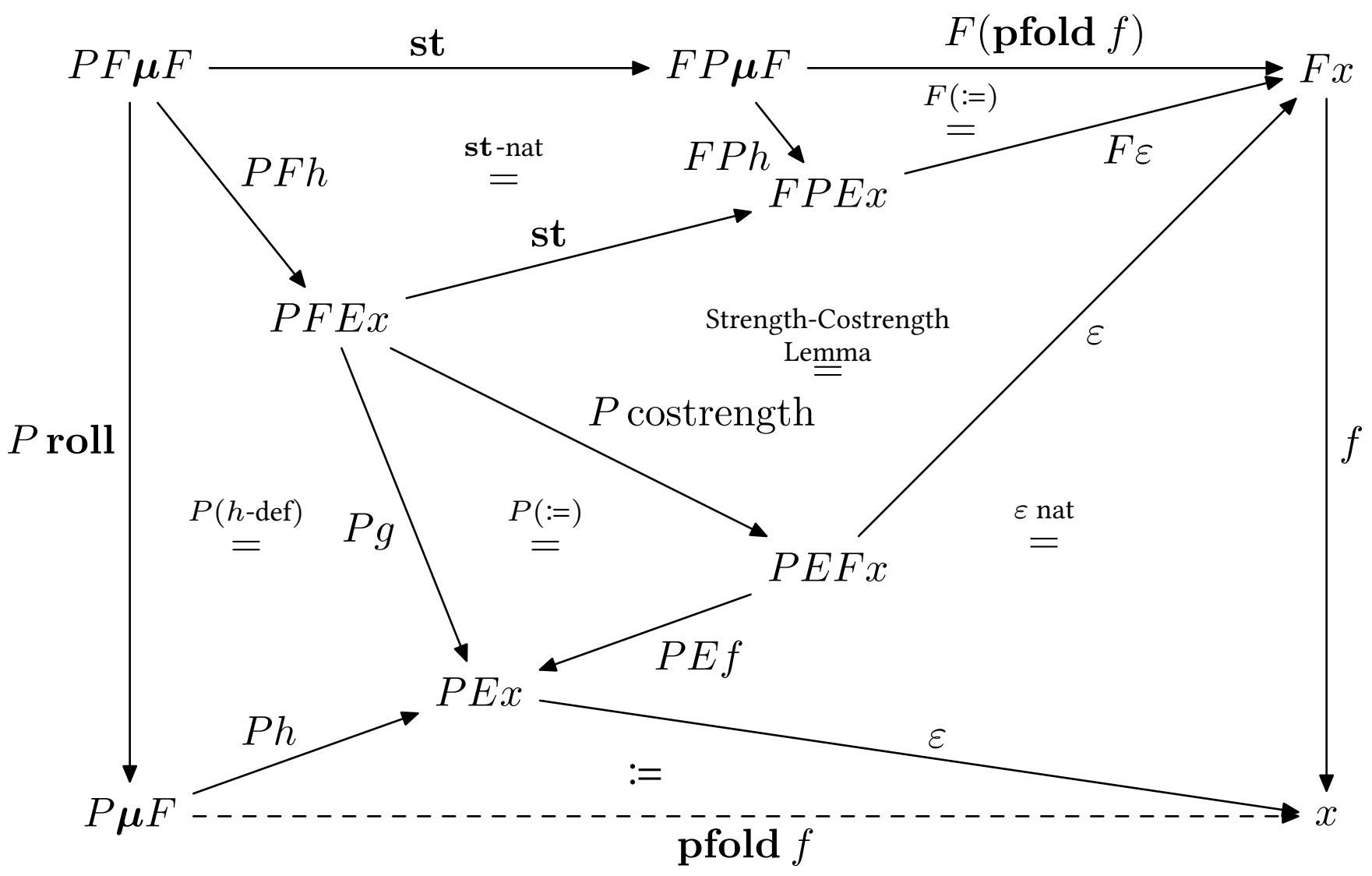
\diagram{39}
  Now consider any morphism $k : P\fix F \to x$ making the following
  diagram commute:
  
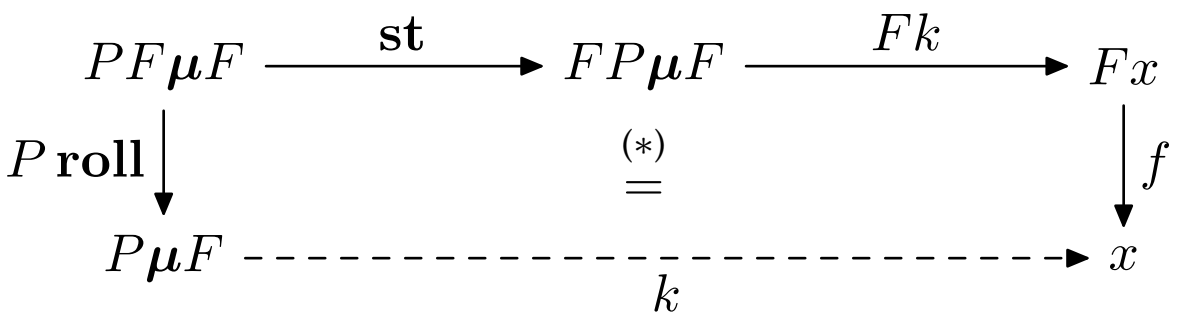
\diagram{40}
  We will show that its mate must be $h$, i.e.: $h : \fix F \xto{\eta}
  EP\fix F \xto{Ek} Ex$. Calculate:
  
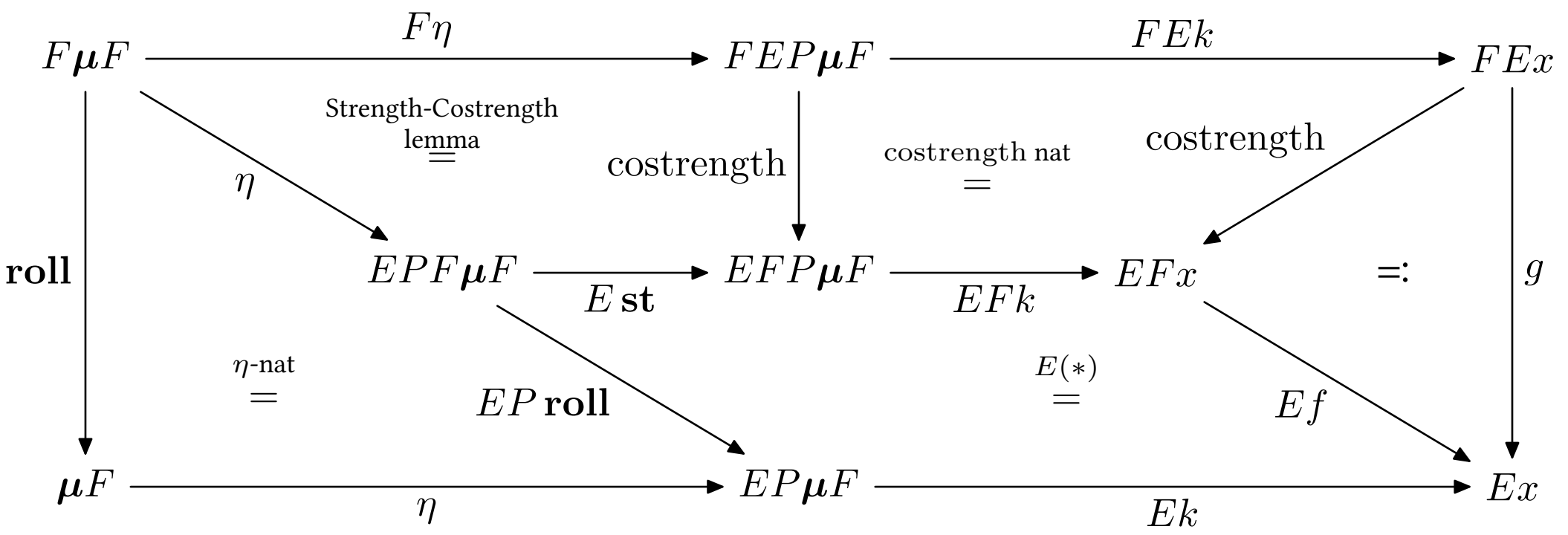
\diagram{41}
  By universality, $h$ is the mate of $k$, and so $k = \pfold f$.
\end{proof}

Therefore, if an endofunctor $F$ over $\carrier\Ctx$-indexed families
has a $(\gast-)$-strength, then its initial algebras are both
$(\gast-)$-initial and $(\gast\gast-)$-initial.

\begin{definition}
  Let $P = \triple{P : \Cat C \to \Cat C}{\eta}{\mu}$ be a monad, and $F :
  \Cat C \to \Cat C$ an endofunctor over $\Cat C$. A \emph{$P$-strength for $F$}
  is a natural transformation $\strength : PF \to FP$ satisfying:
  
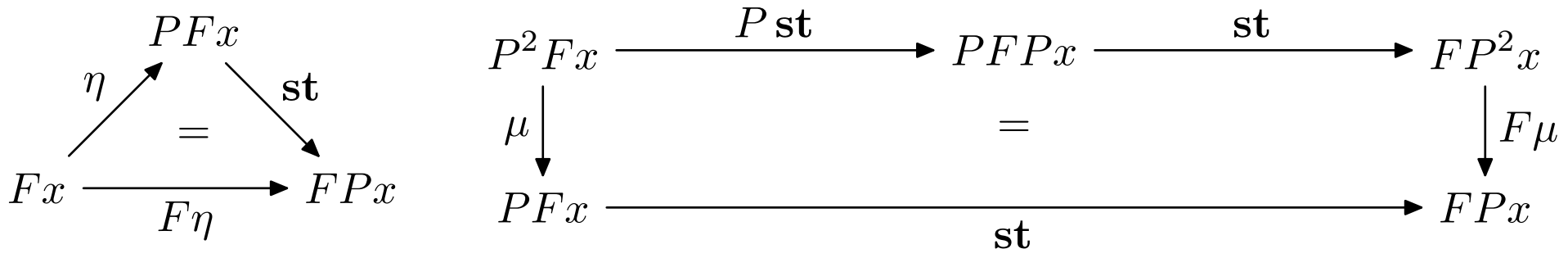
\begin{center}
    \diagram*{42}\qquad
    \diagram*{43}
  \end{center}
\end{definition}
This notion is sometimes called: a \emph{distributive law} between a
monad and a functor; a \emph{module}; or a \emph{monad endomorphism}.

The following notion generalises equivariant $F$-algebras as algebras
of the lifting of $F$ to $P$-algebras:
\begin{definition}
  Let $P = \triple{P : \Cat C \to \Cat C}{\eta}{\mu}$ be a monad over
  $\Cat C$, and $F : \Cat C \to \Cat C$ an endofunctor equipped with a
  $P$-strength $\strength P : PF \to FP$. A \emph{$P$-linear $F$-algebra}
  $\triple{x}{f}{a}$ consists of:
  \begin{itemize}
  \item an $F$-algebra $f : Fx \to x$; and
  \item a $P$-algebra $a : Pa \to a$;
  \end{itemize}
  such that $f$ is $P$-linear $f : \pair{Fx}{a_F} \to \pair{x}{a}$. A
  \emph{homomorphism} $h : \triple xfa \to \triple ygb$ of $P$-linear
  $F$-algebras is a morphism $h : x \to y$ that is simultaneously an
  $F$-algebra homomorphism and a $P$-algebra homomorphism.
\end{definition}

Since a $(\gast-)$-algebra is precisely a $\Ctx$-indexed $\G$-action
and a $(\gast-)$-linear map between them is precisely an equivariant
function, we have that the $(\gast-)$-linear $F$-algebras are
precisely the equivariant $F$-algebras. The capstone of this
development is the following theorem:

\begin{theorem}[general representation theorem]%
  \theoremlabel{general representation}
  Let $P = \triple{P : \Cat C \to \Cat C}{\eta}{\mu}$ be a monad over
  $\Cat C$, and $F : \Cat C \to \Cat C$ an endofunctor equipped with a
  $P$-strength $\strength : PF \to FP$. If an initial $F$-algebra
  $\roll : F\fix F \to \fix F$ is both $P$-initial and $P^2$-initial,
  then it is also the initial $P$-linear $F$-algebra. Its $P$-algebra
  structure is given by $P$-initiality as
  \(
  \alg \definedby \pfold\roll : P\fix F \to \fix F
  \)
  and the unique $F$-homomorphism $\fold f : \fix F \to x$ into any
  other $P$-linear $F$-algebra $\triple xfa$ is $P$-linear.
\end{theorem}

We chose this formulation as its proof follows our proof in the
simply-typed case more closely, which itself follows the argument in
abstract syntax with binding and
substitution~\cite{fiore-plotkin-turi:ast}. Our situation is
semantically well-behaved, and there are alternative theorems we could
use. For example, since $\Ctx$-indexed $\G$-actions are $E \definedby
(\lollybox-)$-algebras, we could use
\citeauthor{szamozvancev:thesis}'s Prop.~3.3.1
instead~\cite{fiore-szamozvancev:soas-theory}. Whichever formulation
we use, it is clear it proves the Indexed Representation
\theoremref{indexed representation}.

\begin{proof}
  We show that:
  \begin{itemize}
  \item $\alg : P\fix F \to \fix F$ is a $P$-algebra;
  \item $\roll$ is $P$-linear by the defining property of $P$-initiality;
  \item each $\fold f$ is $P$-linear.
  \end{itemize}

  We show the $P$-algebra unit axiom:
  
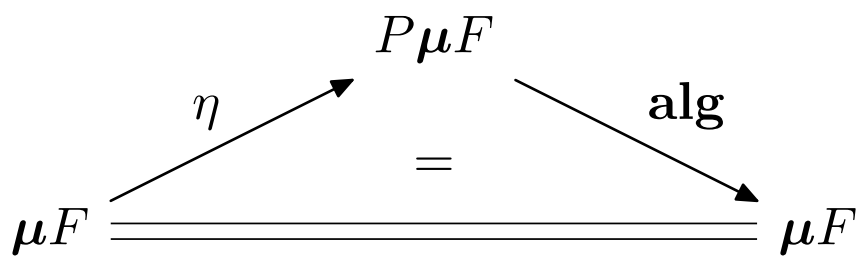
\diagram{44}
  by recourse to mere initiality:
  
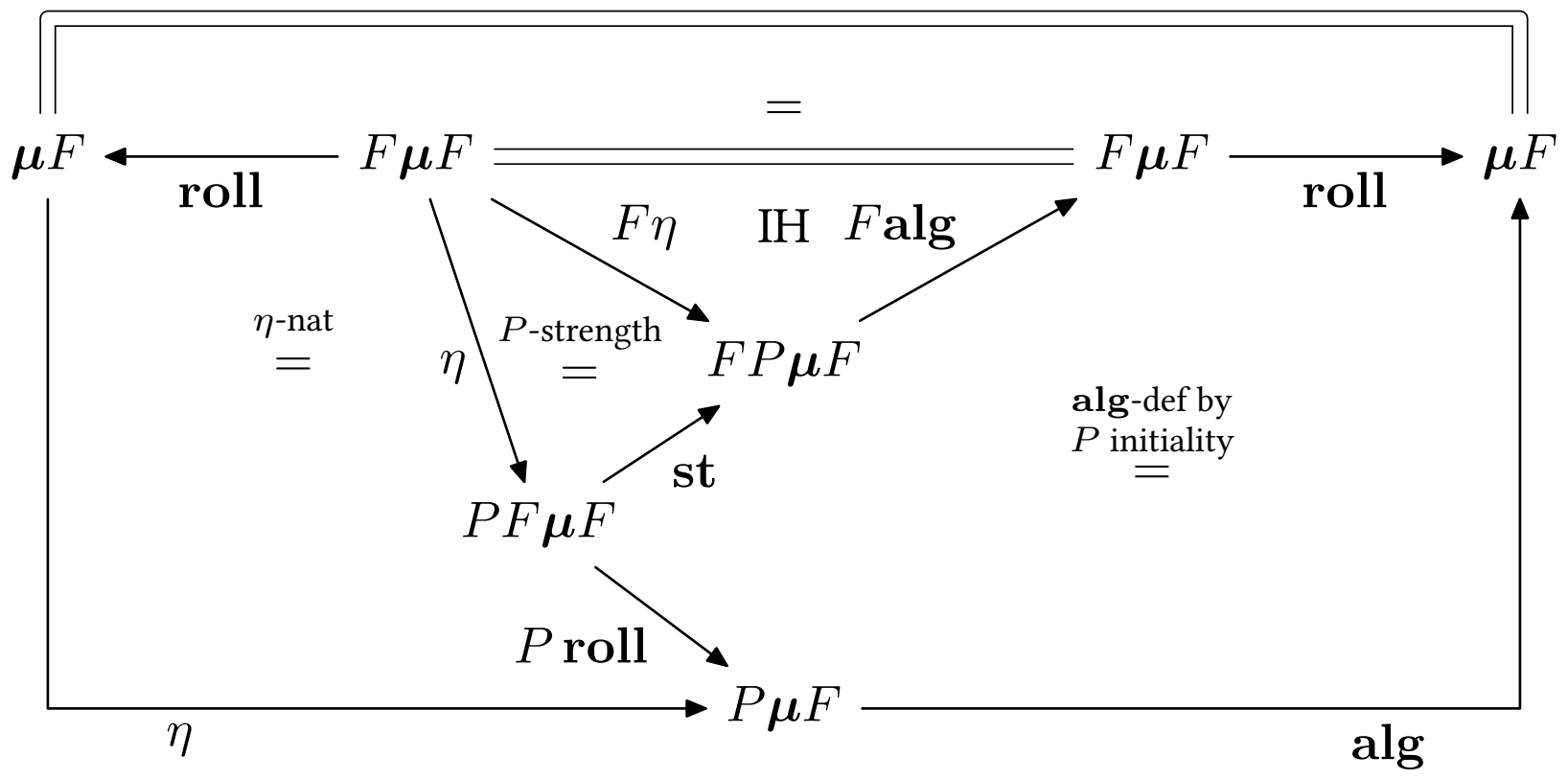
\diagram{45}
  We show the $P$-algebra multiplication axiom:
  
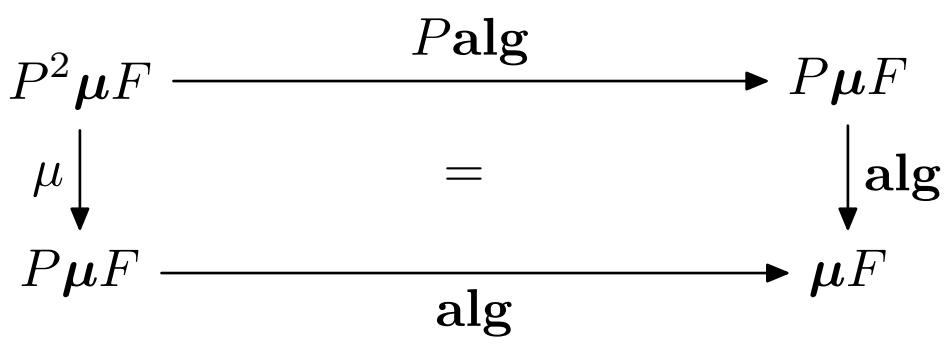
\diagram{47}
  by recourse to $P^2$ initiality:
  
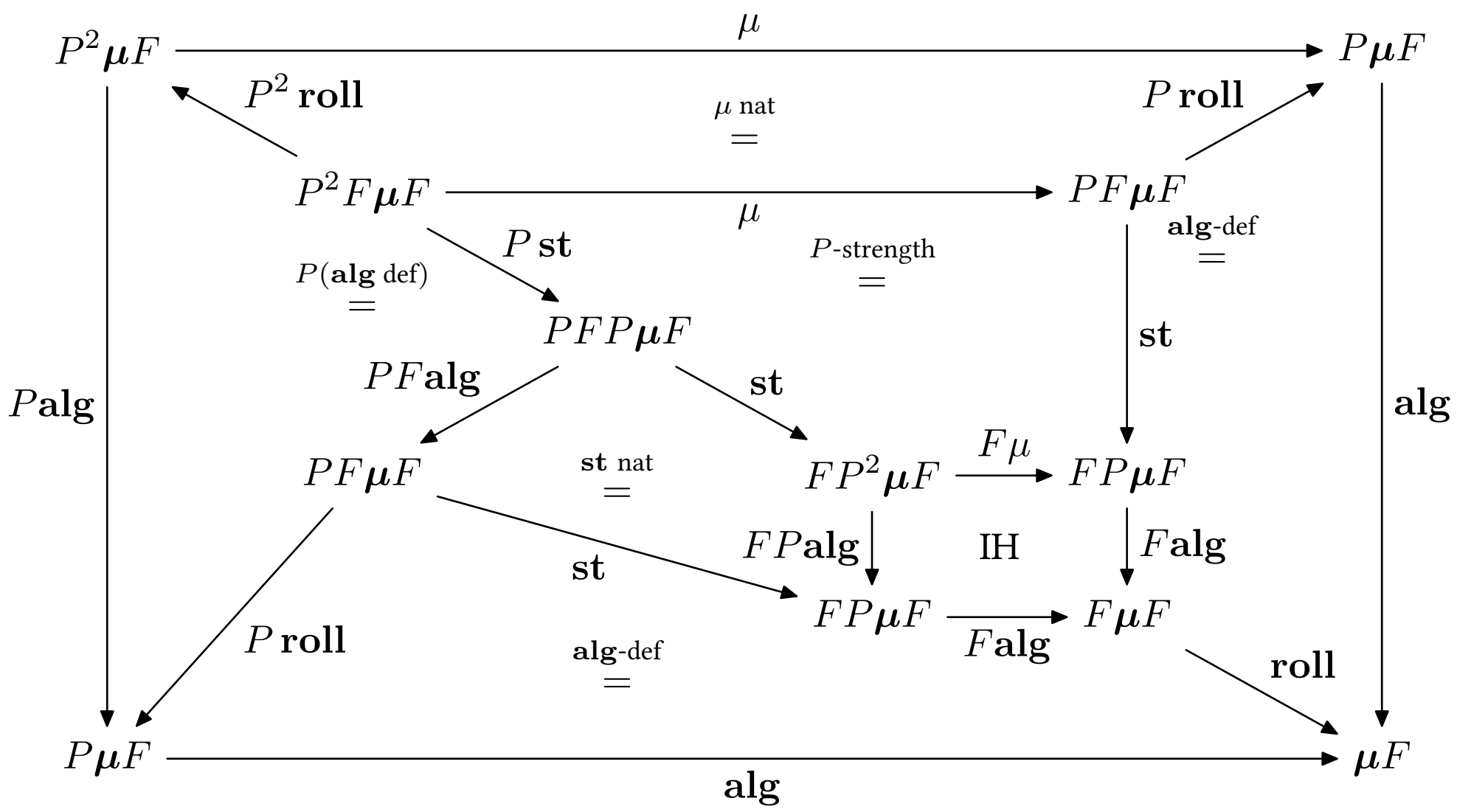
\diagram{46}
  Finally, take any $P$-linear $F$-algebra $\triple xfa$. By
  initiality there is at most one $P$-linear $F$-homomorphism, and
  exactly one $F$-homomorphism, into it. We show:
  
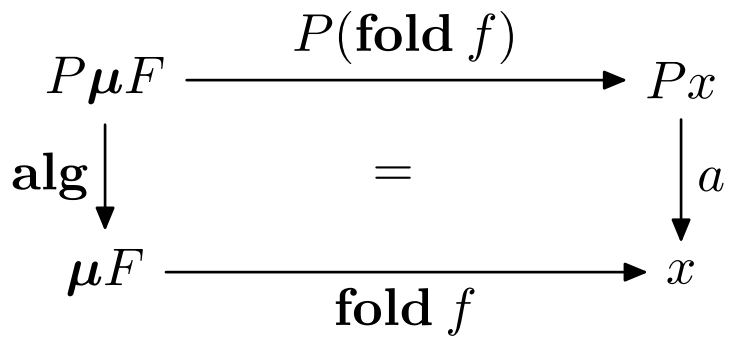
\diagram{48}
  by recourse to $P$-initiality:
  \diagram{49}
  concluding the proof.
\end{proof}

}{}

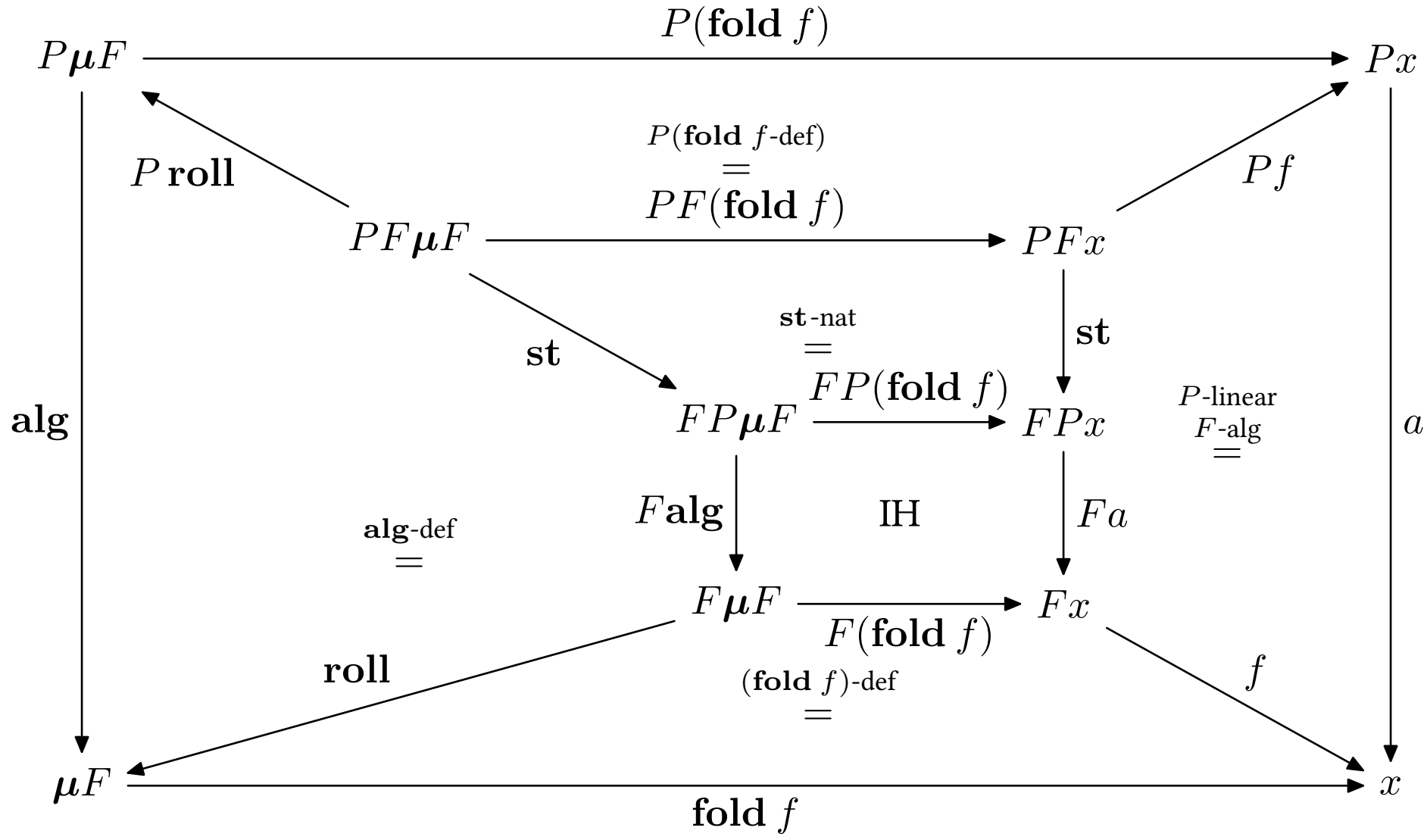
\end{document}